\documentclass[aps,prd,twocolumn,preprintnumbers,showpacs,nofootinbib,amssymb]
{revtex4}
\usepackage{graphicx}
\usepackage{amsmath}
\usepackage{amssymb}
\usepackage{amsfonts}
\usepackage{bm}
\usepackage{color}

\def\be{\begin{equation}}
\def\ee{\end{equation}}
\def\bea{\begin{eqnarray}}
\def\eea{\end{eqnarray}}

\begin{document}

\title{Thermal activation rate of dilute axion stars close to the maximum mass}
\author{Pierre-Henri Chavanis}
\affiliation{Laboratoire de Physique Th\'eorique, Universit\'e de Toulouse,
CNRS, UPS, France}

\begin{abstract}

We compute the thermal activation rate of metastable
self-gravitating Bose-Einstein condensates with attractive
self-interaction (e.g., dilute
axion stars) by using the instanton theory.  Explicit
analytical
results are given close to the maximum mass $M_{\rm max}$ [P.H. Chavanis, Phys.
Rev. D {\bf 84}, 043531 (2011)] by using the normal form of the saddle-node
bifurcation close to that point. We show that the lifetime of metastable states
is extremely long, scaling as
$t_{\rm life}\sim e^N\, t_D$,
where $N$ is
the number of bosons in the system and $t_D$ is the
dynamical time ($N\sim 10^{57}$ and $t_D\sim 10\, {\rm hrs}$
for
typical QCD axion stars; $N\sim 10^{96}$ and $t_D\sim 100\, {\rm
Myrs}$ for the quantum core of a dark matter halo made of ultralight axions).
Therefore,
metastable equilibrium states can be considered as stable equilibrium states in
practice.   We compare our
results with similar results obtained for  Bose-Einstein
condensates in laboratory, globular clusters and self-gravitating
Brownian
particles in astrophysics, the Brownian mean field model (BMF) in statistical mechanics, and bacterial populations in biology. Our
presentation parallels the calculation of the quantum
tunneling rate of dilute axion stars given in a previous paper [P.H. Chavanis,
Phys.
Rev. D {\bf 102}, 083531 (2020)]. These calculations can find application in
various domains of physics and astrophysics.

\end{abstract}


\maketitle

\section{Introduction}
\label{sec_introduction}

The nature of dark matter (DM) is one of the greatest mysteries of modern
cosmology. The DM particle could be a boson such as the axion of quantum
chromodynamics (QCD) with a mass $m\sim 10^{-4}\, {\rm eV}/c^2$ \cite{pq}  or an
ultralight
axion (ULA)    with a
much smaller
mass $m\sim 10^{-22}\, {\rm eV}/c^2$ predicted by string
theory \cite{marshrevue}. This leads to the notion of ``axiverse''
\cite{axiverse}. At
sufficiently low temperatures, bosons form Bose-Einstein
condensates (BECs). Because of their large occupation numbers, they can be
described by a classical scalar field (SF) 
that represents the wavefunction  $\psi({\bf r},t)$ of the BEC. Self-gravitating axion condensates can produce macroscopic objects such as axion
stars
or even DM halos. The mass
of the axion has to
be very small for quantum mechanics to manifest itself at astrophysical
or galactic scales. This leads to the so-called fuzzy dark matter (FDM) model
\cite{barkana}, which is also called the SFDM model \cite{sfdm} or the BECDM
model
\cite{mocz,moczmnras}. In the nonrelativistic
regime appropriate to dilute axion
stars and BECDM halos, the evolution of the condensate wavefunction $\psi({\bf
r},t)$ is governed by the Gross-Pitaevskii-Poisson (GPP) equations with a
$|\psi|^4$ potential taking into account the 
self-interaction  of the
bosons (two-body collisions) \cite{bohmer,prd1,prd2,tanja}. By
making the Madelung \cite{madelung} transformation, one can write the GPP
equations in the form
of hydrodynamic equations involving a quantum potential accounting for the
Heisenberg uncertainty principle and a pressure accounting for the
self-interaction of the bosons
\cite{bohmer,prd1,prd2,tanja} (see, e.g., the reviews \cite{srm,rds,chavanisbook,marshrevue,leerevue,braatenrevue,niemeyer,mg16B,ferreira,
huirevue,visinelliREVUE,khoury,mul,phgtmg,reviewBECDM}
containing an exhaustive list of references).

One particularity of the axion is to have a
negative scattering length ($a_s<0$) corresponding to an attractive $|\psi|^4$ 
self-interaction. The mass-radius relation of nonrelativistic self-gravitating
BECs with an
attractive self-interaction has been determined in \cite{prd1,prd2}. It is
shown that dilute axion stars can exist only below a
maximum mass \cite{prd1,prd2,mg16B} 
\begin{eqnarray}
\label{intro1}
M_{\rm max}^{\rm exact}=1.012\, \frac{\hbar}{\sqrt{Gm|a_s|}}
\end{eqnarray}
corresponding to a radius\footnote{Equivalent expressions of the maxium
mass and the corresponding radius
written in terms of the dimensionless
self-interaction constant $\lambda=8\pi a_smc/\hbar$ or the axion decay constant
$f=(\hbar c^3 m/32\pi|a_s|)^{1/2}$ are
given in \cite{phi6}.}
\begin{eqnarray}
\label{intro2}
(R_{99}^*)^{\rm exact}=5.5\,
\left (\frac{|a_s|\hbar^2}{Gm^3}\right )^{1/2}.
\end{eqnarray}
When $M>M_{\rm max}$ there is no equilibrium
state. When $M<M_{\rm max}$ there
are two possible equilibrium states for the same mass $M$. The solution with
$R>R_{99}^*$ is stable (minimum of energy at fixed mass) while the solution with
$R<R_{99}^*$ is unstable (maximum of energy at fixed mass). The noninteracting
limit corresponds to $R\gg R_{99}^*$ and the nongravitational limit corresponds
to  $R\ll R_{99}^*$. Dilute axion stars can exist only below the maximum
mass $M_{\rm max}$ and above the minimum radius $R_{99}^*$ given by Eqs.
(\ref{intro1}) and (\ref{intro2}). We stress that the
maximum mass of dilute axion stars \cite{prd1,prd2,mg16B} has a nonrelativistic
origin
unlike the maximum mass of boson stars
\cite{kaup,rb,colpi,chavharko}.
The connection between the maximum mass of nonrelativistic dilute axion stars
and the maximum mass of general relativistic boson stars is made in Ref.
\cite{massmaxrel}.

For the QCD axion of mass  $m=10^{-4}\,
{\rm eV}/c^2$ and scattering length $a_s=-5.8\times 10^{-53}\, {\rm
m}$ (yielding $\lambda=-7.39\times
10^{-49}$ and $f=5.82\times 10^{10}\, {\rm GeV}$), the maximum mass $M_{\rm
max}^{\rm
exact}=6.46\times 10^{-14}\,
M_{\odot}=1.29\times 10^{17}\, {\rm kg}=2.16\times 10^{-8}\,
M_{\oplus}$ and the corresponding radius $(R_{99}^*)^{\rm exact}=3.26\times
10^{-4}\,
R_{\odot}=227\, {\rm km}=3.56\times 10^{-2}\, R_{\oplus}$ of axion stars are
very small, much
smaller than galactic sizes. Therefore, QCD axions are expected to form mini
axion stars of the size of asteroids. For that reason, they are
called ``axteroids''.

For an ULA with a mass 
$m=2.19\times 10^{-22}\, {\rm eV}/c^2$  and a very small attractive
self-interaction $a_s=-1.11\times 10^{-62}\, {\rm fm}$  (yielding
$\lambda=-3.10\times 10^{-91}$ and $f=1.97\times 10^{14}\, {\rm GeV}$), one
finds that  $M_{\rm max}=10^8\, M_{\odot}$ and 
$R^*_{99}=1\, {\rm
kpc}$. For smaller absolute values of the scattering length, the maximum
mass can even be larger.\footnote{If the self-interaction vanishes or is
repulsive instead
of being attractive, then there is no maximum mass in the Newtonian
regime \cite{jeansapp}.} Therefore, ULAs can form giant
BECs with the dimensions of DM halos. These objects may correspond either to
ultracompact DM halos of mass  $(M_h)_{\rm min}\sim 10^8\, M_{\odot}$ and
radius $(r_h)_{\rm min}\sim 1\, {\rm kpc}$ like dwarf spheroidal galaxies
(dSphs) or to the quantum core (soliton) of larger DM halos. In the second case,
the quantum core of mass $M_c$ and radius $R_c$ is surrounded by a halo of
scalar radiation arising from the quantum interferences of excited states. This
``core-envelope'' structure, which results from a process of
violent relaxation \cite{lb} and gravitational cooling \cite{seidel94,gul0,gul},
has
been evidenced in direct numerical simulations of
BECDM \cite{ch2,ch3,schwabe,mocz,moczSV,veltmaat,moczprl,moczmnras,veltmaat2,
moczatt}. The quantum core
(ground
state of the GPP equations) stems from the
equilibrium
between the quantum pressure (Heisenberg's uncertainty principle), the self-interaction of the axions, and the
gravitational attraction. On
the other hand, the envelope has an approximately isothermal
\cite{lb,modeldm,wignerPH} or
Navarro-Frenk-White (NFW) profile \cite{nfw} as obtained in classical numerical
simulations of collisionless cold dark matter (CDM). The
Schr\"odinger-Poisson--Vlasov-Poisson correspondance
is discussed in Refs. \cite{moczSV,wignerPH}. The quantum core (soliton) may
solve the core-cusp problem \cite{moore} of the CDM
model and the envelope accounts for
the flat rotation curves of the
galaxies. It is the envelope that
determines the
mass and the size of large DM halos and explains why the halo
radius $r_h$ increases with the halo mass $M_h$ while the core radius
$R_c$ decreases with the core mass $M_c$ (see Appendix L of
\cite{modeldm} for a detailed discussion of this peculiarity). The core mass --
halo
mass relation $M_c(M_h)$ of BECDM halos with an arbitrary self-interaction has
been determined in \cite{mcmh,mcmhbh} from a maximum entropy principle. It is found that
the core mass $M_c$
increases with the halo mass $M_h$. Of
course, for an attractive self-interaction, these core-envelope configurations
are
stable only if
the mass of their core  is smaller than the maximum mass from Eq.
(\ref{intro1}). In
sufficiently large DM halos, the core mass may reach the maximum
mass, become unstable, and undergo gravitational collapse (see
\cite{mcmh,mcmhbh}
for details). 
The  mass of the
quantum core of a given DM halo or the mass of a dilute axion star may
also increase by collisions, mergers and accretion or by the
process of Bose condensation driven by
gravitational interactions or self-interactions
\cite{levkov2,egg,sgbne,kmp,cdlmn,cdlm,kirk}. When a
dilute axion star\footnote{Here, the term ``axion star'' 
generically refers to a self-gravitating BEC of axions or, even more generally,
to a self-gravitating BEC with an attractive self-interaction. It can
correspond to an
isolated object (e.g. an axteroid or a Bose star) or to the quantum core of a DM
halo surrounded by an envelope.} 
reaches the maximum mass it may:  

(i) collapse and form a {\it Dirac peak} (in an idealized scenario)
\cite{bectcoll};

(ii) break into several
stable pieces ({\it axion drops} \cite{davidson}) of mass $M'<M_{\rm max}$
\cite{cotner};

(iii) give rise to a {\it bosenova} phenomenon where the collapse (implosion)
of
the axion star is accompanied by a burst (explosion) of
outgoing relativistic axions
(radiation) produced by
inelastic reactions when the density reaches high values \cite{tkachevprl};

(iv) form a 
{\it black hole} when the mass
of the axion star is sufficiently large or when the self-interaction is
sufficiently weak so that general relativity must be taken into account
 \cite{helfer,moss}.

Actually, when the axion star becomes dense enough, the $|\psi|^4$
approximation is not valid anymore and one has to take into account higher order
terms in the expansion of the SF potential (or, better, consider the exact
axionic self-interaction potential) \cite{pq}. These higher order terms, which
can
be repulsive (unlike the $|\psi|^4$ term for axions), can account for strong
collisions (three-body, four-body...) between axions. In that case, the collapse above $M_{\rm max}$ can
lead to a {\it dense axion star} in which the gravitational attraction and the
attractive $|\psi|^4$ self-interaction are balanced by the quantum
potential and the repulsive $|\psi|^6$
(or higher order) self-interaction \cite{braaten}.
As a result, a stable branch of dense axion stars
\cite{braaten,phi6} appears in
the mass-radius relation of axion stars in addition to the stable branch of
dilute axion stars and the unstable branch of nongravitational
axions stars
\cite{prd1,prd2}. On the branch of dense axion stars, self-gravity is
negligible (except for very large masses) and the density of the star, which
results from the balance between the attractive $|\psi|^4$ self-interaction and
the repulsive $|\psi|^6$ self-interaction, is uniform with value
$\rho=9m^3c^2/(32\pi|a_s|\hbar^2)$ \cite{phi6}. This is a sort of giant
``quantum droplet.''  The
mass-radius relation of axion
stars presents therefore a maximum mass $M_{\rm max}^{\rm dilute}$ and a
minimum mass $M_{\rm min}^{\rm dense}$.\footnote{In Ref. \cite{phi6} we have
argued that, at very large
masses where general relativistic effects are important, the mass-radius
relation of dense axion stars forms a spiral. This implies the existence
of another maximum mass $M_{\rm max,GR}^{\rm
dense}\sim (|a_s|\hbar^2c^4/G^3m^3)^{1/2}$, which has a general relativistic
origin, 
above which a dense axion star collapses towards a black hole.}  
This leads to the possibility of phase transitions between dilute and dense
axion stars \cite{phi6}. Indeed, there exists a transition mass $M_t$ such that
dilute axion stars are fully stable (global minima of energy) for $M<M_t$
and metastable (local  minima of energy) for $M_t<M<M_{\rm max}^{\rm dilute}$.
Correspondingly, dense axion stars are metastable for $M_{\rm min}^{\rm
dense}<M<M_t$ and fully
stable for
$M>M_t$. If a dilute axion star gains mass, for instance by collision, merger,
accretion or condensation, it can overcome the maximum mass $M_{\rm max}^{\rm
dilute}$, collapse
and form a dense axion star (it may also emit a relativistic radiation --
bosenova -- and disappear into scalar waves, or form a black hole, as
discussed
above) \cite{braaten,cotner,bectcoll,tkachevprl,helfer,phi6,moss}.
Conversely, if a dense axion star loses mass -- decaying by
emitting axion
radiation because of relativistic effects -- it can pass below
the minimum mass $M_{\rm min}^{\rm dense}$ and
disperse outwards
(explosion) due to the repulsive kinetic pressure (quantum
potential). This mechanism determines the
lifetime of dense axion stars close to the minimum mass in the nonrelativistic
regime. As noted by Braaten
and Zhang \cite{braatenrevue}, the lifetime of dense axion stars close to the
minimum mass may be too short to be
astrophysically relevant. However, dense axion stars may have an important
cosmological effect by transforming nonrelativistic axions into relativistic
axions. These phase transitions, involving collapses and
explosions, are similar to those studied in
\cite{pt,ijmpb,clm1,clm2,calettre,acf,mg16F} in the case of
self-gravitating fermions at finite temperature. They
also share similarities
with the phase transitions of compact objects (white dwarfs, neutron
stars and black holes) discussed in \cite{htww,shapiroteukolsky} (see Sec.
XI.C of Ref.
\cite{phi6} for a discussion of this analogy).

Visinelli
{\it et al.} \cite{visinelli} and Eby {\it et
al.} \cite{ebyexpansion,ebyclass,elssw} stressed the importance of
relativistic effects on the branch of dense axion stars while
self-gravity is negligible. In this sense, dense axion stars correspond to
``pseudobreathers'', ``oscillons''
\cite{oscillon,cgm}
or ``axitons'' \cite{kt2}, which are
described by the sine-Gordon equation (they can be
viewed as the 3D version of usual 1D ``breathers'' \cite{breather}). For a real
SF, these
objects are known to
be unstable due to
particle number changing processes such as the $3\rightarrow 1$ process and to 
decay via emission of relativistic axions. More precisely, they
are dynamically stable but they decay
rapidly because of relativistic effects. Visinelli
{\it et al.} \cite{visinelli} and Eby {\it et al.} \cite{elssw} argue that the
decay
timescale is much
shorter than any cosmological timescale so that dense axion stars are not
physically relevant.\footnote{By contrast, dilute axion stars are long-lived
with
respect
to decay in photons with a lifetime
far longer than the age of the Universe
\cite{ebylifetime,ebydecay,ebybh,braatenR}.}  This
conclusion is contested
by Braaten and Zhang \cite{braatenrevue} who argue that dense axion stars can
be long-lived, except close to the minimum mass $M_{\rm min}^{\rm dense}$. The
accuracy of the
nonrelativistic approximation may improve as $M$ increases along the dense
branch. It
is
important to stress that
these authors consider axions (like QCD axions) described by a {\it real}
SF for which the particle number is not conserved in the
relativistic regime. This is the reason for their fast decay. Alternatively,
if one considers ULAs described by a {\it complex} SF (like, e.g., in
Refs. \cite{abrilph,playa}) for which the particle number (or charge) is
conserved, then the
dense axion stars (`axion boson stars'' \cite{guerra}) are
stable and long-lived in the
relativistic regime because of charge conservation.

There exists an interesting analogy between the
dynamics of a self-gravitating BEC described by the 
GPP equations and the motion of a particle in a one-dimensional
potential $V(R)$ \cite{prd1}.\footnote{This analogy was first
developed in the case of nongravitational BECs (see the references given in
footnote 10 of \cite{tunnel}) and in simple models of stars with a uniform
density  (see the references given in \cite{wdsD,prd1}).}
Indeed,
by making a Gaussian ansatz for
the wave function of the BEC, one can reduce the study of the GPP equations to
the mechanical problem of a fictive
particle of mass $\alpha M$ and position $R$ in a
potential $V(R)$, where $R$ represents the typical size of
the BEC of mass $M$. In the case of self-gravitating BECs with a purely
$|\psi|^4$
attractive self-interaction, the potential $V(R)$ is unbounded from below
($V(R)\rightarrow -\infty$ as $R\rightarrow 0$). It has a local minimum
corresponding to a metastable state (dilute axion star) and a maximum
corresponding to
an unstable state (nongravitational axion star). If we account for a repulsive
$|\psi|^6$
self-interaction, a second
minimum appears. It corresponds to a dense axion star which can be either
stable
or metastable. Therefore, we
are led back to the study of a particle in a double-well potential with a global
mimimum at $R_S$ (fully stable), a maximum at $R_U$ (unstable) and a local
minimum at $R_M$ (metastable). In practice, we will assume that the metastable
state corresponds to a dilute axion star and that the fully stable state corresponds 
to a dense axion
star (we can also consider the possibility that there is no
dense axion star, i.e., we consider a purely $|\psi|^4$
attractive self-interaction so that $V(R)\rightarrow
-\infty$ as $R\rightarrow 0$).

Suppose that, initially, the system is entirely localized in the metastable
state (local but not global minimum of energy). There are two
mechanisms by which this state can decay to the ground state. At zero
temperature ($T=0$) or when $k_B T\ll M\omega_M^2R_M^2$, the
system would classically behave like a harmonic oscillator with a pulsation
$\omega_M=\sqrt{V''(R_M)/\alpha M}$. However, it is
rendered unstable by
the quantum mechanical process of barrier-penetration (tunnel effect). When
$\hbar\rightarrow 0$, the
tunneling rate of a quantum particle is basically given by the WKB formula
\cite{llquantique}. On the other hand, when $k_B T\gg
M\omega_M^2R_M^2$, classical thermal fluctuations are strong enough to induce
a diffusion process over the barrier, destroying metastability. In that case,
the metastable state is rendered unstable  by thermal
effects. When $M\omega_M^2R_M^2\ll k_B T\ll \Delta V$, where $\Delta
V=V(R_U)-V(R_M)$ is the
potential
barrier,  the thermal activation rate of a Brownian particle is basically given 
by the Kramers formula
\cite{kramers} which is a generalization of the Arrhenius law \cite{arrhenius}.
One can also obtain the tunneling and activation rates for both processes by
using the theory
of
path integrals and instantons. The calculation of the lifetime
$t_{\rm life}$  of a metastable
state, using the
instanton theory, is a beautiful problem of physics that has a long
history. This formalism
was originally developed for a
particle in a one-dimensional potential $V(x)$, either in the context of quantum
mechanics or in the context of  Brownian theory. It was then extended to a scalar field $\varphi({\bf
r},t)$ in quantum field theory and to a density field $\rho({\bf r},t)$ in the statistical
mechanics of phase transitions\footnote{In that case, the lifetime of a metastable state is related to the nucleation of 3D droplets (bubbles)  in
statistical physics and 4D droplets ($3+1$ dimensions including time) in the context of quantum field theory.} (see Appendices \ref{sec_wnlq}-\ref{sec_ass} and
\cite{preparation} for a review of the main developments of this subject
in quantum mechanics and statistical physics and an exhaustive list of
references).

In a previous paper \cite{tunnel}, we determined the quantum tunneling rate of
dilute axion
stars, and their lifetime, by applying the theory of
path integrals and instantons to the GPP equations. By making a Gaussian
ansatz,
we
showed that this problem could be reduced to the simpler problem of the
tunneling rate of a fictive quantum particle in a one-dimensional
potential. Our study followed the works of
\cite{no,stoof,leggett,huepe,fa,skalski,lnp} for
a
nongravitational
metastable BEC with an attractive self-interaction in a confining  harmonic
potential. 
Within this framework, quantum fluctuations are taken into account by replacing
the deterministic equation of motion of the fictive particle by a Schr\"odinger
equation. We can then apply the usual instanton theory
following \cite{stone1,stone2,polyakov,coleman,cc} (see also the reviews
\cite{abc,brandenberger,gkw,stoof}).

In the present paper, we determine the thermal activation rate and the lifetime
of dilute axion stars. We use an analogy with Brownian motion and introduce a
temperature term and a friction term in the GP equation, leading to a form of
dissipative GP equation \cite{ggpp,wignerPH}. We also introduce
thermal fluctuations at the level of the hydrodynamic equations by using the
Landau and Lifshitz
theory of fluctuating hydrodynamics \cite{ll}. Therefore, our
theoretical formalism is relevant both to a classical ($\hbar=0$) or a quantum
($\hbar\neq 0$)  system of
self-gravitating Brownian
particles. We then apply the theory of
path integrals and instantons to these equations. 
By making a  Gaussian ansatz, we
show that this problem can be reduced to the simpler problem of the
thermal activation rate of a fictive Brownian particle in a one-dimensional
potential. Within this framework, thermal fluctuations are taken into account by
replacing the deterministic equation of motion of the fictive particle by
a stochastic Langevin equation or by using a Fokker-Planck 
equation. We can then apply the usual instanton theory following
\cite{brayprl,mk,bray1,bray,lmk,nbk}.

We find that the lifetime of metastable axion stars is
considerable since it scales as $e^N t_D$ (where $t_D$
is the dynamical time) with $N\sim 10^{57}$ and $t_D\sim 10\, {\rm hrs}$ for QCD
axions and with $N\sim 10^{96}$ and $t_D\sim 100\, {\rm
Myrs}$ for ULAs. A similar result
was
previously established in the case of classical self-gravitating systems such as
globular clusters \cite{katzokamoto,lifetime}. Indeed, it is known
that these objects
are metastable,  being local but 
not global maxima of  entropy at fixed mass and energy \cite{antonov,lbw}. In
principle, they are rendered unstable by
thermal effects or by fluctuations due to finite $N$ effects
\cite{katzokamoto,lifetime}. However, their
lifetime is exponentially large \cite{lifetime}, scaling as $e^N t_D$ with
$N\sim
10^6$ for globular clusters, making them astrophysically relevant. 
More generally, systems with long-range
interactions exhibit metastable states whose lifetime scales as
$e^N t_D$ \cite{langerspin,art,lifetime,cdkramers,bmf}.
When the number of
particles $N$ is large, like in astrophysics, this
lifetime is so great that metastable
states can be
considered as
stable states in practice. Nevertheless, the
detailed
computation of the system's lifetime is necessary in order to obtain the
reduction
factor close to the critical
point (e.g., close to the maximum mass $M_{\rm max}$ for dilute axion stars
\cite{prd1,phi6}
or
close to the minimum energy $E_{c}$ for
globular clusters \cite{lifetime}) and determine whether this reduction
factor is relevant or not. We will see that it is not relevant for
axion stars (except possibly at criticality) while it is relevant for
globular clusters. 
In this paper, we explicitly derive the analytical expression
of the thermal activation rate of dilute axion stars close to the maximum mass.
We show that the
reduction factor scales as $(1-M/M_{\rm max})^{3/2}$ in the thermal case instead
of $(1-M/M_{\rm
max})^{5/4}$ in the
quantum case \cite{tunnel}. By developing a finite size scaling
theory close to the maximum mass, we show that the collapse
time for overdamped systems at criticality (i.e. for $M=M_{\rm max}$) scales
algebraically with $N$ as $t_{\rm coll}^{\rm T}\sim N^{1/3}\xi t_D^2$   in
the thermal case instead of being infinite when fluctuations are neglected.
By comparison, in Ref. \cite{tunnel}, we showed that the collapse time for
dissipationless systems at criticality
(i.e. for $M=M_{\rm max}$) scales algebraically
with $N$ as $t_{\rm
coll}^{\rm Q}\sim N^{1/5}t_D$ in the quantum case and as  $t_{\rm
coll}^{\rm T}\sim N^{1/6}t_D$  in the thermal
case. The
collapse time at criticality
is smaller than the age of the universe for QCD axion stars and
larger than the age of the universe for the quantum cores of DM
halos made of ULAs \cite{tunnel}. Our detailed calculation of the
tunneling and  activation rates of dilute
axion stars may also be
useful
if one is able in the future to perform direct $N$-body simulations or
laboratory experiments of self-gravitating BECs with an attractive
self-interaction mimicking dilute axion stars.  In that case, the number of
bosons $N$ will be relatively small, the temperature will be relatively large, and metastability effects should be observed, especially close to
the maximum mass. On the other hand, our theoretical results may be applied to
other systems which possess metastable states with a relatively short lifetime. In
that case, the determination of their lifetime requires a detailed calculation.
We give examples of such systems in the context of globular
clusters \cite{katzokamoto,lifetime}, self-gravitating Brownian particles   \cite{crs,cdkramers}, the Brownian mean field (BMF) model  \cite{bmf},
and
bacterial populations in biology (Keller-Segel model) \cite{ks,sks}. Therefore, our general formalism
can find applications in various domains of physics and astrophysics.  This is
why we develop it in a rather exhaustive manner.

The paper is organized as follows. In Sec. \ref{sec_be}, we consider a model
of dilute axion stars based on the GPP equations with an
attractive $|\psi|^4$ self-interaction. We also introduce the damped GPP
equations taking into account thermal effects and dissipative effects. In Sec.
\ref{sec_gbm}, we discuss the main properties of the damped GPP equations. In
Sec. \ref{sec_ga}, we use a Gaussian ansatz to transform these equations into
the simpler mechanical problem of a fictive particle in a one-dimensional
potential. We also discuss the normal form of the potential close to the maximum
mass (critical point). In Sec. \ref{sec_ttbec}, we take into account thermal
fluctuations in the
BEC  by using an analogy with the theory of Brownian motion and we determine the thermal
activation rate of dilute axion stars from the instanton theory. We give its
general expression and its approximate expression close to the maximum mass. We
consider corrections to the maximum mass of dilute axion
stars due to quantum or thermal
fluctuations and show that they are generally negligible. We emphasize the
very long lifetime of dilute axion stars below the maximum mass. In Sec.
\ref{sec_gc}, we discuss the lifetime of globular clusters and self-gravitating
Brownian particles in the underdamped limit by using a similar approach. In Sec.
\ref{sec_sfbp}, as another application of our general formalism, we consider the
case of classical self-gravitating Brownian particles in the overdamped limit.
Finally, in Sec. \ref{sec_pain}, we provide a summary of the different
equations appearing in our study and propose some extensions. 
The Appendices provide additional results that are useful in the main text.
In Appendix \ref{sec_tcano}, we consider the thermodynamics of self-gravitating Brownian particles in the canonical ensemble. In Appendix \ref{sec_summary}, we  summarize the basic formulas which are valid for a Brownian particle close to a saddle-node bifurcation. In Appendix \ref{sec_collsf}, we determine the collapse time of an overdamped self-gravitating BEC with attractive self-interactions close to the maximum mass and discuss finite size scalings. In Appendix \ref{sec_tf}, we take into account thermal fluctuations in a self-gravitating BEC, derive the stochastic quantum virial theorem, and simplify the dynamics of the system by using a Gaussian ansatz. In Appendix \ref{sec_wnlq}, we derive the tunneling rate of a quantum particle in a metastable state by using the instanton theory. In Appendix \ref{sec_wnl}, we derive the escape rate of a Brownian particle from a metastable state by using the instanton theory. In Appendix \ref{sec_ass},  we discuss the analogy between the Smoluchowski equation and the Schr\"odinger equation in imaginary time (Kac equation) and consider their path integral representations.

\section{BECDM model with an attractive self-interaction: The
maximum mass of dilute axion stars}
\label{sec_be}

Dilute axion stars can be interpreted as nonrelativistic
self-gravitating BECs with an
attractive $|\psi|^4$ self-interaction \cite{phi6}. They are described by the
GPP
equations
\begin{eqnarray}
i\hbar \frac{\partial\psi}{\partial
t}=-\frac{\hbar^2}{2m}\Delta\psi+m\Phi\psi
+\frac{4\pi a_s\hbar^2}{m^2}|\psi|^{2}\psi,
\label{gpp1}
\end{eqnarray}
\begin{equation}
\Delta\Phi=4\pi G |\psi|^2,
\label{gpp2}
\end{equation}
where $\psi({\bf r},t)$ is the complex wave function of the condensate, $\Phi({\bf
r},t)$ is the gravitational potential, and $a_s$ is the scattering length of the
bosons.\footnote{See Secs. II and III of \cite{phi6}, Appendix
A of \cite{tunnel}, and Appendix C of \cite{massmaxrel} for the
derivation of the GPP
equations from the more general Klein-Gordon-Einstein (KGE) equations for a real SF $\varphi$ describing relativistic
axion
stars with the instantonic potential.} The GP
equation (\ref{gpp1}) involves a cubic nonlinearity associated with a quartic
effective potential taking into account two-body collisions (see Sec. \ref{sec_gbm}). 
We assume an attractive self-interaction between the bosons
corresponding to a negative scattering length ($a_s<0$). For example, the
axion has an attractive self-interaction. As a result, dilute
axion stars can exist only below a maximum mass
\cite{prd1,prd2,phi6,mg16B,massmaxrel}\footnote{This maximum mass
was first evidenced in \cite{prd1,prd2}. Its exact expression can be obtained by
solving the eigenvalue problem associated with the GPP equations numerically \cite{prd2}.
An approximate expression can be obtained analytically by using a Gaussian
ansatz \cite{prd1}.}
\begin{eqnarray}
\label{mmax}
M_{\rm max}^{\rm exact}=1.012\, \frac{\hbar}{\sqrt{Gm|a_s|}}.
\end{eqnarray}
When $M>M_{\rm max}^{\rm exact}$, there is no equilibrium state and the star
collapses. When $M<M_{\rm max}^{\rm exact}$, there is a stable equilibrium state
and an unstable equilibrium state. The stable equilibrium state is metastable
(local but not global minimum of free energy). Because of quantum or thermal
fluctuations, the star may cross the potential barrier and collapse. We
have
studied the effect of quantum fluctuations in \cite{tunnel}. Here, we consider
the effect of thermal fluctuations. To that purpose, we develop an analogy
with Brownian theory. We account for finite
temperature effects and dissipative effects through the generalized GPP
equations \cite{ggpp}
\begin{eqnarray}
\label{g19}
i\hbar \frac{\partial\psi}{\partial
t}=-\frac{\hbar^2}{2m}\Delta\psi+m\Phi\psi+\frac{4\pi
a_s\hbar^2}{m^2}|\psi|^{2}\psi\nonumber\\
+2k_B
T\ln|\psi|\psi
-i\frac{\hbar}{2}\xi\left\lbrack \ln\left (\frac{\psi}{\psi^*}\right
)-\left\langle \ln\left (\frac{\psi}{\psi^*}\right
)\right\rangle\right\rbrack\psi,
\end{eqnarray}
\begin{equation}
\label{g2b}
\Delta\Phi=4\pi G |\psi|^2.
\end{equation}
With respect to the standard GP equation (\ref{gpp1}), the  generalized GP
equation (\ref{g19}) includes two new terms presenting a logarithmic
nonlinearity. The first term takes into account thermal effects (with a
temperature $T>0$) and the second term takes into account dissipative effects
(with a friction coefficient $\xi>0$). The interpretation of
these terms will become clear in Sec. \ref{sec_gbm} when we discuss the
hydrodynamic representation of the generalized GPP equations.\footnote{These
terms  give rise to a quantum Euler equation involving an isothermal equation of
state $P=\rho k_B T/m$ and a linear friction force $-\xi {\bf u}$ \cite{pre11}.}
We also explain in
Appendix \ref{sec_tf} how we can include thermal fluctuations in this model. The
effect of these thermal fluctuations is studied in Sec. \ref{sec_ttbec} with a
Gaussian ansatz.

{\it Remark:} The thermal and dissipative terms present in the
generalized GP equation (\ref{g19}) may have
several interpretations: (i) They may take into account the coupling of the
BEC with an external medium playing the role of a thermal
bath fixing the temperature $T$ (more general GP equations could be
considered if the thermal bath has a non standard origin like in Tsallis
thermodynamics \cite{ggpp,preparation}). In that case, they describe a quantum
Brownian gas, at least in an approximate manner \cite{pre11,ggpp,back,nottale2,preparation}. (ii) They can
parametrize the complicated processes of violent relaxation \cite{lb} and
gravitational cooling \cite{seidel94,gul0,gul} experienced by a self-gravitating
BEC. In that case, the generalized GPP equations (\ref{g19}) and (\ref{g2b})
describe the evolution of a self-gravitating BEC on a ``coarse-grained''
scale \cite{modeldm,wignerPH}. The resulting equilibrium state has a
core-envelope structure with a quantum core (soliton) surrounded by an
isothermal atmosphere. This core-envelope structure is
consistent with the observations of DM halos and with direct numerical
simulations of BECDM
\cite{ch2,ch3,schwabe,mocz,moczSV,veltmaat,moczprl,moczmnras,veltmaat2,
moczatt}. This isothermal model has been studied theoretically in
\cite{modeldm}. (iii) The thermal and dissipative terms present  in the
generalized GP equation (\ref{g19}) may account for the
effect of collisions between the bosons. In this context, other types of
generalized GP equations have been proposed in condensed matter physics \cite{gnz}. The
analogies and the differences between these equations and the generalized GP equation  (\ref{g19}) 
of the present paper are
discussed in \cite{preparation}. (iv) The logarithmic term
$-2b\ln |\psi|\psi$ with $b=-k_B T$ may correspond to a certain form of
self-interaction (beyond mean field contributions) or be a fundamental term present in all the wave equations
of quantum mechanics. In that case, $b$ may be interpreted as a fundamental
constant of physics and $T$ may be interpreted as a quantum
temperature (possibly scale dependent) \cite{ggpp}.
 (iv) Finally, the terms of temperature and friction in the  generalized GP equation
(\ref{g19}) arise from a single formalism related to the theory
of scale relativity according to which space-time has a fractal structure giving
rise to quantum mechanics \cite{nottale,epjpnottale,pdunottale}. In this
framework, the effective temperature $T$ may be positive or negative.

\section{Dissipative BEC dark matter model}
\label{sec_gbm}

\subsection{Generalized GPP equations}
\label{sec_ggpp}

We consider a self-gravitating BEC described by the  generalized
GPP equations \cite{ggpp} 
\begin{eqnarray}
\label{g1}
i\hbar \frac{\partial\psi}{\partial t}=-\frac{\hbar^2}{2m}\Delta\psi
+m\Phi\psi
+m\frac{dV}{d|\psi|^2}\psi\nonumber\\
-i\frac{\hbar}{2}\xi\left\lbrack \ln\left (\frac{\psi}{\psi^*}\right
)-\left\langle \ln\left (\frac{\psi}{\psi^*}\right
)\right\rangle\right\rbrack\psi,
\end{eqnarray}
\begin{equation}
\label{g2}
\Delta\Phi=4\pi G |\psi|^2,
\end{equation}
where $\psi({\bf r},t)$ is the wave function of the condensate, $\Phi({\bf
r},t)$ is the gravitational
potential, $V(|\psi|^2)$ is the potential of the BEC, and $\xi$ is a friction
coefficient. The potential of the BEC can be decomposed into $V(|\psi|^2)=V_{\rm
th}(|\psi|^2)+V_{\rm int}(|\psi|^2)$, where $V_{\rm th}(|\psi|^2)$
takes into account thermal effects and $V_{\rm int}(|\psi|^2)$ takes
into account the self-interaction of the bosons. If
the system is coupled to a standard thermal bath, the thermal potential  is the Boltzmann
potential $V_{\rm
th}(|\psi|^2)=\frac{k_B T}{m}|\psi|^2 (\ln|\psi|^2 -1)$. However, if the system is coupled to a thermal bath of a more
general origin, as in the context of nonextensive (Tsallis) thermodynamics,
the thermal potential $V_{\rm th}(|\psi|^2)$ has a different expression
\cite{wg}.

{\it Remark:} For the standard BEC model of Sec. \ref{sec_be} the 
SF potential is
\begin{equation}
\label{g18rem}
V(|\psi|^2)=\frac{k_B T}{m}|\psi|^2 (\ln|\psi|^2 -1)+\frac{2\pi
a_s\hbar^2}{m^3}|\psi|^4.
\end{equation}
It can be written as $V=V_{\rm th}+V_{\rm int}$. The first term $V_{\rm th}(|\psi|^2)=\frac{k_B T}{m}|\psi|^2 (\ln|\psi|^2
-1)$ is a logarithmic potential taking into account the
coupling of the system with an ordinary thermal bath. The second term $V_{\rm
int}(|\psi|^2)=\frac{2\pi a_s\hbar^2}{m^3}|\psi|^4$ is a quartic potential
taking into account two-body collisions between the bosons in a
dilute quantum gas.

\subsection{Hydrodynamic equations}
\label{sec_he}

Making the  Madelung \cite{madelung} transformation
\begin{equation}
\label{g3b}
\psi({\bf r},t)=\sqrt{{\rho({\bf r},t)}} e^{iS({\bf r},t)/\hbar},\quad
\rho=|\psi|^2,\quad {\bf u}=\frac{\nabla S}{m},
\end{equation}
where $\rho({\bf r},t)$ is the density of the BEC, $S({\bf r},t)$ is the action
and
${\bf u}({\bf r},t)$ is the velocity field, it can be shown \cite{ggpp} that the
generalized GPP equations (\ref{g1}) and (\ref{g2})  are equivalent to quantum
hydrodynamic
equations of the
form 
\begin{equation}
\label{g3}
\frac{\partial\rho}{\partial t}+\nabla\cdot (\rho {\bf u})=0,
\end{equation}
\begin{equation}
\label{g3g}
\frac{\partial S}{\partial t}+\frac{1}{2m}(\nabla S)^2+mV'(\rho)+m\Phi+Q+\xi (S-\langle S\rangle)=0,
\end{equation}
\begin{equation}
\label{g4}
\frac{\partial {\bf u}}{\partial t}+({\bf u}\cdot \nabla){\bf
u}=-\frac{1}{\rho}\nabla P-\nabla\Phi-\frac{1}{m}\nabla
Q-\xi{\bf u},
\end{equation}
\begin{equation}
\label{g5}
\Delta\Phi=4\pi G\rho,
\end{equation}
where
\begin{eqnarray}
\label{g7}
Q=-\frac{\hbar^2}{2m}\frac{\Delta
\sqrt{\rho}}{\sqrt{\rho}}=-\frac{\hbar^2}{4m}\left\lbrack
\frac{\Delta\rho}{\rho}-\frac{1}{2}\frac{(\nabla\rho)^2}{\rho^2}
\right\rbrack\nonumber\\
=-\frac{\hbar^2}{8m}({\nabla\ln\rho})^2
-\frac{\hbar^2}{4m}\Delta\ln\rho
\end{eqnarray}
is the quantum potential taking into account the Heisenberg uncertainty
principle and $P({\bf r},t)$ is the pressure arising from the self-interaction
of the bosons. The barotropic equation of state
$P(\rho)$ is determined by the
self-interaction potential $V(\rho)$ through the relation \cite{ggpp}
\begin{equation}
\label{g6}
P(\rho)=\rho
V'(\rho)-V(\rho)=\rho^2\left\lbrack
\frac{V(\rho)}{\rho}\right\rbrack'.
\end{equation}
The squared speed of sound is 
\begin{equation}
\label{g6sound}
c_s^2=P'(\rho)=\rho V''(\rho).
\end{equation}
Conversely, the
self-interaction potential is determined by the equation of state through the relation
\begin{equation}
\label{mad5b}
V(\rho)=\rho\int^{\rho}\frac{P(\rho')}{{\rho'}^2}\, d\rho'.
\end{equation} 
We note the identities
\begin{equation}
\label{mad5bb}
V'(\rho)=\int^{\rho} \frac{P'(\rho')}{\rho'}\, d\rho'\qquad {\rm and}\qquad 
V''(\rho)=\frac{P'(\rho)}{\rho},
\end{equation} 
which result from simple integrations by parts. Equations
(\ref{g3})-(\ref{g5}) are called the  quantum damped Euler-Poisson equations. The first equation is the
equation of continuity, the second equation is the quantum damped 
Hamilton-Jacobi (or Bernoulli) equation, the third equation is the quantum
damped Euler equation, and the fourth equation is 
the Poisson equation.  In
the
conservative
case $\xi=0$, we recover the quantum Euler-Poisson equations \cite{prd1}. In the
strong friction limit $\xi\rightarrow +\infty$, Eqs. (\ref{g3})-(\ref{g5}) lead
to the quantum Smoluchowski-Poisson equations \cite{ggpp}
\begin{equation}
\label{g8}
\xi\frac{\partial\rho}{\partial t}=\nabla\cdot\left (\nabla
P+\rho\nabla\Phi+\frac{\rho}{m}\nabla Q\right ),
\end{equation}
\begin{equation}
\label{g8b}
\Delta\Phi=4\pi G\rho.
\end{equation}
In the following, we will  exclusively use
the
hydrodynamic formalism.\footnote{We refer to
\cite{ggpp} for
the description of
the results in terms of the wave function.} In that case, the normalization
condition on the
wave function is equivalent to
the conservation of the mass of the BEC $M=\int \rho\, d{\bf r}$.

The foregoing equations can be given a
thermodynamical interpretation (see Appendix B.5 of \cite{wignerPH} and Appendices A and D of \cite{back}). The
relation from Eq. (\ref{g6}) is equivalent to the first principle of
thermodynamics $d(u/\rho)=-Pd(1/\rho)+Td(s/\rho)$ for a barotropic gas at zero temperature ($T=0$) provided
that $V(\rho)$ is interpreted as the internal energy density $u(\rho)$. In that
case, $V'(\rho)$ represents the enthalpy density $h(\rho)$. We have the
relations
\begin{equation}
\label{hell1}
u(\rho)=V(\rho),\qquad h(\rho)=V'(\rho),
\end{equation} 
\begin{equation}
\label{hell2}
h'(\rho)=\frac{P'(\rho)}{\rho},\qquad h=\frac{P+u}{\rho}.
\end{equation}
This thermodynamical
interpretation is also valid for an
isothermal gas at fixed
temperature $T$, in which
case the internal energy $u$ in the equations above is replaced by the free energy density
$f=u-T s$ where $s$ is the entropy density and $h$ is replaced by the local
chemical potential by unit of mass $\mu_{\rm loc}/m$ (see \cite{back,nottale2}
for more details). To simplify the notations (and be consistent with our previous papers) it is convenient to write  the free energy density $f$ as $u=u_{\rm int}+u_{\rm th}=u_{\rm int}-T s$ and
the local chemical potential by unit of mass $\mu_{\rm loc}/m$ as $h$, even if it is an abuse of language.

{\it Remark:} For the standard BEC model of Sec. \ref{sec_be},  the SF potential $V=V_{\rm th}+V_{\rm int}$ 
(free energy density) is 
\begin{equation}
\label{g18}
V(\rho)=\frac{k_B T}{m}\rho(\ln\rho-1)+\frac{2\pi a_s\hbar^2}{m^3}\rho^2.
\end{equation}
It can be written as $u=u_{\rm th}+u_{\rm int}=-Ts+u_{\rm int}$, 
where the first term is the entropy term $u_{\rm th}=-Ts$, where $s=-\frac{k_B}{m}\rho\ln\rho$ is the Boltzmann entropy density, and the second term is
the self-interaction term $u_{\rm int}=(2\pi a_s\hbar^2/m^3)\rho^2$.  The associated
equation of state reads
\begin{equation}
\label{g20}
P=\rho \frac{k_B T}{m}+\frac{2\pi a_s\hbar^2}{m^3}\rho^{2}.
\end{equation}
It can be written as $P=P_{\rm th}+P_{\rm int}$. It has a linear part $P_{\rm th}=\rho k_B T/m$ and a quadratic part
$P_{\rm int}=2\pi
a_s\hbar^2\rho^2/m^3$. The linear part
corresponds to
an isothermal equation of state with a temperature $T$.
The quadratic part corresponds to a polytropic equation of state with index
$n=1$, which arises from the self-interaction of the bosons. For an
attractive
self-interaction between the bosons ($a_s<0$), the self-interaction pressure
$P_{\rm int}$ is negative. The
equation
of state (\ref{g20}) defines a composite model of BECDM halos with
a core-envelope
structure. The quadratic equation of state
dominates in the core where the density is high and the isothermal equation of
state dominates in the envelope where the density is low. As a result, the
corresponding BECDM
halos present a quantum (solitonic) core surrounded by an isothermal
envelope. This composite model has been studied in detail in \cite{modeldm}.

\subsection{Equilibrium state}
\label{sec_eq2}

In the hydrodynamic representation, an equilibrium state of the  quantum
damped Euler-Poisson equations (\ref{g3})-(\ref{g5}), obtained
by taking $\partial_t=0$
and ${\bf u}={\bf 0}$, satisfies the relation
\begin{equation}
\label{eq1q}
\nabla P+\rho\nabla\Phi+\frac{\rho}{m}\nabla Q={\bf
0}.
\end{equation}
This equation can be interpreted as a condition of quantum hydrostatic
equilibrium. It describes the balance between the pressure
(due to the self-interaction of the bosons and their coupling to a thermal
bath), the gravitational force, and the
quantum force arising from the Heisenberg
uncertainty principle.  Combining Eq. (\ref{eq1q}) with the Poisson
equation (\ref{g5}), we
obtain the fundamental differential equation of quantum hydrostatic equilibrium
\begin{equation}
\label{bm1}
-\nabla\cdot \left (\frac{\nabla P}{\rho}\right )+\frac{\hbar^2}{2m^2}\Delta
\left (\frac{\Delta\sqrt{\rho}}{\sqrt{\rho}}\right )=4\pi G\rho.
\end{equation}

{\it Remark:} For the standard BEC model of Sec. \ref{sec_be}, Eq. (\ref{bm1})
becomes
\begin{equation}
\label{bm2}
-\frac{k_B T}{m}\Delta\ln\rho-\frac{4\pi
a_s\hbar^2}{m^3}\Delta\rho+\frac{\hbar^2}{2m^2}\Delta
\left (\frac{\Delta\sqrt{\rho}}{\sqrt{\rho}}\right )=4\pi G\rho.
\end{equation}
When $T=0$, this differential equation has
been solved numerically in Ref. \cite{prd2} in the general case of a repulsive 
or an attractive self-interaction. When $T\neq 0$, this differential equation
has
been solved numerically in Ref. \cite{modeldm} for a repulsive
self-interaction ($a_s>0$)  in the Thomas-Fermi (TF) limit ($\hbar=0$) and in \cite{ggpp,pdunottale}
for a noninteracting BEC ($a_s=0$).

\subsection{Free energy}
\label{sec_fe}

The free energy associated with the  quantum
damped Euler equations (\ref{g3})-(\ref{g5}) is given by
\begin{eqnarray}
\label{g9}
F=\Theta_c+\Theta_Q+U+W,
\end{eqnarray}
where $\Theta_c$ is the classical kinetic energy, $\Theta_Q$ is the quantum
kinetic energy, $U$ is the internal energy including self-interaction effects
and thermal effects, and $W$ is the gravitational energy
\cite{ggpp}. The free energy
can be  written explicitly as
\begin{equation}
\label{g10}
F=\int\rho \frac{{\bf u}^2}{2}\, d{\bf r}+\frac{1}{m}\int \rho Q\, d{\bf r}+\int
V(\rho)\, d{\bf r}
+\frac{1}{2}\int\rho\Phi\, d{\bf r}.
\end{equation}
The free energy associated with the quantum Smoluchowski-Poisson equations
(\ref{g8}) and (\ref{g8b}) is
given by 
\begin{eqnarray}
\label{g11}
F=\Theta_Q+U+W.
\end{eqnarray}
It can be  written explicitly as
\begin{eqnarray}
\label{g12}
F=\frac{1}{m}\int
\rho Q\, d{\bf r}+\int
V(\rho)\, d{\bf r}
+\frac{1}{2}\int\rho\Phi\, d{\bf r}.
\end{eqnarray}
It is obtained from Eqs. (\ref{g9}) and (\ref{g10}) by discarding the classical
kinetic term
which is small in the strong friction limit $\xi\rightarrow +\infty$.

The time
derivative of the generalized free energy (\ref{g10}) satisfies the identity
\cite{ggpp}
\begin{eqnarray}
\label{gh1}
\dot F=-\xi\int \rho {\bf u}^2\, d{\bf r}=-2\xi\Theta_c.
\end{eqnarray}
For dissipationless systems ($\xi=0$), Eq. (\ref{gh1}) 
shows that the quantum Euler-Poisson equations  
conserve
the free energy ($\dot F=0$). Using general arguments
\cite{holm}, one concludes that an equilibrium state of the quantum Euler-Poisson equations is stable if,
and only if, it is a minimum of free energy at fixed
mass. For dissipative systems ($\xi>0$), Eq.
(\ref{gh1}) 
shows that the quantum damped Euler-Poisson
equations decrease the free energy ($\dot F\le 0$).
This provides a form of
 $H$-theorem implying that the system relaxes for $t\rightarrow +\infty$ towards a stable  equilibrium
state (minimum of free energy at fixed mass) when it exists. When
$\dot F=0$, we get ${\bf u}={\bf 0}$, leading to the
condition of
quantum hydrostatic equilibrium (\ref{eq1q}). In the
strong friction limit $\xi\rightarrow +\infty$,
the time derivative of the generalized free energy (\ref{g12}) associated with
the quantum Smoluchowski-Poisson 
equations satisfies the $H$-theorem
\begin{eqnarray}
\label{gh2}
\dot F=-\frac{1}{\xi}\int \frac{1}{\rho}\left (\nabla
P+\rho\nabla\Phi+\frac{\rho}{m}\nabla Q\right )^2\,
d{\bf r}\le 0.
\end{eqnarray}
When $\dot F=0$, the term in parenthesis vanishes, leading to the condition
of quantum hydrostatic equilibrium (\ref{eq1q}). A more detailed
discussion of these results is given
in Sec. 3.3 of \cite{ggpp}.

{\it Remark:} For the standard BEC model of Sec. \ref{sec_be}, we have $U=U_{\rm int}+U_{\rm th}=U_{\rm
int}-T
S_B$ where
\begin{eqnarray}
\label{g10b}
S_B=-k_B\int \frac{\rho}{m}(\ln\rho-1)\, d{\bf
r}
\end{eqnarray}
is the Boltzmann entropy and
\begin{eqnarray}
\label{g10bv}
U_{\rm int}=\frac{2\pi
a_s\hbar^2}{m^3}\int \rho^2\,
d{\bf
r}
\end{eqnarray}
is the internal energy of a polytrope of index $n=1$. The Boltzmann entropy
$S_B$ accounts for thermal effects and the internal energy $U_{\rm int}$
accounts for
the self-interaction of the bosons.  On the other hand, the quantum kinetic energy $\Theta_Q$ can be
written in different forms
\cite{epjpnottale,ggpp}. It can be
written as
\begin{eqnarray}
\label{ef5km}
\Theta_Q=\int
\rho \frac{{\bf u}_Q^2}{2}\, d{\bf r},
\end{eqnarray}
where ${\bf u}_Q=(\hbar/2m)\nabla\ln\rho$ is the quantum (osmotic) velocity
introduced by Nelson \cite{nelson} (see also \cite{epjpnottale,axioms}).
Under that form it really looks like a kinetic
energy. It can also be written as
\begin{equation}
\label{ef4}
\Theta_Q=\frac{1}{m}\int \rho Q\, d{\bf r},
\end{equation}
where $Q$ is the quantum potential. Under that form it looks like a potential
energy. Finally, it can be written as
\begin{eqnarray}
\label{ef5}
\Theta_Q=\frac{\hbar^2}{8m^2}\int \frac{(\nabla\rho)^2}{\rho}\, d{\bf r}.
\end{eqnarray}
This functional  was introduced by Madelung \cite{madelungN,madelung} and von
Weizs\"acker \cite{weizsacker}. It is related to the Fisher \cite{fisher}
entropy 
\begin{eqnarray}
\label{fish}
S_F=\frac{1}{m}\int \frac{(\nabla\rho)^2}{\rho}\, d{\bf r}, 
\end{eqnarray}
which was introduced independently from quantum mechanics, by
\begin{eqnarray}
\label{fishrel}
\Theta_Q=\frac{\hbar^2}{8m} S_F.
\end{eqnarray}
It is interesting to note that the damped logarithmic
Schr\"odinger  equation (\ref{g19}) satisfies an $H$-theorem for a free energy
including the Boltzmann entropy  (taking into account thermal effects) and
the Fisher entropy (taking into account quantum effects). This is because our
formalism combines thermodynamics and quantum mechanics. The functional (\ref{ef5}) is also a special case of the Korteweg functional $\Theta_Q=\frac{1}{2}\int K(\rho)(\nabla\rho)^2\, d{\bf r}$  for capillary fluids \cite{vdwc,korteweg}, corresponding to a capillary coefficient  $K(\rho)=\hbar^2/(4m^2\rho)$ \cite{back}.

 \subsection{Variational principle}
\label{sec_eq}

According to the results of the preceding section, a
stable equilibrium state of the quantum Euler-Poisson
equations\footnote{The following results are valid for damped and undamped
systems.} is the
solution of
the minimization problem
\begin{eqnarray}
\label{eq1}
F(M)=\min_{\rho,{\bf u}} \left\lbrace F[\rho,{\bf u}]\quad |\quad M\quad {\rm
fixed}\right\rbrace.
\end{eqnarray}
The
variational principle for the first variations
(extremization) can be written as 
\begin{eqnarray}
\label{eq2}
\delta F-\frac{\mu}{m}\delta M=0,
\end{eqnarray}
where $\mu$ is a Lagrange multiplier (global chemical potential) taking  into
account
the conservation of mass. This variational principle gives
${\bf u}={\bf 0}$ (the equilibrium state is static) and the quantum Gibbs
condition 
\begin{eqnarray}
\label{eq3}
\Phi+V'(\rho)+\frac{Q}{m}=\frac{\mu}{m}
\end{eqnarray}
expressing the fact that the gravitational potential $\Phi$ 
$+$ the enthalpy $h(\rho)=V'(\rho)$ (equal to the local chemical potential by
unit of mass $\mu_{\rm loc}(\rho)/m$) $+$ the
quantum potential $Q/m$ is constant (equal to the global chemical potential  by
unit
of mass $\mu/m$). Taking the gradient of Eq.
(\ref{eq3}) and using Eq. (\ref{mad5bb}), we recover the condition of 
quantum
hydrostatic equilibrium
(\ref{eq1q}). Therefore, an extremum of free energy at fixed mass is an
equilibrium
state of the
quantum  Euler-Poisson equations. Furthermore, it can be
shown (see Appendix B of \cite{jeansapp})
that the self-gravitating BEC
is linearly stable with respect to the quantum Euler-Poisson
equations
if, and
only if,
it is a local minimum of free energy  at fixed mass (a maximum or a saddle
point is linearly unstable) \cite{ggpp}.

{\it Remark:} The series of equilibria of dilute axion stars with an
attractive $|\psi|^4$ self-interaction can be parametrized by their central
density $\rho_0$. It presents a maximum mass $M_{\rm max}^{\rm dilute}$ [see Eq.
(\ref{mmax})] corresponding to a critical central density
$(\rho_0)_{\rm crit}=0.04\, Gm^4/(a_s^2\hbar^2)$  \cite{prd1,prd2}. Using the Poincar\'e turning point criterion
\cite{poincare,katzpoincare},\footnote{See Appendix C.2 of \cite{acf}
for a summary of the Poincar\'e theory of linear series of equilibria.} the
Whitney theorem \cite{whitney}  of catastrophe
(or bifurcation) theory \cite{post,arnold},  the
Derrick theorem \cite{derrick}, or the Wheeler $M(R)$ theorem \cite{htww},
one can show that the
series of equilibria is dynamically stable before
the turning point of mass (i.e. for $\rho_0<(\rho_0)_{\rm crit}$)
and becomes dynamically unstable
afterwards (i.e. for $\rho_0>(\rho_0)_{\rm crit}$) (see, e.g.,
\cite{prd1,prd2,massmaxrel} and
references therein). In the
case of dense axion stars with an
additional repulsive
$|\psi|^6$ potential, the stability is regained after a second turning point of
mass (minimum mass $M_{\rm min}^{\rm dense}$). Phase transitions between dilute
and dense axion
stars are studied in \cite{phi6} by using a Gaussian ansatz.

\subsection{Quantum virial theorem}
\label{sec_qvt}

The time-dependent scalar  virial theorem associated with the  quantum damped
Euler-Poisson equations can be written as (see Appendix G of \cite{ggpp})
\begin{equation}
\label{g15}
\frac{1}{2}\ddot I+\frac{1}{2}\xi\dot I=2(\Theta_c+\Theta_Q)+3\int
P\, d{\bf r}+W,
\end{equation}
where $I=\int\rho r^2\, d{\bf r}$ is the moment of inertia and $W$ is the
gravitational energy (coinciding with the virial of the gravitational force
$W_{ii}$ in $d=3$ dimensions). In
the
strong friction limit $\xi\rightarrow +\infty$, it reduces to
\begin{equation}
\label{g16}
\frac{1}{2}\xi\dot I=2\Theta_Q+3\int P\, d{\bf r}+W.
\end{equation}
At equilibrium, we obtain the quantum scalar virial theorem
\begin{equation}
\label{g17}
2\Theta_Q+3\int P\, d{\bf r}+W=0.
\end{equation}

{\it Remark:} For the standard BEC model of Sec. \ref{sec_be}, using Eqs.
(\ref{g20}) and (\ref{g10bv}) we have 
\begin{equation}
\label{g17b}
\int P\, d{\bf r}=N k_B T+U_{\rm int}.
\end{equation}

\section{Gaussian ansatz}
\label{sec_ga}

\subsection{Free energy}
\label{sec_feg}

We can obtain an approximate analytical solution
of the generalized GPP equations (\ref{g1}) and (\ref{g2}) by developing a
mechanical analogy \cite{prd1}. Making a Gaussian ansatz for the
wavefunction
(see Sec. 8.2 of Ref. \cite{ggpp} for details)
\begin{eqnarray}
\label{gte1a}
\psi({\bf r},t)=\left \lbrack \frac{M}{\pi^{3/2}
R(t)^3}\right \rbrack^{1/2}e^{-r^2/2R(t)^2}e^{imH(t)r^2/2\hbar},
\end{eqnarray}
corresponding to a density profile, an action, and a velocity field of the form 
\begin{eqnarray}
\label{gte1asup}
\rho({\bf r},t)=\frac{M}{\pi^{3/2}
R(t)^3}e^{-r^2/R(t)^2},\nonumber\\
S({\bf r},t)=\frac{1}{2}mH(t)r^2,\qquad {\bf u}({\bf r},t)=H(t){\bf r},
\end{eqnarray}
where $R(t)$ is the typical
radius of the BEC,\footnote{For a
Gaussian density profile, the relation
between the radius $R$ and the
radius $R_{99}$ containing $99\%$
of the mass is $R_{99}=2.38167 R$  \cite{prd1}.} and $H=\dot R/R$ is the
analogue of the Hubble function in
cosmology (see Sec. 8.8 of Ref. \cite{ggpp} for details), we find that the
free energy
functional (\ref{g9}) 
can be written
as a function of $R$ and $\dot R$  (for a fixed mass $M$) as
\begin{eqnarray}
\label{g24}
F=\frac{1}{2}\alpha M\left (\frac{dR}{dt}\right )^2+V(R),
\end{eqnarray}
with $\alpha=3/2$. The first term in Eq. (\ref{g24}) is the classical kinetic energy
$\Theta_c$ while the 
effective potential $V(R)$ comprises the quantum kinetic energy $\Theta_Q$,
the internal energy $U$, and the gravitational energy $W$. Therefore,
$F=\Theta_c+V$ with $V=\Theta_Q+U+W$. With the Gaussian ansatz, the 
$H$-theorem (\ref{gh1}) can be written as
\begin{eqnarray}
\label{g26}
\dot F=-\xi\alpha M\left (\frac{dR}{dt}\right )^2\le 0.
\end{eqnarray}
Taking the time derivative of Eq. (\ref{g24}) and using Eq. (\ref{g26}) we find
that the evolution of the
radius $R(t)$ of
the BEC is determined by the differential equation
\begin{eqnarray}
\label{g25}
\alpha M\frac{d^2R}{dt^2}+\xi\alpha M\frac{dR}{dt}=-\frac{d{V}}{dR}.
\end{eqnarray}
This equation is similar to the equation of motion of a damped
particle of mass $\alpha M$ and position $R$ moving in a one-dimensional
potential $V(R)$. This equation can also be obtained from the quantum
damped virial theorem (\ref{g15}) (see Sec. 8.4 of Ref. \cite{ggpp} for
details). In the
strong friction limit $\xi\rightarrow +\infty$, we have $F=V(R)$ and the
foregoing equations
reduce to
\begin{eqnarray}
\label{g27}
\xi\alpha M\frac{dR}{dt}=-\frac{d{V}}{dR}
\end{eqnarray}
and
\begin{eqnarray}
\label{g28}
\dot F=-\frac{1}{\xi\alpha M}\left (\frac{dV}{dR}\right )^2\le 0.
\end{eqnarray}
In the following, we shall specifically consider the  BEC model of Sec.
\ref{sec_be} with an attractive self-interaction ($a_s<0$). In that case,  the
effective potential reads (see Sec. 8.2 of Ref. \cite{ggpp} for details)
\begin{eqnarray}
\label{g29p}
V(R)=\sigma \frac{\hbar^2M}{m^2R^2}-\zeta\frac{2\pi
|a_s|\hbar^2M^2}{m^3R^{3}}-\nu
\frac{GM^2}{R}
\end{eqnarray}
with the coefficients
\begin{eqnarray}
\label{gte2b}
\sigma=\frac{3}{4},  \quad
\zeta=\frac{1}{(2\pi)^{3/2}}, \quad \nu=\frac{1}{\sqrt{2\pi}}.
\end{eqnarray}

{\it Remark:} We have neglected the temperature $T$ in the
effective  potential $V(R)$ because its effect is weak in the quantum core of
the system that we consider, but we
shall take into account thermal fluctuations in Sec. \ref{sec_ttbec}.

\subsection{Mass-radius relation}
\label{sec_mr}

\begin{figure}[!h]
\begin{center}
\includegraphics[clip,scale=0.3]{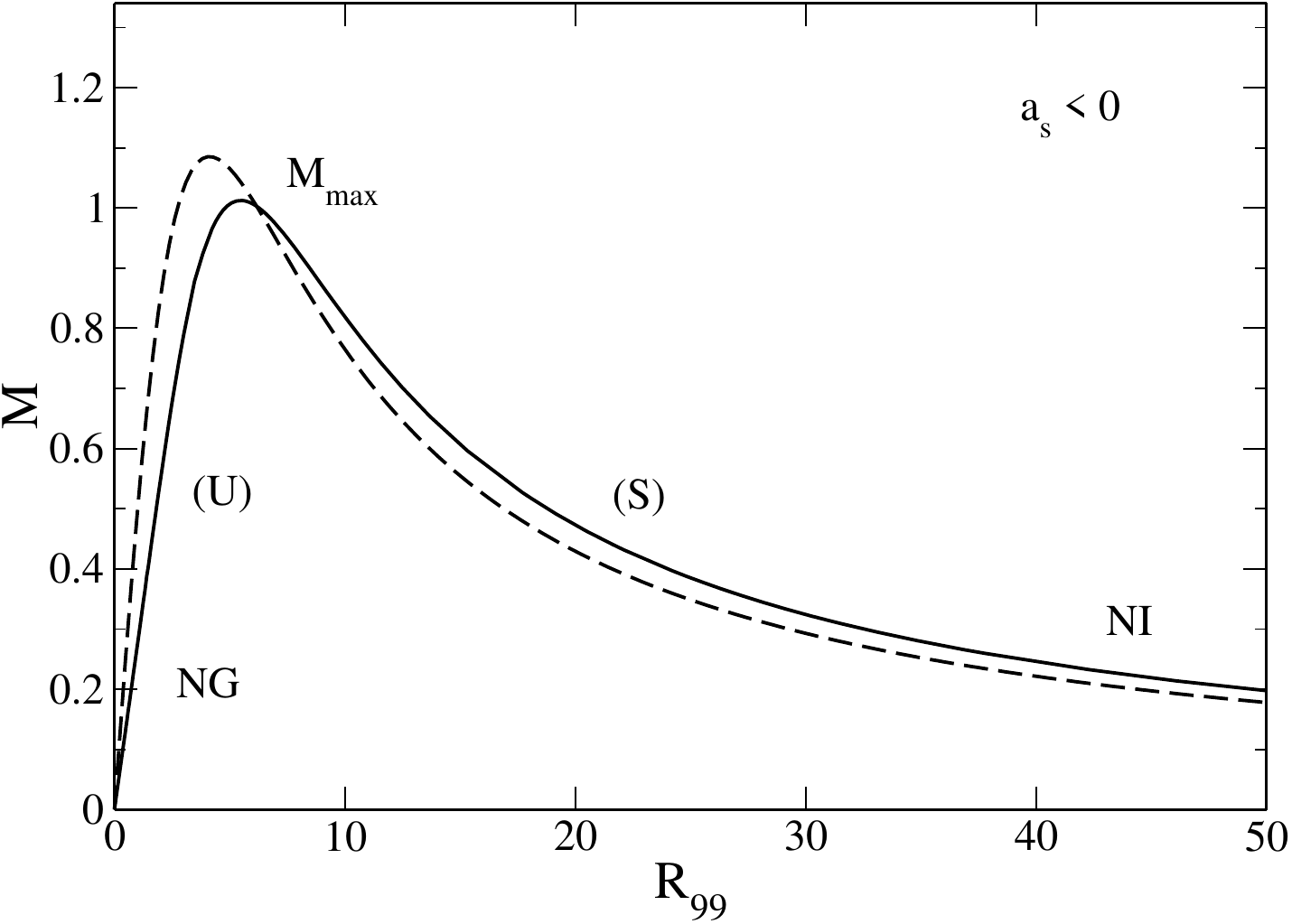}
\caption{Mass-radius relation of dilute
axion stars interpreted as a nonrelativistic self-gravitating BEC with an
attractive
self-interaction ($a_s<0$). We have chosen a normalization such that
$\hbar=G=m=|a_s|=1$ \cite{prd1}. The solid line is the exact mass-radius
relation obtained
by solving the GPP equations numerically \cite{prd2}. The dotted line
corresponds to the approximate analytical mass-radius relation from Eq. 
(\ref{g29}) obtained with the Gaussian ansatz \cite{prd1}.}
\label{MRmodif}
\end{center}
\end{figure}

We have seen in Sec. \ref{sec_eq} that an extremum of free energy $F[\rho,{\bf
u}]$ given by
Eq. (\ref{g10}) at
fixed mass determines an equilibrium state of
the generalized GPP equations (\ref{g1}) and
(\ref{g2}). Furthermore, a  (local)
minimum of free energy is (meta)stable while a maximum or a saddle point is
unstable. Within the Gaussian ansatz, we have to minimize the free
energy $F(R,\dot R)$ given by Eq.
(\ref{g24}) at fixed mass $M$ in order to determine stable equilibrium states.
An extremum corresponds to $dR/dt=0$ and
$V'(R)=0$, leading to the mass-radius relation \cite{prd1,ggpp}
\begin{eqnarray}
\label{g29}
M=\frac{2\sigma\frac{\hbar^2}{m^2R^3}}{\frac{\nu
G}{R^2}+6\pi\zeta\frac{|a_s|\hbar^2}{m^3R^4}}.
\end{eqnarray}
This equation can also be obtained from the equilibrium
quantum virial theorem
(\ref{g17}) (see Sec. 8.3 of \cite{ggpp} for details). The mass-radius relation 
is plotted in Fig.
\ref{MRmodif} in the case of an
attractive self-interaction ($a_s<0$). It displays a 
maximum mass
\cite{prd1}
\begin{eqnarray}
\label{g30}
M_{\rm max}=\left (\frac{\sigma^2}{6\pi\zeta\nu}\right
)^{1/2}\frac{\hbar}{\sqrt{Gm|a_s|}}
\end{eqnarray}
corresponding to a radius
\begin{eqnarray}
\label{g30b}
R_{*}=\left
(\frac{6\pi\zeta}{\nu}\right )^{1/2}\left
(\frac{|a_s|\hbar^2}{Gm^3}\right )^{1/2}.
\end{eqnarray}
The prefactors are $1.085$ and $1.73$ (using the result of footnote 14, the
prefactor of the radius $R_{99}^*$ containing $99\%$ of the mass is
$4.12$).\footnote{The exact values
of the maximum
mass $M_{\rm max}^{\rm
exact}$ and of the
corresponding radius $(R_{99}^*)^{\rm exact}$ [see Eqs.
(\ref{intro1}) and (\ref{intro2})] have been obtained in Ref. \cite{prd2} by
solving the eigenvalue problem associated with the GPP equations (\ref{gpp1})
and (\ref{gpp2}) numerically. The exact prefactors are $1.012$ and $5.5$
respectively. The value of the maximum mass given by the Gaussian ansatz is
relatively accurate.} We have the identity
\begin{eqnarray}
M_{\rm max}=\frac{\sigma}{\nu}\frac{\hbar^2}{Gm^2 R_*}.
\label{mr3}
\end{eqnarray}
There is no equilibrium state with $M>M_{\rm max}$. When $M<M_{\rm max}$, two
equilibrium states exist with the same mass. By studying the
sign of the second
derivatives of the potential $V(R)$ 
one can analytically show \cite{prd1} that the equilibrium
states with $R>R_*$ are stable ($V''>0$) while the equilibrium
states with $R<R_*$ are unstable ($V''<0$). This result can also be recovered
from the identity \cite{bectcoll,ggpp,tunnel,reviewBECDM}
\begin{eqnarray}
\frac{dM}{dR}=-\alpha M\frac{m^2R^3}{2\sigma \hbar^2}\omega^2,
\label{mr3b}
\end{eqnarray}
which relates the squared radial pulsation
of the BEC $\omega^2=V''(R)/\alpha M$ (see Sec. \ref{sec_relt}) to the slope $dM/dR$ of the mass-radius
relation (see
Secs. 8.6 and 8.7 of Ref. \cite{ggpp} for details). An equilibrium
state is stable ($\omega^2>0$) when
$dM/dR<0$ and unstable ($\omega^2<0$) when $dM/dR>0$.
The change of stability occurs at the maximum mass $M_{\rm max}$ (i.e.
$dM/dR=\omega=0$). Therefore, $R_*$ is the minimum radius of
stable equilibrium states. This stability result can also be directly 
established from the topology of the series of equilibria by using
the Poincar\'e turning point criterion
\cite{poincare,katzpoincare}, the Whitney theorem \cite{whitney}, the
Derrick theorem \cite{derrick}, or  the Wheeler
theorem
\cite{htww} stating
that a change of stability can occur only at a turning point
of mass. These general methods are valid beyond the Gaussian
ansatz as discussed in Sec. \ref{sec_eq}.

In the noninteracting limit, corresponding to $R\gg R_*$, the mass-radius
relation reduces to
\begin{eqnarray}
\label{g33}
M\sim \frac{2\sigma}{\nu}\frac{\hbar^2}{Gm^2R}.
\end{eqnarray}
These equilibrium states are stable. In the nongravitational limit,
corresponding to $R\ll R_*$, the mass-radius
relation reduces to
\begin{eqnarray}
\label{g32}
M\sim \frac{\sigma}{3\pi\zeta}\frac{mR}{|a_s|}.
\end{eqnarray}
These equilibrium states are unstable.

{\it Remark:} The above results apply to dilute axion stars. Stable  dilute
axion stars exist only below a maximum mass $M_{\rm max}$ and above a minimum
radius $R_{*}$ given by Eqs. (\ref{g30}) and (\ref{g30b}) with the Gaussian
ansatz \cite{prd1}.\footnote{These results can be generalized by accounting
for the presence of a black hole at the center of the system. The expression of
the maximum mass of a dilute axion star in
the presence of a central black hole, obtained with the Gaussian
ansatz, is given in \cite{epjpbh}.} If we
take into account a $\varphi^6$ repulsion in the potential of
self-interaction (or consider the exact potential of axions), an additional
stable branch appears in the mass-radius relation at small radii, corresponding
to dense axion stars \cite{braaten,phi6} (see also \cite{visinelli,ebyexpansion} for relativistic effects).

\subsection{Normal form of the potential close to the maximum mass}
\label{sec_nfcmm}

\begin{figure}
\begin{center}
\includegraphics[clip,scale=0.3]{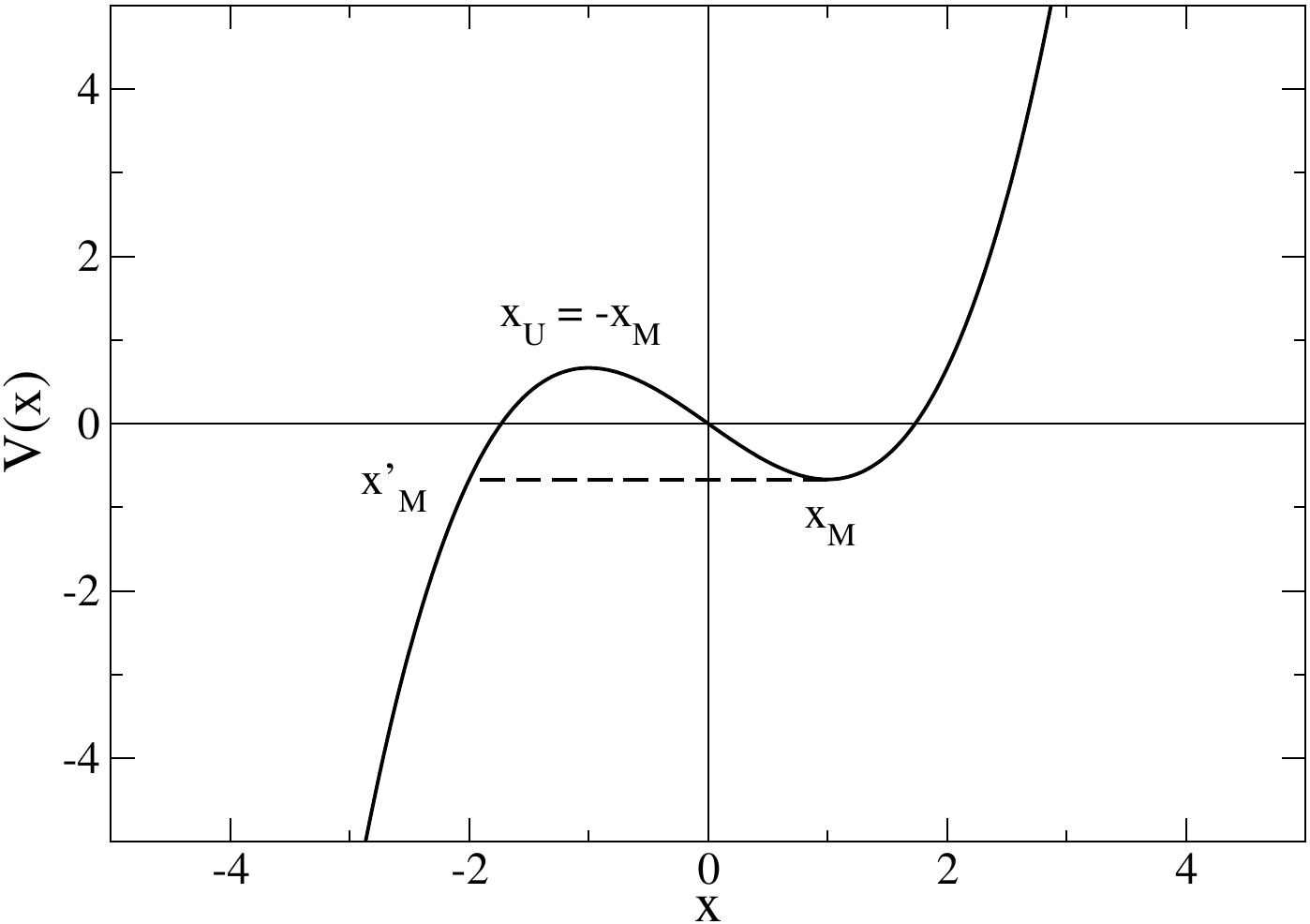}
\caption{Normal form of the potential $V(R)$ close to a saddle-node
bifurcation (we have introduced the notation $x=R-R_*$ of Sec.
\ref{sec_exp}).}
\label{normalform}
\end{center}
\end{figure}

Expanding the effective potential from Eq. (\ref{g29p}) to third order
close to the maximum mass $M_{\rm max}$, we obtain \cite{bectcoll}
\begin{eqnarray}
\label{nfcmm1}
\frac{V(R)}{V_0}&=&\frac{1}{3R_*^3}(R-R_*)^3-\frac{2}{R_*}\left
(1-\frac{M}{M_{\rm max}}\right )(R-R_*)\nonumber\\
&-&\frac{1}{3}+\frac{5}{3}
\left (1-\frac{M}{M_{\rm max}}\right ),
\end{eqnarray}
where
\begin{eqnarray}
V_0=\nu \frac{GM_{\rm
max}^2}{R_*}=\frac{\sigma^2\nu^{1/2}}{(6\pi\zeta)^{3/2}}\frac{\hbar
m^{1/2}G^{1/2}}{|a_s|^{3/2}}.
\label{nfcmm2}
\end{eqnarray}
Equation (\ref{nfcmm1}) is the normal form of a potential $V(R)$
close to a saddle-node bifurcation (see Fig. \ref{normalform}). With this
approximation, the equation of motion (\ref{g25}) of the fictive particle
becomes
\begin{equation}
\label{g25sc}
\alpha M\frac{d^2R}{dt^2}+\xi\alpha
M\frac{dR}{dt}=-\frac{V_0}{R_*^3}(R-R_*)^2+\frac{2V_0}{R_*}\left
(1-\frac{M}{M_{\rm max}}\right ).
\end{equation}

The mass-radius
relation close to  $M_{\rm max}$, obtained from the condition $V'(R)=0$, is
given by
\begin{eqnarray}
\frac{M}{M_{\rm max}}&=&1-\frac{1}{2R_*^2}(R-R_*)^2\nonumber\\
\quad {\rm i.e.}\quad
R-R_*&=&\pm
\sqrt{2}R_*\left (1-\frac{M}{M_{\rm max}}\right )^{1/2}.
\label{nfcmm3}
\end{eqnarray}
The upper sign corresponds to the branch $R>R_*$ and the lower
sign corresponds to the branch $R<R_*$.

The squared radial pulsation of the BEC close to the maxumum mass
is given by
\begin{eqnarray}
\omega^2=\frac{V''(R)}{\alpha M}=\frac{2V_0}{\alpha M_{\rm max}R_*^3}(R-R_*)\nonumber\\
=\pm
\frac{2\sqrt{2}V_0}{\alpha M_{\rm max}R_*^2} \left (1-\frac{M}{M_{\rm
max}}\right
)^{1/2}.
\label{nfcmm4ts}
\end{eqnarray}
This expression
confirms that the branch $R>R_*$ is
stable
($\omega^2>0$) while the branch $R<R_*$ is unstable ($\omega^2<0$). The squared
pulsation can also be written as
\begin{eqnarray}
\omega^2=\pm \frac{2\sqrt{2}}{t_D^2} \left (1-\frac{M}{M_{\rm max}}\right
)^{1/2},
\label{nfcmm4}
\end{eqnarray}
where we have introduced the dynamical time
\begin{eqnarray}
t_D=\left (\frac{\alpha M_{\rm max}R_*^2}{V_0}\right
)^{1/2}&=&\left
(\frac{\alpha}{\nu}\right
)^{1/2}\frac{1}{\sqrt{G\rho_0}}\nonumber\\
&=&\frac{6\pi\zeta}{\nu}\left
(\frac{\alpha}{\sigma}\right
)^{1/2}\frac{|a_s|\hbar}{Gm^2}
\label{nfcmm5}
\end{eqnarray}
constructed with the typical density
\begin{eqnarray}
\label{nfcmm6b}
\rho_{0}=\frac{M_{\rm max}}{R_{*}^3}=\frac{\sigma
\nu}{(6\pi\zeta)^2}\frac{Gm^4}{a_s^2\hbar^2}.
\end{eqnarray}
From Eq.
(\ref{nfcmm3}) we get
\begin{eqnarray}
\label{tsgm}
\frac{dM}{dR}=-\frac{M_{\rm
max}}{R_*^2}(R-R_*).
\end{eqnarray}
Comparing Eq. (\ref{tsgm}) with Eq. (\ref{nfcmm4ts}), we obtain 
the relation
\begin{eqnarray}
\label{nfcmm7}
\frac{dM}{dR}=-\frac{\omega^2}{2}\frac{\alpha M_{\rm
max}^2R_*}{V_0}=-\frac{\omega^2}{2}\frac{M_{\rm max}t_D^2}{R_*}
\end{eqnarray}
that links the stability of the system (through the sign of the squared
pulsation $\omega^2$) to the slope of the mass-radius relation. This is the
expression of the Poincar\'e turning point criterion [see Eq. (\ref{mr3b})]
close to the maximum mass
(see Sec. 8.7 of Ref. \cite{ggpp}).

\begin{figure}
\begin{center}
\includegraphics[clip,scale=0.3]{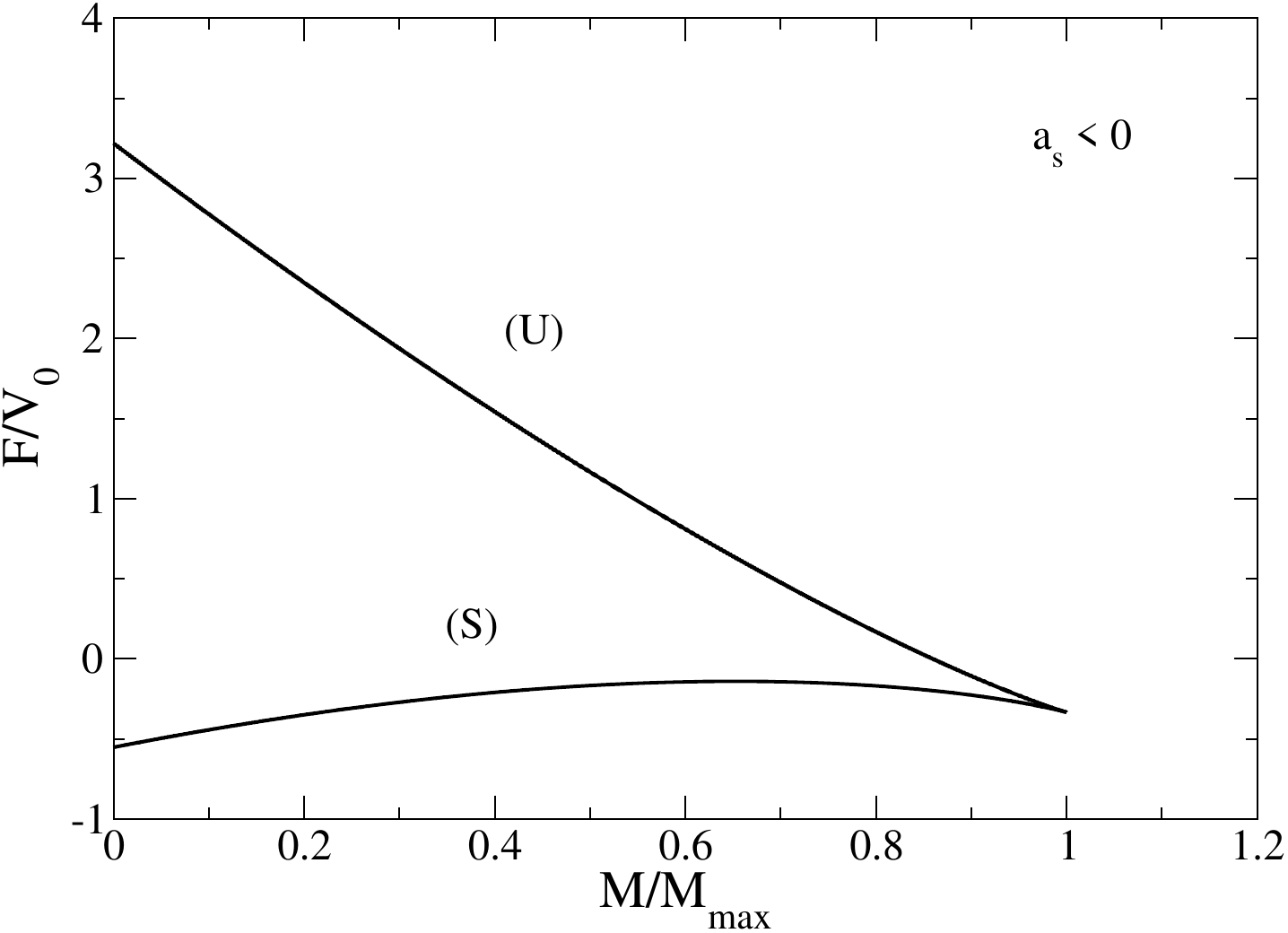}
\caption{Free energy of the equilibrium states as a function of the mass. The
curve $F(M)$ presents a spike
at the maximum mass $M_{\rm max}$. This is because $\delta F-\frac{\mu}{m}\delta
M=0$ for an equilibrium state [see Eq. (\ref{eq2})] so that $M$ and $F$ achieve
their extrema at the same point.}
\label{freeenergy}
\end{center}
\end{figure}

Another manner to investigate the
stability of an equilibrium state is to compute its free energy. The
free energies of the
equilibrium states close to the maximum mass are
\begin{eqnarray}
\label{nfcmm8}
\frac{F}{V_0}=\frac{V}{V_0}&=&-\frac{1}{3}+\frac{5}{3}\left
(1-\frac{M}{M_{\rm max}}\right
)\nonumber\\
&\mp&\frac{4}{3}\sqrt{2}\left (1-\frac{M}{M_{\rm max}}\right
)^{3/2}.
\end{eqnarray}
The free energy of the stable state ($R>R_*$, upper sign) is lower
than the free energy of the unstable state ($R<R_*$, lower sign) for the same
mass
$M$ (see Fig. \ref{freeenergy}). This is expected since an equilibrium state is
a minimum of free energy. The
barrier of free energy is
\begin{eqnarray}
\frac{\Delta F}{V_0}=\frac{\Delta V}{V_0}=\frac{8}{3}\sqrt{2}\left
(1-\frac{M}{M_{\rm max}}\right
)^{3/2}.
\label{bouley2}
\end{eqnarray}

\subsection{Period of oscillations and relaxation time when $M<M_{\rm max}$}
\label{sec_relt}

When $M<M_{\rm max}$, there exist equilibrium states. To
study the linear
dynamical stability of an equilibrium state of Eq.
(\ref{g25}), we make a small perturbation about that state
and write $R(t)=R+\epsilon(t)$ where $R$ is the equilibrium radius and
$\epsilon(t)\ll R$ is the perturbation. Using $V'(R)=0$ and keeping only
terms
that are linear in $\epsilon$, we obtain the equation
\begin{eqnarray}
\label{a27}
\frac{d^2\epsilon}{dt^2}+\xi \frac{d\epsilon}{dt}+\omega^2\epsilon=0,
\end{eqnarray}
where
\begin{eqnarray}
\label{pulse2}
\omega^2=\frac{{V}''(R)}{\alpha M}
\end{eqnarray}
is the squared pulsation.\footnote{We generically call it ``squared pulsation''
but it actually represents the squared pulsation of the BEC only in the
dissipationless case.} Looking for solutions of Eq. (\ref{a27})
under the form
$\epsilon\sim e^{\lambda
t}$, we obtain the quadratic equation $\lambda^2+\xi\lambda+\omega^2=0$ yielding
$\lambda_{\pm}=(-\xi\pm\sqrt{\xi^2-4\omega^2})/2$. If $\omega^2>\xi^2/4$, the
two
modes ($\pm$) are damped at a rate $-\xi/2<0$ and oscillate with a pulsation
$\pm\sqrt{4\omega^2-\xi^2}/2$. If $\omega^2=\xi^2/4$, the
perturbation decays as $\epsilon=(At+B) e^{-\xi t/2}$. If $0<\omega^2<\xi^2/4$,
the two modes are damped
at a rate $(-\xi\pm\sqrt{\xi^2-4\omega^2})/2<0$. If $\omega^2<0$, one mode is
damped at a rate $(-\xi-\sqrt{\xi^2-4\omega^2})/2<0$ and the other one grows
at a rate $(-\xi+\sqrt{\xi^2-4\omega^2})/2>0$.  For dissipationless
systems
($\xi=0$), the perturbation oscillates with a pulsation $\pm\omega$ if
$\omega^2>0$ and grows at a rate  $\sqrt{-\omega^2}>0$ if $\omega^2<0$ (the
other mode is damped at a rate $-\sqrt{-\omega^2}<0$). For overdamped systems
($\xi\rightarrow +\infty$), the perturbation is damped at a rate
$-\omega^2/\xi<0$ if $\omega^2>0$ and grows at a rate  $-\omega^2/\xi>0$ if
$\omega^2<0$. From that discussion, we can see that the equilibrium state of Eq.
(\ref{g25}) is
linearly stable if, and only
if, $\omega^2>0$ that is to say if, and only if, it is a (local) minimum of
the potential
energy $V(R)$, i.e., $V''(R)>0$. The condition of marginal stability (critical
point) corresponds to $\omega=0$, i.e., $V''(R)=0$.

In the dissipationless case ($\xi=0$), close
to the maximum mass, the
period of
oscillations $T_{\rm osc}=2\pi/\omega$ about the stable equilibrium state is
given
by [see Eq. (\ref{nfcmm4})]
\begin{eqnarray}
\frac{T_{\rm osc}}{t_D}=\frac{2\pi} {8^{1/4}}\left (1-\frac{M}{M_{\rm
max}}\right
)^{-1/4}\quad (M\rightarrow M_{\rm max}^-).
\label{voy1}
\end{eqnarray}
When $M\rightarrow M_{\rm max}^{-}$, the period of oscillations diverges at the
critical
point
as  $T_{\rm osc}/t_D\sim
3.73600... (1-M/M_{\rm max})^{-1/4}$.

In the overdamped limit ($\xi\rightarrow +\infty$), close
to the maximum mass, the
relaxation (damping) time $t_{\rm
relax}=1/\gamma=\xi/\omega^2$ towards the stable equilibrium state   is given
by [see Eq. (\ref{nfcmm4})]
\begin{eqnarray}
\frac{t_{\rm relax}}{\xi t_D^2}=\frac{1}{2\sqrt{2}} \left (1-\frac{M}{M_{\rm
max}}\right
)^{-1/2}\quad (M\rightarrow M_{\rm max}^-).
\label{voy2}
\end{eqnarray}
When $M\rightarrow M_{\rm max}^{-}$, the
relaxation time diverges
at the
critical
point
as 
$t_{\rm relax}/(\xi t_D^2)\sim 0.353553...(1-M/M_{\rm
max})^{-1/2}$.

\subsection{Collapse time when $M>M_{\rm max}$}
\label{sec_ce}

When $M>M_{\rm max}$, there is no equilibrium state. In that case, the
dilute axion
star is expected to collapse. Under our idealistic assumptions (nonrelativistic
treatment $+$ purely attractive
self-interaction $+$ spherical collapse), it should ultimately form a classical
singularity
(Dirac peak).

The collapse time has been investigated in the dissipationless
case ($\xi=0$) in \cite{bectcoll} by
using a Gaussian ansatz. Close to the maximum mass, the
collapse time is given by 
\begin{equation}
\label{nfcmm9}
\frac{t_{\rm coll}}{t_D}\sim 2.90178... \left (\frac{M}{M_{\rm max}}-1\right
)^{-1/4} \quad (M\rightarrow M_{\rm max}^+).
\end{equation}
When
$M\rightarrow M_{\rm max}^+$, the collapse time diverges at the
critical
point
as $t_{\rm coll}/t_D\sim 2.90178... (M/M_{\rm max}-1)^{-1/4}$. This
is the same
scaling as for the divergence of the period of oscillations $T_{\rm osc}/t_D\sim
3.73600... (1-M/M_{\rm max})^{-1/4}$
close
to the critical
point when $M\rightarrow M_{\rm max}^{-}$ [see Eq. (\ref{voy1})]. The prefactor
is, however,
different being
$2.90178...$ for $t_{\rm coll}$ ($M>M_{\rm max}$) and $3.73600...$
for $T_{\rm osc}$ ($M<M_{\rm max}$).

In the overdamped limit ($\xi\rightarrow
+\infty$), the collapse time close to the maximum mass is given by (see Appendix
\ref{sec_collsf})
\begin{equation}
\label{louane}
\frac{t_{\rm coll}}{\xi t_D^2}\sim \frac{\pi}{2\sqrt{2}}\left
(\frac{M}{M_{\rm
max}}-1\right )^{-1/2} \quad (M\rightarrow M_{\rm max}^+).
\end{equation}
When $M\rightarrow M_{\rm max}^+$, the collapse
time diverges at
the
critical
point
as $t_{\rm coll}/(\xi t_D^2)\sim 0.785398... (M/M_{\rm
max}-1)^{-1/2}$. This
is the same
scaling as for the divergence of the relaxation (damping) time
$t_{\rm relax}/(\xi t_D^2)\sim 0.353553...(1-M/M_{\rm
max})^{-1/2}$
close
to the critical
point when $M\rightarrow M_{\rm max}^{-}$ [see Eq. (\ref{voy2})]. They
differ by a factor $\pi$.

{\it Remark:} A general expression of the collapse time for a mechanical system described by Eq. (\ref{g25}) with $\xi=0$ or $\xi\rightarrow +\infty$ is given in \cite{preparation}.

\subsection{Gravitational cooling}
\label{sec_gcoo}

When $M<M_{\rm max}$ there are two possible equilibrium states for the same
mass $M$ with radius $R_S>R_*$ and $R_U<R_*$. The equilibrium state $R_S$ is
stable
(S) and the equilibrium state $R_U$ is unstable (U).
If slightly perturbed, an unstable axion star
can either collapse towards a Dirac peak  ($R\rightarrow 0$)
or migrate towards a stable dilute axion star  by
gravitational cooling ($R\rightarrow R_S$) \cite{seidel94,gul0,gul}.\footnote{See \cite{bectcoll,tunnel,glennon} for a detailed discussion depending on how $M$ compares to the
critical mass $M_c=(\sqrt{3}/2)M_{\rm max}$.} This is a
dissipative process similar to violent relaxation \cite{lb} during which the
star undergoes damped oscillations and emits a SF
radiation.\footnote{The friction term in the generalized GP equation (\ref{g19})
may account heuristically for this dissipative process \cite{wignerPH}.} Through
this process, it loses free energy (and mass) and settles down on a stable
equilibrium
state (S) with a larger radius and a lower free energy than the initial
configuration
(U).

In the following, we shall consider a stable axion star with a mass close
to $M_{\rm max}$. If it can reach the unstable
state (U) by thermal activation, thereby reducing its
radius, it can
then collapse towards a Dirac peak.

\section{Thermal activation rate of the BEC}
\label{sec_ttbec}

In Sec. \ref{sec_ga}, by making a Gaussian ansatz, we have reduced the original
problem (solving the generalized GPP equations (\ref{g1}) and (\ref{g2})) to the
simpler mechanical problem of a damped particle of mass $\alpha
M$ in a one-dimensional potential $V(R)$ governed by the deterministic
Newton equation
(\ref{g25}). When $M<M_{\rm max}$, the potential $V(R)$ has two equilibrium
states (see Fig.
\ref{normalform}): a
stable equilibrium state at $R_M>R_*$ (local minimum) and an unstable
equilibrium state at $R_U<R_*$ (local maximum). Since the potential 
 $V(R)$ has no global minimum for a purely attractive $\varphi^4$
self-interaction
(it tends to $-\infty$ when $R\rightarrow 0$),
the stable equilibrium state at $R_M$ is actually metastable. This
metastable equilibrium state represents a dilute axion star. In principle,
because of thermal fluctuations, the metastable BEC can decay towards a more
stable state -- a dense axion star if we take into account the repulsive
$\varphi^6$ term in the
self-interaction potential -- or collapse to $R=0$. In this section, we compute
the activation rate of the BEC and the thermal  lifetime of the metastable state
by making an analogy with Brownian motion and using the
instanton theory (a pedagogical exposition of this theory is
presented in \cite{preparation}). This path integral formulation lends itself
naturally to the study of the $T\rightarrow 0$ limit via a steepest-descent
approach. To simplify the presentation, we consider below the overdamped
limit.

\subsection{Stochastic Langevin equation}

In the  strong friction limit $\xi\rightarrow +\infty$,
the equation of motion of the fictive particle representing the BEC reads [see
Eq. (\ref{g27})]
\begin{eqnarray}
\label{g62}
\xi\alpha M\frac{dR}{dt}=-\frac{d{V}}{dR}.
\end{eqnarray}
At $T=0$, the fictive particle can be in equilibrium in the local
minimum $R_M$
of the potential $V(R)$. If slightly displaced from its equilibrium position,
it will return to equilibrium with an exponential damping rate
$\gamma=\omega^2/\xi=V''(R)/\xi\alpha M$. 
However, because of thermal fluctuations ($T\neq 0$), this equilibrium state is
metastable and the particle can overcome the
potential barrier and escape (implying that the BEC will collapse). In the
present formalism, using an analogy with Brownian theory, thermal fluctuations
can be taken into account by replacing the deterministic equation (\ref{g62})
for the fictive particle by the stochastic Langevin equation\footnote{As shown in Appendix \ref{sec_tf}, the Langevin equation (\ref{ttr0})
can be obtained from the stochastic quantum
Smoluchowski-Poisson equations (\ref{qs1}) and (\ref{qes2b})  by making a
Gaussian ansatz.}
\begin{eqnarray}
\label{ttr0}
\xi\alpha
M\frac{dR}{dt}=-\frac{dV}{dR}+\sqrt{2\xi\alpha M k_B T}\, \eta(t),
\end{eqnarray} 
where $\eta(t)$ is a Gaussian white noise with $\langle \eta(t)\rangle=0$ and
$\langle \eta(t)\eta(t')\rangle=\delta (t-t')$. The strength of the noise
is measured by the temperature $T$. The noise is responsible for a diffusive
motion with a diffusion coefficient given by the Einstein relation $D=
k_B T/\xi\alpha M$ (see Appendix \ref{sec_wnl}). This accounts for the
fluctuation-dissipation theorem. The probability
density $P(R,t)$ of finding the fictive Brownian particle in $R$ at time $t$ is
determined by the Fokker-Planck (Smoluchowski) equation (see Appendix
\ref{sec_wnl})
\begin{eqnarray}
\label{b5txt}
\frac{\partial P}{\partial t}=\frac{\partial }{\partial
R}\left \lbrack D\frac{\partial P}{\partial R}+\frac{1}{\xi \alpha M} P
V'(R)\right
\rbrack.
\end{eqnarray}
The equilibrium distribution is the Boltzmann distribution
\begin{eqnarray}
\label{kmaro}
P_{\rm eq}(R)=A\, e^{-\beta V(R)}.
\end{eqnarray}

In the Gaussian approximation (see Appendix
\ref{sec_summary}), the average value of the radius $\langle R\rangle$
coincides with its most probable value $R_e$ (corresponding to the minimum of
the potential $V(R)$ determined by $V'(R_e)=0$ and
$V''(R_e)>0$) and its
variance is
\begin{eqnarray}
\label{mar1}
\langle (\Delta R)^2\rangle=\frac{1}{\beta V''(R_e)}=\frac{1}{\beta \alpha M \omega^2},
\end{eqnarray}
where $\Delta R=R-R_e$ is the fluctuation of the radius about its equilibrium
value. Combining Eq. (\ref{mar1}) with Eq. (\ref{mr3b}), we obtain the
relation 
\begin{eqnarray}
\frac{dM}{dR}&=&-\alpha M\frac{m^2R_e^3}{2\sigma
\hbar^2}\omega^2=-\frac{m^2R_e^3}{2\sigma
\hbar^2}V''(R_e)\nonumber\\
&=&-\frac{m^2R_e^3}{2\sigma \hbar^2}\frac{1}{\beta
\langle (\Delta R)^2\rangle},
\label{mar2}
\end{eqnarray}
which is analogous to the fluctuation-dissipation theorem in thermodynamics (see
Appendix \ref{sec_tcano}). Close to the maximum mass $M_{\rm max}$, using  Eqs.
(\ref{nfcmm4}) and (\ref{mar1}), we have for the stable equilibrium state
\begin{eqnarray}
\omega^2&=&\frac{2\sqrt{2}}{t_D^2} \left (1-\frac{M}{M_{\rm max}}\right
)^{1/2},\nonumber\\ 
\langle (\Delta
R)^2\rangle&=&\frac{t_D^2}{2\sqrt{2}\beta\alpha M} \left (1-\frac{M}{M_{\rm
max}}\right
)^{-1/2}.
\label{mar3}
\end{eqnarray}
When $M\rightarrow M_{\rm max}^-$, the squared pulsation vanishes as 
$\omega^2\propto \left (1-{M}/{M_{\rm max}}\right )^{1/2}$ and the variance
diverges as $\langle (\Delta R)^2\rangle\propto  \left (1-{M}/{M_{\rm
max}}\right )^{-1/2}$. This is similar to the phenomenon of critical opalescence
in the theory of liquids \cite{opal1,opal2} (see also \cite{paper5}). In the
Gaussian approximation, one can show that for $\omega^2 t/\xi \gg 1$, i.e., in
the stationary state, the temporal correlation function of the fluctuations
$\Delta R(t)$ reads (see Appendix \ref{sec_summary})
\begin{eqnarray}
\langle \Delta R(0)\Delta R(t)\rangle=\frac{k_B T}{\alpha M\omega^2}e^{-\omega^2 t/\xi}.
\label{mar4}
\end{eqnarray}
Close to the maximum mass $M_{\rm max}$ where $\omega\rightarrow 0$, the
variance 
of the fluctuations $\langle (\Delta R)^2\rangle={k_B T}/{\alpha M
\omega^2}$ (prefactor), and the correlation time $t_{\rm corr}=\xi/\omega^2$
(exponent), which coincides with the relaxation (damping) time $t_{\rm relax}$, 
both
diverge as $\omega^{-2}$.

\subsection{General expression of the thermal activation rate}

In the weak noise limit $T\rightarrow 0$,
the thermal activation (escape) rate of the BEC is given by (see
Appendix \ref{sec_wnl})
\begin{eqnarray}
\label{g65}
\Gamma\sim A e^{-B/k_B T},
\end{eqnarray}
where the prefactor $A$ is specified below and the
exponent $B$ is equal to
\begin{eqnarray}
\label{g65brc}
B=S[R_c],
\end{eqnarray}
where
\begin{eqnarray}
\label{g66}
S[R]=\frac{\xi \alpha M}{4}\int_{-\infty}^{+\infty} \left\lbrack
{\dot
R}+\frac{V'(R)}{\xi \alpha M}\right\rbrack^2 \, {d} t
\end{eqnarray}
is the Onsager-Machlup \cite{onsagermachlup} functional of the fictive particle
representing the
BEC. It plays the role of the action in quantum mechanics.\footnote{The thermal
activation process discussed in this section is similar to the quantum tunneling
process discussed in \cite{tunnel} (see Appendices
\ref{sec_wnlq}-\ref{sec_ass}).}
The
trajectory $R_c(t)$ occuring in $B$ is the one that makes
the Onsager-Machlup functional  (\ref{g66}) extremal. This is the so-called
instanton
solution. Therefore, the exponent $B$ is equal to the
value of the Onsager-Machlup functional   evaluated along this optimal
trajectory (instanton). The instanton solution gives the dominant
contribution to the transition rate for a weak noise (see
Appendix \ref{sec_wnl}).

To determine the escape rate $\Gamma$ from Eq. (\ref{g65}) we thus
have to compute the action $S[R_c]$ of the optimal path. 
The least action principle $\delta S=0$ leads to the equation
(see Appendix \ref{sec_wnl})
\begin{eqnarray}
\label{aaa}
E=\frac{1}{2}\alpha M\dot R_c^2-\frac{1}{2\xi^2\alpha M}\lbrack
V'(R_c)\rbrack^2,
\end{eqnarray}
where $E$ is a constant. Since the extremal path $R_c(t)$ satisfies $V'(R_c)=0$
and $\dot R_c=0$ at the equilibrium state $R_c=R_M$ (attractor), this solution
has zero ``energy''
($E=0$). Therefore, the extremal path  satisfies the differential
equation
\begin{eqnarray}
\label{g67}
{\dot R_c}=\pm \frac{V'(R_c)}{\xi \alpha M}
\end{eqnarray}
with the boundary conditions $R_c(-\infty)=R_M$ and
$R_c(+\infty)=0$ (we assume
that the potential has no global minimum so that $R_c(t)$ tends to zero for
$t\rightarrow +\infty$). The
solution $R_c(t)$ satisfying these conditions is an instanton solution of the
theory. We call it the ``thermal instanton''. Now, Eq. (\ref{g67}) is the
deterministic equation of motion of an
overdamped particle in a
potential $\pm V(R)$. Therefore, the optimal path
corresponds to the deterministic gradient driven dynamics (\ref{g62}) with a
sign
$\pm$. Because of the $\pm$ sign, we have
two types of
solutions:

(i) The solution with the minus sign
is
\begin{eqnarray}
\label{b28}
\dot R_c=- \frac{V'(R_c)}{\xi \alpha M} \quad \Rightarrow
\quad
\int^{R_c(t)}\frac{dR}{V'(R)}=-\frac{t}{\xi \alpha M}.
\end{eqnarray}
It corresponds to the deterministic motion of the overdamped particle in the
original potential $V(R)$ [see Eq. (\ref{g62})], i.e., to the motion
described by  Eq. (\ref{ttr0}) when the noise is switched off or averaged out. 
Since $V(R_c(t))$
decreases monotonically with time along this path\footnote{Indeed,
$dV(R_c(t))/dt=V'(R_c(t))dR_c/dt=-(1/\xi\alpha M)V'(R_c(t))^2\le 0$.}  this
solution is called the ``downhill'' solution: The particle descends the
potential $V(R)$. Substituting Eq. (\ref{b28}) into Eq.
(\ref{g66}), we see that this solution has a vanishing action:
$S[R_c(t)]=0$. Since the Onsager-Machlup
functional is positive definite, the downhill solution corresponds
to the global minimum of the action. As a result, the
deterministic trajectory corresponds to the most probable path of the Brownian
particle
provided that the
initial and final positions can be connected by this
path (see Appendix \ref{sec_wnl}). This is also
the average trajectory of the Brownian particle.\footnote{Considering
the solution with the sign $-$, which corresponds to the downhill solution,
we see that the most probable path (instanton) coincides with
the ensemble average path, i.e., it corresponds to the deterministic equation
(\ref{g62}) obtained by averaging the stochastic Langevin equation
(\ref{ttr0}) over the noise. This path has a zero action ($S=0$). As a
result, the deterministic
equation (\ref{g62}) -- the average path -- can be obtained by
minimizing the Onsager-Machlup functional
(\ref{g66}). This is a form of least action principle for a dissipative system.
Similarly, one can show from the Feynman path integral formulation of quantum
mechanics (see Appendix \ref{sec_wnlq}) that the classical trajectory of a
particle minimizes the action
($\delta S=0$). This justifies the least action principle
of Maupertuis \cite{maupertuis1,maupertuis2} and Hamilton
\cite{hamilton} for a conservative system.}
However, when the
initial and final positions cannot be connected by this path there is another
optimal trajectory [solution (ii)] that minimizes the action.

(ii) The solution with the plus sign
is
\begin{eqnarray}
\label{b29}
\dot R_c=\frac{V'(R_c)}{\xi
\alpha M}\quad \Rightarrow \quad
\int^{R_c(t)}\frac{dR}{V'(R)}=\frac{t}{\xi \alpha M}.
\end{eqnarray}
It corresponds to the deterministic motion of the overdamped particle in
the reversed potential,
i.e., in the potential $-V(R)$ instead of $V(R)$. Since
$V(R_c(t))$ increases monotonically with time along this path (see footnote 22
with
$V\rightarrow -V$) this solution is called the ``uphill''
solution: The particle climbs the
potential $V(R)$. This solution has a nonzero action (see below).

\begin{figure}
\begin{center}
\includegraphics[clip,scale=0.3]{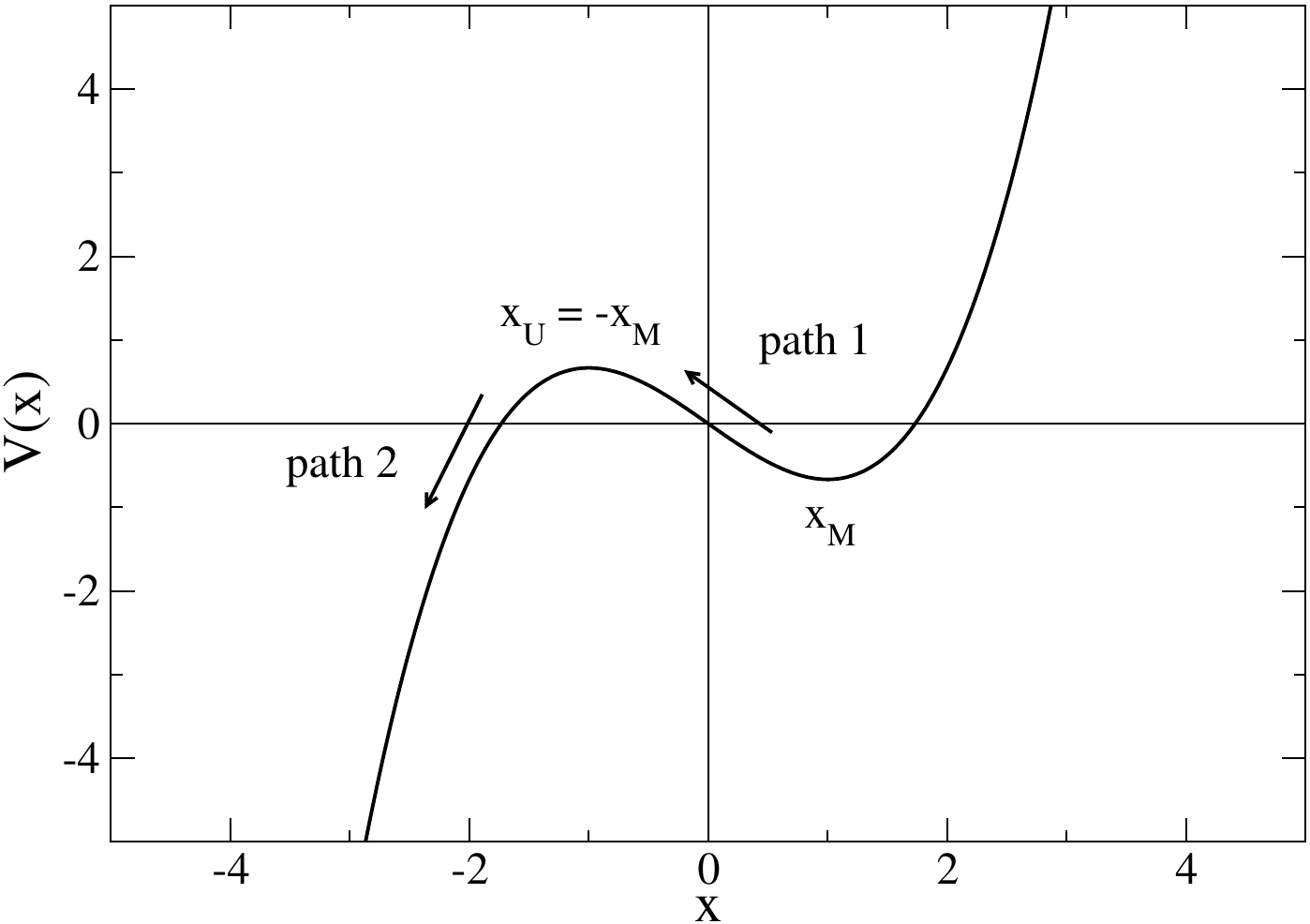}
\caption{The rate of escape $\Gamma$ from the
metastable state is governed by the action $S$ of the instanton associated
with the ``uphill'' path $1$: $\Gamma\sim A \, {\rm exp}(-S_c/k_B T)$. The
instanton
associated with the ``downhill''
path $2$ has zero action. Therefore, the rate of
escape $\Gamma$
from the metastable state is
governed by the action $S[R_c]=B=\Delta V$ of the thermal instanton
$R_c(t)$ associated with
the path labeled 1 in the figure connecting the metastable minimum of the
potential to the
unstable maximum (uphill path).  This is
in accord with physical expectation: the
downhill path $2$
corresponds to
a free descent, i.e., it does not require any external noise to drive it. By
contrast, the uphill path $1$ does require external noise. This is a rare event.
Therefore, the escape
rate is determined solely by the uphill path. This yields the Arrhenius law
$\Gamma\sim e^{-\Delta V/k_B T}$ where $\Delta V$ is the potential
barrier.
}
\label{path}
\end{center}
\end{figure}

Now,
the evolution of the optimal trajectory (instanton) from the metastable state
past the potential barrier
is clear (see Fig. \ref{path}). The uphill solution (\ref{b29}) corresponds to a
path which starts at $R_M$, ascends the potential $V(R)$ (or descends the
potential $-V(R)$) and finishes at $R_U$ (path $1$), whereas the downhill
solution (\ref{b28}) corresponds to a path which starts at $R_U$ and
freely descends the
potential $V(R)$ towards $R=0$ (path
$2$). For this latter solution the action $S[R_c]$ is equal to zero. Thus
$S[R_c]$ is determined solely by the uphill
solution. As a result, the equation of the thermal instanton (uphill solution) which
starts from $R_c=R_M$ for
$t\rightarrow -\infty$ and reaches $R_c=R_U$  for
$t\rightarrow
+\infty$ is
\begin{eqnarray}
\label{b29b}
\int_{R_*}^{R_c(t)}\frac{dR}{V'(R)}=\frac{t}{\xi\alpha M}.
\end{eqnarray}
The physical interpretation of Eq.
(\ref{b29b}) is the following. Starting from the metastable state
$R_{M}$, the most
probable path follows the time-reversed deterministic dynamics against the
potential gradient up to the maximum $R_{U}$; beyond the maximum
point, it follows the
forward-time deterministic dynamics down to the stable state $R_{S}$ (or $R=0$
in the present case).\footnote{In the limit of weak noise, and for a
stochastic process that
obeys a fluctuation-dissipation relation, 
it can be shown that the most
probable
path between the metastable state and the stable state must necessarily pass
through the maximum  $R_U$. Once
the system reaches the maximum it may either return to the initial
metastable state or reach the stable state. In the latter
case, it has
crossed the potential barrier.} According to Eqs.
(\ref{g66}) and (\ref{b28}), the action of the most
probable path corresponding to the transition from the unstable state
$R_U$ to the
stable state $R_S$ or $R=0$ (downhill solution corresponding to Eq.
(\ref{g67}) with
the
sign $-$) is zero  while the action of the most probable path  corresponding to
the transition from the metastable state $R_M$ to the unstable state
$R_U$ (uphill solution
corresponding to Eq. (\ref{g67}) with the sign $+$) is nonzero. This is to be
expected since the descent from the maximum to the stable state is a
``free'' descent that does not require thermal noise; it thus gives the
smallest possible value of zero of the action. By contrast, the rise from the
metastable state to the maximum is a rare event that requires thermal
noise.

It is now
easy to obtain a
closed expression
for the action of the instanton and  compute the escape rate. Using Eq.
(\ref{b29}) the exponent
\begin{eqnarray}
\label{ru1}
B=\frac{\xi \alpha
M}{4}\int_{-\infty}^{+\infty} \left\lbrack
{\dot
R_c}+\frac{V'(R_c)}{\xi \alpha M}\right\rbrack^2 \, {d} t
\end{eqnarray}
can be written
under the equivalent forms 
\begin{eqnarray}
\label{g68}
B=\frac{1}{\xi \alpha M}\int_{-\infty}^{+\infty} \left\lbrack
V'(R_c(t))\right\rbrack^2 \, {d} t
\end{eqnarray}
or
\begin{eqnarray}
\label{g69}
B=\xi \alpha M\int_{-\infty}^{+\infty} {\dot
R_c}^2 \, {d} t.
\end{eqnarray}
The second integral can be rewritten as
\begin{eqnarray}
\label{b33}
B=\int_{-\infty}^{+\infty} {\dot
R_c} V'(R_c(t)) \, {d} t=\int_{R_M}^{R_U} V'(R_c) \,
{d}R,
\end{eqnarray}
leading to
\begin{eqnarray}
\label{g70}
B=\Delta
V=V(R_U)-V(R_M),
\end{eqnarray}
where $\Delta V=V(R_U)-V(R_M)$ is the barrier of potential between the
metastable state and the unstable
state. If we use Eq. (\ref{g68}) or Eq.
(\ref{g69})
to compute $B$, we have to explicitly determine the trajectory of the
instanton (see below). It we use Eq. (\ref{g70}), this is not necessary. We just
need to
know the barrier of potential $\Delta V$. This leads to the following
expression of the thermal activation rate, i.e., the escape rate of an
overdamped Brownian particle over a
one-dimensional potential barrier\footnote{As mentioned in Appendix \ref{sec_tf}, this result can also be directly obtained from the stochastic quantum Smoluchowski-Poisson equations (\ref{qs1}) and (\ref{qes2b}) giving $\Gamma\propto e^{-\Delta F/k_B T}$, where $\Delta F=F[\rho_U]-F[\rho_M]$ is the difference of free energy 
between the unstable state and the metastable state. Using a Gaussian ansatz, this expression reduces to Eq. (\ref{b35}).}
\begin{eqnarray}
\label{b35}
\Gamma\sim A\, e^{-\Delta V/k_B T}.
\end{eqnarray}
The exponential term in Eq.
(\ref{b35}) was obtained a long time ago by
Arrhenius \cite{arrhenius} from an empirical analysis of chemical reaction
rates. This is the celebrated Arrhenius law.\footnote{Actually, Arrhenius \cite{arrhenius} credits this law to Van't Hoff  \cite{hoff}  so it is sometimes called the Van't Hoff-Arrhenius law \cite{hanggi}.} It states that the
transition rate
is inversely proportional to the exponential of the barrier of potential
divided by $k_B T$.\footnote{This formula can be
simply
obtained as follows. The equilibrium probability of observing the particle at
$R$ is given by the Boltzmann distribution
$\propto e^{-\beta V(R)}$. Therefore, the probability for the particle
initially prepared in the metastable state $R_M$ to reach the potential maximum
(unstable state $R_U$) and collapse is $\propto
e^{-\beta [V(R_U)-V(R_M)]}$. The typical lifetime of a metastable state
may then be estimated by $t_{\rm life}\sim e^{\beta\Delta V}$, where $\Delta
V=V(R_U)-V(R_M)$ is the potential barrier between the
metastable state and the unstable state.}

In many applications, the exponential
behavior of the activation rate is sufficient. The calculation
of the prefactor $A$ is more involved. It requires the determination of a
fluctuation determinant which was obtained
by Luckock and
McKane \cite{lmk} by using
a general method due to Dreyfuss and Dym \cite{dd}. This leads to the following
expression of the prefactor
\begin{eqnarray}
\label{qtr15}
A=\frac{\omega_M|\omega_U|}{2\pi\xi},
\end{eqnarray}
where $\omega_M^2=V''(R_M)/\alpha M$ and $\omega_U^2=V''(R_U)/\alpha M$ are the squared pulsations of the
fictive particle at the
metastable minimum and at the maximum of the potential (and the notation
$|\omega_U|$ means $\sqrt{-\omega_U^2}$). The complete expression of the
activation rate
including the prefactor is
therefore 
\begin{eqnarray}
\label{g65b}
\Gamma\sim \frac{\omega_M|\omega_U|}{2\pi\xi} e^{-\Delta V/k_B T}
\end{eqnarray}
or, equivalently,
\begin{eqnarray}
\Gamma=\frac{\sqrt{V''(R_M)|V''(R_U)|}}{2\pi\xi\alpha M} e^{-\Delta V/k_B T}.
\end{eqnarray}
This expression, which is valid in the weak noise limit $T\rightarrow 0$ (more
precisely when $k_B T\ll \Delta V$), coincides with the formula obtained by
Kramers
\cite{kramers} who
calculated the escape rate of an overdamped Brownian particle over a
one-dimensional potential barrier by using  the Fokker-Planck 
equation (reducing to the Smoluchowski equation in the overdamped limit). This
is the so-called Kramers formula. Finally, the escape time
or the typical lifetime of the
metastable state can be estimated by 
\begin{equation}
\label{qtr17b}
t_{\rm life}\sim \Gamma^{-1}
\sim \frac{2\pi\xi}{\omega_M|\omega_U|} e^{\Delta V/k_B
T}
\end{equation}
or, equivalently, by
\begin{eqnarray}
\label{b36b}
t_{\rm life}\sim \Gamma^{-1}\sim \frac{2\pi\xi\alpha
M}{\sqrt{V''(R_M)|V''(R_U)|}} e^{\Delta V/k_B T}.
\end{eqnarray}
This is the characteristic time (e.g. the ``mean first passage
time'') taken
by the Brownian particle initially located at the local minimum of $V(R)$ to
escape over the potential barrier.

{\it Remark:} Equation (\ref{g65b}) is valid for an
overdamped Brownian particle, i.e., for $\xi\rightarrow +\infty$. When inertial
effects are taken into account, the exponential
term remains the same but the prefactor is different (see Sec. V of
\cite{tunnel}). The calculation of the
escape rate of a Brownian particle over a potential barrier for an arbitrary
value of the friction coefficient has been performed by Kramers \cite{kramers} 
using a Fokker-Planck equation in phase space (the Kramers equation)
and by Newman {\it et al.} \cite{nbk}  (in certain limits) using path integrals and instanton theory.  If
we ignore the
pre-exponential factor, the escape rate of a Brownian particle over a
one-dimensional potential barrier in the weak noise limit $k_B T\ll \Delta V$
is given in all cases by the Arrhenius law $e^{-\Delta V/k_B T}$ where
$\Delta V$ is the height of the potential barrier. This result is valid for
a Gaussian white noise. The case of a  colored (correlated) noise gives a
different result. It has been
considered by \cite{brayprl,mk,bray1,bray,lmk,nbk} using path integrals and
instanton theory.

\subsection{Expression of the thermal activation rate valid close to the maximum
mass}
\label{sec_exp}

In this section, we determine the thermal activation rate of
the BEC close to
the maximum mass $M_{\rm max}$ by using the normal form of the
potential close to a saddle-node bifurcation
given in Sec. \ref{sec_nfcmm}. It is
convenient to set $x=R-R_*$. In that
case, the Langevin equation (\ref{ttr0}) can be rewritten as
\begin{eqnarray}
\label{ttr0bis}
\xi\alpha
M\frac{dx}{dt}=-\frac{dV}{dx}+\sqrt{2\xi\alpha M k_B T}\, \eta(t),
\end{eqnarray} 
and the potential (\ref{nfcmm1})  as 
\begin{eqnarray}
\label{qtr18}
V(x)=\frac{1}{3}ax^3-bx,
\end{eqnarray}
where the constants $a$ and $b$ are given by
\begin{eqnarray}
a=\frac{V_0}{R_*^3}\quad {\rm and} \quad b=\frac{2V_0}{R_*}\left
(1-\frac{M}{M_{\rm
max}}\right ).
\label{qtr19}
\end{eqnarray}
For simplicity, we have taken the additional constant in the potential equal
to
zero.\footnote{This is possible without restriction of generality since only
differences
of potential occur in our problem.} The potential (\ref{qtr18})
presents a local minimum and a local maximum (see Fig. \ref{normalform}). The
local minimum of $V(x)$ is located at
\begin{eqnarray}
x_M=\sqrt{\frac{b}{a}}
\label{xm}
\end{eqnarray}
and the value of the potential at that point is
$V(x_M)=-(2/3)ax_M^{3}$. The maximum of $V(x)$ is located at $x_U=-x_M$ and
the value of the potential at that point is $V(x_U)=(2/3)ax_M^{3}$. Let
us call $x'_M$ the point where $V(x'_M)=V(x_M)$. We find that $x'_M=-2x_M$. In
the quantum tunneling
problem \cite{tunnel}, this point is very important; it is called the
bouncing (or escape) point. In the present case, this point will not
play any role but it can be used to simplify some formulae (see below).
With these notations, the potential (\ref{qtr18}) can be
rewritten as
\begin{equation}
\label{qtr20}
V(x)-V(x_M)=a\left (\frac{1}{3}x^3-x_M^2x+\frac{2}{3}x_M^3\right ).
\end{equation}
The roots of the third degree equation defined by the term in parenthesis in
Eq. (\ref{qtr20}) are
$x_M$ (double root) and $x'_M$ (single root). We then find that the
potential (\ref{qtr18}) can be written as
\begin{equation}
\label{qtr22}
V(x)-V(x_M)=\frac{a}{3}(x-x_M)^2(x-x'_M).
\end{equation}
Finally, we note that $V(x)=0$ for $x=0$ and for $x=\pm\sqrt{3b/a}$. On the
other hand, the squared pulsations ($\omega^2=V''(x)/\alpha M$) of the
fictive particle around  the metastable and unstable equilibrium positions are
\begin{eqnarray}
\label{bttr6}
\omega_M^2=\frac{2}{\alpha M}\sqrt{ab}\qquad {\rm and}\qquad
\omega_U^2=-\frac{2}{\alpha M}\sqrt{ab}.
\end{eqnarray} 
When $T\rightarrow 0$, the thermal activation rate of
the BEC is given by Eq. (\ref{g65}). We propose
two methods to compute the exponent $B$ appearing in the exponential
factor using respectively the Kramers
formula and the instanton solution. We also compute the prefactor $A$ of the
activation rate.

\subsubsection{The potential barrier}

When $T\rightarrow 0$, the activation rate
of the
particle can be estimated
by the Arrhenius law (\ref{b35}) or by the Kramers formula (\ref{g65b}), where
$B$
is equal to the potential
barrier $\Delta V=V(x_U)-V(x_M)$ between the metastable state and the
unstable state [see Eq. (\ref{g70})]. When the
potential is given by Eq. (\ref{qtr22}), the potential barrier is given by
\begin{eqnarray}
\label{b37}
B&=&\Delta V=\frac{4}{3}ax_M^3=\frac{4}{3}a\left (\frac{b}{a}\right
)^{3/2}\nonumber\\
&=&\frac{4V_0}{3}\left\lbrack 2\left
(1-\frac{M}{M_{\rm max}}\right )\right\rbrack^{3/2}.
\end{eqnarray}
We stress that this method does not require the explicit expression of the
instanton solution $x_c(t)$.

\subsubsection{The instanton solution}

\begin{figure}
\begin{center}
\includegraphics[clip,scale=0.3]{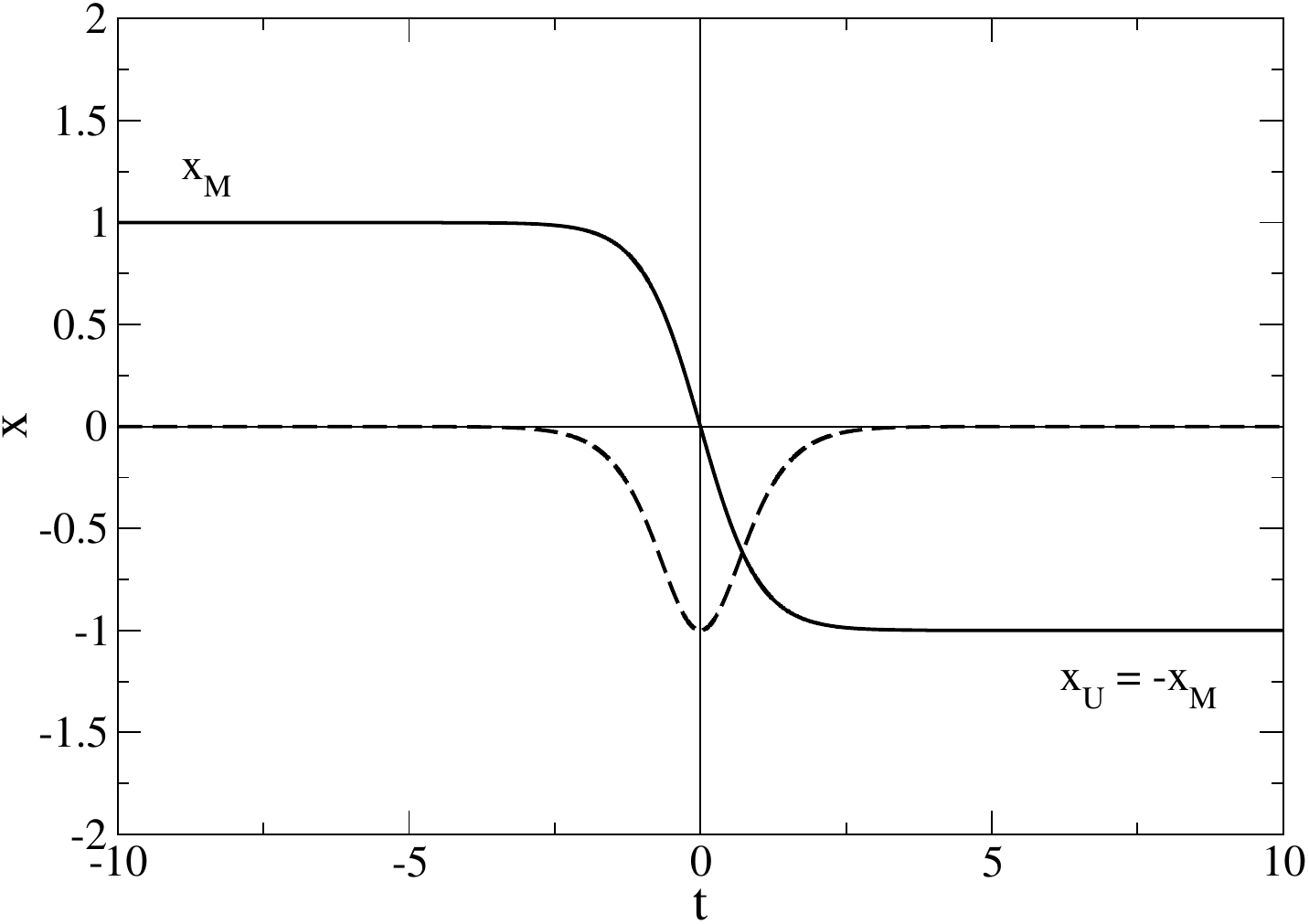}
\caption{Thermal instanton $x_c(t)$ close to a saddle-node bifurcation
(the dashed  line
corresponds to the velocity ${\dot x}_c(t)$).}
\label{instantonT}
\end{center}
\end{figure}

The expression of
$B$ can also be obtained
from Eq. (\ref{g68}) or
(\ref{g69}) by explicitly calculating the instanton solution. The uphill
trajectory (the one that has a nonzero contribution to the action) is
determined by Eq. (\ref{b29}). When the potential is
given by Eq. (\ref{qtr22}), this equation becomes 
\begin{equation}
\label{b41}
\frac{dx_c}{dt}=\frac{a}{\xi\alpha M}(x_c^2-x_M^2)\, \Rightarrow
\, \int^{x_c(t)} \frac{{d}x}{x_M^2-x^2}=-\frac{at}{\xi \alpha M}.
\end{equation}
Writing
\begin{equation}
\label{b42}
\frac{1}{x_M^2-x^2}=\left (\frac{1}{x_M-x}+\frac{1}{x_M+x}\right
)\frac{1}{2x_M}
\end{equation}
and integrating over $x$, we obtain after simplification
\begin{equation}
\label{b43}
x_{c}(t)=-x_M\tanh\left (\frac{a x_M t}{\xi \alpha M}\right ).
\end{equation}
This is the thermal instanton solution (see Fig. \ref{instantonT}). We have
chosen
the origin of time such that
the instanton center is at $x=0$ at $t=0$. For
$t\rightarrow -\infty$, $x_c(t)\rightarrow x_M$ and for
$t\rightarrow +\infty$, $x_c(t)\rightarrow x_U=-x_M$. It is
precisely this type of solutions, which
approach a
static limit in the distant past and future, that are refered to as
``instantons''. The
velocity of the instanton is
\begin{equation}
\label{b44}
\dot x_{c}(t)=-\frac{ax_M^2}{\xi \alpha M}\frac{1}{\cosh^2\left (\frac{a x_M
t}{\xi
\alpha M}\right )}.
\end{equation}
Substituting this expression into Eq. (\ref{g69}), we obtain 
\begin{equation}
\label{b45}
B=a x_M^{3}\int_{-\infty}^{+\infty}\frac{{d}x}{\cosh^4(x)}.
\end{equation}
Using the identity
\begin{equation}
\label{b46}
\int_{-\infty}^{+\infty}\frac{{d}x}{\cosh^4(x)}
=\frac{4}{3},
\end{equation}
we recover Eq. (\ref{b37}). Of course, the same result is found if we
substitute Eq. (\ref{b43}) into Eq. (\ref{g68}).

\subsubsection{The prefactor}

To obtain the prefactor of the activation rate given by Eq. (\ref{qtr15}) we
recall that
$\omega^2=V''(x)/\alpha M$ and note that $V''(x_M)=2ax_M$ and  $V''(x_U)=-2ax_M$.
Therefore, the prefactor of the activation rate is
\begin{equation}
\label{b38m}
A=\frac{ax_M}{\pi\xi \alpha M}=\frac{\sqrt{ab}}{\pi\xi \alpha M}=\frac{V_0}{\pi\xi\alpha MR_*^2} \left\lbrack 2\left
(1-\frac{M}{M_{\rm max}}\right )\right\rbrack^{1/2}.
\end{equation}

\subsubsection{The Kramers formula}

Combining Eqs. (\ref{g65}), (\ref{b37}) and (\ref{b38m}), or directly using Eq. (\ref{g65b}), we find that the
complete expression
of the thermal activation rate of the BEC close to the maximum mass is given by
\begin{equation}
\label{b38}
\Gamma\sim\frac{ax_M}{\pi\xi \alpha M}e^{-\frac{4ax_M^3}{3k_B
T}}\sim\frac{\sqrt{ab}}{\pi\xi \alpha
M}e^{-\frac{4a(b/a)^{3/2}}{3k_B T}}.
\end{equation}
Returning to the original variables, we get
\begin{equation}
\label{g75ts1}
\Gamma\sim \frac{V_0}{\pi\xi\alpha MR_*^2} \left\lbrack 2\left
(1-\frac{M}{M_{\rm max}}\right )\right\rbrack^{1/2}
e^{-\frac{4V_0}{3k_B T}\left\lbrack
2\left
(1-\frac{M}{M_{\rm max}}\right )\right\rbrack^{3/2}}
\end{equation}
or, equivalently,
\begin{equation}
\label{g75}
\Gamma\sim \frac{1}{\pi\xi t_D^2} \left\lbrack 2\left
(1-\frac{M}{M_{\rm max}}\right )\right\rbrack^{1/2} e^{-\frac{4}{3}\left\lbrack
2\left
(1-\frac{M}{M_{\rm max}}\right )\right\rbrack^{3/2}\nu\eta_{\rm max} N},
\end{equation}
where we have introduced the dynamical
time from
Eq. (\ref{nfcmm5}), the particle number
$N=M/m$, and the dimensionless inverse temperature
\begin{eqnarray}
\label{g75b2}
\eta_{\rm max}=\frac{GM_{\rm max} m}{R_* k_B T}.
\end{eqnarray}
We recall that Eq. (\ref{g75}) is valid for $M\rightarrow M_{\rm max}$. The
lifetime of the metastable state is $t_{\rm life}\sim 1/\Gamma$. We note that
the thermal activation rate exponent (equal to the potential barrier $\Delta V$)
scales as $B\propto
\left
(1-{M}/{M_{\rm max}}\right
)^{3/2}$ and the prefactor as $A\propto \left (1-{M}/{M_{\rm
max}}\right
)^{1/2}$. These are the same scalings as those obtained in
Ref. \cite{huepe} for the thermal activation rate of nongravitational
BECs\footnote{Actually, the scaling of their prefactor is
different, $A\propto \left (1-{M}/{M_{\rm
max}}\right
)^{1/4}$ instead of $A\propto \left (1-{M}/{M_{\rm
max}}\right
)^{1/2}$,  because they considered the  weak friction limit while we consider the overdamped limit (see
Sec. V of
\cite{tunnel} for a detailed discussion).}  and in
Ref. \cite{lifetime} for the metastable lifetime of globular
clusters in astrophysics (see Sec. \ref{sec_gcs}).  In comparison, the
quantum
tunneling rate
exponent
scales as $B\propto
\left
(1-{M}/{M_{\rm max}}\right
)^{5/4}$ and the prefactor as $A\propto \left (1-{M}/{M_{\rm
max}}\right
)^{7/8}$ \cite{tunnel}. These are the same scalings as those obtained in
Refs. \cite{leggett,huepe} for the quantum tunneling rate of
nongravitational
BECs. These scalings are
universal because they only depend on the normal form of the potential close to
a saddle-node bifurcation (see Appendix \ref{sec_summary}). 
We recall, however, that the Kramers formula and the WKB formula assume 
$B_T/k_B T\gg 1$ (i.e. $\Delta V/k_B T\gg 1$) and
$B_Q/\hbar\gg 1$ (i.e. $\frac{2}{\hbar}\int_{R'_M}^{R_M}\sqrt{2\alpha M\lbrack V(R)-V(R_M)\rbrack}\, dR\gg 1$) respectively, so
the above expressions may not be valid if we are too close to the maximum mass
$M_{\rm max}$ where $B\rightarrow 0$.

{\it Remark:} The prefactor in Eq. (\ref{g75}) is valid in the strong friction
limit $\xi\rightarrow +\infty$. More general expressions of the  thermal
activation rate, which are valid for an arbitrary value of the friction
coefficient, are given in Sec. V of \cite{tunnel} based on the Kramers theory (see footnote 29). As
we have already indicated, the value of the friction  affects only the
prefactor of the thermal activation rate, not the exponential term, which is the
leading term.

\subsection{Correction of the critical mass due to
thermal
fluctuations}
\label{sec_scm}

Let us summarize the preceding results. A self-gravitating BEC with
an attractive self-interaction can exist only
below a maximum mass $M_{\rm max}$ \cite{prd1,prd2}. For $M<M_{\rm max}$
and $R>R_*$ the BEC is in a metastable state (local but not global minimum
of free energy) corresponding to a dilute axion star. Because
of
thermal
fluctuations\footnote{The case of quantum fluctuations has been discussed in
\cite{tunnel}.} it can overcome the energy barrier
 and decay into a more
stable state (dense axion star) if we account for the repulsive
self-interaction
between the bosons, or collapse if there is no repulsive
self-interaction.  The lifetime of the
metastable
state due to thermal fluctuations can be estimated by $t_{\rm life}^{\rm T}\sim
1/\Gamma_{\rm T}$ where $\Gamma_{\rm T}$ is the thermal activation rate of the
BEC. According to
Eqs. (\ref{b38}) and (\ref{g75}), close to the maximum mass, we have 
\begin{equation}
\label{b38bis}
t_{\rm life}^{\rm T}\sim\frac{\pi\xi
\alpha
M}{\sqrt{ab}} e^{\frac{4a(b/a)^{3/2}}{3k_B T}}
\end{equation}
or, with the original notations,
\begin{eqnarray}
\label{k1}
\frac{t_{\rm life}^{\rm T}}{\xi t_D^2}\sim \frac{\pi}{\sqrt{2}}
\left
(1-\frac{M}{M_{\rm max}}\right )^{-1/2} e^{\frac{4}{3}\left\lbrack
2\left
(1-\frac{M}{M_{\rm max}}\right )\right\rbrack^{3/2}\nu\eta_{\rm max} N}\nonumber\\
(M\rightarrow M_{\rm max}^-).\qquad
\end{eqnarray}
We see that the time is naturally scaled by $\xi t_D^2$ in
agreement with Eq. (\ref{qtr17b}). When $M\rightarrow M_{\rm max}^{-}$, the
prefactor
behaves as
$2.221441...(1-M/M_{\rm
max})^{-1/2}$. This is the same scaling as for the
relaxation time (\ref{voy2}) when $M\rightarrow M_{\rm max}^{-}$ (they differ by factor $2\pi$) and as for the
collapse time
(\ref{louane}) when $M\rightarrow M_{\rm max}^{+}$ (they differ by a factor $2$). We symbolically have ``$t_{\rm
life}^{\rm T}=2t_{\rm coll}=2\pi t_{\rm relax}$'' (see Appendix
\ref{sec_summary}). The thermal
lifetime of dilute
axion stars
scales as
\begin{eqnarray}
\label{k2}
\frac{t_{\rm life}^{\rm T}}{\xi t_D^2}\sim e^{N}
\end{eqnarray}
except at, or very close
to, the critical point. This exponential scaling is characteristic of
systems with long-range interactions \cite{lifetime}. Since $N$ is very large
($N=7.21\times
10^{56}$ for QCD
axions
and $N=5.09\times 10^{95}$ for ULAs), the
lifetime of
a metastable state is considerable \cite{phi6}. As a matter of fact, metastable
states can
be considered as stable states. Only extraordinarily close to the maximum mass
will their
lifetime decrease because the barrier of potential is reduced.\footnote{Exactly at
the maximum mass $M_{\rm max}$,
the collapse time taking into
account thermal fluctuations scales as $t_{\rm coll}^{\rm T}\propto N^{1/3}\,
\xi t_D^2$ in the strong friction limit (see Appendix \ref{sec_size}) instead
of being infinite according to Eq. (\ref{louane}) if we ignore the
fluctuations (note that in the dissipationless case, the
collapse time taking into
account thermal fluctuations scales as $t_{\rm coll}^{\rm T}\propto N^{1/6}\,
t_D$ \cite{tunnel}). In comparison, the collapse time  taking into
account quantum  fluctuations
scales as $t_{\rm coll}^{\rm Q}\propto N^{1/5}\,
t_D$ at  $M=M_{\rm max}$ \cite{tunnel}.}
In principle, the BEC
will collapse at a mass 
$M_{\rm crit}$ smaller than $M_{\rm max}$. The mass
at which the BEC collapses because of thermal activation can
be estimated by writing that the exponent of the exponential term in Eq.
(\ref{k1}) is of order unity.
This gives
\begin{equation}
\label{k5}
M_{\rm crit}^{\rm T}\sim M_{\rm max}\left (1-0.762\, N^{-2/3}\right
),
\end{equation}
where we have used Eq. (\ref{gte2b}) and
taken $\eta_{\rm max}\sim 1$ for simplicity. This critical mass displays the
scaling
$N^{-2/3}$. In comparison, the analogous formula for the critical mass
due to quantum tunneling displays the scaling $N^{-4/5}$ \cite{tunnel}. For
large values
of $N$, which is the case
for axion stars, this correction is
extremely
small and can be neglected. Therefore, the value of the maximum mass of axion
stars \cite{prd1,prd2} is essentially unaffected by thermal activation (and
quantum
tunneling). However, corrections due to thermal activation (and quantum
tunneling) could be
observed in laboratory
experiments and in numerical simulations attempting to mimic ``axion stars''
because, in these situations, the number of particles will be necessarily
smaller than in astrophysics.

{\it Remark:} The temperature $T$ may be due to an external environment inducing a stochastic forcing on the axion star. If the temperature $T$ is large, then $\eta_{\rm max}$ can be much smaller than unity and the thermal lifetime of the axion star is reduced. Using Eqs. (\ref{g30}), (\ref{g30b}) and  (\ref{g75b2}), we find that
\begin{equation}
\label{etalou}
\eta_{\rm max}\sim \frac{Gm^2}{|a_s|k_B T}.
\end{equation}
Therefore, the reduction of the thermal lifetime of the axion star is important ($\eta_{\rm max}\ll 1$) if $k_B T\gg Gm^2/|a_s|$. For QCD axions, this leads to the condition $T\gg 10^{-15}\, {\rm K}$ and for ULAs to the condition $T\gg 10^{-25}\, {\rm K}$ (using the numerical values of the introduction). If $T\sim 1\, {\rm K}$, we have a reduction of a factor $10^{-15}$--$10^{-25}$ but this does not change our main conclusions because of the enormous value of $N$. Thermal activation would become important ($\eta_{\rm max} N\sim 1$) when $k_B T\sim \hbar\sqrt{Gm}/|a_s|^{3/2}$ giving  $T\sim 10^{42}\, {\rm K}$ for QCD axions and $T\sim 10^{70}\, {\rm K}$ for ULAs. These high temperatures are not realistic, even in the early universe, because they are above the Planck temperature $T_P=(\hbar c^5/G)^{1/2}/k_B=1.42\times 10^{32}\, {\rm K}$.

\section{Globular clusters and self-gravitating Brownian particles in the
underdamped limit}
\label{sec_gc}

\subsection{Globular clusters}
\label{sec_gcs}

\begin{figure}
\begin{center}
\includegraphics[width=0.45\textwidth]{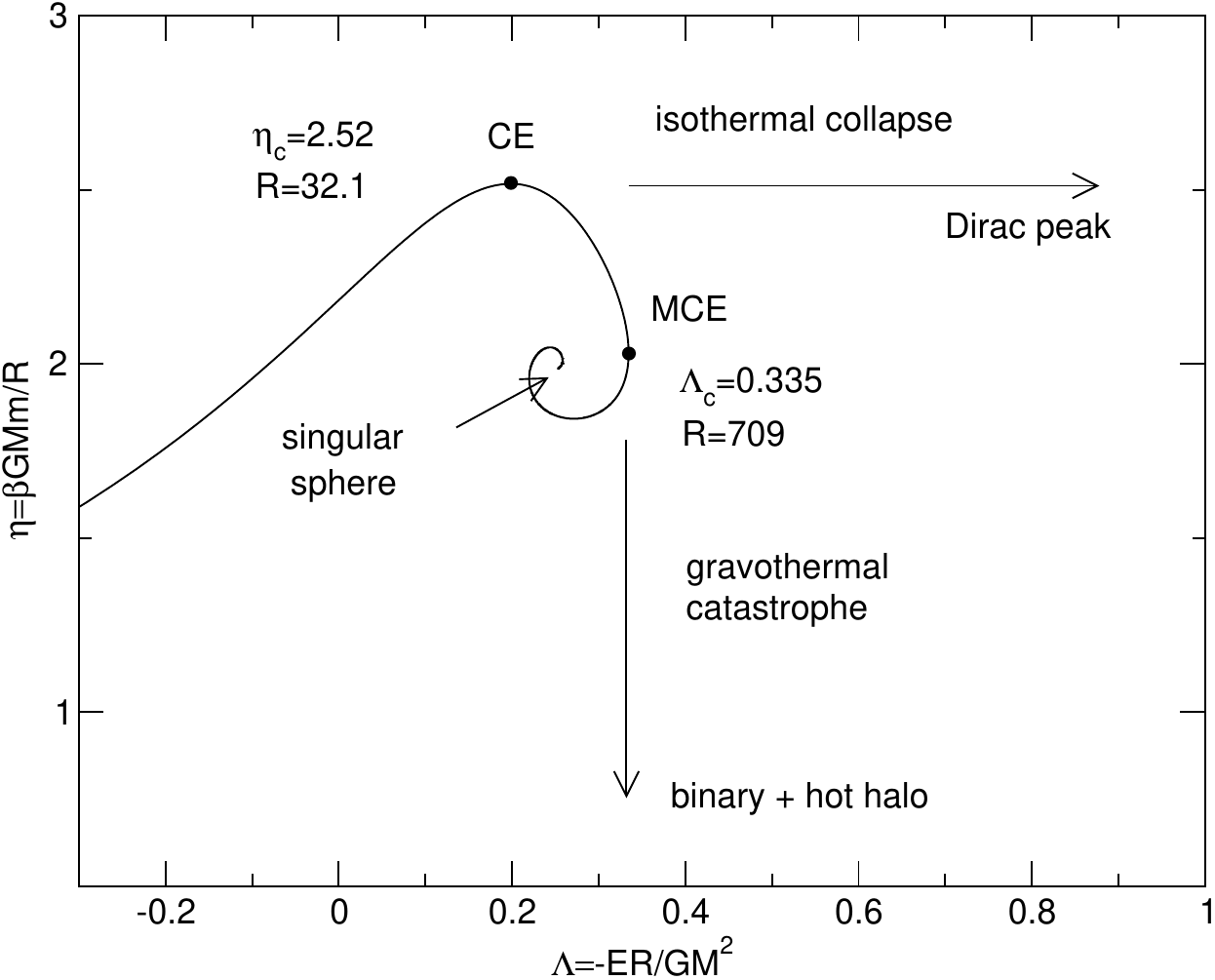}
\end{center}
\caption[]{Series
of equilibria (caloric curve) for classical self-gravitating systems in a
spherical box. For
$\Lambda=-ER/GM^2>\Lambda_{c}$ in the microcanonical ensemble or for 
$\eta=\beta GMm/R>\eta_{c}$ in the canonical ensemble, there is no equilibrium
state and the system undergoes a
gravitational collapse (gravothermal catastrophe in the microcanonical ensemble
and isothermal collapse in the 
canonical ensemble).} \label{etalambda}
\end{figure}

The previous results are similar to those obtained by Katz and Okamoto
\cite{katzokamoto} and Chavanis \cite{lifetime} 
for classical self-gravitating systems such as globular clusters in
astrophysics. It is
well-known since the works of Antonov \cite{antonov} and Lynden-Bell and Wood
\cite{lbw} that self-gravitating systems at statistical equilibrium  in
the microcanonical ensemble are, at
most, metastable (local but not global maxima of entropy at fixed mass and
energy).
Furthermore, metastable equilibrium states exist only above a minimum energy
$E_{c}=-0.335\, GM^2/R$ (see Fig. \ref{etalambda}).\footnote{Here $R$
represents the radius of the spherical ``box'' in which the
system has to be artificially confined in order to prevent its
evaporation \cite{antonov,lbw,paddy,found,ijmpb}. It
may be assimilated with the tidal radius of more realistic models based on the
King distribution function \cite{clm1,clm2}.} When
$E<E_c$
the system
undergoes a ``gravothermal catastrophe'' (core collapse) \cite{lbw}. These
results are valid in a proper thermodynamic limit $N\rightarrow +\infty$, in which the mean field approximation is exact. However, because of
fluctuations (finite
$N$ effects), the system should collapse at an energy $E_l$ slightly
larger than $E_{c}$. This critical energy can be estimated as follows.

Let
us assume that the self-gravitating 
system is in the metastable state (M) at $E>E_{c}$. It is characterized
by a distribution function $f_{M}({\bf r},{\bf v})$. Now, the probability of a
fluctuation  $f({\bf r},{\bf v})$ in the microcanonical ensemble is given by
\cite{ijmpb,lifetime}
\begin{equation}
\label{k5ein}
P[f]\sim W[f] \delta(E[f]-E)\sim e^{(S[f]-S[f_M])/k_B} \delta(E[f]-E).
\end{equation}
To obtain this expression, we have used  the Einstein formula $W[f]=e^{S[f]/k_B}$
\cite{einsteinentropy} giving the probability density of the macrostate $f$ in terms of the entropy $S[f]$.
It can be seen as the reversed expression of the celebrated Boltzmann
formula for the entropy $S[f]=k_B \ln W[f]$, where $W[f]$ is the number of
microstates (complexions) corresponding to the macrostate $f({\bf
r},{\bf v})$.\footnote{As is
well-known, Boltzmann never wrote this
formula explicitly. It was actually formulated by Planck \cite{planck} in his
fundamental paper on the blackbody radiation in which he introduced both the
Planck constant $h$ and the Boltzmann constant $k_B$. However, on the basis of
the kinetic theory of gases, Boltzmann had discovered that, aside from a
constant factor, entropy is equivalent to the logarithm of the ``probability''
of the state under consideration \cite{einsteinQT}.}

The system will
collapse if the fluctuations take it
to the unstable configuration (U) with distribution function $f_{U}({\bf r},{\bf
v})$ corresponding to an entropy
minimum (actually a saddle point) with the same energy $E$ and the same mass
$M$. 
Therefore, the lifetime
of the metastable state in the microcanonical ensemble can be estimated
by\footnote{This timescale represents the typical 
lifetime of a globular cluster provided that the escape of stars are inhibited
(for example by placing the system in a box). In practice, because of
evaporation, a globular cluster evolves dynamically along the series of
equilibria until it
experiences the gravothermal catastrophe at $E_{c}$. In that case, the effect of
fluctuations is to shift the critical energy to a larger value $E_l$.} 
\begin{equation}
\label{k6}
t_{\rm life}^{\rm GC}\sim e^{\Delta S[f]/k_B},
\end{equation}
where $\Delta S=S[f_M]-S[f_U]$ is the barrier of entropy between the metastable
state and the stable state \cite{lifetime}. Since the entropy
scales as $N$, we can write
\begin{equation}
\label{k7}
t_{\rm life}^{\rm GC}\sim e^{N\Delta s/k_B},
\end{equation}
where $\Delta s$ is the barrier of entropy per star.\footnote{Physically, in
order to
cross the entropic barrier, the system has to spontaneously create a dense
nucleus (similar to a ``germ'' or a ``critical droplet'' in the language of phase
transitions). This  nucleation
process is a rare event, leading to small transition rates or long
metastable state lifetimes (see below).}
Therefore, the lifetime of globular clusters scales as
$t_{\rm life}^{\rm
GC}\sim e^{N}$ \cite{lifetime}. Since
$N$ is large, this leads to the
conclusion that metastable states of globular clusters are very robust and that
they are
long-lived. However, close to the minimum energy $E_c$, their lifetime is reduced because
the barrier of entropy $\Delta
s$ decreases. It is therefore easier for the system to cross the barrier of
entropy and collapse. In the thermodynamic limit $N\rightarrow +\infty$, where the mean field approximation becomes exact, the barrier
of entropy $\Delta s$ can be calculated
analytically close
to the minimum energy and leads to the result\footnote{The
prefactor is difficult to calculate exactly but it is expected to be of the
order of the dynamical time $t_D$.}
\begin{equation}
\label{k8}
t_{\rm life}^{\rm GC}\sim e^{2\lambda' N (\Lambda_{c}-\Lambda)^{3/2}},
\end{equation}
where $\lambda'=0.863159...$, $\Lambda=-ER/GM^2$ and
$\Lambda_{c}=0.335$ \cite{lifetime}. From this formula, writing that the term
in the exponential is of order unity, we find that
the correction to the collapse energy due to finite $N$ effects is
\begin{equation}
\label{k9}
E_l \sim E_{c}  (1-2.077\, N^{-2/3}).
\end{equation}
An equivalent expression has been obtained in \cite{katzokamoto}. Eqs.
(\ref{k8}) and (\ref{k9}) are similar to Eqs.
(\ref{k1})-(\ref{k5}) derived for
self-gravitating BECs. In particular, the scaling exponents are the
same.\footnote{As discussed in Sec. \ref{sec_exp}, this universality arises
because we are just using the normal form of the thermodynamic potential for a
saddle-node
bifurcation close to a critical point.} For large clusters of stars with
$N\sim 10^6$ the
correction to the critical energy is negligible. For small
clusters with
$N\sim 10^3$ the correction becomes significant. This is particularly visible on
the corresponding density contrast ${\cal
R}_l=709\, (1-6.01 N^{-1/3})$ which is reduced from $709$ when $N\rightarrow
+\infty$ to ${\cal
R}_l=283$ when $N\sim 10^3$
\cite{katzokamoto,lifetime}. Fluctuations can
thus explain why observations reveal that a greater number
of globular clusters than is normally believed may already be in an advanced
stage of core collapse. Finally, for small values of $N$ such as those achieved
in early $N$-body simulations of self-gravitating systems, the correction
is drastic. 
Similarly, quantum tunneling and thermal activation for self-gravitating BECs
with attractive self-interaction are negligible for real axion stars but
they could be relevant in future numerical simulations and laboratory
experiments aiming at mimicking axion stars since they will necessarily
involve a small number of bosons.

\subsection{Self-gravitating Brownian particles in the
underdamped limit}
\label{sec_gcb}

Let us consider a system of self-gravitating Brownian particles at
temperature $T$ enclosed within
a spherical box of radius $R$. The statistical
equilibrium states of this system in the canonical ensemble are, at
most, metastable (local but not global minima of free energy at fixed mass).
Furthermore, metastable equilibrium states exist only above a minimum
temperature
$k_B T_{c}=GMm/(2.52\, R)$ (see Fig. \ref{etalambda}). When
$T<T_c$
the system
undergoes an ``isothermal collapse''  \cite{aaISO,crs}. These
results are valid in a proper thermodynamic limit $N\rightarrow +\infty$, in which the mean field approximation is exact. 
However, because of
fluctuations (finite
$N$ effects), the system should collapse at a temperature $T_l$ slightly
larger than $T_c$. This critical temperature can be estimated as
follows (we use arguments similar to those developed in the previous section).

Let us assume that the system is in the metastable state
(M) at $T>T_c$. It is characterized
by a distribution function $f_{M}({\bf r},{\bf v})$. Now, the probability of a
fluctuation  $f({\bf r},{\bf v})$ in the canonical ensemble is given by
\cite{ijmpb,lifetime}
\begin{eqnarray}
\label{k5eincan}
P[f]&\sim& e^{-E[f]/k_B T} W[f]\sim e^{S[f]/k_B} e^{-E[f]/k_B T} \nonumber\\
&\sim&
e^{-(F[f]-F[f_M])/k_BT},
\end{eqnarray}
where $F[f]=E[f]-TS[f]$ is the Helmholtz free energy. The system will collapse
if the fluctuations take it
to the unstable configuration (U) with distribution function $f_{U}({\bf r},{\bf
v})$ corresponding to a free energy maximum (actually a saddle point) with the
same temperature $T$ and the same mass
$M$. 
Therefore, the lifetime
of the metastable state can be estimated in the canonical ensemble by
\begin{equation}
\label{k6can}
t_{\rm life}^{\rm SGB}\sim e^{\Delta F/k_B T},
\end{equation}
where $\Delta F=F[f_U]-F[f_M]$ is the barrier of free energy between the
metastable
state and the stable state \cite{lifetime}.  This is the
thermodynamical version of the Kramers formula
\cite{kramers}. For systems with long-range
interactions, the free
energy scales as $N$ so the typical lifetime of a metastable state scales
as 
\begin{equation}
\label{k7can}
t_{\rm life}^{\rm SGB}\sim e^{N\Delta f/k_B T},
\end{equation}
where $\Delta f$ is the barrier of free energy per particle. Therefore, we find
that, in general, the lifetime of self-gravitating Brownian particles scales as
$t_{\rm life}^{\rm SGB}\sim e^{N}$ \cite{lifetime}. Since
$N$ is large, this leads to the
conclusion that metastable states of  self-gravitating Brownian particles are
very robust and that
they are
long-lived (see footnote 35).
However, close to the minimum temperature $T_c$, their lifetime is reduced
because the barrier of free energy $\Delta
f$ decreases. It is therefore easier for the system to cross the barrier of
free energy and collapse.

A detailed derivation of the Kramers formula (\ref{k6can}) for self-gravitating
Brownian particles in
the underdamped limit $\xi\rightarrow 0$ is provided in
\cite{lifetime,n2d3} starting from the $N$-body Kramers equation. In the weak
friction limit, it can be shown that the evolution of the
distribution of energies $P(E,t)$ is governed by the Fokker-Planck equation 
\cite{lifetime}
\begin{eqnarray}
\label{kq3} 
\frac{\partial P}{\partial t}=\frac{\partial}{\partial E}\left\lbrack D(E)\left
(\frac{\partial P}{\partial E}+\beta P F'(E)\right )\right\rbrack,
\end{eqnarray}
where $F(E)=E-TS(E)$ is the Helmholtz free energy and the expression of the
diffusion coefficient $D(E)$ is detailed in  \cite{lifetime,n2d3}. Close to the
minimum temperature $T_c$ the free energy is given by the normal form
\cite{n2d3} 
\begin{equation}
\label{e7} 
F(E)=F_0-\frac{\beta_c-\beta}{\beta_c}(E-E_0)-\frac{1}{6}\frac{
\beta''(E_0)}{\beta_c}(E-E_0)^3,
\end{equation}
where $E_0$ is the energy at $T_c$ and $F_0=E_0-T_cS(E_0)$ is the corresponding free energy. The equilibrium states determined by the condition $F'(E_e)=0$ are given by $E_e=E_0\pm \sqrt{2/|\beta''(E_0)|}(\beta_c-\beta)^{1/2}$ with $+$ for the stable state and $-$ for the  unstable state (see Appendix \ref{sec_tcano}). The stochastic
Langevin equation governing the evolution of the fluctuating energy $E(t)$ and
leading to the Fokker-Planck equation
(\ref{kq3}) reads 
\begin{eqnarray}
\label{lan1}
\frac{dE}{dt}=-D_0\beta\frac{d{F}}{dE}+\sqrt{2D_0}
\eta(t),
\end{eqnarray}
where $\eta(t)$ is a Gaussian white noise satisfying $\langle
\eta(t)\rangle=0$ and $\langle \eta(t)\eta(t')\rangle=\delta(t-t')$.\footnote{We
have
replaced $D(E)$ by $D_0=D(E_0)$ in order to simplify the discussion. This is
possible
because we are close to the critical point.} This stochastic equation is
similar to the
equation of motion of an overdamped Brownian
particle 
of unit mass with ``position'' $E$ evolving in a potential $F(E)$. Therefore,
we can directly apply the results of the previous sections (see
Appendix \ref{sec_summary} for a summary) to the present
situation by setting $x=E-E_0$, $\xi=1/(D_0\beta)$, $\alpha M=1$, $V=F$,
$a=-\beta''(E_0)/(2\beta_c)$ and
$b=(\beta_c-\beta)/\beta_c$.

The lifetime of a metastable state close to the critical point is given by
\cite{lifetime,n2d3}\footnote{There is a difference by a factor
$2$ with the
results of \cite{n2d3}. This is related to the Remark A1 of
\cite{n2d3}.}
\begin{eqnarray}
\label{lcp3} 
t_{\rm
life}^{\rm
T}=\frac{\pi}{3}\sqrt{\frac{2}{-\eta''(\Lambda_0)}}\frac{\eta_c}
{{\cal D}_0}(\eta_c-\eta)^{-1/2}\nonumber\\
\times e^{\frac{4}{3}N\sqrt{\frac{2}{
-\eta''(\Lambda_0)}}(\eta_c-\eta)^{3/2}}\frac{t_K}{N},
\end{eqnarray}
where we have introduced the
notations $\eta=\beta GMm/R$, $\Lambda=-ER/GM^2$, $t_K=1/\xi$, and ${\cal
D}_0=RD_0/3DGM^3$ with $D=k_B T\xi/m$ (see \cite{n2d3} for details).  
We see that the time is naturally scaled by $t_K/N$ in
agreement with Eq. (\ref{qtr17b}).

In the thermodynamic limit $N\rightarrow +\infty$, where the mean field approximation becomes exact,  the value of the coefficients are
$\eta_c=2.517551...$,
$\Lambda_0={1}/({2\eta_c})=0.1986057...$, and
$-\eta''(\Lambda_0)={\eta_c^3}/({\eta_c-2})=30.8306$ \cite{lifetime}. From Eq.
(\ref{lcp3}), writing that the term in the exponential is of order unity, we
find that the correction to the collapse temperature due to
finite $N$ effects is
\begin{eqnarray}
\label{ln5} 
\eta_l=2.517551\left ( 1-0.816043\,  N^{-2/3}\right ).
\end{eqnarray}
This returns the  scaling $N^{-2/3}$  obtained in
\cite{lifetime}. We can also show that the correction to the critical
density
contrast of collapse is
${\cal R}_l=32.1255\, \left (1-4.47068\,  N^{-1/3}\right )$
\cite{n2d3}.
This returns the $N^{-1/3}$ scaling obtained in
\cite{lifetime} (the prefactor
is different because the critical density contrast is defined and calculated
in
a different manner).

It is also possible to determine relaxation time above $T_c$ and the collapse
time below $T_c$ \cite{n2d3}. 
Close to the critical point, the
relaxation time towards the stable state is
\begin{eqnarray}
\label{rel6}
t_{\rm relax}\sim \frac{1}{6}\sqrt{\frac{2}{-\eta''(\Lambda_0)}}
(\eta_c-\eta)^{-1/2}\frac{\eta_c}{{\cal D}_0}\, \frac{t_K}{N}.
\end{eqnarray}
For $\eta\rightarrow\eta_c^-$, the relaxation time
diverges as $t_{\rm relax}\sim (\eta_c-\eta)^{-1/2}$.  Close to the critical point, the collapse time
is
given by
\begin{eqnarray}
\label{coll3}
t_{\rm coll}\sim \frac{\pi}{6}\sqrt{\frac{2}{-\eta''(\Lambda_0)}}
(\eta-\eta_c)^{-1/2}\frac{\eta_c}{{\cal D}_0}\, \frac{t_K}{N}.
\end{eqnarray}
For $\eta\rightarrow\eta_c^+$, the collapse time
diverges
as $t_{\rm coll}\sim (\eta-\eta_c)^{-1/2}$. This
is the same
scaling as for the divergence of the relaxation (damping) time close
to the critical
point [see Eq. (\ref{rel6})]. The
prefactors
differ by a factor $\pi$. These prefactors are also similar to
the prefactor in front of the
exponential term in the metastable state lifetime  [see Eq. (\ref{lcp3})]. They
differ by a factor $2\pi$ and $2$ respectively. We symbolically
have ``$t_{\rm
life}^{\rm T}=2t_{\rm coll}=2\pi t_{\rm relax}$'' (see Appendix
\ref{sec_summary}). Exactly at the minimum temperature $T_c$, the collapse time taking into
account thermal fluctuations scales as 
 \begin{equation}
\label{size3b}
t_{\rm coll}\propto N^{1/3}\, \frac{t_K}{N} \qquad
(\eta=\eta_{c}).
\end{equation}
These
expressions are similar to those obtained in Eqs. (\ref{k1})-(\ref{k5}) for
self-gravitating BECs. In particular, the exponent of $N$ is the
same. These scalings are ``universal'' because they are based on the normal form
of the potential close to a critical point (see Appendix \ref{sec_summary}).

\section{Self-gravitating Brownian particles in the overdamped limit}
\label{sec_sfbp}

\subsection{Stochastic Smoluchowski-Poisson equations}
\label{sec_ssp}

In this section, as another application of our general formalism, we consider a
system of classical self-gravitating Brownian particles in the strong friction
limit $\xi\rightarrow +\infty$ \cite{crs} and take thermal fluctuations into
account. The
evolution of the spatial density is governed by the stochastic
Smoluchowski-Poisson equations \cite{paper5,longshort,cdkramers,entropy1,entropy2,back}\footnote{These equations are isomorphic to the stochastic Keller-Segel model \cite{sks} describing the chemotaxis of bacterial populations in biology. They are also similar to the stochastic Fokker-Planck equations describing the BMF model \cite{bmf}.}
\begin{equation}
\label{ssp1}
\xi\frac{\partial\rho}{\partial t}=\nabla\cdot\left (\nabla
P+\rho\nabla\Phi\right)
+\nabla\cdot \left \lbrack \sqrt{2\xi k_B T\rho}\, {\bf R}({\bf r},t)\right \rbrack,
\end{equation}
\begin{equation}
\label{ssp2}
\Delta\Phi=4\pi G\rho.
\end{equation}
The free energy associated with these equations is 
\begin{eqnarray}
\label{ssp3}
F=\int u(\rho)\, d{\bf r}
+\frac{1}{2}\int\rho\Phi\, d{\bf r},
\end{eqnarray}
where the internal energy density $u(\rho)$ is determined by the equation
of state $P(\rho)$ from Eq. (\ref{mad5b}).\footnote{In this
section, we use the
notation $u(\rho)$ instead of $V(\rho)$.} This corresponds to the equations of
Sec. \ref{sec_gbm} and Appendix \ref{sec_tf} with $\xi\rightarrow
+\infty$ and $Q=0$. In the absence of thermal
fluctuations, an extremum of free energy at fixed mass is an equilibrium state.
We consider an equation of state $P(\rho)$ such that, above a critical
temperature $T_c$, the system possesses two equilibrium states, one stable (free
energy minimum) and one unstable (saddle point or maximum of free energy). In
the presence of thermal fluctuations, the stable equilibrium state is actually
metastable
and we want to determine its typical lifetime. By writing the solution of the
stochastic 
Smoluchowski-Poisson equations (\ref{ssp1}) and (\ref{ssp2}) in the form of a
path integral characterized 
by the Onsager-Machlup functional, and by applying the theory of instantons, 
we have shown in \cite{cdkramers,entropy2} that the thermal lifetime of the
metastable equilibrium state is given by a Kramers-like formula of the form
\begin{equation}
\label{ssp4}
t_{\rm life}\propto e^{\Delta F/k_B T},
\end{equation}
where $\Delta F=F[\rho_U]-F[\rho_M]$ is the difference of free energy 
between the unstable state and the metastable state.  Below, we consider a
particular equation of state with the above-mentioned properties and we simplify
the problem by making an $f$-ansatz. This leads to a one-dimensional
stochastic Langevin equation with an effective potential $V(R)$. We apply the
instanton theory directly on this mechanical problem and we recover the result
of Eq. (\ref{ssp4})  from this simplified approach with an explicit expression
of the
prefactor. We also analyze the behavior of the system close to the critical
point.

\subsection{Equation of state}

We consider an equation of state of the form
\begin{eqnarray}
P=\rho_* \frac{k_B T}{m}\left (\sqrt{1+\rho/\rho_*}-1\right
)^2.
\label{ssp5}
\end{eqnarray}
For $\rho\gg\rho_*$, it reduces to the isothermal equation
of state $P=\rho k_B T/m$. For $\rho\ll \rho_*$, it reduces to the
polytropic equation of state $P=K\rho^2$ with index $\gamma=2$ and polytropic
constant $K=k_B T/(4\rho_*m)$. When coupled to gravity, the equation of state (\ref{ssp5}) leads to equilibrium configurations with an isothermal core surrounded by a polytropic envelope. The enthalpy density is given by [see Eq. (\ref{hell2})]
\begin{eqnarray}
h(\rho)=2\frac{k_B T}{m} \ln \left ( 1+\sqrt{1+\rho/\rho_*}\right ).
\label{ssp6}
\end{eqnarray}
It behaves as $h(\rho)\simeq \frac{k_B
T}{m}\ln(\rho/\rho_*)$ for $\rho\gg \rho_*$ and as 
$h(\rho)\simeq 2\frac{k_B T}{m}\ln(2)+\frac{k_B
T}{m}\frac{\rho}{2\rho_*}$ for $\rho\ll \rho_*$. The internal energy
density vanishing when $\rho=0$
is given by [see Eq. (\ref{hell1})]
\begin{eqnarray}
u(\rho)&=&\rho_* \frac{k_B T}{m}\biggl\lbrack 2
(\sqrt{1+\rho/\rho_*}-1)\nonumber\\
&+&2(\rho/\rho_*) \ln \left
( 1+\sqrt{1+\rho/\rho_*}\right )-\rho/\rho_*\biggr\rbrack.\qquad
\label{ssp7}
\end{eqnarray}
It behaves as $u(\rho)\simeq \rho\frac{k_B
T}{m}\ln(\rho/\rho_*)-\rho\frac{k_B T}{m}$ for $\rho\gg \rho_*$ and as
$u(\rho)=2\rho \frac{k_B T}{m}\ln(2)+\frac{1}{4}\rho_* \frac{k_B
T}{m}(\rho/\rho_*)^2$ for $\rho\ll \rho_*$.

Self-gravitating systems described by the equation of state (\ref{ssp5}) have
been studied in detail in \cite{pomeau0,pomeau,pomeauL} in relation to a model of
supernova. At
high densities, this equation of state coincides with the  isothermal equation
of state. As a result, the caloric curve $\beta(E)$ has a
snail-like (or
spiral) structure \cite{pomeau0,pomeau,pomeauL} similar to the one reported in Fig. \ref{etalambda}
for box-confined systems.
Note that the polytropic equation of state at low densities serves as an
envelope that confines the isothermal core without having to introduce an
artificial box as in the case of purely isothermal systems
\cite{antonov,lbw,paddy,found,ijmpb}.
This is more satisfactory on a physical point of view. This system presents  a
saddle-node bifurcation at a critical temperature $T_c$
\cite{pomeau0,pomeau,pomeauL}. Above $T_c$,
there exists one stable and one or several unstable states (close to $T_c$ only
one unstable state remains). At $T=T_c$, the stable and the unstable states
merge. Below $T_c$, there is no equilibrium state. In that case, the system
undergoes a gravitational collapse. In the canonical ensemble (fixed temperature
$T$), this is called the isothermal collapse \cite{aaISO}.

\subsection{$f$-ansatz}

To simplify the problem and make it analytically tractable, we use an
$f$-ansatz as explained in Appendix G.2 of \cite{jeansapp}. This allows us to
reduce
the stochastic Smoluchowski-Poisson equations (\ref{ssp1}) and (\ref{ssp2}) with
the equation of state (\ref{ssp5}) to a simple Langevin equation of the
form (see Appendix \ref{sec_tf})
\begin{eqnarray}
\label{fg25}
\xi\alpha
M\frac{dR}{dt}=-\frac{d{V}}{dR}+\sqrt{2\xi\alpha M k_B T}\, \eta(t)
\end{eqnarray}
with the effective potential 
\begin{eqnarray}
V(R)=\chi u\left (\frac{3M}{4\pi R^3}\right )\frac{4}{3}\pi
R^3-\nu\frac{GM^2}{R},
\label{eos7}
\end{eqnarray}
where $\alpha$, $\chi$ and $\nu$ are coefficients of order unity that depend on the
$f$-ansatz. For a uniform sphere, we get $\alpha=3/5$, $\chi=1$ and $\nu=3/5$.
For
the internal energy density from Eq. (\ref{ssp7}) we  explicitly obtain\footnote{We use a uniform sphere ansatz instead of a Gaussian ansatz because it makes the internal energy simpler to compute  (see Appendix G.2 of \cite{jeansapp}).}
\begin{eqnarray}
V(R)&=&\frac{8}{3}\chi\pi R^3 \rho_* \frac{k_B T}{m} \left
(\sqrt{1+\frac{3M}{4\pi\rho_*
R^3}}-1\right )\nonumber\\
&+&2\chi\frac{k_B T}{m} M \ln \left (
1+\sqrt{1+\frac{3M}{4\pi\rho_*
R^3}}\right
)\nonumber\\
&-&\chi Nk_B T-\nu\frac{GM^2}{R}.
\label{sil}
\end{eqnarray}
Eq. (\ref{fg25}) is similar to Eq. (\ref{ttr0}) so we can apply the same
arguments as those developed in Secs. \ref{sec_ga} and \ref{sec_ttbec}.

\subsection{Temperature-radius relation}

\begin{figure}[!h]
\begin{center}
\includegraphics[clip,scale=0.3]{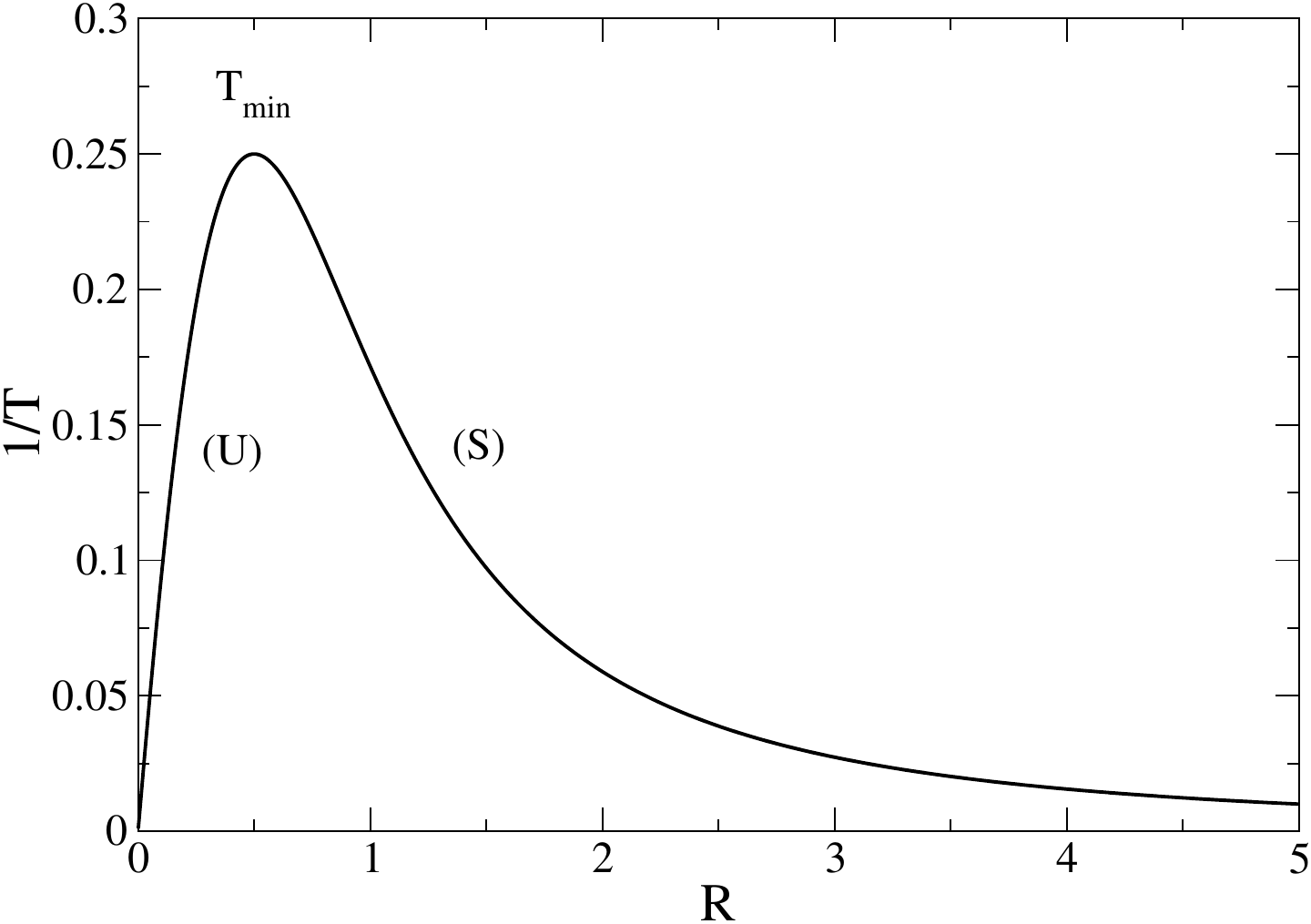}
\caption{Temperature-radius relation
of
self-gravitating Brownian particles described by the equation of state
(\ref{ssp5}). It displays a minimum temperature $T_{\rm min}$ below which
no equilibrium state exists. We have normalized the radius by $(3M/4\pi
\rho_*)^{1/3}$ and the
temperature by $(\nu GMm/3\chi k_B)(4\pi \rho_*/3M)^{1/3}$.}
\label{USTR}
\end{center}
\end{figure}

An equilibrium state is determined by the condition $V'(R)=0$. Using
\begin{eqnarray}
V'(R)&=&8\chi\pi R^2 \rho_* \frac{k_B T}{m} \left
(\sqrt{1+\frac{3M}{4\pi\rho_*
R^3}}-1\right )\nonumber\\
&-&3\chi\frac{k_B T}{m} \frac{M}{R}+\nu\frac{GM^2}{R^2},
\label{sil1}
\end{eqnarray}
we obtain the temperature-radius
relation
\begin{equation}
\frac{m}{k_B T}=\frac{3\chi R}{\nu GM}
-\frac{8\chi\pi R^4 \rho_*}{\nu GM^2}  \left
(\sqrt{1+\frac{3M}{4\pi\rho_*
R^3}}-1\right ).
\label{sil2}
\end{equation}
The curve $(1/T)(R)$  is
represented in
Fig. \ref{USTR}. It is similar to the mass-radius
relation of
dilute axion stars in Fig. \ref{MRmodif}. There is a minimum
temperature
\begin{eqnarray}
k_B T_{\rm min}=\frac{4\nu G M m}{3\chi}\left (\frac{4\pi\rho_*}{3M}\right
)^{1/3}
\label{eos7b}
\end{eqnarray}
at
\begin{eqnarray}
R_*=\frac{1}{2}\left (\frac{3M}{4\pi\rho_*}\right )^{1/3}.
\label{eos7bk}
\end{eqnarray}
We have the identity
\begin{eqnarray}
k_B T_{\rm min}=\frac{2\nu GMm}{3\chi R_*}.
\label{eos7c}
\end{eqnarray}
Equilibrium solutions exist only for $T>T_{\rm min}$. The series of equilibria
displays two branches of
solutions.\footnote{The $f$-ansatz is not able to account for all the unstable
states (saddle points) of the original model where
the temperature-radius relation forms a spiral \cite{pomeau0,pomeau,pomeauL}.} As we
shall show
below, the
right branch ($R>R_*$) corresponds to stable states (S) [actually, metastable (M) states] and the
left branch  ($R<R_*$) corresponds to unstable (U) states. When $R\rightarrow 0$ we can neglect the contribution of the polytropic envelope
and we
get ${m}/{k_B T}\sim {3\chi R}/{\nu GM}$. When $R\rightarrow +\infty$ we
can neglect the contribution of the isothermal core and we get ${m}/{k_B T}\sim
{9\chi}/({16\pi\rho_*\nu GR^2})$.

\subsection{Squared pulsation and Poincar\'e turning point criterion}
\label{sec_pt}

The squared pulsation around an equilibrium state is given by (see Sec.
\ref{sec_relt})
\begin{eqnarray}
\omega^2=\frac{V''(R)}{\alpha M}
\label{silw}
\end{eqnarray}
with
\begin{eqnarray}
V''(R)&=&16\chi\pi R\rho_* \frac{k_B T}{m} \left
(\sqrt{1+\frac{3M}{4\pi\rho_*
R^3}}-1\right )\nonumber\\
&-&9M\chi \frac{k_B
T}{m}\frac{1}{\sqrt{1+\frac{3M}{4\pi\rho_*R^3}}}\frac{1}{R^2}\nonumber\\
&+&3\chi\frac{k_B T}{m} \frac{M}{R^2}-2\nu\frac{GM^2}{R^3}.
\label{sil3}
\end{eqnarray}
Taking the derivative of Eq. (\ref{sil2}) with respect to $R$, we obtain
\begin{eqnarray}
\frac{d}{dR}\left (\frac{m}{k_B T}\right )=\frac{3\chi}{\nu GM}
+\frac{9\chi}{\nu GM}\frac{1}{\sqrt{1+\frac{3M}{4\pi\rho_*
R^3}}}\nonumber\\
-\frac{32\chi\pi R^3 \rho_*}{\nu GM^2}  \left
(\sqrt{1+\frac{3M}{4\pi\rho_*
R^3}}-1\right ).
\label{sil4}
\end{eqnarray}
Using Eq. (\ref{sil2}), we can rewrite Eq. (\ref{sil3}) as
\begin{eqnarray}
\frac{m}{k_B T} V''(R)=32\chi\pi R\rho_*  \left
(\sqrt{1+\frac{3M}{4\pi\rho_*
R^3}}-1\right )\nonumber\\
-9M\chi \frac{1}{\sqrt{1+\frac{3M}{4\pi\rho_*R^3}}}\frac{1}{R^2}-3\chi
\frac{M}{R^2}.
\label{sil5}
\end{eqnarray}
Comparing Eqs. (\ref{sil4}) and  (\ref{sil5}) we obtain the identity
\begin{equation}
\frac{d}{dR}\left (\frac{1}{k_B T}\right )=-\frac{R^2}{\nu GM^2 k_B T}V''(R)=-\frac{\alpha R^2}{\nu GM k_B T}\omega^2(R),
\label{sil6}
\end{equation}
which can be interpreted as a form of Poincar\'e turning point criterion
\cite{poincare}. We explicitly check that the pulsation vanishes ($\omega=0$) at the
turning point of temperature ($T'(R)=0$). More precisely, this equation relates
the stability of the equilibrium solution to the slope of the temperature-radius
relation. The solutions on the right branch of Fig. \ref{USTR}
are stable ($\omega^2>0$) since the inverse temperature is a decreasing function
of the radius ($d(1/T)/dR<0$), while the solutions on the left branch are
unstable ($\omega^2<0$) since the inverse temperature is an increasing function
of the radius ($d(1/T)/dR>0$).

\subsection{Normal form of the potential close to the
maximum mass}

Expanding the effective potential from Eq. (\ref{sil}) to
third order
close to the minimum temperature $T_{\rm min}$, we obtain
\begin{eqnarray}
\frac{V(R)}{V_0}=\frac{1}{3R_*^3}
(R-R_*)^3\nonumber\\
-\frac{3}{2R_*}\left
(1-\frac{T_{\rm min}}{T}\right )(R-R_*)
-2+4\ln(2),
\label{weos18}
\end{eqnarray}
where
\begin{eqnarray}
V_0=\chi N k_B T_{\rm min}=\frac{2\nu GM^2}{3R_*}.
\label{eos18by}
\end{eqnarray}
Equation (\ref{weos18}) is the normal form of a potential $V(R)$
close to a saddle-node  bifurcation (see Fig.
\ref{normalform}). With this
approximation, the equation of motion (\ref{fg25}) of the fictive particle
becomes
\begin{eqnarray}
\label{g25scst}
\xi\alpha
M\frac{dR}{dt}=-\frac{V_0}{R_*^3}(R-R_*)^2+\frac{3V_0}{2R_*}\left
(1-\frac{T_{\rm min}}{T}\right )\nonumber\\
+\sqrt{2\xi\alpha M k_B T}\, \eta(t).
\end{eqnarray}
Setting $x=R-R_*$, the potential
from Eq. (\ref{weos18}) can be written as in Eq.
(\ref{qtr18}) with
\begin{eqnarray}
a=\frac{V_0}{R_*^3}\quad {\rm and} \quad b=\frac{3V_0}{2R_*}\left
(1-\frac{T_{\rm
min}}{T}\right ).
\label{qtr19pom}
\end{eqnarray}
By using this correspondence we can establish the following results directly from the general relations obtained in the previous sections, which are valid close to a saddle-node
bifurcation (see Appendix \ref{sec_summary} for a summary).

The radii of the equilibrium states ($V'(R)=0$) close to the critical point are
given by 
\begin{eqnarray}
\frac{T_{\rm min}}{T}&=&1-\frac{2}{3R_*^2}
(R-R_*)^2\nonumber\\
{\rm i.e.} \qquad R-R_*&=&\pm\sqrt{\frac{3}{2}}R_*\left (1-\frac{T_{\rm
min}}{T}\right )^{1/2}.
\label{eos18}
\end{eqnarray}
The upper sign corresponds to the right  branch ($R>R_*$) and the lower
sign corresponds to the left branch ($R<R_*$).

According to Eqs. (\ref{silw}), (\ref{weos18}) and (\ref{eos18}) the squared
radial pulsation around an equilibrium state is given by
\begin{eqnarray}
\omega^2&=&\frac{V''(R)}{\alpha M}=\frac{2\chi k_B T_{\rm min}}{\alpha m
R_*^3}(R-R_*)\nonumber\\
&=&\pm\frac{\sqrt{6}\chi k_B T_{\rm min}}{\alpha m R_*^2}\left (1-\frac{T_{\rm
min}}{T}\right )^{1/2}.
\label{nfcmm4b}
\end{eqnarray}
This expression, which is valid close to the minimum temperature, confirms
that the branch $R>R_*$ is stable ($\omega^2>0$) while the branch $R<R_*$ is
unstable ($\omega^2<0$). The squared pulsation close to the critical point can
be rewritten as
\begin{eqnarray}
\omega^2=\pm \frac{2\sqrt{6}}{3t_D^2} \left (1-\frac{T_{\rm min}}{T}\right
)^{1/2}
\label{nfcmm4h}
\end{eqnarray}
with the dynamical time
\begin{eqnarray}
t_D&=&\left (\frac{\alpha}{\nu}\right )^{1/2}\frac{1}{\sqrt{G\rho_0}}=\left
(\frac{\alpha R_*^3}{\nu GM}\right )^{1/2}\nonumber\\
&=& \left (\frac{2\alpha M R_*^2}{3V_0}\right )^{1/2}
=\left (\frac{2\alpha m R_*^2}{3\chi k_B T_{\rm min}}\right )^{1/2}
\label{tdynj}
\end{eqnarray}
constructed with the typical density
\begin{eqnarray}
\label{nfcmm6}
\rho_{0}=\frac{M}{R_{*}^3}.
\end{eqnarray}

Taking the derivative of Eq. (\ref{eos18}) with respect to $R$, we find that
\begin{eqnarray}
\frac{d}{dR}\left (\frac{1}{k_BT}\right )&=&-\frac{4}{3R_*^2}(R-R_*)\frac{1}{k_B
T_{\rm min}}\nonumber\\
&=&\mp\frac{2\sqrt{6}}{3R_*}\frac{1}{k_B T_{\rm min}}\left
(1-\frac{T_{\rm min}}{T}\right )^{1/2}.\qquad
\label{we2}
\end{eqnarray}
Comparing Eqs. (\ref{nfcmm4b}) and (\ref{we2}), we obtain the relation
\begin{eqnarray}
\frac{d}{dR}\left (\frac{1}{k_BT}\right
)&=&-\frac{2}{3}R_*\frac{m}{\chi M(k_B T_{\rm
min})^2}V''(R)\nonumber\\
&=&-\frac{2}{3}R_*\frac{\alpha m}{\chi (k_B T_{\rm
min})^2}\omega^2
\label{we2b}
\end{eqnarray}
that links the stability of the system (through the sign of the squared
pulsation $\omega^2$) to the slope of the temperature-radius relation. This is
the expression of the Poincar\'e turning point criterion [see Eq. (\ref{sil6})]
close to the minimum
temperature.

Another manner to investigate the
stability of an equilibrium state is to compute its free energy. The
free energies of the (stable and unstable) equilibrium states close to the
minimum temperature are
\begin{equation}
\frac{F}{V_0}=\frac{V}{V_0}=\mp\left (\frac{3}{2}\right )^{1/2}\left
(1-\frac{T_{\rm
min}}{T}\right )^{3/2}-2+4\ln(2).
\label{eos7bg}
\end{equation}
We see that the free energy of the stable state ($R>R_*$, upper sign) is lower
than the free energy of the unstable state ($R<R_*$, lower sign) for the same temperature
$T$. This is expected since an equilibrium state is a minimum of free energy.
The barrier of free energy is
\begin{eqnarray}
\frac{\Delta F}{V_0}=\frac{\Delta V}{V_0}=2\left (\frac{3}{2}\right
)^{1/2}\left
(1-\frac{T_{\rm
min}}{T}\right )^{3/2}.
\label{bouley1}
\end{eqnarray}
The evolution of the free energy as a 
function of the temperature is similar to the one reported in Fig.
\ref{freeenergy} for self-gravitating BECs (see also Fig. 6 in \cite{pt}).

\subsection{Relaxation time when $T>T_{\rm min}$}

If we neglect the noise term in Eq. (\ref{fg25}) the
equation of motion reduces
to
\begin{eqnarray}
\label{g25a}
\xi\alpha
M\frac{dR}{dt}=-\frac{d{V}}{dR}.
\end{eqnarray}
When $T>T_{\rm min}$, there are equilibrium states. Let us make a small
perturbation $R(t)=R_e+\delta R(t)$ about an equilibrium
state and linearize Eq. (\ref{g25a}). This yields
\begin{eqnarray}
\label{g25b}
\frac{d\delta R}{dt}+\frac{\omega^2}{\xi}\delta R=0.
\end{eqnarray}
The solution of this equation is
\begin{eqnarray}
\label{g25c}
\delta R(t)=\delta R(0)e^{-\frac{\omega^2}{\xi}t}.
\end{eqnarray}
The perturbation evolves exponentially rapidly with a rate 
$\gamma=-\omega^2/\xi$. The perturbation is damped when $\omega^2>0$ (stable
case) and grows when $\omega^2<0$ (unstable case). In the stable case, the
relaxation time of the system towards the equilibrium state is $t_{\rm
relax}=-1/\gamma=\xi/\omega^2$. Close to
the critical point, using Eqs. (\ref{nfcmm4b}) and (\ref{nfcmm4h}), we obtain
\begin{eqnarray}
t_{\rm relax}&\sim&\frac{\xi\alpha m R_*^2}{\sqrt{6}\chi k_B T_{\rm
min}}\left (1-\frac{T_{\rm min}}{T}\right )^{-1/2}\nonumber\\
&\sim&\frac{3}{2\sqrt{6}}\xi
t_D^2\left
(1-\frac{T_{\rm min}}{T}\right
)^{-1/2}.
\label{sb}
\end{eqnarray}
The relaxation time diverges like $t_{\rm
relax}/(\xi t_D^2)\sim
0.612372...(1-T_{\rm
min}/T)^{-1/2}$ when $T\rightarrow T_{\rm min}^+$.

\subsection{Collapse time when $T<T_{\rm min}$}

When $T<T_{\rm min}$ there is no equilibrium state and the system collapses. The
collapse time  close to the critical point is given by (see Appendix
\ref{sec_collsfa})
\begin{eqnarray}
t_{\rm coll}&\sim& \frac{\xi\alpha m\pi R_*^2}{\sqrt{6}\chi k_B T_{\rm
min}}\left (\frac{T_{\rm
min}}{T}-1\right )^{-1/2}\nonumber\\
&\sim& \frac{3\pi}{2\sqrt{6}}\xi t_D^2 \left (\frac{T_{\rm
min}}{T}-1\right )^{-1/2}.\label{str}
\end{eqnarray}
The collapse time diverges like $t_{\rm coll}/(\xi
t_D^2)\sim 1.92382... (T_{\rm
min}/T-1)^{-1/2}$ when $T\rightarrow T_{\rm min}^-$. This
is the same
scaling as for the divergence of the relaxation (damping) time
$t_{\rm relax}/(\xi t_D^2)\sim  0.612372...(1-T_{\rm
min}/T)^{-1/2}$
close
to the critical
point when $T\rightarrow T_{\rm min}^+$ [see Eq. (\ref{sb})]. The prefactors
differ by a factor $\pi$.

\subsection{Variance and correlation of the fluctuations}
\label{sec_corr}

We now take fluctuations into account in Eq. (\ref{fg25}). The evolution
of the probability density $P(R,t)$ is governed by the
Fokker-Planck (Smoluchowski) equation (\ref{b5txt}) and the equilibrium
distribution is given by the Boltzmann distribution from Eq. (\ref{kmaro}).
In the Gaussian approximation (see Appendix
\ref{sec_summary}), the average value of the radius $\langle R\rangle$
coincides with its most probable value $R_e$ (corresponding to the
minimum of the potential $V(R)$ determined by $V'(R_e)=0$ and $V''(R_e)>0$) and
its
variance is
\begin{eqnarray}
\label{tmar1}
\langle (\Delta R)^2\rangle=\frac{1}{\beta V''(R_e)}=\frac{1}{\beta \alpha M \omega^2},
\end{eqnarray}
where $\Delta R=R-R_e$ is the fluctuation of the radius about its equilibrium
value. Combining Eq. (\ref{tmar1}) with Eq. (\ref{sil6}), we obtain the
relation 
\begin{eqnarray}
\frac{d}{dR}\left (\frac{1}{k_B T}\right )&=&-\frac{R_e^2}{\nu GM^2 k_B
T}V''(R_e)
=-\frac{\alpha R_e^2}{\nu GM k_B T}\omega^2\nonumber\\
&=&-\frac{R_e^2}{\nu GM^2}\frac{1}{\langle (\Delta R)^2\rangle},
\label{tmar2}
\end{eqnarray}
which is analogous to the fluctuation-dissipation theorem in thermodynamics (see Appendix \ref{sec_tcano}). Close to the minimum temperature $T_{\rm min}$, using  Eqs. (\ref{nfcmm4h}) and (\ref{tmar1}), we have for the stable equilibrium state
\begin{eqnarray}
\omega^2&=&\frac{2\sqrt{6}}{3t_D^2} \left (1-\frac{T_{\rm min}}{T}\right
)^{1/2},\nonumber\\
\langle (\Delta R)^2\rangle&=&\frac{3t_D^2}{2\sqrt{6}\beta
\alpha M} \left (1-\frac{T_{\rm min}}{T}\right
)^{-1/2}.
\label{tmar3}
\end{eqnarray}
When $T\rightarrow T_{\rm min}^+$, the squared pulsation vanishes as 
$\omega^2\propto \left (1-T_{\rm min}/T\right )^{1/2}$ and the variance diverges
as $\langle (\Delta R)^2\rangle\propto  \left (1-T_{\rm min}/T\right )^{-1/2}$.
This is similar to the phenomenon of critical opalescence in the theory of
liquids \cite{opal1,opal2} (see also \cite{paper5}). In the  Gaussian
approximation, one can show that for $\omega^2 t/\xi \gg 1$, i.e., in the
stationary state, the temporal correlation function of the fluctuations $\Delta
R(t)$ reads (see Appendix \ref{sec_summary})
\begin{eqnarray}
\langle \Delta R(0)\Delta R(t)\rangle=\frac{k_B T}{\alpha M\omega^2}e^{-\omega^2 t/\xi}.
\label{tmar4}
\end{eqnarray}
Close to the minimum temperature $T_{\rm min}$ where 
$\omega\rightarrow 0$, the variance of the fluctuations $\langle (\Delta
R)^2\rangle={k_B T}/{\alpha M
\omega^2}$ (prefactor), and the correlation
time $t_{\rm corr}=\xi/\omega^2$ (exponent), which coincides with the relaxation
(damping) time $t_{\rm relax}$, both diverge as $\omega^{-2}$.

\subsection{Thermal activation rate}

When $T>T_{\rm min}$ the stable equilibrium state is in fact metastable. The
lifetime of the
metastable state close to the minimum temperature ($T\rightarrow T_{\rm
min}^+$) is given by (see Sec. \ref{sec_ttbec} and Appendix \ref{sec_summary})
\begin{equation}
\label{g75ts}
\frac{t_{\rm life}^{\rm T}}{\xi t_D^2}\sim \frac{3\pi}{\sqrt{6}}\left
(1-\frac{T_{\rm min}}{T}\right )^{-1/2}
e^{\frac{4}{\sqrt{6}}\left
(1-\frac{T_{\rm min}}{T}\right )^{3/2}\nu\eta_{\rm max} N},
\end{equation}
where we have introduced the particle number
$N=M/m$ and the dimensionless inverse temperature
\begin{eqnarray}
\label{g75b}
\eta_{\rm max}=\frac{GMm}{R_*k_B T_{\rm min}}.
\end{eqnarray}
We see that the time is naturally scaled by $\xi t_D^2$ in
agreement with Eq. (\ref{qtr17b}). The prefactor behaves as $3.84765...\, (1-T_{\rm
min}/T)^{-1/2}$. This is the same scaling as for the
relaxation time from Eq. (\ref{sb}) when $T\rightarrow T_{\rm min}^+$ (they differ by factor $2\pi$) and as for the
collapse time from Eq.
(\ref{str}) when $T\rightarrow T_{\rm min}^-$ (they differ by a factor $2$). We
symbolically have ``$t_{\rm
life}^{\rm T}=2t_{\rm coll}=2\pi t_{\rm relax}$'' (see Appendix
\ref{sec_summary}). The thermal
lifetime of the metastable state
scales as
\begin{eqnarray}
\label{k2bruel}
\frac{t_{\rm life}^{\rm T}}{\xi t_D^2}\sim e^{N}
\end{eqnarray}
except at, or very close
to, the critical point.\footnote{We recall that the Kramers
formula assumes $\Delta V/k_B T\gg 1$
so the above expression of $t_{\rm life}^{\rm T}$  may not be valid if we
are too close to the
collapse temperature $T_c$ where $\Delta V\rightarrow 0$.}

Exactly at the minimum temperature $T_{\rm min}$,
the collapse time taking into
account thermal fluctuations scales as
\begin{eqnarray}
\frac{t_{\rm coll}^{\rm T}}{\xi t_D^2}\propto N^{1/3}
\end{eqnarray}
in the strong friction limit (see Appendix \ref{sec_size}) instead
of being infinite according to Eq. (\ref{str}) if we ignore the
fluctuations.

From Eq. (\ref{g75ts}), writing that the term in the exponential is of order
unity, we
find that the correction  to the collapse temperature due to
finite $N$ effects is
\begin{eqnarray}
\label{ln5sup} 
T_l=T_{\rm min}\left (1-\frac{1}{6^{1/3}\chi^{2/3}}N^{-2/3}\right )^{-1}.
\end{eqnarray}

{\it Remark:} The expressions obtained in this section are similar to those
obtained in Secs. \ref{sec_gbm}-\ref{sec_ttbec} for self-gravitating BECs and in
Sec. \ref{sec_gc}
for globular clusters and self-gravitating Brownian particles in the
underdamped limit. In particular, the exponents are the
same. These scalings are ``universal'' because they are based on the normal form
of the potential close to a critical point (see Appendix \ref{sec_summary}).

\section{BMF model}
\label{sec_othex}

\begin{figure}[!h]
\begin{center}
\includegraphics[clip,scale=0.3]{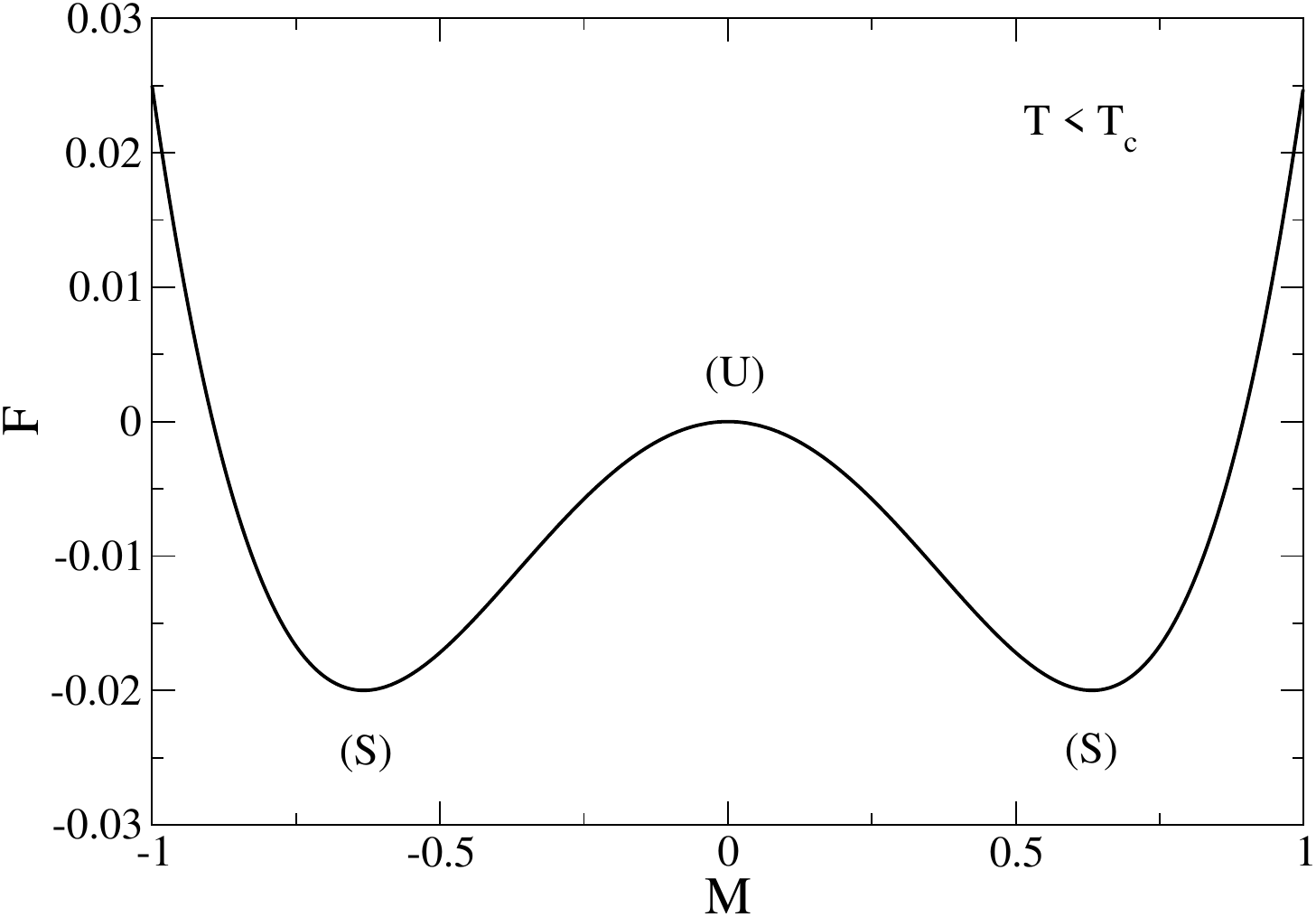}
\caption{Normal form of the free energy per particle for $T<T_c$ close to a pitchfork bifurcation showing a bistable behavior (we have taken $T=0.4$).}
\label{free}
\end{center}
\end{figure}

The previous results can be extended to a system of overdamped Brownian particles moving on a circle and interacting via a cosine binary potential (BMF model) \cite{bmf}. In that case, the evolution of the density $\rho(\theta,t)$ is governed by the stochastic Smoluchowski equation \cite{bmf}
\begin{equation}
\label{ep1}
\xi\frac{\partial\rho}{\partial t}=\frac{\partial}{\partial\theta}\left (T\frac{\partial\rho}{\partial\theta}+\rho\frac{\partial\Phi}{\partial\theta}\right )+\frac{1}{\sqrt{N}}\frac{\partial}{\partial\theta}\left \lbrack\sqrt{2\xi T\rho(\theta,t)}\, R(\theta,t)\right\rbrack
\end{equation}
with the mean field potential
\begin{equation}
\label{ep2}
\Phi(\theta,t)=\int \left\lbrack 1-\cos(\theta-\theta')\right\rbrack\rho(\theta',t)\, d\theta'.
\end{equation}
This system exhibits a pitchfork bifurcation at a critical temperature $T_c=1/2$. For $T>T_c$ the equilibrium state is homogeneous and for $T<T_c$ it is inhomogeneous. In the weakly inhomogeneous (magnetized) phase slightly below the critical temperature $T_c=1/2$, one can show \cite{bmf} that the magnetization $M(t)\equiv M_x(t)=\int_{0}^{2\pi} \rho(\theta,t) \cos\theta\, d\theta$ (we impose $M_y(t)=0$) satisfies a stochastic Langevin equation of the form 
\begin{equation}
\label{ep3}
\xi\frac{dM}{dt}=-\frac{1}{2}F'(M)+\sqrt{\frac{\xi T}{N}}\, \eta(t)
\end{equation}
with the free energy per particle 
\begin{equation}
\label{ep4}
F(M)=F_0(T)+(T-T_c)M^2+\frac{1}{8}M^4
\end{equation}
playing the role of a potential. In this  bistable system (see Fig. \ref{free}) the magnetization $M(t)$ undergoes random transitions between the equilibrium values $M_M=2(T_c-T)^{1/2}$ and $M_M=-2(T_c-T)^{1/2}$ corresponding to $F'(M)=0$ (the equilibrium state $M_U=0$ is stable for $T>T_c$ and unstable for $T<T_c$). Such random transitions have been studied for a system of one-dimensional self-gravitating Brownian particles in a box \cite{cdkramers}. They can also be studied with the methods presented in the present paper.

\begin{figure}[!h]
\begin{center}
\includegraphics[clip,scale=0.3]{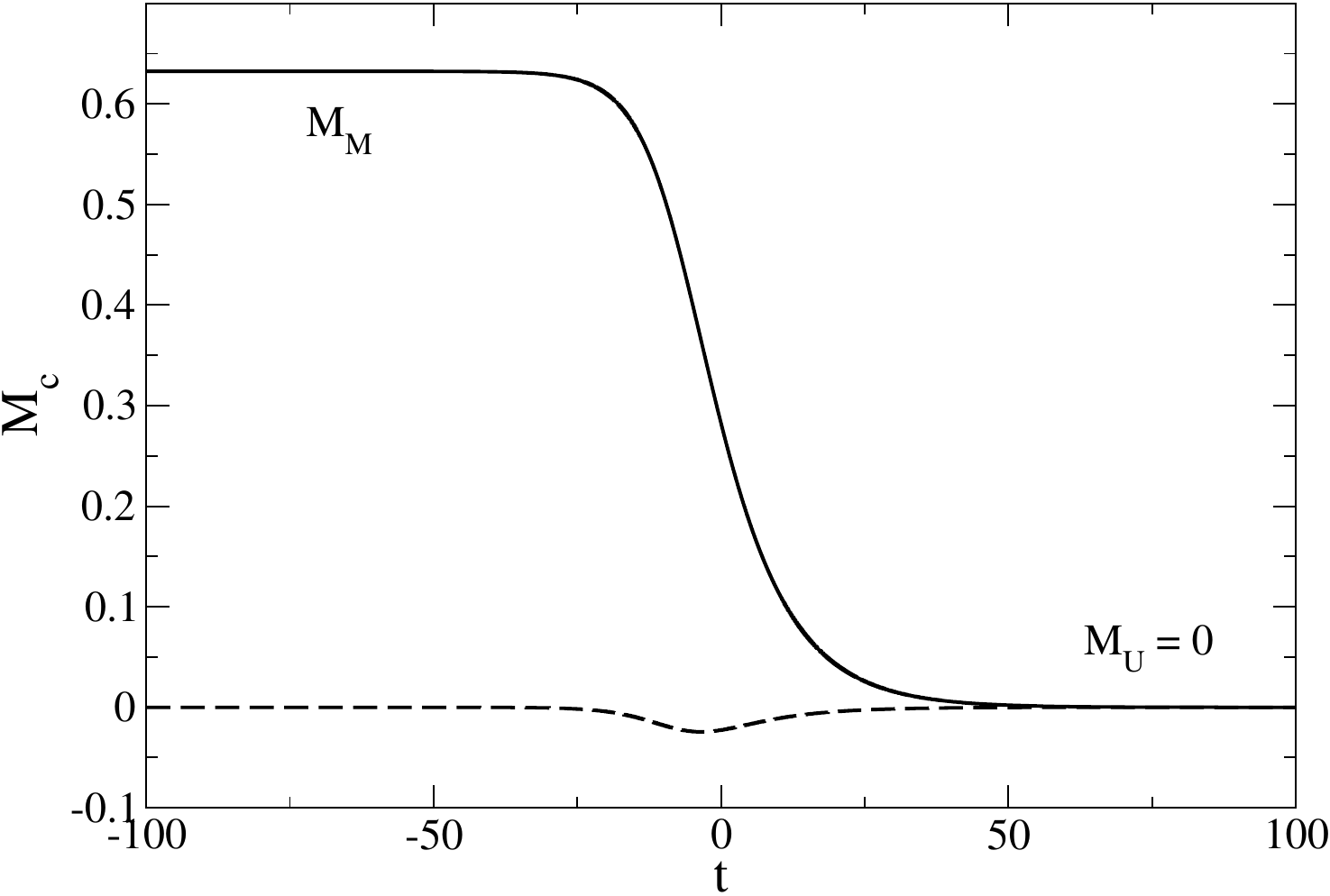}
\caption{Instanton $M_c(t)$ close to a pitchfork bifurcation. The dashed  line corresponds to the velocity $\dot M_c$.}
\label{instantonBMF}
\end{center}
\end{figure}

The escape rate from a metastable state (minimum of free energy) is given by the Kramers formula [see Eq. (\ref{g65b})]
\begin{equation}
\label{ep5}
\Gamma=\frac{\omega_M |\omega_U|}{2\pi\xi}e^{-N\Delta F/T},
\end{equation}
where $\Delta F=F_U-F_M$ is the barrier of free energy. Using  $\omega_M^2=F''(M_M)/2=2(T_c-T)$, $\omega_U^2=F''(M_U)/2=-(T_c-T)$ and $\Delta F=2(T_c-T)^2$ \cite{bmf}, we obtain
\begin{equation}
\label{ep6}
\Gamma=\frac{T_c-T}{\sqrt{2}\pi\xi}e^{-2N(T_c-T)^2/T}.
\end{equation}
This result can also be obtained with the instanton theory. The escape rate is given by Eq. (\ref{g65}) with the exponent 
\begin{equation}
\label{ep7}
B=\frac{1}{2}N\xi\int_{-\infty}^{+\infty} \left \lbrack\dot M_c+\frac{F'(M_c)}{2\xi}\right \rbrack^2\, dt,
\end{equation}
where $M_c(t)$ is the instanton determined by the equation (uphill solution)
\begin{equation}
\label{ep8}
\xi\frac{dM_c}{dt}=\frac{1}{2}F'(M_c)=(T-T_c)M_c+\frac{1}{4}M_c^3.
\end{equation}
The solution of this equation is
\begin{equation}
\label{ep9}
M_c(t)=\frac{2(T_c-T)^{1/2}}{\sqrt{1+4e^{2(T_c-T)(t-t_0)/\xi}}},
\end{equation}
where $t_0$ is arbitrary (see Fig. \ref{instantonBMF}). The instanton $M_c(t)$ ascends the potential from $M_M$ to $M_U=0$ (afterwards, the fictive particle descends freely to the symmetric equilibrium state $-M_M$). Its velocity is
\begin{equation}
\label{ep10}
\dot M_c=-\frac{8}{\xi} (T_c-T)^{3/2}
\frac{e^{2(T_c-T)(t-t_0)/\xi}}{\left\lbrack 1+4e^{2(T_c-T)(t-t_0)/\xi}\right\rbrack^{3/2}}.
\end{equation}
Substituting this expression into Eq. (\ref{ep7}) and using the instanton equation  (\ref{ep8}) we obtain 
\begin{equation}
\label{ep11}
B=2N\xi\int_{-\infty}^{+\infty} {\dot M}^2 \, dt=2N(T_c-T)^2=N\Delta F,
\end{equation}
which is equal to the barrier of free energy. On the other hand, the prefactor in Eq. (\ref{g65})  is
\begin{equation}
\label{ep12}
A=\frac{\omega_M |\omega_U|}{2\pi\xi}=\frac{T_c-T}{\sqrt{2}\pi\xi}.
\end{equation}
This returns the Kramers escape rate from Eq. (\ref{ep5}).

\begin{figure}[!h]
\begin{center}
\includegraphics[clip,scale=0.3]{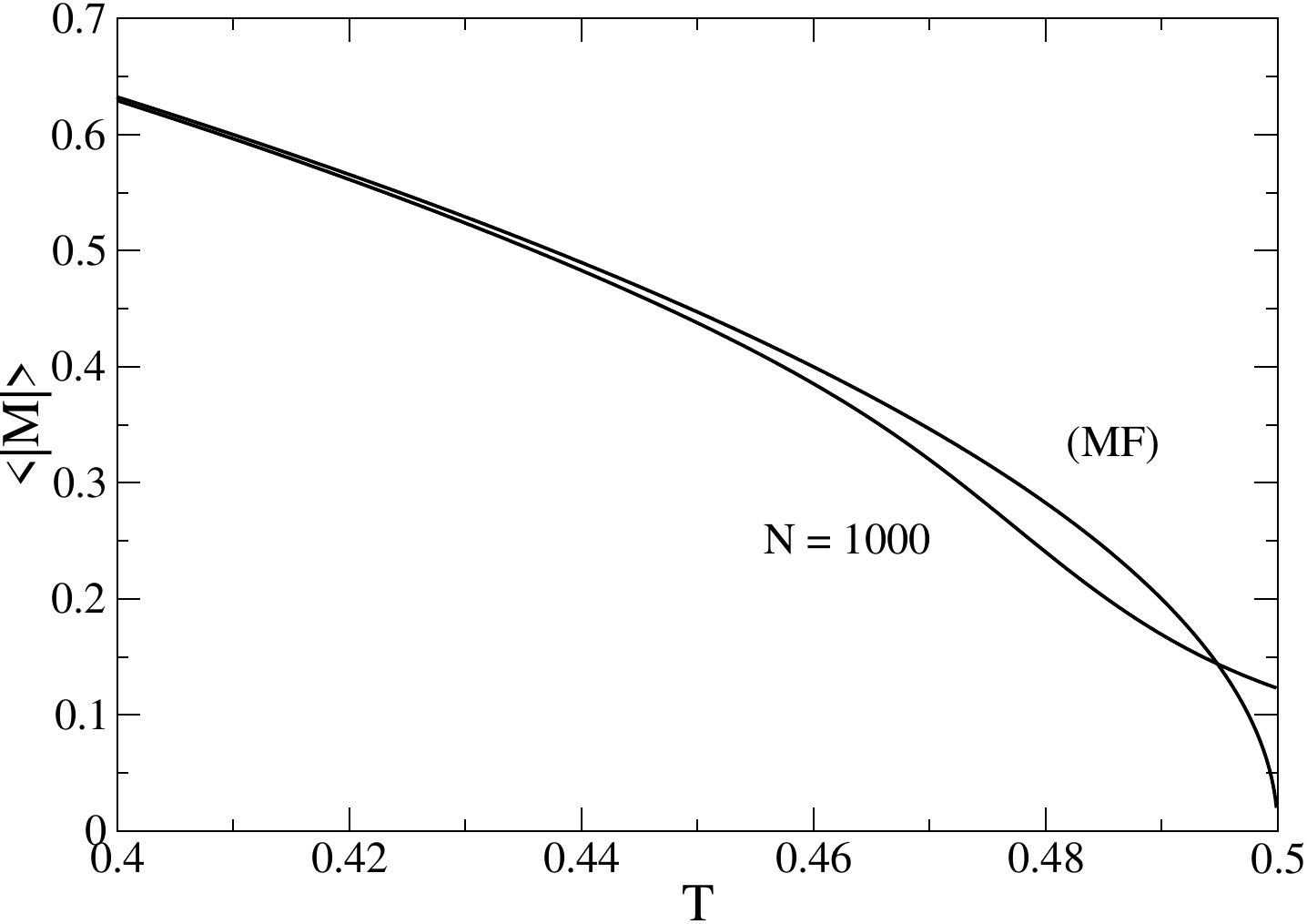}
\caption{Finite size effects close to the critical point of the BMF model (pitchfork bifurcation). We have compared the case $N=1000$ [see Eq. (\ref{ep15})] for which  $d\langle |M|\rangle/dT$ is finite at $T=T_c$ to the mean field (MF) result $|M|_M=2(T_c-T)^{1/2}$ valid for $N\rightarrow +\infty$ for which $d|M|_M/dT\rightarrow +\infty$ when $T\rightarrow T_c$.}
\label{tm}
\end{center}
\end{figure}

The equilibrium distribution of the magnetization at the temperature $T$ is $P(M)=\frac{1}{Z}e^{-NF(M)/T}$ with $Z=\int e^{-NF(M)/T}\, dM$ \cite{bmf}. For a bistable behavior, in order to distinguish  the concentration of the particles around $\theta=0$ ($M=1$) and  $\theta=\pi$ ($M=-1$), we consider the average modulus of the magnetization $\langle |M|\rangle=\int |M| e^{-NF(M)/T}\, dM/\int  e^{-NF(M)/T}\, dM$ (the average magnetization is zero since the free energy is even). We want to study finite size effects close to the critical point $T_c=1/2$. We consider the scaling $T=T_c-x/N^a$ with $x=O(1)$, so that we are just below $T_c$ when $N\gg 1$. The free energy is then given by $F(M)=-(x/N^a) M^2+M^4/8$. From the mean field theory, we know that $|M|\sim (T_c-T)^{1/2}\sim 1/N^{a/2}$, so we introduce the variable $t= N^{a/2} M$ in terms of which $F(M)=(-x t^2+t^4/8)/N^{2a}$. When $N\rightarrow +\infty$, we must have $NF(M)\sim 1$  which imposes $a=1/2$. Therefore, the average modulus of the magnetization close to the critical point is given by 
\begin{eqnarray}
\langle |M|\rangle=\frac{1}{N^{1/4}}f\left \lbrack N^{1/2}(T_c-T)\right \rbrack
\label{ep15}
\end{eqnarray}
with
\begin{eqnarray}
f(x)=\frac{\int_{0}^{+\infty} t e^{2x t^2-t^4/4}\, dt}{\int_{0}^{+\infty} e^{2x t^2-t^4/4}\, dt}.
\label{ep15b}
\end{eqnarray}For $x\rightarrow +\infty$, using the saddle point approximation, we find that $f(x)\sim 2\sqrt{x}$. Therefore, for $T<T_c$ and $N\rightarrow +\infty$, we recover the mean field result $|M|_M=2(T_c-T)^{1/2}$.  For $T>T_c$ (homogeneous phase) and $N\rightarrow +\infty$, the fluctuations of the magnetization are gaussian \cite{bmf} with variance $\langle M^2\rangle=1/[4N(T-T_c)]$  and average modulus $\langle |M|\rangle=1/\sqrt{2\pi N  (T-T_c)}$ (for $x\rightarrow -\infty$ we have  $f(x)\sim 1/\sqrt{2\pi |x|}$). At the critical point $T=T_c$, we have $\langle |M|\rangle=f(0)/N^{1/4}$ with $f(0)=\sqrt{\pi}/[2\sqrt{2}\Gamma(5/4)]=0.691367...$. For $N$ finite, there is no divergence of the slope $d\langle |M|\rangle/dT$ of the magnetization curve $\langle |M|\rangle (T)$ (which is similar to the specific heat $C=d\langle E\rangle/dT$ associated with a caloric curve $\langle E\rangle(T)$) and of the fluctuations at the critical point (see Fig. \ref{tm}).

\section{Summary of the different equations}
\label{sec_pain}

In this section, we summarize the different equations derived in this paper
and propose some extensions.

By making a Gaussian ansatz, we can reduce the
stochastic quantum damped Euler-Poisson
equations (\ref{qes1})-(\ref{qes2}) into
a stochastic Langevin equation of
the form (see Appendix \ref{sec_tf})
\begin{eqnarray}
\label{pain1}
\alpha M\frac{d^2R}{dt^2}+\xi\alpha M\frac{dR}{dt}=-\frac{d{V}}{dR}+\sqrt{2\xi
k_B
T \alpha M}\eta(t),\quad
\end{eqnarray}
where $V(R)$ is an effective potential and $\eta(t)$ is a Gaussian white noise
satisfying $\langle
\eta(t)\rangle=0$ and $\langle \eta(t)\eta(t')\rangle=\delta(t-t')$. Let us
assume that the system displays a critical point associated with a
maximum mass as in the case of dilute axion stars (see
Secs. \ref{sec_be}-\ref{sec_ttbec}).\footnote{It could also be a
minimum energy as in the case of globular clusters (see Sec. \ref{sec_gcs}),
a minimum temperature as in the case of self-gravitating Brownian particles
(see Secs. \ref{sec_gcb} and \ref{sec_sfbp}), or any other control parameter.}
For a saddle-node bifurcation, the normal
form of the potential close to the maximum mass can be
written as (see Sec. \ref{sec_nfcmm})
\begin{equation}
\label{pain2}
V(R)=V_0+\frac{1}{3}a(R-R_*)^3-b_0\left (1-\frac{M}{M_{\rm max}}\right
)(R-R_*),
\end{equation}
where $a$ and $b_0$ are positive coefficients. If we substitute the normal form
of the potential close to the maximum mass from Eq. (\ref{pain2}) into Eq.
(\ref{pain1}) we obtain
\begin{eqnarray}
\label{pain3}
\alpha M\frac{d^2R}{dt^2}+\xi\alpha
M\frac{dR}{dt}=-a(R-R_*)^2\nonumber\\
+b_0\left
(1-\frac{M}{M_{\rm max}}\right )+\sqrt{2\xi
k_B
T \alpha M}\eta(t).
\end{eqnarray}
This stochastic equation
has been used in Sec. V of \cite{tunnel} to determine the
thermal lifetime of the
metastable state of a self-gravitating BEC with an attractive self-interaction
close to the
maximum mass for arbitrary values of $\xi$. Below, we consider particular cases
of this equation and give different examples of application.

\subsection{Dissipationless case without noise}

In the dissipationless case ($\xi=0$), and in the absence of noise ($\eta=0$),
Eq. (\ref{pain3}) reduces to
\begin{eqnarray}
\label{pain4}
\alpha M\frac{d^2R}{dt^2}=-a(R-R_*)^2+b_0\left
(1-\frac{M}{M_{\rm max}}\right ).
\end{eqnarray}
This equation can be derived from the deterministic (quantum) Euler-Poisson
equations. It
has been used in \cite{bectcoll} to determine the collapse time
of a self-gravitating BEC with an attractive self-interaction close to the
maximum mass $M_{\rm max}$. It could be used similarly to determine the collapse
time of a 
self-gravitating classical gas described by the equation of state (\ref{ssp5}) close to the
minimum temperature $T_{\rm min}$ \cite{pomeau0,pomeau,pomeauL}.

\subsection{Dissipationless case with noise}

In the dissipationless case ($\xi=0$), Eq.
(\ref{pain3}) reduces to
\begin{eqnarray}
\label{pain5}
\alpha M\frac{d^2R}{dt^2}=-a(R-R_*)^2+b_0\left
(1-\frac{M}{M_{\rm max}}\right )\nonumber\\
+\sqrt{2\xi
k_B
T \alpha M}\eta(t).\quad
\end{eqnarray}
This equation can be derived from the  stochastic (quantum) Euler-Poisson
equations. It
has been used in \cite{tunnel} to determine the thermal lifetime of the
metastable state of a self-gravitating BEC with an attractive self-interaction
close to the
maximum mass $M_{\rm max}$. It could be used similarly to
determine the lifetime of the metastable state of a self-gravitating
classical gas described by 
the
equation of state (\ref{ssp5}) close to the minimum
temperature $T_{\rm min}$ \cite{pomeau0,pomeau,pomeauL}.

\subsection{Overdamped case without noise}

In the strong friction limit ($\xi\rightarrow +\infty$), and in the absence of
noise
($\eta=0$), Eq.
(\ref{pain3}) reduces to
\begin{eqnarray}
\label{pain6}
\xi\alpha
M\frac{dR}{dt}=-a(R-R_*)^2+b_0\left
(1-\frac{M}{M_{\rm max}}\right ).
\end{eqnarray}
This equation can be derived from the deterministic (quantum)
Smoluchowski-Poisson
equations. It has
been used in Appendix \ref{sec_collsf} to determine the
collapse time
of an overdamped self-gravitating BEC with an attractive self-interaction close
to the maximum mass $M_{\rm max}$, and in Sec. \ref{sec_sfbp} to determine
the collapse time
of an overdamped self-gravitating classical Brownian gas described by the
equation of
state (\ref{ssp5})
close to the minimum temperature $T_{\rm min}$. This
equation has also been used in \cite{blowup}  to determine the
collapse time of a self-gravitating classical isothermal  Brownian gas
confined within a box close to the minimum temperature $T_c$. In that case, it
was
derived  by
making a systematic
expansion of the Smoluchowski-Poisson equations close to
the minimum temperature.\footnote{We cannot obtain the normal form of the potential from a Gaussian ansatz because of the presence of the box.} Eq. (\ref{pain6}) (with a different
interpretation of the
parameters) also governs the average 
evolution of the energy $E$ (in place of $R$)  [see Eq. (\ref{lan1}) with
$\eta=0$] of 
an underdamped virialized
self-gravitating classical isothermal Brownian gas confined within a box \cite{lifetime,n2d3}.\footnote{In \cite{lifetime,n2d3} it is assumed that the system has reached a state of mechanical equilibrium (virialized state) and that it evolves towards a state of thermodynamical equilibrium. In that case, the quantity of interest is $P(E,t)$. By contrast, in \cite{pomeau0,pomeau,pomeauL}, it is assumed that the system has reached a state of local thermodynamical equilibrium (LTE) and that it evolves towards a state of mechanical equilibrium. In that case, the quantities of interest are $\rho({\bf r},t)$ and ${\bf u}({\bf r},t)$. Therefore, although the two systems are underdamped ($\xi\rightarrow 0$), they have a very different evolution. Overdamped systems ($\xi\rightarrow +\infty$) are also in a situation when they have reached a LTE and evolve towards a state of mechanical equilibrium. In that case, the quantity of interest is 
$\rho({\bf r},t)$.} It
has been
used in \cite{n2d3} to determine the collapse time of this system close to the
minimum temperature $T_c$.

\subsection{Overdamped case with noise}

In the strong friction limit ($\xi\rightarrow +\infty$), Eq.
(\ref{pain3}) reduces to
\begin{eqnarray}
\label{pain7}
\xi\alpha
M\frac{dR}{dt}=-a(R-R_*)^2+b_0\left
(1-\frac{M}{M_{\rm max}}\right )\nonumber\\
+\sqrt{2\xi
k_B
T \alpha M}\eta(t).
\end{eqnarray}
This equation can be derived from the stochastic
(quantum) Smoluchowski-Poisson equations.
It has been used in Sec. \ref{sec_ttbec} to determine the thermal lifetime of
the
metastable state of an overdamped self-gravitating BEC with an attractive
self-interaction
close to the
maximum mass $M_{\rm max}$, and in Sec. \ref{sec_sfbp}  to
determine the thermal lifetime of the metastable state of an overdamped
self-gravitating classical
Brownian gas described by the
equation of state (\ref{ssp5})  close to the minimum
temperature $T_{\rm min}$. Eq. (\ref{pain7}) (with a different interpretation of
the
parameters) also governs the stochastic evolution of the
energy $E$ (in place of $R$) [see Eq. (\ref{lan1})]  of an underdamped
virialized
self-gravitating classical isothermal Brownian gas confined within a box \cite{lifetime,n2d3}. It
has been used in
\cite{n2d3} to determine the thermal lifetime of the metastable state of this
system close to the minimum temperature $T_c$.

\subsection{Stochastic damped Painlev\'e equation}

Let us finally consider the case where the mass $M$ of
the system (e.g. an axion star) slowly
increases with time by accreting matter around it. We  assume an
evolution
of the form $M(t)=M_{\rm
max}(1+Kt)$, where the origin of time is
chosen when $M=M_{\rm max}$. Therefore, $M(t)\le M_{\rm max}$  when $t\le 0$ and
$M(t)=M_{\rm max}$ when
$t=0$. In this manner, an initially stable system ($t<0$) becomes unstable
at
$t=0$
and collapses. The equation of motion (\ref{pain3}) can then be
written as 
\begin{eqnarray}
\label{pain8}
\alpha M\frac{d^2R}{dt^2}+\xi\alpha
M\frac{dR}{dt}=-a(R-R_*)^2-b_0K t\nonumber\\
+\sqrt{2\xi
k_B
T \alpha M}\eta(t).
\end{eqnarray}
It can be viewed  as a  stochastic damped Painlev\'e I equation.

In the dissipationless case ($\xi=0$), and in the absence of noise ($\eta=0$),
Eq.
(\ref{pain8}) reduces to the ordinary (deterministic) Painlev\'e I equation
\begin{eqnarray}
\label{pain9}
\alpha M\frac{d^2R}{dt^2}=-a(R-R_*)^2-b_0K
t.
\end{eqnarray}
This equation has
been studied in detail  in a model of supernova \cite{pomeau0,pomeau,pomeauL} where the star
is treated as a  self-gravitating classical gas described by the equation of
state
(\ref{ssp5}) which displays a saddle-node bifurcation close to a minimum
temperature
$T_{\rm min}$. In that
case, the temperature $T(t)$ is assumed to decrease with time until it reaches
its critical value.
The Painlev\'e I equation (\ref{pain9}) exhibits
a striking transition between a slow dynamics and a fast dynamics, which
is marked by the
implosion of the core and the explosion of the halo. The implosion-explosion
of the star is identified with the supernova phenomenon. The Painlev\'e I
equation (\ref{pain9}) could also be
applied to a
self-gravitating BEC with an attractive self-interaction close to the maximum
mass by assuming that the mass of the BEC (axion star) slowly
increases because of a continuous inflow of particles (accretion).

In the strong friction limit ($\xi\rightarrow +\infty$), and  in the absence of
noise ($\eta=0$), Eq. (\ref{pain8}) reduces to 
\begin{eqnarray}
\label{pain10}
\xi\alpha
M\frac{dR}{dt}=-a(R-R_*)^2-b_0K t.
\end{eqnarray}
This equation describes a situation where the system remains for a long time in the quasistable state  until the time $t=0$ at which the barrier of potential disappears and the system undergoes a fast collapse.  For $t<0$, the system is at ``equilibrium'' in the minimum of the slowly changing potential. Its radius is $R_e(t)=R_*+\sqrt{-b_0Kt/a}$.  If we consider a perturbation of the form $R(t)=R_e(t)+\delta R(t)$ with $\delta R\ll R_e$, we obtain the differential equation $\delta \dot R+(2a/\xi \alpha M)\sqrt{-b_0Kt/a}\,\delta R=(1/2)\sqrt{-b_0K/at}$ whose solution is 
\begin{eqnarray}
\label{pain10j}
\delta R(t)=e^{\frac{4a}{3\xi\alpha M}\sqrt{\frac{b_0K}{a}}(-t)^{3/2}}\nonumber\\
\times \frac{1}{2}\sqrt{\frac{b_0K}{a}}\int_{-\infty}^t   e^{-\frac{4a}{3\xi\alpha M}\sqrt{\frac{b_0K}{a}}(-\tau)^{3/2}}\, \frac{d\tau}{\sqrt{-\tau}}.
\end{eqnarray}
At $t=0$, we get  $\delta R(0)=(3b_0K\xi\alpha M/4a^2)^{1/3}\Gamma(4/3)$.  For $t\gtrsim 0$, the radius decreases slowly as $R(t)=R_*-b_0Kt^2/(2\xi\alpha M)$ according to the differential equation $\xi\alpha
M\frac{dR}{dt}=-b_0K t$. Then, the system collapses and displays a  finite time singularity. For $t\rightarrow t_{\rm coll}$, the radius behaves as $R(t)=R_*-\frac{\xi\alpha M}{a}(t_{\rm coll}-t)^{-1}$, which is the solution of the differential equation $\xi\alpha M\frac{dR}{dt}=-a(R-R_*)^2$. Eq. (\ref{pain10}) could be applied to an overdamped self-gravitating
classical Brownian gas described by the
equation of state (\ref{ssp5}), to an overdamped self-gravitating classical
isothermal
Brownian gas confined within a box 
\cite{blowup}, to an overdamped self-gravitating BEC
with an attractive self-interaction close to the maximum mass, or to an
underdamped virialized self-gravitating classical isothermal Brownian gas
enclosed within a
box (with $E$ in place
of $R$) by
assuming, in each case, that the mass or the temperature of the system slowly
evolves with time up to the critical point. It would also be interesting to study how the fluctuations (noise)
present in the stochastic Painlev\'e I equation (\ref{pain8}) affect the
evolution of the system.

\section{Conclusion}
\label{sec_conclusion}

In this paper, we have computed the thermal activation rate of dilute axion stars close to the maximum mass $M_{\rm max}$ \cite{prd1}. We have shown that the exponent (energy barrier) vanishes as $(1-M/M_{\rm max})^{3/2}$ and  the amplitude as $(1-M/M_{\rm max})^{1/2}$ in the overdamped limit and as $(1-M/M_{\rm max})^{1/4}$ in the weak friction limit. The same scalings were previously obtained in the case of globular clusters close to the point of gravitational collapse \cite{lifetime}. By contrast, in the quantum case, for the tunneling rate,  the bounce exponent  vanishes as $(1-M/M_{\rm max})^{5/4}$ and  the amplitude as $(1-M/M_{\rm max})^{7/8}$ \cite{tunnel} as previously found in the case of laboratory BECs confined by  a harmonic potential \cite{stoof,leggett,huepe}. These scalings reflect the universal form of the potential close to a saddle-node bifurcation. Despite these attenuation factors, the lifetime of dilute axion stars generically scales as $e^{N}$. In the case of axion stars, the number of bosons is of the order $N\sim 10^{50}-10^{100}$ implying that the lifetime of dilute axion stars is considerable. As a matter of fact, these metastable states can be considered as stable states \cite{phi6}, except extremely close to the critical point.\footnote{At the critical point ($M=M_{\rm max}$), using finite size scaling arguments, we have shown that the collapse time of overdamped systems scales as $t_{\rm coll}^{\rm T}\sim N^{1/3}\xi t_D^2$   in the thermal case. By comparison,  the collapse time of
dissipationless systems at criticality scales as $t_{\rm
coll}^{\rm Q}\sim N^{1/5}t_D$ in the quantum case and as  $t_{\rm
coll}^{\rm T}\sim N^{1/6}t_D$  in the thermal
case. The
collapse time at criticality
is smaller than the age of the universe for QCD axion stars and
larger than the age of the universe for the quantum cores of DM
halos made of ULAs \cite{tunnel}.} Barrier activation and quantum tunneling are notoriously slow processes. Similar results regarding the very long lifetime of metastable states have been previously obtained in the case of systems with long-range interactions \cite{langerspin,art,lifetime,cdkramers,bmf},  neutron stars \cite{htww}, quantum field theory at zero or nonzero temperature in the early universe \cite{linde77,linde80,linde81,linde83,gw,cook,billoire,steinhardt81,witten,hm,chh}, laboratory BECs \cite{stoof,leggett,huepe}, globular clusters \cite{katzokamoto,lifetime} etc. In all these examples, quantum and thermal tunneling processes are very rare events. They can, however, have important physical consequences.  They  can be observed in laboratory BECs \cite{stoof,leggett,huepe} and in numerical simulations  of systems with long-range interactions \cite{langerspin,art,lifetime,cdkramers,bmf}. They can also advance the onset of gravitational collapse of globular clusters in astrophysics \cite{katzokamoto,lifetime}.\footnote{The onset of collapse is shifted by a factor proportional to $N^{-2/3}$ in the thermal case and by a factor proportional to $N^{-4/5}$ in the quantum case \cite{tunnel}.}

Our results show that the thermal activation rate and the tunneling rate of dilute axion stars are usually negligible. This strengthens the validity of our former results \cite{prd1,bectcoll,phi6} where these effects were neglected from the start. This is because, in axion stars, the number of bosons $N$ is huge. However, our results may be useful to interpret future numerical simulations and laboratory experiments trying to reproduce  ``axion stars'' with a limited number of particles. In that case, thermal and quantum fluctuations will be more important and possibly measurable. On the other hand, our results may apply to systems of classical Brownian particles in interaction described by the stochastic Smoluchowski equation: self-gravitating Brownian particles \cite{cdkramers}, the BMF model \cite{bmf}, the Keller-Segel \cite{ks,sks} model of chemotaxis... For these systems, the number of particles can be relatively small so that metastability and random transitions can be observed. This is illustrated in Ref. \cite{cdkramers} for a 1D model of self-gravitating Brownian particles or bacterial colonies. For these systems, a  detailed derivation of the  thermal activation rate is required. Since our general formalism covers a wide range of physical  situations, we have presented our results in a  rather exhaustive manner.

To obtain our results we started from the generalized GPP equations (\ref{g1}) and (\ref{g2}) which possess a temperature term and a friction term. We transformed them into hydrodynamic equations [see Eqs. (\ref{g3})-(\ref{g5})] and noted the analogy with the hydrodynamic equations describing classical self-gravitating Brownian particles (they differ by the presence of the quantum potential $Q$ which may be interpreted as arising from correlations between the Brownian particles due to short-range interactions) \cite{back}. Using this analogy, we introduced a stochastic term in the quantum Euler (and Smoluchowski) equations as in the case of classical Brownian particles in interaction \cite{paper5} (see Appendix  \ref{sec_tf}).  We can then apply to these equations the instanton formalism based on the Onsager-Machlup functional in order to obtain the activation rate, like in our previous papers \cite{cdkramers,entropy2} (we just have to account for the presence of the quantum kinetic energy $\Theta_Q$ in the free energy functional). However, in the present paper, we proceeded differently. By making a Gaussian ansatz, we transformed the stochastic hydrodynamic equations into a stochastic Langevin equation for a fictive particle in a one-dimensional potential.\footnote{This equation can also be introduced in a more direct manner by introducing a friction term and a noise term in the equation of motion obtained from the GPP equations by making a Gaussian ansatz.} We then applied the instanton theory directly to this equation in order to obtain the activation rate.

In the present paper, we have taken into account thermal fluctuations but not quantum fluctuations. By contrast, in our previous paper \cite{tunnel}, we have taken into account only quantum fluctuations. Of course, a more general approach should take into account both quantum and thermal fluctuations.\footnote{Preliminary results are given in Appendix G of \cite{nottale2} in the high temperature limit.} In this respect, we note that our treatment of thermal fluctuations based on the stochastic quantum Smoluchowski equation  is not standard, and at most approximate. It is exact in the classical case but not in the quantum case. Indeed, the generalized GP equation (\ref{g1}) from which the quantum Smoluchowski equation (\ref{g8}) is derived does not account for thermal effects in a rigorous manner.\footnote{For example, the quantum Smoluchowski equation with a fixed external potential does not relax towards the Wigner distribution (or toward the Bloch distribution in the case of a harmonic potential) predicted by quantum statistical mechanics \cite{nottale2}.}.  Since we consider  only thermal effects in the present paper  (quantum effects are only used to stabilize the star against gravitational collapse), our treatment should be reliable. However, it would be interesting to compare our results with those obtained from more rigorous formalisms. This will be considered in future works.

\appendix

\section{Thermodynamics of self-gravitating Brownian particles in the canonical ensemble}
\label{sec_tcano}

In this Appendix, we consider the thermodynamics of a system of self-gravitating Brownian particles in the canonical ensemble (see Sec. 
\ref{sec_gcb} and \cite{n2d3} for a more detailed treatment). The results of
this Appendix are actually very general and can be applied to systems different
from self-gravitating Brownian particles. At statistical equilibrium, the 
distribution of energy is given by\footnote{In the canonical ensemble, the
$N$-body distribution is $P_N=\frac{1}{Z}e^{-\beta H}$ where $H$ is the
Hamiltonian and $Z=\int e^{-\beta H}\,   \prod_{i=1}^N d{\bf r}_id{\bf v}_i$ is
the partition function. Introducing the entropy $S(E)=k_B \ln g(E)$, where
$g(E)=\int \delta (E-H)\, \prod_{i=1}^N d{\bf r}_id{\bf v}_i$ is the density of
states, we can write the partition function as  $Z=\int e^{-\beta E}g(E)\,
dE=\int e^{-\beta E+S(E)/k_B}\, dE$, leading to Eqs. (\ref{tcano1}) and
(\ref{tcano2}).} 
\begin{eqnarray}
P(E)=\frac{1}{Z}e^{-\beta F(E)}, \qquad Z=\int e^{-\beta F(E)}\, dE,
\label{tcano1}
\end{eqnarray} 
with
\begin{eqnarray}
 F(E)=E-TS(E).
 \label{tcano2}
\end{eqnarray} 
This is the analogous of the Boltzmann distribution from Eq. (\ref{kra3}), where $E$ plays the role of $x$ and $F(E)$ the role of $V(x)$. The control parameter is the temperature $T$ and the critical point corresponds to $T_c$ (see Sec. 
\ref{sec_gcb}).  The most probable energy $E_S$ at temperature $T$ corresponds to the minimum  of $F(E)$, i.e., $F'(E_S)=0$ and $F''(E_S)>0$. The moments of the energy are
\begin{eqnarray}
\langle E^n\rangle=\int P(E)E^n\, dE=(-1)^n\frac{1}{Z}\frac{d^nZ}{d\beta^n}.
\label{tcano9}
\end{eqnarray} 
The first moment $\langle E\rangle$ is the average energy. In general $\langle
E\rangle\neq E_S$.

An equilibrium state corresponds to $F'(E_e)=0$. Using Eq. (\ref{tcano2}) we
obtain the   temperature-energy relation
\begin{eqnarray}
\frac{1}{T}=S'(E_e).
\label{tcano3}
\end{eqnarray} 
This equation determines the series 
of equilibria (caloric curve). The equilibrium state is stable (most probable)
if $F''(E_e)>0$ (minimum of free energy) and unstable (least probable) if
$F''(E_e)<0$ (maximum of free energy).  Taking the derivative of Eq.
(\ref{tcano3}) with respect to $E_e$ and using the relation  $F''(E)=-TS''(E)$
resulting from Eq. (\ref{tcano2}) we obtain the identity
\begin{eqnarray}
\frac{dT}{dE_e}=TF''(E_e).
\label{tcano4}
\end{eqnarray} 
This is a form of Poincar\'e turning point criterion. It relates the slope of the caloric curve (specific heat $C=dE_e/dT$) to the stability of the system (second variations of $F(E)$). In the canonical ensemble, the equilibrium state is stable ($F''>0$) when the specific heat is positive ($C>0$) and unstable ($F''<0$) when the specific heat is negative ($C<0$). The change of stability corresponds to the turning point of temperature ($dT=0$) where the  specific heat diverges ($C\rightarrow +\infty$).

For self-gravitating Brownian particles in the underdamped limit, the evolution
of the distribution of energy $P(E,t)$ is governed by the Fokker-Planck equation
(\ref{kq3}).   At statistical equilibrium, we recover the canonical distribution
from Eq. (\ref{tcano1}). Considering the equation of motion (\ref{lan1}), we
introduce the notation 
\begin{eqnarray}
\omega^2=F''(E_e),
\label{tcano5}
\end{eqnarray}
which is analogous to the squared pulsation of a material particle in the
ordinary mechanical problem (here, however, $\omega$ has not the dimension
of a pulsation).

Expanding the free energy for small fluctuations around an equilibrium state,
writing $F(E)=F(E_e)+F'(E_e)(E-E_e)+\frac{1}{2}F''(E_e)(E-E_e)^2+...$, and using
$F'(E_e)=0$, we obtain the Gaussian distribution 
\begin{eqnarray}
P_e(E)=\sqrt{\frac{\beta F''(E_e)}{2\pi}}\, e^{-\frac{1}{2}\beta F''(E_e)(E-E_e)^2}.
\label{tcano6}
\end{eqnarray}
In the Gaussian approximation, the average value of the energy $\langle
E\rangle$ coincides with its most probable value $E_e$ (corresponding to the
minimum of the free energy $F(E)$ determined by $F'(E_e)=0$ and $F''(E_e)>0$)
and its
variance is 
\begin{eqnarray}
\langle (\Delta E)^2\rangle=\frac{1}{\beta F''(E_e)}=\frac{1}{\beta\omega^2},
\label{tcano7}
\end{eqnarray}
where $\Delta E=E-E_e$ is the fluctuation of energy about equilibrium.
Combining Eq. (\ref{tcano7}) with Eq. (\ref{tcano4}), we obtain the relation
\begin{eqnarray}
\frac{dT}{dE_e}=\frac{1}{C}=T F''(E_e)=T \omega^2=\frac{k_B T^2}{\langle (\Delta E)^2\rangle},
\label{tcano8}
\end{eqnarray}
which corresponds to the fluctuation-dissipation theorem in thermodynamics. 
Although this relation has been obtained here in the Gaussian approximation, the
fluctuation-dissipation theorem turns out to be exact. It can be directly
derived from Eq. (\ref{tcano1}) by calculating the variance $\langle (\Delta
E)^2\rangle=\langle E^2\rangle-\langle E\rangle$ and the specific heat
$d\langle E\rangle/T$ from Eq. (\ref{tcano9}) (see, e.g., \cite{n2d3} for
details). This leads to the Gibbs-Einstein \cite{gibbs,einsteinfd} relation 
\begin{eqnarray}
C\equiv \frac{d\langle E\rangle}{dT}=k_B\beta^2 \langle (\Delta E)^2\rangle.
\label{tcano11}
\end{eqnarray}

Close to the critical point $T_c$, we can use the normal form of the free energy from Eq. (\ref{e7}).  The energies of the metastable ($+$) and unstable ($-$) equilibrium states are given by
\begin{eqnarray}
E_e&=&E_0\pm \sqrt{\frac{2}{-\beta''(E_0)}}(\beta_c-\beta)^{1/2}\nonumber\\
 {\rm i.e.}\qquad \beta&=&\beta_c+\frac{1}{2}\beta''(E_0)(E_e-E_0)^2.
\label{tcano12}
\end{eqnarray}
For the stable equilibrium state, we have
\begin{eqnarray}
C&=&\frac{\beta_c^2}{2}\sqrt{\frac{2}{-\beta''(E_0)}}
(\beta_c-\beta)^{-1/2},\nonumber\\
 \omega^2&=&\frac{2}{\beta_c}\sqrt{\frac{-\beta''(E_0)}{2}}
(\beta_c-\beta)^{1/2},\nonumber\\
   \langle (\Delta E)^2\rangle&=&\frac{1}{2}\sqrt{\frac{2}{-\beta''(E_0)}}
(\beta_c-\beta)^{-1/2}.
\label{tcano13}
\end{eqnarray}
Therefore, at the critical point, the specific heat diverges as
$(\beta_c-\beta)^{-1/2}$, 
the squared pulsation vanishes as $(\beta_c-\beta)^{1/2}$, and the variance of
the energy diverges as $(\beta_c-\beta)^{-1/2}$. This is similar to the
phenomenon of critical opalescence in the theory of liquids \cite{opal1,opal2}
(see also \cite{paper5}). Using Eqs. (\ref{e7}) and (\ref{tcano12}), we find
that the barrier of free energy between the metastable state and the unstable
state is given by
\begin{eqnarray}
\Delta F=\frac{4}{3\beta_c}\sqrt{\frac{2}{-\beta''(E_0)}}(\beta_c-\beta)^{3/2}.
\label{deltaf}
\end{eqnarray}

In the Gaussian approximation, one can show that for $D_0\omega^2 t/k_B T
\gg 1$, i.e., in the stationary state, the temporal correlation function of
the fluctuations of energy $\Delta E(t)$ reads (see Appendix \ref{sec_summary})
\begin{eqnarray}
\langle \Delta E(0)\Delta E(t)\rangle=\frac{k_B T}{\omega^2}e^{-D_0\omega^2 t/k_B T}.
\label{tcano14}
\end{eqnarray}
Close to the critical point  $T_c$ where $\omega\rightarrow 0$, 
the variance of the fluctuations  $\langle (\Delta
E)^2\rangle={k_B T}/{\omega^2}$ (prefactor), and the correlation time
$t_{\rm corr}=k_B T/D_0\omega^2$ (exponent),
which coincides with the relaxation
(damping) time $t_{\rm relax}$, both diverge as $\omega^{-2}$.

\section{Summary of the basic formulas valid close to a saddle-node bifurcation}
\label{sec_summary}

In this Appendix, we summarize the basic formulas which are valid for a Brownian
particle close to a saddle-node bifurcation.

We consider the stochastic Langevin
equation 
\begin{eqnarray}
m\frac{d^2x}{dt^2}+\xi m\frac{dx}{dt}=-\frac{dV}{dx}+\sqrt{2\xi m
k_B T}\, \eta(t),
\label{lan}
\end{eqnarray} 
where $\eta(t)$ is a Gaussian white noise such that $\langle \eta(t)\rangle=0$
and
$\langle \eta(t)\eta(t')\rangle=\delta (t-t')$. We assume that the potential
has the form
\begin{eqnarray}
\label{qqtr18}
V(x)=\frac{1}{3}ax^3-bx,
\end{eqnarray}
where $a>0$ and $b$ can be of any sign. This is
the normal form of a potential $V(x)$ close to a saddle-node
bifurcation (see Fig. \ref{normalform}). The stochastic Langevin equation (\ref{lan}) then becomes
\begin{equation}
m\frac{d^2x}{dt^2}+\xi m\frac{dx}{dt}=-ax^2+b+\sqrt{2\xi m
k_B T}\, \eta(t).
\label{lanexp}
\end{equation} 
The control parameter is $b$ and the critical point corresponds to
$b_c=0$.

We first neglect the fluctuations in Eq. (\ref{lan}) by taking $T=0$. An
equilibrium state corresponds to $V'(x_e)=0$. When $b<0$ the
potential $V(x)$ is monotonically increasing and there is no equilibrium state.
When $b>0$ there are two equilibrium states (extrema of $V$) given by
\begin{eqnarray}
x_e=\pm\sqrt{\frac{b}{a}}.
\label{qxm}
\end{eqnarray}
The upper sign corresponds to the metastable equilibrium state (local minimum)
$x_M$ and the lower sign corresponds to the unstable state (local maximum)
$x_U$. The curve $x_e(b)$ defines a saddle-node bifurcation. The inverse 
curve $b(x_e)$ reads 
\begin{eqnarray}
b=a x_e^2.
\label{qxmc}
\end{eqnarray}
The squared pulsation around an equilibrium state $x_e$
(extremum of the potential) is given by
\begin{eqnarray}
\omega^2=\frac{V''(x_e)}{m}=\pm\frac{2}{m}\sqrt{ab}.
\label{spu}
\end{eqnarray}
Taking the derivative of Eq. (\ref{qxmc}) with respect to
$x_e$, we get
\begin{eqnarray}
\frac{db}{dx_e}=\pm 2\sqrt{ab}.
\label{qxmd}
\end{eqnarray}
Combining Eqs. (\ref{spu}) and (\ref{qxmd}), we obtain
\begin{eqnarray}
\omega^2=\frac{1}{m}\frac{db}{dx_e}.
\label{poi}
\end{eqnarray}
This formula can be interpreted as a form of Poincar\'e turning point
criterion. The
pulsation vanishes ($\omega=0$) at the
turning point of the control parameter ($b'(x_e)=0$). In that case, the equilibrium is marginally stable. More precisely, this
equation
relates
the stability of the equilibrium solutions to the slope of the series of equilibria $b(x_e)$. The equilibrium states $x_e>0$
are stable ($\omega^2>0$) since $b(x_e)$ increases ($db/dx_e>0$) and the
equilibrium states $x_e<0$
are unstable ($\omega^2<0$) since $b(x_e)$ decreases ($db/dx_e<0$). Therefore,
if we follow the series of equilibria $b(x_e)$, the change of stability occurs
at the turning point of the control parameter $b$. Finally, the value of the
potential at the equilibrium states is
\begin{eqnarray}
V_e=\mp \frac{2}{3}a\left (\frac{b}{a}\right )^{3/2}.
\label{ve}
\end{eqnarray}
The barrier of potential between the metastable state and the unstable
state is
\begin{eqnarray}
\Delta V=V(x_U)-V(x_M)=\frac{4}{3}a\left (\frac{b}{a}\right )^{3/2}.
\label{diffve}
\end{eqnarray}

If we take fluctuations into account in Eq. (\ref{lan}), the probability density
$P(x,v,t)$ of finding the particle in $x$ with velocity $v$ at time $t$ is
governed by the Kramers (or Klein-Kramers) \cite{klein,kramers} equation
\begin{eqnarray}
\frac{\partial P}{\partial t}+v\frac{\partial P}{\partial
x}-\frac{V'(x)}{m}\frac{\partial P}{\partial v}=\frac{\partial}{\partial
v}\left ({\cal D}\frac{\partial P}{\partial v}+\xi P v\right ).\quad
\label{kra1}
\end{eqnarray}
The diffusion coefficient in velocity space is given by an Einstein relation of the form \cite{chandrarevue}
\begin{eqnarray}
{\cal D}=\frac{\xi k_B T}{m}.
\label{kra2}
\end{eqnarray}
At statistical equilibrium, we have the Boltzmann distribution 
\begin{eqnarray}
P_e(x,v)&=&A'\, e^{-\beta \left \lbrack m \frac{v^2}{2}+V(x)\right
\rbrack}\nonumber\\
 \Rightarrow \qquad P_e(x)&=&A\, e^{-\beta V(x)}.
\label{kra3}
\end{eqnarray}
The most probable position of the particle at equilibrium corresponds to the minimum $x_M$ of $V(x)$. Expanding the potential for small fluctuations around an equilibrium state, writing $V(x)=V(x_e)+V'(x_e)(x-x_e)+\frac{1}{2}V''(x_e)(x-x_e)^2+...$, and using $V'(x_e)=0$, we obtain the Gaussian distribution 
\begin{eqnarray}
P_e(x)=\sqrt{\frac{\beta V''(x_e)}{2\pi}}\, e^{-\frac{1}{2}\beta V''(x_e)(x-x_e)^2}.
\label{kra4}
\end{eqnarray}
It coincides with the Boltzmann distribution for a 
particle submitted to a harmonic potential
$V(x)=V(x_e)+\frac{1}{2}m\omega^2 (x-x_e)^2$. The average value of the
position $\langle x\rangle$
coincides with its most probable value $x_e$ (corresponding to the minimum of
the potential $V(x)$ determined by $V'(x_e)=0$ and $V''(x_e)>0$) and its
variance is 
\begin{eqnarray}
\langle (\Delta x)^2\rangle=\frac{1}{\beta V''(x_e)}=\frac{1}{\beta m
\omega^2},
\label{kra5}
\end{eqnarray}
where $\Delta x=x-x_e$ is the fluctuation about equilibrium. Combining Eq.
(\ref{kra5}) with Eq. (\ref{poi}) we obtain the relation
\begin{eqnarray}
\frac{db}{dx_e}=m \omega^2=V''(x_e)=\frac{1}{\beta \langle (\Delta
x)^2\rangle},
\label{kra6}
\end{eqnarray}
which is analogous to the fluctuation-dissipation theorem 
in thermodynamics (see Appendix \ref{sec_tcano}). Close to the critical point
$b_c=0$, using Eqs. (\ref{spu}) and (\ref{kra5}), we have for the stable
equilibrium state
\begin{eqnarray}
\omega^2=\frac{1}{m}2\sqrt{ab},\qquad  \langle (\Delta
x)^2\rangle=\frac{1}{2\beta\sqrt{ab}}.
\label{kra7}
\end{eqnarray}
When $b\rightarrow 0^+$, the squared pulsation vanishes as $b^{1/2}$ and the
variance of the fluctuations diverges as $b^{-1/2}$. This is similar to the
phenomenon of critical opalescence in the theory of liquids \cite{opal1,opal2}
(see also \cite{paper5}).

In the strong friction limit $\xi\rightarrow +\infty$, by making as before  a
Gaussian approximation, i.e. 
by approximating the potential by a harmonic potential
$V(x)=V(x_e)+\frac{1}{2}m\omega^2 (x-x_e)^2$, the Langevin equation
(\ref{lan}) becomes
\begin{eqnarray}
\frac{dx}{dt}=-\frac{\omega^2}{\xi}(x-x_e)+\sqrt{\frac{2k_B T}{\xi m}}\,
\eta(t).
\label{lan2}
\end{eqnarray} 
This is an Ornstein-Uhlenbeck \cite{uo} process for the variable $\Delta x(t)=x(t)-x_e$. 
One can show that for $\omega^2 t/\xi \gg 1$, i.e., in the stationary
state, the temporal correlation function of the fluctuations $\Delta x(t)$ reads
\cite{risken}
\begin{eqnarray}
\langle \Delta x(0)\Delta x(t)\rangle=\frac{k_B T}{m\omega^2}e^{-\omega^2
t/\xi}.
\label{kra8}
\end{eqnarray}
The prefactor corresponds to the variance  $\langle (\Delta x)^2\rangle$ 
of the equilibrium distribution [see Eq. (\ref{kra5})]  and the decorrelation
time (exponent) corresponds to the relaxation time $t_{\rm relax}=\xi/\omega^2$
(see Sec. \ref{sec_relt}). This is a manifestation of the regression of fluctuations
\cite{risken}.  At the critical point $b_c=0$, where $\omega\rightarrow 0$, the
variance of the fluctuations and the relaxation time diverge as $\langle (\Delta
x)^2\rangle=k_BT/m\omega^2\rightarrow +\infty$ and
$t_{\rm relax}=\xi/\omega^2\rightarrow +\infty$.

{\it Remark:} For a more general potential of the form $V(x)=f(x)-bg(x)$, the foregoing equations become
\begin{eqnarray}
b=\frac{f'(x_e)}{g'(x_e)},
\label{st2}
\end{eqnarray}
\begin{eqnarray}
\omega^2=\frac{V''(x_e)}{m}=\frac{g'(x_e)f''(x_e)-f'(x_e)g''(x_e)}{m
g'(x_e)},
\label{st3}
\end{eqnarray}
\begin{eqnarray}
\frac{db}{dx_e}=\frac{f''(x_e)g'(x_e)-g''(x_e)f'(x_e)}{g'(x_e)^2},
\label{st4}
\end{eqnarray}
\begin{eqnarray}
\omega^2=\frac{g'(x_e)}{m}\frac{db}{dx_e},
\label{st5}
\end{eqnarray}
\begin{eqnarray}
g'(x_e)\frac{db}{dx_e}=m\omega^2=V''(x_e)=\frac{1}{\beta\langle (\Delta
x)^2\rangle}.
\label{st6}
\end{eqnarray}
This returns the results of Secs. \ref{sec_ga}-\ref{sec_sfbp} with suitable expressions of $f(x)$ and $g(x)$.

\subsection{Dissipationless systems}

For dissipationless systems ($\xi=0$), the stochastic Langevin equation (\ref{lan})
reduces to
\begin{eqnarray}
m\frac{d^2x}{dt^2}=-\frac{dV}{dx}+\sqrt{2\xi m k_B T}\, \eta(t).
\end{eqnarray} 

We have the following results:

(i) The period of the oscillations of the particle about the stable
equilibrium
state when $T=0$ and $b>0$ is given
by (see Eq. (\ref{voy1}) in Sec. \ref{sec_relt} and Eq. (64) of \cite{bectcoll})
\begin{eqnarray}
T_{\rm osc}=\frac{2\pi}{\omega_M}=2\pi\left
(\frac{m^2}{4ab}\right )^{1/4}.
\label{osc}
\end{eqnarray} 
When $b\rightarrow 0^+$, the period of the
oscillations diverges at the critical point as $b^{-1/4}$.

(ii) The collapse time of the particle starting from $x_0=0$ without initial velocity ($\dot x_0=0$) when $T=0$ and
$b<0$  is given
by (see Eq. (115) of \cite{bectcoll})
\begin{eqnarray}
t_{\rm coll}=B\left (\frac{2m^2}{a|b|}\right )^{1/4}
\label{bona}
\end{eqnarray}
with
\begin{eqnarray}
B=\left (\frac{3}{8}\right
)^{1/4}\int_{-\infty}^0\frac{dy}{\sqrt{-y-y^3}}=2.90178...
\end{eqnarray}
When $b\rightarrow 0^-$, the collapse time diverges at the
critical point
as $|b|^{-1/4}$. Interestingly, this
is the same
scaling as for the divergence of the period of the oscillations $T_{\rm osc}\sim
b^{-1/4}$
close
to the critical
point when $b>0$ [see Eq. (\ref{osc})]. The
prefactor is, however, different being $2\pi=6.28319...$ for $T_{\rm osc}$ and
$8^{1/4}B=4.88019...$ for $t_{\rm coll}$. We
symbolically have ``$t_{\rm coll}=0.776707...T_{\rm osc}$''.

(iii) The quantum  lifetime of the  metastable equilibrium state when
$T=0$ and $b>0$  is given
by (see Eq. (98)
of \cite{tunnel})
\begin{equation}
t_{\rm life}^{\rm Q}\sim \frac{1}{12}\left
(\frac{\pi^2m\hbar^2}{8a^3}\right
)^{1/4} (b/a)^{-7/8} e^{\frac{24}{5\hbar}\sqrt{2ma}(b/a)^{5/4}}.
\label{giu1}
\end{equation}
The quantum tunneling rate exponent (equal to the WKB
transmission amplitude) scales as $b^{5/4}$ and the prefactor as $b^{-7/8}$.

(iv) The thermal lifetime of the  metastable equilibrium state when
$\hbar=0$ and $b>0$ is
given
by (see Eq. (97) of \cite{tunnel})
\begin{equation}
t_{\rm life}^{\rm T}\sim 2\pi\left (\frac{m^2}{4ab}\right
)^{1/4} e^{\frac{4a(b/a)^{3/2}}{3k_B T}}.
\label{giu2}
\end{equation}
The thermal activation rate exponent (equal to the potential barrier
$\Delta V$)
scales as $b^{3/2}$ and the prefactor as $b^{-1/4}$. The
scaling of the prefactor is the same as for the period of the oscillations from
Eq. (\ref{osc}) when $b>0$ (they 
coincide)\footnote{Actually, the prefactor of Eq. (\ref{giu2}), corresponding to
the result $2\pi/\omega_M$ of the transition state theory, is not correct. A
more accurate result has been established by Kramers \cite{kramers} (see the
discussion in Sec. V of \cite{tunnel}).} and as for the collapse time from Eq.
(\ref{bona}) when $b<0$ (they differ by a factor $1.28749...$).
We symbolically have
``$t_{\rm life}^{\rm T}=T_{\rm osc}=1.28749\, t_{\rm coll}$''.

(v) Writing that the term in the exponential of Eqs. (\ref{giu1}) and
(\ref{giu2}) is of order unity, we find that
the corrections to the values of the critical points due to quantum and thermal
fluctuations are (see Sec. VI of \cite{tunnel}) 
\begin{equation}
\label{bcorr}
b_c^{\rm Q}=a\left (\frac{25\hbar^2}{1152 m a}\right )^{2/5}\quad
{\rm and}\quad b_c^{\rm T}= a\left (\frac{3k_B T}{4a}\right )^{2/3}.
\end{equation}

(vi)  Using finite size
scaling arguments  we find that the collapse times
at the critical point due to quantum and thermal fluctuations are not infinite
but equal to  (see Sec. VII of \cite{tunnel}) 
\begin{eqnarray}
\label{bona3}
t_{\rm coll}^{\rm Q}\sim B\left (\frac{2m^2}{a^2}\right )^{1/4}\left
(\frac{24}{5\hbar}\right )^{1/5}(2ma)^{1/10} \nonumber\\
 {\rm and} \qquad
t_{\rm coll}^{\rm T}\sim B\left (\frac{2m^2}{a^2}\right )^{1/4}\left
(\frac{4a}{3k_B T}\right )^{1/6}.
\end{eqnarray}

\subsection{Dissipative systems in the strong friction limit}

In the strong friction limit $\xi\rightarrow +\infty$, the stochastic Langevin
equation (\ref{lan}) reduces to
\begin{eqnarray}
\label{qttr0bis}
\xi m\frac{dx}{dt}=-\frac{dV}{dx}+\sqrt{2\xi m k_B T}\, \eta(t).
\end{eqnarray}

We have the following results:

(i) The relaxation (damping) time of the particle towards the stable equilibrium
state when $T=0$ and 
$b>0$ is given by (see Eq. (\ref{voy2}) in Sec. \ref{sec_relt})
\begin{eqnarray}
\label{qbttr6ab}
t_{\rm relax}=\frac{\xi}{\omega_M^2}=\frac{\xi m}{2\sqrt{ab}}. 
\end{eqnarray} 
When $b\rightarrow 0^+$, the relaxation time
diverges at the critical point as $b^{-1/2}$.

(ii) The collapse time of the particle starting from $x_0=0$ when $T=0$ and 
$b<0$ is given by (see Eq. (\ref{tay1}) in Appendix \ref{sec_collsf})
\begin{eqnarray}
t_{\rm coll}=\frac{\pi\xi m}{2\sqrt{a|b|}}.
\label{qtay1}
\end{eqnarray}
When $b\rightarrow 0^-$, the collapse time diverges at the
critical point
as $|b|^{-1/2}$. Interestingly, this
is the same
scaling as for the divergence of the relaxation time $t_{\rm
relax}\sim b^{-1/2}$ close
to the critical
point when  $b>0$ [see Eq. (\ref{qbttr6ab})]. They
differ by a factor $\pi$. We symbolically have ``$t_{\rm coll}=\pi t_{\rm
relax}$''.

(iii) The thermal  lifetime of the metastable equilibrium state when $\hbar=0$
and $b>0$ is given
by (see Eq.
(\ref{b38bis}) in Sec. \ref{sec_ttbec})
\begin{equation}
\label{qb38bis}
t_{\rm life}^{\rm T}\sim\frac{\pi\xi m}{\sqrt{ab}} e^{\frac{4a(b/a)^{3/2}}{3k_B
T}}.
\end{equation}
The thermal activation rate exponent (equal to the potential barrier
$\Delta V$)
scales as $b^{3/2}$ and the prefactor as $b^{-1/2}$. The scaling of the prefactor is the same as
for the
relaxation time from Eq. (\ref{qbttr6ab}) when $b>0$ (they differ by factor $2\pi$) and as for the
collapse time from Eq. (\ref{qtay1}) when $b<0$ (they differ by a factor $2$). We
symbolically have
``$t_{\rm
life}^{\rm T}=2t_{\rm coll}=2\pi t_{\rm relax}$''.

(iv) Writing that the term
in the exponential of Eq. (\ref{qb38bis}) is of order unity, we find that
the correction to the value of the critical point due to thermal fluctuations is
\begin{equation}
\label{bcorr2}
b_c^{\rm T}=\left (\frac{3k_B T}{4a}\right )^{2/3}a.
\end{equation}

(v)  Using finite size
scaling arguments (see Appendix \ref{sec_size}) we find that the collapse time
at the critical point due to thermal fluctuations is not infinite but equal to 
\begin{equation}
\label{bona2}
t_{\rm coll}^{\rm T}\sim \frac{\pi\xi m}{2a}\left (\frac{4a}{3k_B
T}\right )^{1/3}.
\end{equation}

\section{Dissipative self-gravitating BECs in the strong friction limit}
\label{sec_collsf}

In Refs. \cite{bectcoll,tunnel} we have established the expression of the
collapse time of a dissipationless self-gravitating BEC with attractive self-interactions close to the
maximum mass, and discussed finite size scalings. In
this Appendix, we extend these results to the case of a dissipative
self-gravitating BEC in the strong friction limit.

\subsection{Collapse time of self-gravitating BECs close to the maximum mass}
\label{sec_collsfa}

\begin{figure}
\begin{center}
\includegraphics[clip,scale=0.3]{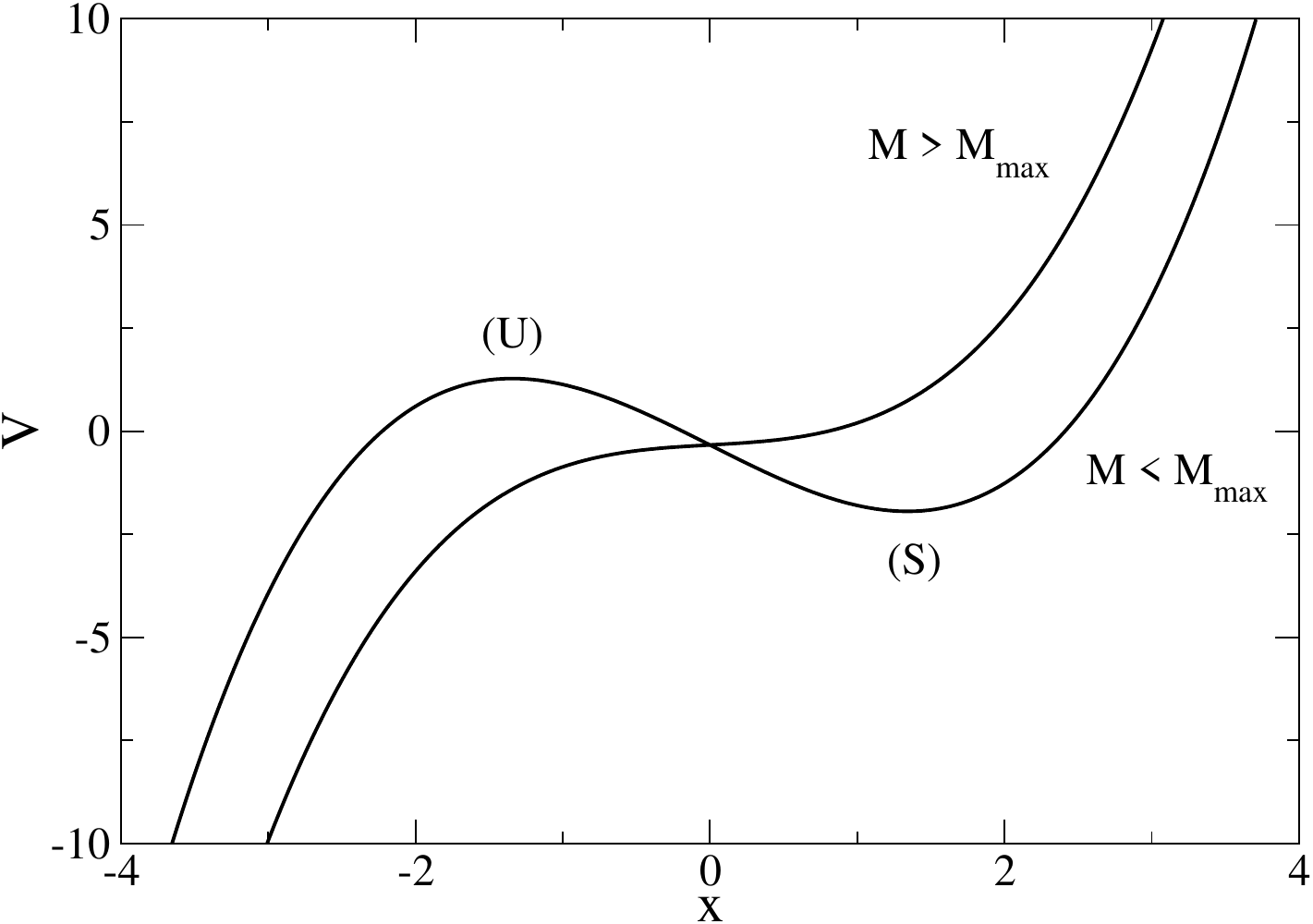}
\caption{Normal form of the effective potential close to the critical point,
corresponding to a saddle-node bifurcation, for $M<M_{\rm max}$ and $M>M_{\rm
max}$.}
\label{saddlenode}
\end{center}
\end{figure}

Let us first determine the collapse time of a BEC with an attractive self-interaction
close to the maximum mass in the strong friction limit $\xi\rightarrow +\infty$. We make a Gaussian
ansatz and adapt the
calculations of \cite{bectcoll} to this situation. We consider the collapse of
the BEC when its mass $M$ is slightly above the maximum mass $M_{\rm
max}$ and when its initial radius $R_0$ is close to the critical radius $R_*$.
Since we are close to the critical point, corresponding to a saddle-node
bifurcation,  we can approximate the potential $V(R)$ by its normal
form from Eq. (\ref{nfcmm1}). Using Eq. (\ref{g27}),  setting $x=R-R_*$, and introducing
the notations from Eqs. (\ref{qtr18}) and (\ref{qtr19}), we obtain the equation of
motion 
\begin{eqnarray}
\xi\alpha M\frac{dx}{dt}=-V'(x)=-ax^2+b.
\label{saddle2}
\end{eqnarray}
This corresponds to Eq. (\ref{ttr0bis}) without the noise term. Since $M\ge
M_{\rm max}$ we have $b\le 0$ (while $a>0$). Calling  $x_0$ the value of $x$ at
$t=0$, we can write the solution of Eq. (\ref{saddle2}) as
\begin{eqnarray}
\int_{x(t)}^{x_0} \frac{dx}{a x^2-b}=\frac{t}{\xi\alpha M}.
\label{saddle5}
\end{eqnarray}
The fictive particle released from $x_0$  runs down the potential in the
direction of increasingly negative values
of $x$ (see Fig. \ref{saddlenode}). Its
trajectory
is given by
\begin{equation}
\tan^{-1}\left (\sqrt{\frac{a}{|b|}}x_0\right )-\tan^{-1}\left
(\sqrt{\frac{a}{|b|}}x\right )=\sqrt{a|b|}\frac{t}{\xi\alpha M}.
\label{saddle5u}
\end{equation}
In particular, when $R_0=R_*$ (i.e. $x_0=0$), we obtain
\begin{eqnarray}
x(t)=-\sqrt{\frac{|b|}{a}}\tan\left (\sqrt{a|b|}\frac{t}{\xi\alpha
M}\right ).
\label{saddle5v}
\end{eqnarray}
The collapse time, corresponding to $x(t_{\rm coll})\rightarrow
-\infty$, is generally given by
\begin{eqnarray}
t_{\rm coll}(M,R_0)=\xi\alpha M\int_{-\infty}^{x_0}
\frac{dx}{a x^2-b}.
\label{saddle5b}
\end{eqnarray}
If the fictive particle starts from $R_0=R_*$ (i.e. $x_0=0$) at $t=0$, the collapse time diverges when $M\rightarrow
M_{\rm max}^+$
(i.e. $b\rightarrow 0^-$) because the
effective potential given by Eq. (\ref{nfcmm1}) presents an inflexion
point
at $R=R_*$ when $M=M_{\rm max}$. Therefore, it remains an infinite time at this position. From Eq. (\ref{saddle5b}), we obtain
\begin{eqnarray}
t_{\rm coll}(M,R_0=R_*)=\xi\alpha M\int_{-\infty}^{0}
\frac{dx}{a x^2-b}.
\label{saddle5c}
\end{eqnarray}
Making the change of variables $y=\sqrt{a/|b|}\, x$, we can rewrite this expression as
\begin{eqnarray}
t_{\rm coll}(M,R_0=R_*)=\xi\alpha M\frac{B}{\sqrt{a|b|}}
\label{tay1}
\end{eqnarray}
with
\begin{eqnarray}
B=\int_{-\infty}^0
\frac{dy}{1+y^2}=\frac{\pi}{2}=1.5708...
\label{saddle11}
\end{eqnarray}
This result can also be directly deduced from Eq. (\ref{saddle5v}).
Returning to the original
variables, we find that the collapse
time is given,  when  $M\rightarrow M_{\rm max}^+$, by
\begin{equation}
t_{\rm coll}(M,R_0=R_*)\sim \xi \alpha M \frac{\pi}{2\sqrt{2}} \frac{R_*^2}{V_0}
\left (\frac{M}{M_{\rm
max}}-1\right )^{-1/2}.
\label{saddle10ts}
\end{equation}
Introducing the dynamical time from Eq. (\ref{nfcmm5}), we get 
\begin{equation}
t_{\rm coll}(M,R_0=R_*)\sim \xi t_D^2 \frac{\pi}{2\sqrt{2}}\left
(\frac{M}{M_{\rm
max}}-1\right )^{-1/2}.
\label{saddle10}
\end{equation}
When $M>M_{\rm max}$, the collapse time diverges at the
critical
point
as $t_{\rm coll}/(\xi t_D^2)\sim 0.785398... (M/M_{\rm
max}-1)^{-1/2}$. This
is the same
scaling as for the divergence of the relaxation (damping) time
$t_{\rm relax}/(\xi t_D^2)\sim 0.353553...(1-M/M_{\rm
max})^{-1/2}$
close
to the critical
point when $M<M_{\rm max}$ [see Eq. (\ref{voy2})]. The two prefactors
differ by a factor $\pi$.

{\it Remark:}  When $M=M_{\rm max}$, corresponding to $b=0$, the trajectory of
the overdamped fictive particle in the potential $V(x)=\frac{1}{3}ax^3$
starting from $x_0\neq 0$ is
\begin{eqnarray}
x=\frac{x_0}{1+\frac{ax_0t}{\xi\alpha M}}.
\label{bz1}
\end{eqnarray}
When $x_0<0$, the particle collapses ($x\rightarrow -\infty$) in a finite time
\begin{eqnarray}
t_{\rm coll}=\frac{\xi\alpha M}{a|x_0|}.
\label{bz2}
\end{eqnarray}
When $x_0>0$, the particle does not collapse. It reaches the inflexion point $x=0$ of the potential in infinite time.\footnote{The conservative case ($\xi=0$) is treated in Sec. IV.B of \cite{bectcoll}. In that case, the particle starting from $x_0>0$ can collapse  ($x\rightarrow -\infty$) in a finite time because it comes to the inflexion point $x=0$ with a nonzero velocity and therefore continues its motion towards $x\rightarrow -\infty$.} However,  the presence of fluctuations ($T\neq 0$) allows the particle to collapse even when $x_0>0$ because, in that case, the particle can cross the inflexion point and continue its motion towards $x\rightarrow -\infty$. 

\subsection{Finite size scaling close to the maximum mass}
\label{sec_size}

In the previous section, we have studied the collapse time of a
self-gravitating BEC with an attractive self-interaction
when $M>M_{\rm max}$ in the strong friction limit. For $M\rightarrow M_{\rm
max}^+$ (with $R_0=R_*$) we have found that [see Eq. (\ref{saddle10})]
\begin{equation}
\label{size1}
\frac{t_{\rm coll}}{\xi t_D^2}\sim \frac{0.785398... }{(M/M_{\rm
max}-1)^{1/2}}.
\end{equation}
This study neglects thermal fluctuations. As a result, it
predicts that the collapse time is infinite
when $M=M_{\rm max}$. Actually, because of thermal fluctuations, the collapse
time should be
large but finite at $M=M_{\rm max}$. This is because thermal fluctuations can displace the fictive particle from the inflexion point and facilitate its collapse. We develop below an argument, similar to
the one developed in Sec. VII of \cite{tunnel}, to estimate
the finite size scaling of the collapse time close to
the maximum mass and its finite value at $M=M_{\rm max}$. We
consider
thermal fluctuations in an overdamped BEC contrary to Ref.
\cite{tunnel} where we considered quantum and thermal fluctuations in a
dissipationless
BEC.

For $M<M_{\rm max}$, we have found that the thermal lifetime of a
self-gravitating BEC with an attractive self-interaction is
given by Eq. (\ref{k1}). Finite size effects enter in the expression of
the metastable state lifetime in the combination $N(1-M/M_{\rm max})^{3/2}$. If
we assume that a similar combination enters in the expression of the collapse
time for $M>M_{\rm max}$, we expect a scaling of the form
\begin{equation}
\label{size2}
\frac{t_{\rm coll}}{\xi t_D^2}\propto (M/M_{\rm max}-1)^{-1/2} F\left\lbrack
N(M/M_{\rm max}-1)^{3/2}\right\rbrack
\end{equation}
with $F(x)\rightarrow 1$ for $x\rightarrow +\infty$ in order to recover Eq.
(\ref{size1}) when $N\rightarrow +\infty$. At $M=M_{\rm max}$, the singular
factor $(M/M_{\rm max}-1)^{-1/2}$ must cancel out implying that $F(x)\sim
x^{1/3}$ for $x\rightarrow 0$. Therefore, at $M=M_{\rm max}$, the collapse time
taking into account thermal fluctuations scales as
 \begin{equation}
\label{size3}
\frac{t_{\rm coll}^{\rm T}}{\xi t_D^2}\propto N^{1/3} \qquad (M=M_{\rm
max}).
\end{equation}
This differs from the scalings $t_{\rm coll}^{\rm Q}\propto
N^{1/5}\, t_D$ (quantum) and $t_{\rm coll}^{\rm T}\propto
N^{1/6}\, t_D$ (thermal) obtained in \cite{tunnel} in the dissipationless case.

\section{Thermal fluctuations in a self-gravitating BEC}
\label{sec_tf}

The generalized GPP equations (\ref{g1}) and (\ref{g2}) describe the
mean evolution of a dissipative self-gravitating BEC with a self-interaction
potential $V(|\psi|^2)$ at
$T=0$.\footnote{Thermal effects can be taken into account by adding a
thermal potential
$V_{\rm th}(|\psi|^2)=\frac{k_B T}{m}|\psi|^2 (\ln|\psi|^2
-1)$ in addition to the self-interaction potential $V_{\rm int}(|\psi|^2)$
as explained in Sec. \ref{sec_be} and in the Remark at the end of Sec.
\ref{sec_ggpp}.}
Therefore, they
ignore thermal
fluctuations. These equations  may have several
equilibrium states, e.g., a stable one (global minimum of free energy), an
unstable one (maximum or saddle point of free energy), and a
metastable one (local minimum of free energy). When $N$ is finite and
$T\neq 0$,
the fluctuations can induce a thermal activation from the metastable 
equilibrium state to the stable state. If we want to describe this process, we
need to take thermal fluctuations into account. To that purpose, we shall use
the analogy with the theory of Brownian particles with long-range interactions
developed in \cite{paper5,longshort,cdkramers,entropy1,entropy2,back} and include fluctuations at the level of the
hydrodynamic equations (\ref{g3})-(\ref{g5}) associated with the generalized GPP equations (\ref{g1})
and (\ref{g2}). This corresponds to the concept of fluctuating hydrodynamics.

\subsection{Fluctuating hydrodynamics}
\label{sec_qes}

By using the general theory of fluctuations developed by Landau and
Lifshitz (see Ref. \cite{ll}, Chap. XVII), we can
obtain the stochastic quantum damped Euler-Poisson equations (see \cite{wg} and Appendix B of
\cite{paper5}) 
\begin{equation}
\label{qes1}
\frac{\partial\rho}{\partial t}+\nabla\cdot (\rho {\bf u})=0,
\end{equation}
\begin{eqnarray}
\label{qes3}
\frac{\partial {\bf u}}{\partial t}+({\bf u}\cdot \nabla){\bf
u}=-\frac{1}{\rho}\nabla P-\nabla\Phi-\frac{1}{m}\nabla
Q\nonumber\\
-\xi{\bf u}-\sqrt{\frac{2\xi k_B
T}{\rho}}\, {\bf R}({\bf r},t),
\end{eqnarray}
\begin{equation}
\label{qes2}
\Delta\Phi=4\pi G\rho,
\end{equation}
where ${\bf R}({\bf r},t)$ is a Gaussian white noise satisfying
$\langle {\bf R}({\bf r},t)\rangle={\bf 0}$ and $\langle 
{R}_i({\bf
r},t){R}_j({\bf r}',t')\rangle=\delta_{ij}\delta({\bf r}-{\bf
r}')\delta(t-t')$. Introducing the free energy functional from Eq.
(\ref{g10}) we can
rewrite the stochastic quantum
damped  Euler equation (\ref{qes3}) under the form 
\begin{equation}
\label{qes5}
\frac{\partial {\bf u}}{\partial
t}+({\bf u}\cdot \nabla){\bf u}=-\nabla \frac{\delta
F}{\delta {\rho}}
-\xi\,  {\bf u}-\sqrt{\frac{2\xi k_B
T}{\rho}}\, {\bf R}({\bf r},t).
\end{equation}
The stochastic quantum damped Euler-Poisson equations (\ref{qes1})-(\ref{qes2})
describe the evolution of the fluctuating density and velocity field while
the deterministic quantum  damped Euler-Poisson equations (\ref{g3})-(\ref{g5})
describe the evolution of the average density and velocity field. When
$T\rightarrow 0$ or $N\rightarrow +\infty$, the noise term in Eqs. (\ref{qes3}) and (\ref{qes5}) tends to zero and
the fluctuations can be neglected.

In the strong friction limit $\xi\rightarrow +\infty$, we obtain the
stochastic quantum Smoluchowski-Poisson equations  \cite{paper5,longshort,cdkramers,entropy1,entropy2,back} \begin{eqnarray}
\label{qs1}
\xi\frac{\partial\rho}{\partial t}=\nabla\cdot\left (\nabla
P+\rho\nabla\Phi+\frac{\rho}{m}\nabla Q\right
)\nonumber\\
+\nabla\cdot \left (\sqrt{2\xi k_B T\rho}\, {\bf R}({\bf r},t)\right ),
\end{eqnarray}
\begin{equation}
\label{qes2b}
\Delta\Phi=4\pi G\rho.
\end{equation}
Introducing the free energy functional from Eq. (\ref{g12}), the stochastic
quantum
Smoluchowski equation can be written as\footnote{We stress that
this equation is fundamentally different from the Dean equation \cite{dean}, which
governs the evolution of the exact density field $\rho_{d}({\bf r},t)=\sum_i m
\delta({\bf r}-{\bf r}_i(t))$ made of a sum of Dirac distributions (see
\cite{paper5,entropy2} for a detailed discussion).}
\begin{eqnarray}
\label{qs2}
\frac{\partial {\rho}}{\partial t}=\nabla\cdot \left (
\frac{1}{\xi} {\rho}\nabla\frac{\delta F}{\delta {\rho}}\right
)+\nabla\cdot \left (\sqrt{\frac{2k_{B}T {\rho}}{\xi}}\, {\bf R}\right ).
\end{eqnarray}
The stochastic quantum Smoluchowski-Poisson equations (\ref{qs1}) and
(\ref{qes2b})
describe the evolution of the fluctuating density field while
the deterministic quantum Smoluchowski-Poisson equations (\ref{g8}) and 
(\ref{g8b})
describe the evolution of the average density field. Eq. (\ref{qs2}) can be viewed as a Langevin equation for the fluctuating density field $\rho({\bf r},t)$. We note that the noise
appearing in Eq. (\ref{qs1}) is multiplicative since it
depends on the density. The time-dependent distribution $P[\rho({\bf r},t),t]$
of the density satisfies a functional Fokker-Planck equation of the form
\begin{eqnarray}
\frac{\partial P}{\partial t}[\rho,t]
&=&-\frac{1}{\xi}\int d{\bf r}\, \frac{\delta}{\delta\rho({\bf
r})}\nonumber\\
&\times&\left\lbrace \nabla\cdot  \rho\nabla \left\lbrack k_B
T\frac{\delta}{\delta\rho({\bf r})}+\frac{\delta F}{\delta\rho({\bf
r})}\right\rbrack P[\rho,t]\right\rbrace.\nonumber\\
\label{fpfonc}
\end{eqnarray}
This Fokker-Planck equation relaxes towards  the Gibbs canonical distribution 
\begin{eqnarray}
\label{gibbs}
P_{\rm eq}[\rho({\bf r})]\propto e^{-\beta F[\rho]}.
\end{eqnarray}
This distribution determines the equilibrium probability density of the macrostate $\rho({\bf r})$. The most probable macrostate $\rho({\bf r})$ at statistical equilibrium minimizes the free energy $F[\rho]$ at fixed mass. When
$T\rightarrow 0$ or $N\rightarrow +\infty$, the noise term in Eqs. (\ref{qs1}) and (\ref{qs2}) tends to zero
and the fluctuations can be neglected. In that case, we recover the
deterministic dynamics.

{\it Remark:} The thermal lifetime of a metastable state of the
functional $F[\rho]$ has been computed in \cite{cdkramers,entropy2} by writing
the solution of the stochastic 
Smoluchowski-Poisson equations (\ref{ssp1}) and (\ref{ssp2}) as a path
integral involving the Onsager-Machlup action, and by applying the theory
of instantons. This procedure leads to the Kramers-like formula from Eq.
(\ref{ssp4}).

\subsection{Density correlation function}

We can use the stochastic quantum Smoluchowski-Poisson equations (\ref{qs1}) and
(\ref{qes2b}) to derive the equilibrium correlation
function of the fluctuations of the density about
a homogeneous distribution of particles with density $\rho$. The detailed
calculations will be reported 
elsewhere \cite{preparation}. They extend those developed in
\cite{paper5,newdean} for classical Brownian particles in interaction. In
Fourier space, we find that the density fluctuation spectrum
$S(k,\omega)\sim \langle |\delta{\hat
\rho}_{k\omega}|^2\rangle$ is given by
\begin{equation}
\label{fd10}
S({k},\omega)=\frac{2\xi k_B T \rho
k^2}{\left\lbrack \frac{\hbar^2k^4}{4m^2}+c_s^2 k^2+ (2\pi)^d{\hat
u}(k)\rho k^2\right\rbrack^2+\xi^2\omega^2}.
\end{equation}
Here $c_s^2=P'(\rho)=\rho V''(\rho)$ is the squared speed 
of sound [see Eq. (\ref{g6sound})] and ${\hat u}(k)$ is the Fourier transform of
the potential of interaction.\footnote{It could be the true potential of
interaction in the case of long-range interactions like gravity ($u=u_{\rm
LR}^{\rm exact}$)  or an effective potential of interaction taking into account
the correlations between the particles in the case of short-range interactions
($u=u_{\rm eff}\neq u_{\rm SR}^{\rm exact}$) \cite{preparation}. On the other
hand,  a nonlinear
equation of state $P(\rho)$ can  take into account correlations between the
particles due to short-range interactions \cite{preparation}. It can also take
into account the
interaction of the system with a nonstandard thermostat like in Tsallis
generalized thermodynamics \cite{ggpp,wg} (see Sec. \ref{sec_ggpp}).} The
structure
function (\ref{fd10}) is a  Cauchy (or Lorentz) distribution in pulsation
$\omega$. Taking the inverse Fourier transform in time of Eq. (\ref{fd10}) and
using the Cauchy residue theorem, we find that the temporal correlation function
of the Fourier transform of the density fluctuations $S(k,t)\sim \langle
\delta{\hat \rho}_{k}(0)\delta{\hat \rho}^*_{k}(t)\rangle$  is
given by 
\begin{eqnarray}
\label{tcf3}
S({k},t)=\frac{k_B T \rho k^2}{\frac{\hbar^2k^4}{4m^2}+c_s^2 k^2+(2\pi)^d\hat{u}(k)\rho
k^2}\nonumber\\
\times e^{-\left\lbrack
\frac{\hbar^2k^4}{4m^2}+c_s^2 k^2+(2\pi)^d\hat{u}(k)\rho k^2\right\rbrack t/\xi}.
\end{eqnarray}
Integrating the Cauchy (or Lorentz) distribution from Eq. (\ref{fd10})
over $\omega$, or directly setting $t=0$ in Eq. (\ref{tcf3}), we obtain 
the static structure factor $S(k)\sim \langle |\delta{\hat
\rho}_{k}|^2\rangle$:
\begin{eqnarray}
\label{skd}
S(k)=\frac{k_B T\rho k^2}{\frac{\hbar^2k^4}{4m^2}+c_s^2 k^2+(2\pi)^d\hat{u}(k)\rho k^2}.
\end{eqnarray}
Eq. (\ref{tcf3}) is the counterpart of the velocity auto-correlation function
$\langle v(0)v(t)\rangle=(D/\xi)e^{-\xi t}$ with $D/\xi=k_B T/m$ of the 
Ornstein-Uhlenbeck process \cite{risken} (see also Eq. (\ref{kra8})). It can be
written as  $S(k,t)=S(k)\, e^{\gamma(k)t}$ where $S(k)=\xi k_B T\rho
k^2/\gamma(k)$ is the static structure factor and
$\gamma(k)=[\frac{\hbar^2k^4}{4m^2}+c_s^2 k^2+(2\pi)^d\hat{u}(k)\rho
k^2]/\xi$ is the exponential rate. 
The correlation function decays with time in the stable case ($\gamma(k)<0$)
and grows with time
in the unstable case ($\gamma(k)>0$).\footnote{When
$\gamma(k)<0$, the prefactor in Eq. (\ref{tcf3}) corresponds to
the variance  $\langle |\delta\rho_k|^2\rangle$ 
of the equilibrium distribution of the density fluctuations and the
decorrelation
time (exponent) coincides with the relaxation time $t_{\rm
relax}(k)=1/\gamma(k)$ of a fluctuation $\delta\rho_k(t)$ \cite{preparation}.}
At the critical
point where $\gamma(k)\rightarrow 0$, the static structure factor  $S(k)\propto
1/\gamma(k)$
(variance of the fluctuations) and the correlation (or relaxation) time
$\tau(k)= 1/\gamma(k)$
diverge.  This corresponds to the phenomenon of critical opalescence
in the theory of liquids \cite{opal1,opal2} (see also \cite{paper5}). In that
case, the linear  
theory breaks down and one must consider nonlinear effects or use
renormalization group technics.

\subsection{Stochastic quantum virial theorem}

In the absence of noise, the quantum virial theorem associated with the
generalized GPP equations (\ref{g1}) and (\ref{g2}) has been derived in
Appendix G of \cite{ggpp} from their hydrodynamic representation. It is
summarized in Sec. \ref{sec_qvt}. If we take the noise term into account, we
obtain by the same procedure an equation of the form
\begin{equation}
\label{sqv1}
\frac{1}{2}\ddot I+\frac{1}{2}\xi\dot I=2(\Theta_c+\Theta_Q)+3\int P\, d{\bf
r}+W+A(t),
\end{equation}
where
\begin{eqnarray}
\label{sqv2}
A(t)=-\int
\sqrt{2\xi k_B T\rho}\, {\bf r}\cdot {\bf R}({\bf r},t)\, d{\bf
r}
\end{eqnarray}
is the virial of the stochastic force. The stochastic term $A(t)$ has a zero
average $\langle A(t)\rangle=0$ and a correlation function 
\begin{eqnarray}
\label{sqv4}
&&\langle A(t)A(t')\rangle
=2\xi k_B T\nonumber\\
&\times&\int d{\bf
r} d{\bf
r}'\, \sqrt{\rho({\bf
r},t)\rho({\bf r}',t')}
 x_i x'_j 
\langle R_i({\bf r},t)R_j({\bf r}',t')\rangle\nonumber\\
&=&2\xi k_B T \delta(t-t')\int \rho r^2  
\, d{\bf
r}
=2\xi k_B T I \delta(t-t').\qquad
\end{eqnarray}
Therefore, it is statistically equivalent to a fluctuation of the form
\begin{eqnarray}
\label{sqv5}
A(t)=\sqrt{2\xi k_B T I}\eta(t),
\end{eqnarray}
where $\eta(t)$ is a Gaussian white noise such that $\langle\eta(t)\rangle=0$
and 
$\langle\eta(t)\eta(t')\rangle=\delta(t-t')$. In conclusion, the stochastic
quantum virial theorem associated with the stochastic
quantum damped Euler-Poisson equations (\ref{qes1})-(\ref{qes2}) can be written
as
\begin{equation}
\label{sqv6}
\frac{1}{2}\ddot I+\frac{1}{2}\xi\dot I=2(\Theta_c+\Theta_Q)+3\int P\, d{\bf
r}+W+\sqrt{2\xi k_B T I}\eta(t).
\end{equation}

In the strong friction limit $\xi\rightarrow +\infty$,  the stochastic quantum
virial theorem associated with the stochastic quantum Smoluchowski-Poisson
equations
(\ref{qs1}) and (\ref{qes2b}) takes the form
\begin{eqnarray}
\label{qs3}
\frac{1}{2}\xi\dot I=2\Theta_Q+3\int P\, d{\bf
r}+W+\sqrt{2\xi k_B T I}\eta(t).
\end{eqnarray}

{\it Remark:} To the best of our knowledge, this stochastic generalization of the virial theorem has not been
given
previously.

\subsection{Gaussian ansatz}
\label{sec_sga}

We can simplify the foregoing equations by using a Gaussian ansatz for the
condensate wave function. In the absence of noise, the Gaussian ansatz
associated with the generalized GPP equations (\ref{g1}) and (\ref{g2}) has
been discussed in Sec. 8 of \cite{ggpp}. It is summarized
in Sec. \ref{sec_ga}. If we take
into account the noise term and repeat the calculations of Sec. 8 of
\cite{ggpp} we find that
the stochastic quantum virial theorem (\ref{sqv6}) with the
Gaussian ansatz reduces to the stochastic damped Langevin
equation of motion
\begin{equation}
\label{sg2}
\alpha M\frac{d^2R}{dt^2}+\xi\alpha M\frac{dR}{dt}=-\frac{d{V}}{dR}+\sqrt{2\xi
k_B
T \alpha M}\eta(t).
\end{equation}
In the strong friction limit $\xi\rightarrow +\infty$, we obtain
\begin{eqnarray}
\label{qs4}
\frac{dR}{dt}=-\frac{1}{\xi\alpha M}\frac{d{V}}{dR}+\sqrt{\frac{2k_B
T}{\xi\alpha M}}\eta(t).
\end{eqnarray}
These
equations describe the stochastic damped and overdamped motion of a fictive
particle with
mass
$\alpha M$ and position $R$ in a potential $V(R)$. 
Therefore, the radius of the BEC obeys the equation of motion of a Brownian
particle with mass $\alpha M$ in a potential $V(R)$. Since the Langevin equation
contains
fluctuations, it may be used to describe the destabilization of the metastable
state and the thermal activation of the BEC when the self-interaction
between the bosons is attractive
($a_s<0$).
Indeed, thermal fluctuations can allow the condensate to overcome the free
energy
barrier between the metastable state and the unstable state and collapse (see
Appendix \ref{sec_wnl}). For
ordinary self-gravitating BECs, the effect of fluctuations is weak because $N$
is very large (see Sec. \ref{sec_scm}). Without the noise term, we recover the
deterministic equations (\ref{g25}) and (\ref{g27}).

\section{Derivation of the tunneling
rate of a quantum particle in a
metastable state by
using the
instanton theory}
\label{sec_wnlq}

In this Appendix, we use the instanton theory to derive the tunneling rate of a
quantum particle in a metastable state through a potential barrier. We call it
the quantum tunneling rate. We follow the seminal works
of \cite{polyakov,coleman,cc,stone1,stone2} (see also the reviews
\cite{abc,brandenberger,gkw,stoof}) but we focus on the basics. A more detailed
treatment and an early history of instantons in quantum mechanics is given in
\cite{preparation}.

\subsection{The Schr\"odinger equation}
\label{sec_trqBse}

We consider a particle of mass $m$ evolving in a one-dimensional
potential $V(x)$. We assume that the potential $V(x)$ has a local minimum at
$x_M$ and a local maximum at $x_U$. It may also possess a global minimum at
$x_S$ or be unbounded from below. We shall consider this second possibility so
as to
be in the situation of Fig. \ref{normalform}. The difference of potential
between the
local minimum and the local maximum creates a barrier of height $\Delta
V=V(x_U)-V(x_M)$. 
In classical mechanics, the motion of the particle is described by
Newton's equation
\begin{eqnarray}
\label{qm1}
m\frac{d^2x}{dt^2}=-\frac{d{V}}{dx}.
\end{eqnarray}
In quantum mechanics, the particle is represented by a complex wave function 
$\psi(x,t)$ whose evolution is governed by the Schr\"odinger equation
\begin{eqnarray}
\label{qm2}
i\hbar \frac{\partial\psi}{\partial
t}=-\frac{\hbar^2}{2m}\frac{\partial^2\psi}{\partial x^2}
+V(x)\psi.
\end{eqnarray}

Suppose that, at the initial time, the particle is localized in the
local minimum of the potential. According to the classical laws of motion [see
Eq. (\ref{qm1})]
it cannot
go away from it. This minimum is a stable equilibrium state in
which the particle is at rest. For a small perturbation, the classical
particle oscillates with a pulsation $\omega_M=\sqrt{V''(x_M)/m}$. The left of
the barrier is
inaccessible. By contrast, in quantum
mechanics [see Eq. (\ref{qm2})], the 
local minimum becomes unstable through
barrier-penetration.   In semiclassical
language, the particle penetrates the potential barrier by tunnel effect
 and
materializes at the
escape point $x'_M$ (see Fig. \ref{normalform}) with zero kinetic energy, after
which it
propagates classically. Therefore, the particle can escape from the local
minimum of the potential by quantum-mechanical tunneling (quantum fluctuations).
In this sense, the
local minimum is called a metastable state (or false vacuum in quantum field
theory).
Our
aim is to compute the decay rate
of this metastable state or, equivalently, the mean time taken by a particle
initially in the nonglobal minimum $x_M$ of the potential $V(x)$ to
escape through the potential barrier by quantum-mechanical tunneling
transfers.

Calculations of tunneling rates (or mean times of escape) have
traditionally been accomplished by solving
the Schr\"odinger equation in the semiclassical limit $\hbar\rightarrow
0$, leading to the WKB formula from Eq. (\ref{q54}) below
\cite{llquantique}. Instead, our starting
point
will be the expression of the  probability amplitude as a path integral
\cite{fh}. We
shall treat quantum fluctuations in a semiclassical approximation by using
functional integral methods. We present below a brief summary of how the
semiclassical approximation to the tunneling rate of a quantum particle
through a one-dimensional potential barrier is
derived by path integral methods (instanton theory) without going
into technical
details.

\subsection{Feynman path integral}
\label{sec_matrix}

Dirac \cite{dirac33,dirac35,dirac45} suggested that there is
an interesting
relation
between the quantum propagator $\langle x|e^{-i{ H}t/\hbar}|x_0\rangle$, which
gives the probability amplitude of the particle in $x$ at time $t$ assuming
that it was located in $x_0$ at time $t_0=0$,  and the
classical action of a particle of mass $m$ in a potential $V(x)$:
\begin{equation}
\label{q10}
S_{\rm class}[x(t)]=\int \left\lbrack \frac{1}{2}m{\dot
x}^2-V(x(t))\right\rbrack \, {d} t=\int L_{\rm class}(x,\dot
x)
\, {d} t,
\end{equation}
where
\begin{eqnarray}
\label{q18bq}
L_{\rm class}(x,\dot x)=\frac{1}{2}m{\dot
x}^2-V(x)
\end{eqnarray}
is the classical Lagrangian.
Developing Dirac's
ideas, Feynman \cite{feynman}  showed that the formal
relation between
these quantities can be written as a functional (path) integral
of the form
\begin{eqnarray}
\label{q8}
\langle x|e^{-i{ H}t/\hbar}|x_0\rangle=\int{\cal
D}x(t)\,
e^{iS_{\rm class}[x(t)]/\hbar}
\end{eqnarray}
with the boundary conditions $x(t_0=0)=x_0$ and $x(t)=x$. He showed by a direct
calculation that the
wavefunction $\psi(x,t)\equiv\langle x|e^{-i{ H}t/\hbar}|x_0\rangle$
expressed as a path integral by Eq.
(\ref{q8}) satisfies the Schr\"odinger equation
(\ref{qm2}).
Therefore,
Eq. (\ref{q8}) provides the exact solution of Eq.
(\ref{qm2}) corresponding to the initial condition
$\psi(x,0)=\delta(x-x_0)$. This is
the Green function of the Schr\"odinger  equation. 
The quantum
propagator describes the reversible evolution of a quantum
probability amplitude. According to Feynman's formulation of quantum mechanics,
the probability amplitude $\psi(x,t)$  is equal to the sum over all paths joining the world
points
$(t_0=0,x_0)$ and $(t,x)$ with an exponential weight $e^{iS_{\rm
class}[x(t)]/\hbar}$.
In Eq. (\ref{q8}) the action $S_{\rm class}[x(t)]$ must be viewed as a
functional of all possible trajectories starting in $x_0$ at time $t_0=0$ and
ending in $x$ at time $t$.  Therefore, the probability amplitude to find
a quantum particle in $x$ at time $t$ is given by a linear superposition of
states that contribute
with a phase factor $e^{iS_{\rm class}[x(t)]/\hbar}$ corresponding to
a trajectory joining $x_0$ to $x$ during the time
$t$. This leads to a sum
of oscillating exponentials. When $\hbar\rightarrow 0$, the exponent in Eq.
(\ref{q8}) oscillates very
rapidly and its positive and negative contributions nearly cancel. The path 
which contributes most strongly to the integral is the one for which
the phase of the exponent varies the less rapidly, i.e., for which the phase
suffers no first order change on varying the path (stationary
phase) \cite{feynman}. This justifies   the least action principle $\delta
S_{\rm class}=0$ introduced heuristically by Maupertuis \cite{maupertuis1,maupertuis2} and Hamilton
\cite{hamilton}, and
leading to the classical path $x_{\rm class}(t)$  [see Eq.
(\ref{qm1})] determined by the  Lagrangian
or
Hamiltonian equations of motion.

{\it Remark:} The type of integrals which
is used in Feynman's
theory was new in theoretical physics but not in probability theory. In
fact, the integral over trajectories introduced by Feynman \cite{feynman} is
almost the same as the one used by Wiener \cite{wiener} in studying the
trajectories which arise in Brownian motion (see Appendix \ref{sec_wnl}). The connection
between Wiener's path integral for Brownian motion and Feynman's path integrals in quantum theory was first made by 
Kac \cite{kac49,kac},
Montroll \cite{montroll}, Gel'fand and Yaglom \cite{gy}, Saito and
Namiki \cite{saito}, and Brush \cite{brush}.

\subsection{Euclidean action}
\label{sec_ea}

The functional integral (\ref{q8}) requires delicate integrations on
oscillatory
functions. To overcome this problem, it is convenient to study its
analytical continuation to purely imaginary times for which the corresponding
Feynman-Kac functional integral is well-defined. To this end we make the substitution
$t\rightarrow
-it$ (rotation of $\pi/2$ in the complex plane of time).
In the literature, the transition to an imaginary time is called the Wick
rotation. In that case, we say that we work with the euclidean formulation of
quantum mechanics.\footnote{This terminology is related to the fact that in
special
relativity the change $t\rightarrow -it$ transforms the Minkowski metric into
the usual euclidean metric.} We come back to the quantum propagator by the
inverse Wick rotation. In the euclidean formulation, the quantum
propagator can be
written as 
\begin{eqnarray}
\label{q16}
\langle x|e^{-{H}t/\hbar}|x_0\rangle=\int{\cal
D}x(t)\,
e^{-S[x(t)]/\hbar}, 
\end{eqnarray}
where $S[x(t)]$ is the
euclidean action
\begin{eqnarray}
\label{q17}
S[x(t)]=\int \left\lbrack \frac{1}{2}m{\dot
x}^2+V(x(t))\right\rbrack \, {d} t=\int L(x,\dot x)\, {d}t\quad
\end{eqnarray}
and
\begin{eqnarray}
\label{q18b}
L(x,\dot x)=\frac{1}{2}m{\dot
x}^2+V(x)
\end{eqnarray}
is the euclidean Lagrangian. It is important to note that the euclidean action
and the euclidean Lagrangian
differ from those of classical mechanics by the sign of the potential. Indeed,
Eq. (\ref{q18b}) is the Lagrangian of the particle with the ``wrong'' sign
($+V$ instead of $-V$). The potential energy changes sign
$V(x)\rightarrow -V(x)$ under the Wick rotation and a minimum (resp.
maximum) transforms into
a maximum (resp. minimum).  The
wavefunction $\psi(x,t)\equiv\langle x|e^{-{H}t/\hbar}|x_0\rangle$
expressed as a path integral by Eq.
(\ref{q16}) satisfies the Schr\"odinger equation in imaginary time (also
called the Kac equation):
\begin{eqnarray}
\label{qm2im}
-\hbar \frac{\partial\psi}{\partial
t}=-\frac{\hbar^2}{2m}\frac{\partial^2\psi}{\partial x^2}
+V(x)\psi.
\end{eqnarray}

The path integral formalism can be used to compute the propagator
$\langle x|e^{-{H}t/\hbar}|x_0\rangle$ of the quantum particle
from one state to the other in general situations. 
Our task is therefore to evaluate this path integral. In general this cannot
be done exactly. However, in the semiclassical approximation
$\hbar\rightarrow 0$, we can use the method of steepest descent (also called the
saddle point approximation) to evaluate the
propagator. The leading behavior of the propagator is
given by the trajectory which minimizes the euclidean action.
In that case the entire integral in Eq. (\ref{q16}) is accumulated from
regions near the minimum of $S[x(t)]$. To determine the dominant contribution
to the path integral we thus have to minimize the action $S[x(t)]$. The paths
corresponding to the least
action, which we denote by $x_c(t)$, are known in the literature as extremal
paths or as stationary points. In the semiclassical limit $\hbar\rightarrow
0$, the functional integral is dominated by the stationary points of $S[x(t)]$
so the dominant contribution to the transition amplitude comes from these
stationary points. If there is just one extremal path $x_c(t)$ then
\begin{eqnarray}
\label{q20}
\langle x|e^{-{H}t/\hbar}|x_0\rangle\propto
e^{-S[x_c(t)]/\hbar}.
\end{eqnarray}
This expression provides an approximate solution of the Schr\"odinger equation
(\ref{qm2})
after coming back to real time by making an inverse Wick
rotation.\footnote{Eq. (\ref{q20}) is the dynamical generalization of the WKB formula \cite{llquantique}, which gives  the stationary solution for $\hbar\rightarrow 0$.}
If there are several stationary points, we have to sum the
contribution of all the stationary points.

\subsection{Quantum escape rate}

We now specifically apply the path integral formalism to the computation of the
escape rate of a quantum particle out of a metastable state. As discussed
at the beginning of this Appendix, we assume that the potential
$V(x)$ has a local minimum at $x_M$ (metastable equilibrium), a maximum at
$x_U$ (unstable equilibrium) and a global minimum at $x_S$ (stable
equilibrium), or no global minimum at all (see Fig. \ref{normalform}). We wish
to calculate the tunneling escape rate $\Gamma$ from the metastable equilibrium state at
$x_M$. We have seen that the classical trajectory $x_{\rm class}(t)$ determined
by Eq. (\ref{qm1}) is a minimum of the classical action ($\delta S_{\rm
class}=0$). According to classical mechanics, a particle at
a minimum of the potential $V(x)$ is stable and remains at this place for ever,
so that $x_{\rm class}(t)=x_M$.
However, there can be other optimal trajectories that minimize the
action ($\delta S=0$). They
can  be seen most clearly by working in imaginary time.  These optimal
trajectories account for
quantum tunneling. For $\hbar\rightarrow 0$, it can be shown that the 
escape rate out of the metastable state is given by
\begin{eqnarray}
\label{q18c}
\Gamma\propto
e^{-S[x_c]/\hbar},
\end{eqnarray}
where $S[x_c]$ is the euclidean action of the most probable
path
that connects the metastable state $x_M$ to the escape point $x'_M$ in the
inverted potential $-V(x)$. This is called a bounce (see Fig. 5
in \cite{tunnel}). In the present case, we
will call it a ``quantum instanton'' \cite{tunnel}. It is obtained by
cancelling the first order
variations of the euclidean action: $\delta S=0$.  The precise
justification of Eq.
(\ref{q18c}) and the evaluation of
the
preexponential factor requires integrating
over the Gaussian fluctuations around the extremal path $x_c(t)$, using
collective variables to deal with the invariance of the action $S[x(t)]$ under
time translation of the instanton (the presence of a zero eigenvalue), and
including multi-instanton contributions. This is a difficult calculation that is
not detailed here (see the references given at the beginning of this Appendix).
Duru {\it et al.} \cite{duru} managed to determine the
prefactor explicitly by showing that the 
fluctuation determinant can  be evaluated exactly with the
Gelfand-Yaglom \cite{gy} method.

\subsection{The semiclassical limit and the instanton}
\label{sec_inst}

In the semiclassical approximation $\hbar\rightarrow 0$,
we have to determine
the trajectory
$x_c(t)$ that minimizes the euclidean action $S[x(t)]$
defined by Eq. (\ref{q17}). The
stationary points of $S[x(t)]$ can be obtained in different manners.

{\it First method:} The extremal condition $\delta S/\delta x=0$ (least
action principle) applied to Eq.
(\ref{q17})  yields the Euler-Lagrange equation
\begin{eqnarray}
\label{q21}
\frac{d}{dt}\left (\frac{\partial L}{\partial \dot x}\right
)-\frac{\partial
L}{\partial x}=0. 
\end{eqnarray}
Using Eq. (\ref{q18b}), we obtain
\begin{equation}
\label{q22}
m\frac{{d}^2x_c}{{d}t^2}=V'(x_c).
\end{equation}
Equation (\ref{q22}) has the same form
as the classical equation of motion of a point
particle with mass $m$  in the inverted
potential $-V(x)$ (the minus sign is due to the fact that the euclidean
formulation is considered).  The first integral of motion is
\begin{eqnarray}
\label{q23}
E=\frac{1}{2}m\dot x_c^2-V(x_c),
\end{eqnarray}
where the energy $E$ is a constant determined by the boundary
conditions.  A nontrivial solution $x_c(t)$ of Eq. (\ref{q23}) is called
an instanton (a soliton in the time variable). An instanton 
$x_c(t)$ corresponds to an optimal
path, i.e., a path that minimizes the euclidean action $S[x(t)]$. Mechanical
analogs such as Eqs. (\ref{q22}) and (\ref{q23}) are
common in instanton physics.

{\it Second method:} The minimization of the action leads to the
Hamilton equations, which are equivalent to the Newton equation (\ref{q22}).
Since the Lagrangian (\ref{q18b}) does not explicitly depend
on
time, the Hamiltonian 
\begin{equation}
\label{q24}
H= \dot x\frac{\partial L}{\partial {\dot x}}-L
\end{equation}
is conserved. Using Eq. (\ref{q18b}), we directly obtain Eq. (\ref{q23}).

\subsection{WKB formula}

We can now repeat the calculations detailed in Sec. IV of
\cite{tunnel} and
find that the quantum tunneling rate  (decay probability per
unit time) is given by
\begin{eqnarray}
\label{q54}
\Gamma\sim\sqrt{\frac{m\omega_Mv_M^2}{\pi\hbar}}\,
e^{-\frac{2}{\hbar}\int_{x'_M}^{x_M} \sqrt{2m\lbrack V(x)-V(x_M)\rbrack}\,
{d}x},
\end{eqnarray}
where $\omega_M^2=V''(x_M)/m$ is the squared pulsation of the particle at the
metastable minimum of the potential  and $v_M$, which depends on the details of
the potential, is
determined by
the asymptotic behavior of the bounce solution via the formula
\begin{equation}
\label{q56}
x_{c}(t)\simeq {x_M}-\frac{v_M}{\omega_M}e^{-\omega_M|t|} \qquad (t\rightarrow
\pm\infty).
\end{equation}
Finally, the quantum escape time,  or the typical lifetime of the
metastable state, can be estimated by 
\begin{equation}
\label{q53b}
t_{\rm life}\sim \Gamma^{-1}\sim\sqrt{\frac{\pi\hbar}{m\omega_Mv_M^2}}
e^{\frac{2}{\hbar}\int_{x'_M}^{x_M} \sqrt{2m\lbrack V(x)-V(x_M)\rbrack}\,
{d}x}.
\end{equation}
This expression, which is valid in the semi-classical approximation $\hbar\rightarrow
0$ (more precisely for a large exponent), can also be obtained by using the WKB method to find the
transmission
amplitude across the potential barrier \cite{llquantique}. Therefore, it is
oftentimes
called the WKB transmittivity formula. In \cite{tunnel}, this expression is simplified close to a saddle-node bifurcation, leading to Eq. (\ref{giu1}).

\subsection{Early literature on quantum tunneling}
\label{sec_relbib}

The problem of escape from metastable states by quantum mechanical
tunneling\footnote{The term ``tunnel
effect'' ({\it wellenmechanische Tunneleffekt}) was introduced by Walter Schottky in 1931 and its first English use appears in the book of J. Frenkel, {\it Wave
Mechanics, Elementary Theory} \cite{merzbacher}.}  occurs in many domains of physics. The probability that
a quantum particle remains in a metastable state decreases exponentially rapidly
with time as
\begin{equation}
\label{bib1}
P(t)\sim e^{-\Gamma t}\sim e^{-t/t_{\rm life}},
\end{equation}
where $\Gamma$ is the rate coefficient and  $t_{\rm life}\sim 1/\Gamma$ is the typical lifetime of the particle in the metastable state.

Gamow \cite{gamow} and Gurney and Condon \cite{gurney1,gurney2,condon} invoked
quantum tunneling to explain the exponential radioactive decay of nuclei (see
also an alternate approach by Born \cite{bornt}).  They showed that when
$\hbar\rightarrow 0$ the rate coefficient is given by the WKB formula
\cite{wentzel1926,brillouin1926,kramers1926}. Their studies followed previous
works by Hund \cite{hund1,hund2} who demonstrated that quantum tunneling
describes the intramolecular rearrangements in pyramidal molecules, Oppenheimer
\cite{opp1,opp2} who employed it for the description of the ionization of atoms
in intense electric fields, and Nordheim \cite{nordheim} and Fowler and Nordheim
\cite{fono} who used it for the electric field emission of electrons from cold
metals. Mandelstam and Leontovich \cite{mandelstam} studied the behavior of the
wave function through a potential barrier, Eckart \cite{eckart} considered the
penetration of a potential barrier by electrons, Rice \cite{rice} discussed the
analogy between the quantum mechanical theory of radioactivity and the
dissociation by rotation of diatomic molecules, and MacColl \cite{coll}
discussed the transmission and reflection of wave packets by potential barriers.
The connection of quantum tunneling to reaction rates was made by Bourgin
\cite{bourgin} and Wigner \cite{wignertunnel} who calculated quantum tunneling
corrections to activation escape rates up to order $\hbar^2$.

The theory of instantons was first developed in the context of quantum field
theory  in order to determine the decay rate of a metastable state by quantum
tunneling in a cosmological context.

Let us consider the quantum field theory of a  scalar field
$\varphi({\bf x},t)$ evolving in an effective potential $V(\varphi)$ which possesses   two spatially homogeneous local minima with different energy density. This occurs is some unified electroweak and grand unified
theories.  The local minimum is called the ``false vacuum'' and the global minimum is called the ``true
vacuum''.\footnote{This terminology was introduced by Coleman \cite{coleman}.} As argued by Lee and Wick
\cite{lw}, the false vacuum is rendered unstable by quantum effects, in
particular by barrier penetration. As a result,  there is a
finite probability for quantum tunneling from the false vacuum to
the true vacuum of lower energy. To estimate the lifetime of the false vacuum, which in some cases turns
out to be very large, it is necessary to develop a theory of decay of the metastable vacuum state. The quantity to compute is the decay probability of the false vacuum per unit time and per unit volume $\Gamma/V$.

A theory of the decay of false vacuum in spontaneously broken theories at
zero temperature has been first suggested by Voloshin {\it et al.}
\cite{voloshin} and then considerably developed by Coleman and Callan
\cite{coleman,cc} (the expansion of the Universe and gravitational
effects on the vacuum decay have
been considered by Coleman and De Luccia \cite{cdl}). Before treating the case
of a scalar field $\varphi({\bf
x},t)$, Coleman and Callan
\cite{coleman,cc}  considered the simpler problem of a quantum particle in a
potential $V(x)$ presenting a local minimum and
a global minimum. In his first
paper, inspired by Banks {\it et
al.} \cite{banks}, Coleman \cite{coleman} started from
the WKB formula $\Gamma=A e^{-B/\hbar}$  [see Eq. (\ref{q53b})].\footnote{In ordinary quantum mechanics \cite{ll},
beside being used to calculate energy levels, WKB methods were employed to
calculate the lifetime of metastable
states \cite{benderwu}.} He observed that the value of $B$ in the exponential term is equal to
$S[x_c]$ where $S[x]$ is the euclidean action of the particle (with the
potential $-V(x)$ instead of $+V(x)$) and $x_c(t)$ is the
optimal trajectory
that extremizes $S[x]$. He called it the ``bounce''.
He then extended this approach to quantum field theory by
replacing the euclidean action of the particle $S[x]$ by the euclidean
action of the field $S[\varphi]$.  In a second
paper, Callen and Coleman \cite{cc}  rederived these
results in a more satisfactory manner from Feynman's path integral approach and
determined the prefactor
$A$ appearing in the expression of the decay rate. They treated
quantum fluctuations in a
semiclassical approximation
using functional integral methods. After quantum tunneling, the further
evolution of the field is determined by solving the classical field
equations. Very similar results were derived simultaneously and independently by
Stone \cite{stone1,stone2} and Polyakov \cite{polyakov}  using path integral
methods from the
start. They used the term ``instanton'' to designate the optimal
path.\footnote{To our knowledge, the
term ``instanton'' (a soliton in the time variable) was
coined by 't Hooft (see Ref. \cite{hooft} and the footnote P. 430 of Polyakov
\cite{polyakov}).
Polyakov \cite{polyakovPP,polyakov} proposed the name ``pseudoparticle'' which
can also be found in the
literature.  Instantons
describe processes by which a metastable state decays. They provide the
dominant part of the decay amplitude. This give a role to previously rejected
unstable solitons \cite{stone2}.} Some reviews were given in
\cite{abc,brandenberger,gkw,stoof}.

In the field theory context, starting from the false  vacuum state, Stone \cite{stone2} and Coleman
\cite{coleman}  showed that quantum
fluctuations may spontaneously generate a ``bubble'' whose interior is built
upon the true vacuum state. There is a critical
size: Below this size the bubble shrinks to nothing but above it there is a
classical instability with respect to rapid and unlimited expansion. The critical
bubble $\varphi_c({\bf
x},t)$ corresponds to the
instanton soliton of the field theory. The instanton (or bounce) looks like a
large
four-dimensional spherical bubble of radius $R$
with a thin wall separating the false vacuum without from the true vacuum
within.

This process is exactly analogous to the nucleation of droplets in
statistical physics like the
boiling of a superheated fluid. The false vacuum corresponds to the superheated
fluid phase (liquid) and the true vacuum to the vapor phase (gas). This nucleation
process has been studied by Langer \cite{langer} using a field theory
approach involving a free energy functional similar to the
euclidean action in quantum mechanics. The extremal solution is called a
``critical droplet''. This corresponds to the
instanton solution in field theory.\footnote{Coleman \cite{coleman}
mentioned the work of
Langer \cite{langer} on the droplet model in statistical physics as a {\it Note
added
in proof}. It is apparently the work of Langer \cite{langer} that induced
Callan and Coleman \cite{cc} to develop a path integral approach
instead of using WKB methods (see footnote 2 of their paper).  The work
of Langer \cite{langer} to the problem of first order phase transitions
in statistical mechanics was also mentioned by Stone \cite{stone2} at the
end of his paper.} Langer  \cite{langer} reduced the calculation of escape rates to an analytical continuation of the free energy.

This scenario for the cosmological evolution of a first order phase
transition in the early universe was very popular in the early 1980s (in relation to the theory of inflation described by a scalar field). But it was soon realized that
the nucleation of bubbles, thermal or quantum, was a very rare event (as
anticipated in \cite{voloshin}). The
barrier between the two minima can be very high implying that the tunneling
amplitude is
not significant enough. Indeed, numerical investigations of first order phase
transitions in a number of models showed that the tunneling probability is
extremely small. As a result, the lifetime of the metastable vacuum state in
gauge theories is usually extremely large
\cite{linde77,linde80,linde81,linde83,gw,cook,billoire,steinhardt81,witten,
hm,chh}, much larger than the age of the Universe. Barrier penetration is a notoriously slow process.\footnote{The fact that quantum tunneling is very improbable and can be neglected in many (but not all) cases of physical interest was pointed out since the beginning  by Hund \cite{hund1,hund2} and Nordheim \cite{nordheim} (see the discussion in \cite{merzbacher}).}

\section{Derivation of the escape rate
of a Brownian particle from a
metastable state by
using the instanton theory}
\label{sec_wnl}

In this Appendix, we use the instanton theory to derive the escape rate of a
Brownian particle from a metastable state across a potential barrier. We call it
the thermal activation rate. We
follow the seminal works of \cite{brayprl,mk,bray1,bray,lmk,nbk} but we focus on the
basics. The early history
of instantons in statistical
mechanics is retraced in \cite{preparation}. In our presentation, we keep a
certain parallel with the quantum
problem described
in Appendix \ref{sec_wnlq}. The relation between the thermal problem and the
quantum problem is
made explicit in Appendix \ref{sec_ass}.

\subsection{The Smoluchowski equation}
\label{sec_trqB}

We consider an overdamped particle of mass $m$  and
mobility $\mu=1/\xi m$ (where $\xi$ is the friction coefficient) evolving in
a one-dimensional potential $V(x)$. We assume that the potential $V(x)$ has a
local minimum at $x_M$ and a local maximum at $x_U$. It may also possess a
global minimum at $x_S$
or be unbounded from below. We shall consider this second possibility so as to
be in the situation of Fig. \ref{normalform}.  The difference of potential
between the
local minimum and the local maximum creates a barrier of height  $\Delta
V=V(x_U)-V(x_M)$. In the absence of external perturbation, the motion of the
particle is
described by the deterministic equation
\begin{eqnarray}
\label{b0}
\frac{{d}x}{{d}t}=-\frac{1}{\xi m}V'(x).
\end{eqnarray}
If the particle is a Brownian particle submitted to a noise $\eta(t)$ caused by
its interaction with a surrounding medium playing the role of a thermal bath,
its
motion is described by the stochastic Langevin
equation \cite{langevin} 
\begin{eqnarray}
\label{b1}
\frac{{d}x}{{d}t}=-\frac{1}{\xi m}V'(x)+\sqrt{2D}\, \eta(t),
\end{eqnarray}
where $D$ (diffusion coefficient) is a constant measuring the amplitude of the
noise. The noise
$\eta(t)$ is assumed to be Gaussian with zero mean so that it is
completely specified by its second moment. Here, we consider a Gaussian white
noise such
that
its correlation function is proportional to a
$\delta$-function: 
\begin{eqnarray}
\label{b2}
\langle\eta(t)\rangle=0,\qquad \langle\eta(t)\eta(t')\rangle=\delta(t-t').
\end{eqnarray} The probability
density $P(x,t)$ of finding the Brownian particle in $x$ at time $t$ is
determined by a Fokker-Planck equation \cite{fokker,planck1917} of the
Smoluchowski form
\cite{smoluchowski}:
\begin{eqnarray}
\label{b5}
\frac{\partial P}{\partial t}=\frac{\partial }{\partial
x}\left \lbrack D\frac{\partial P}{\partial x}+\frac{1}{\xi m} P V'(x)\right
\rbrack.
\end{eqnarray}
Identifying the 
equilibrium state of the Smoluchowski equation with the canonical Boltzmann
distribution 
\begin{eqnarray}
\label{b6}
P_{\rm eq}(x)=\frac{1}{Z} e^{-\beta V(x)},
\end{eqnarray}
where $\beta=1/k_B T$ is
the
inverse temperature of the bath and $Z$ a normalization factor, we find that the
diffusion
coefficient, the friction
coefficient and the temperature are related to each other by the Einstein
relation \cite{einstein}
\begin{eqnarray}
\label{b6b}
D=\frac{k_B T}{\xi m}.
\end{eqnarray}
According to Eqs. (\ref{b1}) and (\ref{b6b}), the strength of
the noise is controlled by the temperature $T$. The Fokker-Planck equation
(\ref{b5}) satisfies an $H$-theorem $\dot F\le 0$ for the Boltzmann free energy
$F=E-TS=\int \rho V\, dx+k_B T\int P\ln P\, dx$.  Furthermore, the equilibrium
Boltzmann distribution (\ref{b6}) minimizes $F$ under the normalization
condition $\int P\, dx=1$. For a free Brownian particle
($V=0$), Eq. (\ref{b1}) reduces to the Wiener process \cite{wiener} and
Eq. (\ref{b5}) reduces to the diffusion equation.

Suppose that, at the initial time, the particle is localized in the
local minimum of the potential. According to the deterministic laws of motion
[see Eq. (\ref{b0})]
it cannot go away from it. This minimum is a stable equilibrium state in
which the particle is at rest. For a small perturbation, the overdamped
particle returns to equilibrium with a damping rate
$\gamma_M=V''(x_M)/\xi m$. The left of the barrier is inaccessible. By
contrast, if the
particle
is a Brownian particle submitted to an external perturbation (noise) [see Eqs.
(\ref{b1})-(\ref{b5})], the local minimum
becomes unstable through
barrier-crossing. Because of the noise (thermal fluctuations), the particle
can
overcome the
potential barrier and reach the maximum of the potential, after which it can
descend the
potential on the left deterministically (freely). Therefore, the particle can
escape
from the
local minimum of the potential by thermal activation transfers. In this sense,
the
local minimum is called a metastable state. Our aim is to compute
the decay rate
of this metastable state or, equivalently, the mean time taken by a Brownian
particle
initially in the nonglobal minimum $x_M$ of the potential $V(x)$ to
escape over the potential barrier by thermal activation transfers.

This calculation is
most easily carried out by using the Fokker-Planck (Smoluchowski) equation
(\ref{b5}). The
result in the weak-noise limit $D\rightarrow 0$ is the classical Kramers
formula [see Eq. (\ref{gth}) below] which returns the Arrhenius law. Instead,
our
starting point will be the expression of the  probability density as a path
integral. We shall treat thermal fluctuations in the weak noise approximation
by using
functional integral methods. We present below a brief summary of how the
weak-noise
approximation to the escape rate of a Brownian particle  over a
one-dimensional potential barrier is derived by
path integral methods (instanton theory) without going into technical
details.

\subsection{Wiener path integral and Onsager-Machlup functional}
\label{sec_wom}

The distribution of a Gaussian white noise $\eta(t)$ is given by
\begin{eqnarray}
\label{b10}
P[\eta(t)]\propto e^{-\frac{1}{2}\int_{t_i}^{t_f} \eta^2(t)\, dt}.
\end{eqnarray}
The probability of the path $x(t)$  is then obtained by
expressing $\eta(t)$ in terms of $x(t)$ using Eq. (\ref{b1}). This gives
\begin{eqnarray}
P[x(t)]\propto e^{-S[x(t)]/k_B T},
\label{in1b}
\end{eqnarray}
where 
\begin{eqnarray}
\label{b12}
S[x(t)]=\frac{\xi m}{4}\int \left\lbrack
{\dot
x}+\frac{V'(x(t))}{\xi m}\right\rbrack^2 \, {d} t
\end{eqnarray}
is the so-called Onsager-Machlup functional \cite{onsagermachlup}. 
The functional $S$ may be called an action by analogy with the
path integral
formulation of quantum mechanics (the temperature $T$ plays the role of the
Planck constant $\hbar$ in quantum mechanics) \cite{feynman}.  It can be written
as $S=\int L\, dt$ where $L$ is the corresponding Lagrangian. The close analogy between the 
Onsager-Machlup functional in Brownian theory and the classical action of a particle is discussed 
in Appendix \ref{sec_ass}.

The probability density to observe the particle in $x$
at time $t$ provided  that it was in $x_0$ at time $t_0$ is 
given by the path integral
\begin{eqnarray}
\label{in3}
P(x,t|x_0,t_0)= \int {\cal D}x(t)\, 
e^{-S[x]/k_B T},
\end{eqnarray}
where the integral runs over all paths satisfying $x(t_0)=x_0$ and
$x(t)=x$. The weighting factor consists in an exponential of the
Onsager-Machlup
functional. For a given initial condition $x=x_0$ at $t=t_0$, one can show
that
the
probability density $P(x,t)\equiv P(x,t|x_0,t_0)$ expressed as a path integral
by Eq. (\ref{in3}) satisfies the Smoluchowski equation
(\ref{b5}).\footnote{This
can be shown by a direct calculation  or  by using the analogy between the Smoluchowski equation
and the Schr\"odinger equation in imaginary time (see Appendix \ref{sec_ass}),
and applying the results of Feynman \cite{feynman}
in quantum mechanics.} Therefore,
Eq. (\ref{in3}) provides the exact solution of Eq.
(\ref{b5}) corresponding to the initial condition
$P(x,t_0|x_0,t_0)=\delta(x-x_0)$. This is the Green function of the
Smoluchowski equation.  For a
free Brownian particle ($V=0$), Eq. (\ref{in3}) reduces to the Wiener path
integral (or Wiener measure) \cite{wiener} associated with the diffusion equation.

The path integral formalism can be used to compute the transition probability
$P(x,t|x_0,t_0)$ of the Brownian particle from one state to the other in general
situations. Our task is therefore to evaluate this path integral. In general
this cannot be done exactly. However, in the weak noise limit $T\rightarrow 0$,
the path integral (\ref{in3}) can be evaluated by the method of steepest descent
(or saddle point approximation). To leading order, the integral (\ref{in3}) is
dominated by the most probable path $x_c(t)$, i.e., the path that minimizes the
Onsager-Machlup functional $S[x(t)]$ (action). The typical paths explored by the
particle are concentrated close to the most probable path. We note that the
Onsager-Machlup functional vanishes for the deterministic
trajectory determined by Eq. (\ref{b0}).\footnote{The deterministic trajectory
(\ref{b0}) can be viewed as
the average of the stochastic process defined by Eq. (\ref{b1}).} This is
therefore the global minimum of $S[x(t)]$. As a result, the  most probable
path of the Brownian particle corresponds to the deterministic (average)
trajectory of the particle. Therefore, the deterministic
equation (\ref{b0}) -- the average path -- can be obtained by
minimizing the Onsager-Machlup functional (\ref{b12}). This provides 
a sort of least action principle $\delta S=0$ for an overdamped particle, which
is the counterpart of the least action principle of Maupertuis
\cite{maupertuis1,maupertuis2} and Hamilton
\cite{hamilton} for a frictionless particle. These variational principles can be justified from the path integral formalism of Brownian and quantum theories in the limits $T\rightarrow 0$ and $\hbar\rightarrow 0$, respectively.
 If there is just
one extremal path
$x_c(t)$, then
\begin{eqnarray}
\label{in4}
P(x,t|x_0,t_0)\simeq e^{-S[x_c]/k_B T}.
\end{eqnarray}
This expression provides an approximate solution of the 
Smoluchowski equation (\ref{b5}). If there are several stationary points, we have to sum the
contribution of all the stationary points.

\subsection{Thermal escape rate}

We now specifically apply the path integral formalism to the computation of the
escape rate
of a Brownian particle out of a metastable state. As discussed
at the beginning of this Appendix, we assume that the potential
$V(x)$ has a local minimum at $x_M$ (metastable equilibrium), a maximum at
$x_U$ (unstable equilibrium) and a global minimum at $x_S$ (stable
equilibrium), or no global minimum at all (see Fig. \ref{normalform}). We wish
to calculate the escape rate $\Gamma$ from the metastable equilibrium state at
$x_M$. We have seen that the deterministic trajectory $x_{\rm average}(t)$
determined by
Eq. (\ref{b0}) is the global minimum of the Onsager-Machlup functional.
According to this deterministic mechanics, a particle at
a minimum of the potential $V(x)$ is stable and remains at this place for
ever, so that $x_{\rm average}(t)=x_M$. However, there can be other optimal
trajectories that minimize the action
($\delta S=0$). These optimal trajectories account for thermal activation. For
$T\rightarrow 0$, it can be shown that the 
escape rate out of the metastable state is given by
\begin{equation}
\label{b25}
\Gamma\propto e^{-S[x_c]/k_B T},
\end{equation}
where $S[x_c]$ is the action of the most probable path $x_c(t)$ that connects the metastable state $x_M$ to the
stable state $x_S$ (see Fig. \ref{path}). Generally speaking,
the most probable
path $x_c(t)$ that connects two attractors is called an ``instanton''
\cite{instanton}. It is obtained by cancelling the first order
variations of the action: $\delta S=0$. In the
present case, we will call it a
``thermal instanton''. The precise justification of Eq. (\ref{b25}) and the
evaluation of the preexponential factor involve difficult calculations, similar
to those developed in the quantum mechanics case, that are not
detailed here (see the references given at the beginning of this Appendix). They require 
allowing for multi-instantons and including small fluctuations around the extremal path.
Luckock and McKane \cite{lmk} managed to determine the
prefactor explicitly by showing that the 
fluctuation determinant can  be evaluated exactly with the method of Dreyfuss
and Dym \cite{dd}.

\subsection{The weak noise limit and the instanton}
\label{sec_instwn}

In the weak noise limit $T\rightarrow 0$, we have to determine the trajectory
$x_c(t)$ that minimizes the Onsager-Machlup functional $S[x]$
defined by Eq. (\ref{b12}). As indicated previously, the Onsager-Machlup
functional (\ref{b12}) can be viewed as an action $S[x(t)]$. It can be written
as
\begin{eqnarray}
\label{b15}
S=\int L(x,\dot x)\, {d}t,
\end{eqnarray}
where
\begin{eqnarray}
\label{b16}
L(x,\dot x)=\frac{\xi m}{4}\left\lbrack
{\dot
x}+\frac{V'(x)}{\xi m}\right\rbrack^2
\end{eqnarray}
is the Lagrangian. We thus have to minimize the
action $S[x(t)]$. The minimization
problem $\delta S=0$ can be solved by different methods.

{\it First method:} The extremal condition $\delta S/\delta x=0$ applied to Eq.
(\ref{b15}) yields the Euler-Lagrange equation (\ref{q21}).
Using Eq. (\ref{b16}) we obtain 
\begin{eqnarray}
\label{b18}
\ddot x_c=\frac{1}{\xi^2m^2}V'(x_c)V''(x_c).
\end{eqnarray}
This equation can be rewritten as
\begin{eqnarray}
\label{b19}
m\ddot x_c=U'(x_c)
\end{eqnarray}
with
\begin{eqnarray}
\label{b20}
U(x_c)=\frac{1}{2\xi^2m}\lbrack V'(x_c)\rbrack^2.
\end{eqnarray}
Equation (\ref{b19}) is equivalent to the classical equation of motion of a
particle of unit mass moving in a potential $-U(x)$.  The first integral of
motion is
\begin{eqnarray}
\label{b21}
E=\frac{1}{2}m\dot x_c^2-\frac{1}{2\xi^2m}\lbrack V'(x_c)\rbrack^2,
\end{eqnarray}
where $E$ (energy) is a constant determined by the boundary conditions. A
nontrivial solution $x_c(t)$ of Eq. (\ref{b21}) is called
an instanton (a soliton in the time variable). In
the present context, it characterizes the thermal instanton.

{\it Second method:} The minimization of the action leads
to the
Hamilton equations, which are equivalent to the Newton equation (\ref{b19}).
 Since the Lagrangian (\ref{b16}) does not explicitly depend on
time, the Hamiltonian defined by Eq. (\ref{q24})
is conserved. For the Lagrangian
(\ref{b16}) we obtain
\begin{equation}
\label{b23}
H=\frac{\xi m}{4}\left\lbrack
{\dot
x_c}+\frac{V'(x_c)}{\xi m}\right\rbrack \left\lbrack
{\dot
x_c}-\frac{V'(x_c)}{\xi m}\right\rbrack,
\end{equation}
which is equivalent to Eq. (\ref{b21}) provided that we set
$H=\frac{1}{2}\xi E$.

\subsection{Kramers formula}

We can now repeat the calculations detailed in Sec. \ref{sec_ttbec} and find
that the thermal activation rate  (decay probability per
unit time) is given by
\begin{eqnarray}
\label{gth}
\Gamma\sim \frac{\omega_M|\omega_U|}{2\pi\xi} e^{-\Delta V/k_B T},
\end{eqnarray}
where $\omega_M^2=V''(x_M)/m$ and
$\omega_U^2=-V''(x_U)/m$ are the squared pulsations of the
fictive particle at the
metastable minimum and at the maximum of the potential (the notation
$|\omega_U|$ means $\sqrt{-\omega_U^2}$), and $\Delta V$ is the height of the
potential barrier. Finally, the thermal escape time, or the typical lifetime of
the
metastable state, can be estimated by 
\begin{equation}
\label{qtr17bx}
t_{\rm life}\sim \Gamma^{-1}
\sim \frac{2\pi\xi}{\omega_M|\omega_U|} e^{\Delta V/k_B
T}.
\end{equation}
This expression, which is valid in the weak-noise limit $T\rightarrow
0$ (more precisely for $k_B T\ll\Delta V$), can also be obtained by
using the Kramers method 
\cite{kramers}  to find the stationary
current accross the potential barrier from the Smoluchowski equation.
Therefore, it is oftentimes
called the Kramers formula. 
The exponential factor  $e^{-\Delta V/k_B T}$ corresponds to the 
Arrhenius law \cite{arrhenius}. In the present paper, we have simplified this expression close to a saddle-node bifurcation, leading to Eq. (\ref{qb38bis}).

\subsection{Early  literature on thermal activation}
\label{sec_relfd}

The problem of escape from metastable states under the effect of thermal noise occurs in many domains of physics. The probability that a Brownian particle remains in a metastable state decreases exponentially rapidly with time as 
\begin{equation}
\label{bib}
P(t)\sim e^{-\Gamma t}\sim e^{-t/t_{\rm life}},
\end{equation}
where $\Gamma$ is the rate coefficient and  $t_{\rm life}\sim 1/\Gamma$ is the typical lifetime of the particle in the metastable state.

The discipline of rate theory started when Van't Hoff \cite{hoff} and Arrhenius \cite{arrhenius} discovered empirically that the rate of escape follows the law from Eq. (\ref{b35}).

Early theoretical works were made by Farkas \cite{farkas} and Becker and D\"oring \cite{bedo} in the context of homogeneous nucleation of supersaturated vapors, and by Polanyi and Wigner \cite{powi} and  Eyring \cite{eyring} in chemical physics. In particular, Eyring \cite{eyring} developed the transition state theory.

A major step in reaction rate theory was made by Kramers \cite{kramers} within the framework of Brownian theory. He introduced the Kramers equation and analyzed the cases of strong and weak frictions. His work on the weak friction was pioneering while his results on the strong friction (Smoluchowski limit), which were originally the most widely appreciated, had been previously obtained by Farkas \cite{farkas},  Kaischew and Stranski \cite{kais}, Becker and D\"oring \cite{bedo} via the calculation of the flux of particles across the potential barrier, and by Pontryagin {\it et al.} \cite{pont} via the calculation of the mean first passage time.

The calculation of thermal activation rates for overdamped Brownian particles  in the weak noise limit from the path integral formalism based on the Onsager-Machlup functional was performed by Caroli {\it et al.} \cite{ccr0,ccr} and Weiss \cite{weiss} by using the analogy with quantum tunneling. It was also developed by Bray and McKane \cite{brayprl,mk,bray1,bray,lmk,nbk}  who considered both white noises and exponentially correlated (colored) noises, and also treated the case of a finite friction. It has been extended to the (generalized) stochastic Smoluchowski equation by Chavanis \cite{cdkramers,entropy2}  who recovered a form of Kramers formula in this more general context (see also Appendix C of \cite{tunnel} for the stochastic Ginzburg-Landau and Cahn-Hilliard equations). The instanton theory has been formalized by Freidlin and Wetzel \cite{fw} in relation to the theory of large deviations.

\subsection{Relation to Onsager's linear thermodynamics}
\label{sec_rel}

In this section we make the connection between the previous results and the linear thermodynamics of Onsager \cite{onsager1,onsager2,onsagermachlup}. Actually, the theory of Brownian motion is the simplest example of application of Onsager's linear thermodynamics.\footnote{We can better see the analogy with thermodynamics if we interpret the position $x$ as the energy $E$ and the potential $V(x)$ as the free energy $F(E)$ as in Sec. \ref{sec_gcb} and Appendix \ref{sec_tcano}.} We follow a presentation similar to the one developed in \cite{paper5,entropy2} in a more general framework (i.e., for stochastic partial differential equations instead of stochastic differential equations).

Let us write the equation of evolution as
\begin{eqnarray}
\label{rel1}
\dot x=J,
\end{eqnarray}
where $J$ plays the role of a current in more general situations. At equilibrium, the particle is at the minimum of the potential $V(x)$, corresponding to a vanishing force $F_e=-V'(x_e)=0$. Following Onsager, we assume that close to equilibrium the current $J$ is proportional to the force $F=-V'(x)$ so that it vanishes at equilibrium: $J_e=0$. Writing 
\begin{eqnarray}
\label{rel2}
J=-\frac{1}{\xi m}V'(x),
\end{eqnarray}
and substituting Eq. (\ref{rel2}) into Eq. (\ref{rel1}), we obtain the deterministic equation (\ref{b0}). This formulation implies a form of $H$-theorem. Indeed, the rate of potential energy is
\begin{eqnarray}
\label{rel3}
\dot V=J V'(x).
\end{eqnarray}
Using Eq. (\ref{rel2}), we obtain
\begin{eqnarray}
\label{rel4}
\dot V=-\frac{1}{\xi m}\left\lbrack V'(x)\right\rbrack^2=-\xi m J^2\le 0,
\end{eqnarray}
which implies that $V(t)$ decreases monotonically (provided that $\xi>0$) until
it reaches  its minimum value ($J_e=0$).

The current from Eq. (\ref{rel2}) can be obtained from a variational principle called the maximum potential energy dissipation principle. It was introduced in a more elaborate form by Lord Rayleigh \cite{rayleigh1,rayleigh2} and further developed by Onsager \cite{onsager1,onsager2}. It can be justified as follows. At equilibrium, we know that the variable $x_e$ minimizes the potential $V(x)$. Similarly, out-of-equilibrium, we state that the optimal current $J$ maximizes the rate of potential energy dissipation $-\dot V(J)$ under suitable constraints (see below). This principle describes the most probable course of an irreversible process. It can be viewed a a variational formulation of Onsager's linear thermodynamics. To see that, let us introduce the dissipation function
\begin{eqnarray}
\label{rel5}
E_d=\xi m\frac{J^2}{2}.
\end{eqnarray}
Using Eqs. (\ref{rel3}) and (\ref{rel5}), the maximum of potential energy dissipation $-\dot V(J)$ under the constraint of a fixed dissipation function $E_d$ is determined by the variational principle $\delta(-\dot V-E_d)=0$, yielding Eq. (\ref{rel2}). This is a maximum because $\delta^2(-\dot V-E_d)=-\xi m<0$. Therefore, the deterministic equation (\ref{b0}) maximizes the dissipation of potential energy under the above mentioned constraint.  We can also introduce the dual dissipation function
\begin{eqnarray}
\label{rel6b}
E_d^*=\frac{1}{2\xi m}\left\lbrack V'(x)\right\rbrack^2.
\end{eqnarray}
We note that $E_d^*(x)$ is a function of the position $x$ while $E_d(J)$ is a function of the current $J$. Using Eq. (\ref{rel4}), we obtain the relation
\begin{eqnarray}
\label{rel7}
\dot V=-2E_d=-2E_d^*.
\end{eqnarray}  
Therefore, the dissipation functions equal half the rate of the dissipation of potential energy.

We now derive a simple relation that establishes a direct equivalence between dynamical
and thermodynamical\footnote{As stated in footnote 70, we can view $V(x)$ as a thermodynamical potential.} stability for the deterministic equation (\ref{b0}). Let us consider a small
perturbation $\delta x$ around an equilibrium state $x_e$ so that $x(t)=x_e+\delta x(t)$.  We linearize Eq. (\ref{b0}) and write  the time dependence of the perturbation as $\delta x(t)\propto e^{\lambda t}$. The perturbation is damped exponentially rapidly when $\lambda<0$ (stable equilibrium) or increases exponentially rapidly when $\lambda>0$ (unstable equilibrium). The linearized deterministic equation can thus be written as
\begin{eqnarray}
\label{rel8}
\delta\dot x=\lambda\delta x=\delta J=-\frac{1}{\xi m}V''(x_e)\delta x.
\end{eqnarray}  
From this equation we obtain
\begin{eqnarray}
\label{rel9}
\lambda=-\frac{1}{\xi m}V''(x_e),
\end{eqnarray}  
which shows that the equilibrium state is stable ($\lambda<0$) if, and only if, it is a minimum of the potential $V(x)$, i.e., $V''(x_e)>0$. We can write the relation (\ref{rel9}) in a different manner. The second variations of potential energy around equilibrium are $\delta^2V=\frac{1}{2}V''(x_e)(\delta x)^2$. On the other 
hand, according to Eq. (\ref{rel4}), the second variations of the rate of potential energy  around equilibrium are $\delta^2\dot V=-\xi m (\delta J)^2=-\xi m (\delta\dot x)^2=-\xi m \lambda^2(\delta x)^2$, and they are clearly negative. Combining these relations, we can rewrite Eq. (\ref{rel9}) as
\begin{eqnarray}
\label{rel10}
2\lambda \delta^2V=\delta^2\dot V\le 0.
\end{eqnarray} 
This nice relation has a very
general validity (see, e.g., \cite{gen,kingen,nfp}). It shows that an
equilibrium state of the deterministic equation (\ref{b0}) is linearly
dynamically stable ($\lambda<0$) if, and only if, it is a (local) minimum of
potential energy ($\delta^2V>0$). Therefore, dynamical and thermodynamical
stability coincide.

We now take fluctuations into account and derive the stochastic Langevin equation (\ref{b1}) from the theory of fluctuations developed by Landau and Lifshitz (see Chapter XVII in \cite{ll} and Appendix B of \cite{paper5}). To that purpose, we write the current as
\begin{eqnarray}
\label{rel11}
\dot x=J=-\frac{1}{\xi m}V'(x)+y(t).
\end{eqnarray} 
The first term in Eq. (\ref{rel11}) is the deterministic current [see Eq. (\ref{rel2})] and the second term $y(t)$ is a stochastic term that takes fluctuations into account. The problem at hand consists in characterizing the stochastic term $y(t)$. According to the general theory \cite{ll}, we have
\begin{eqnarray}
\label{rel12}
\dot x=-\gamma X+y.
\end{eqnarray} 
The $X$ term can be obtained from the expression of the rate of dissipation of potential energy. According to the general theory \cite{ll}, the rate of dissipation of potential energy is given by
\begin{eqnarray}
\label{rel13}
\dot V=k_B T X \dot x.
\end{eqnarray} 
Comparing Eq. (\ref{rel13}) with Eq. (\ref{rel3}), we find that the $X$ term is given by
\begin{eqnarray}
\label{rel14}
X=\frac{V'(x)}{k_B T}.
\end{eqnarray} 
It is now easy to find the expression of the coefficient $\gamma$ that appears in Eq. (\ref{rel12}). Comparing Eqs. (\ref{rel11}), (\ref{rel12}) and (\ref{rel14}), we find that
\begin{eqnarray}
\label{rel15}
\gamma=\frac{k_B T}{\xi m}=D,
\end{eqnarray} 
where $D$ is the diffusion current from Eq. (\ref{b6b}).
Now, the general theory of fluctuations \cite{ll} gives
\begin{eqnarray}
\label{rel16}
\langle y(t)y(t')\rangle=2\gamma \delta(t-t').
\end{eqnarray} 
Using Eq. (\ref{rel15}) and writing $y(t)=\sqrt{2D}\, \eta(t)$, we obtain the stochastic Langevin equation (\ref{b1}). The Einstein relation (\ref{b6b}) is automatically included in the formalism.

It is interesting to discuss the relation between the instanton theory and the principle of maximum
dissipation of potential energy. The Onsager-Machlup
functional (\ref{b12}) can be expanded into
\begin{eqnarray}
\label{rel17}
S=\frac{1}{2}\int (E_d+E_d^*+\dot V)\, dt,
\end{eqnarray}
where $E_d$, $E_d^*$  and $\dot V$ are the dissipation functions and the rate of potential energy introduced previously. Eq. (\ref{rel17}) is the counterpart of Eq. (4-18) of
Onsager and Machlup \cite{onsagermachlup}. The minimization of the Onsager-Machlup functional  $S$ with
respect to variations on $J$ is therefore equivalent to
the maximization of $-\dot V-E_d$. This returns the principle of maximum
dissipation of potential energy leading to the deterministic 
equation (\ref{b0}). This corresponds to the downhill instanton
solution for which we
have $\dot V=-2E_d=-2E_d^*$ and $S=0$. On the other hand, for the uphill
instanton solution corresponding to the deterministic equation
(\ref{b0}) with the opposite sign, we have $\dot V=2E_d=2E_d^*$ and
$S=\int \dot V\, dt=\Delta V$.

Let us recapitulate the different variational principles that appear in the theory of Brownian motion and in Onsager's linear thermodynamics. We have seen in Appendix \ref{sec_trqB} that the probability of the position $x$ of the Brownian particle at statistical equilibrium is given by the Boltzmann distribution
\begin{eqnarray}
\label{bom1}
P_{\rm eq}(x)\propto e^{-\beta V(x)} \qquad ({\rm
Boltzmann}),
\end{eqnarray}
where $V(x)$ is the potential energy. On the other hand, we
have seen in Appendix \ref{sec_wom} that, out-of-equilibrium,  the probability
of the path $x(t)$ is given by
\begin{eqnarray}
\label{bom2}
P[x(t)]\propto e^{-\beta S[x(t)]}  \qquad ({\rm
Onsager-Machlup}),
\end{eqnarray}
where $S[x(t)]$ is the Onsager-Machlup functional (action) which can
be expressed in terms of the rate of potential energy and dissipation
functions. The Onsager-Machlup theorem (\ref{bom2}) is
analogous to the Boltzmann principle (\ref{bom1}).
The Boltzmann principle tells
the
probability of an equilibrium state in terms of its potential energy. The Onsager-Machlup theorem tells the probability of a
temporal succession of states in terms of the rate of potential energy and dissipation functions.
The most probable equilibrium position $x_e$ is obtained by minimizing
the potential energy $V(x)$. This is the equilibrium position of the deterministic equation (\ref{b0}). The most probable path is obtained
by minimizing the Onsager-Machlup functional $S$, or equivalently by maximizing
the rate of production of potential energy $-\dot V-E_d$ with
respect to variations on $J$. This leads to the deterministic equation (\ref{b0}). Therefore, the
Onsager-Machlup principle can be viewed as the out-of-equilibrium version of the
Boltzmann principle of statistical equilibrium. When $T\rightarrow 0$, the trajectory of the particle is the deterministic trajectory from Eq. (\ref{b0}) and the equilibrium position of the particle is $x_e$.

\section{Analogy between the Smoluchowski
equation and the Schr\"odinger equation}
\label{sec_ass}

In this Appendix, we discuss the analogy between the equations of Brownian
theory and the equations of quantum mechanics. In particular, we show that we can transform the Smoluchowski equation into a Schr\"odinger equation in imaginary time (Kac equation). This transformation allows us to make a connection between the Onsager-Machlup functional in Brownian theory and the euclidean action in quantum mechanics.

\subsection{From the Smoluchowski equation to the Schr\"odinger equation in
imaginary time}
\label{sec_assC}

Let us consider an overdamped Brownian particle of mass $m$ evolving in a
one-dimensional
potential $V(x)$. The probability density $P(x,t)$ of finding the Brownian
particle in $x$ at time $t$ is governed by the Smoluchowski equation
\begin{eqnarray}
\label{ass17}
\frac{\partial P}{\partial t}=\frac{\partial }{\partial x}\left\lbrack D\left
(\frac{\partial P}{\partial x}+\beta P V'(x)\right )\right\rbrack, 
\end{eqnarray}
where the diffusion coefficient $D$ is given by the Einstein relation from Eq.
(\ref{b6b}). Making the change of variables
\begin{eqnarray}
\label{ass19}
P(x,t)=e^{-\frac{\beta}{2}(V(x)-V(x_0))}\psi(x,t),
\end{eqnarray}
we find that $\psi(x,t)$ satisfies an equation of the form
\begin{eqnarray}
\label{ass20}
\frac{\partial \psi}{\partial t}=D\frac{\partial^2\psi}{\partial
x^2}-\frac{U(x)}{2Dm}\psi,
\end{eqnarray}
where the potential $U(x)$ is related to the potential $V(x)$
by
\begin{eqnarray}
\label{ass21}
U(x)=\frac{1}{2\xi^2m}V'(x)^2-\frac{D}{\xi}V''(x).
\end{eqnarray}
In the present situation, $\psi(x,t)$ is real.
Now, the Schr\"odinger equation for a particle of mass $m$
in a potential $U(x)$ reads
\begin{eqnarray}
\label{ass22}
i\hbar \frac{\partial\psi}{\partial
t}=-\frac{\hbar^2}{2m}\frac{\partial^2\psi}{\partial x^2}
+U(x)\psi.
\end{eqnarray}
Making the Wick rotation $t\rightarrow -it$, we can transform Eq. (\ref{ass22})
into a Schr\"odinger
equation in imaginary time of the form
\begin{eqnarray}
\label{ass23}
\frac{\partial\psi}{\partial
t}=\frac{\hbar}{2m}\frac{\partial^2\psi}{\partial x^2}
-\frac{U(x)}{\hbar}\psi.
\end{eqnarray}
This is the so-called Kac equation. Comparing Eqs. (\ref{ass20}) and
(\ref{ass23}) we see that Eq.
(\ref{ass20}) can be interpreted as a
Schr\"odinger equation in imaginary time for a particle of mass $m$ in a
potential $U(x)$ provided that we make the
correspondance\footnote{This ``quantum'' diffusion coefficient has been introduced several times in physics with different interpretations (see a detailed list of references in the Remark at the end of this section). It is sometimes called the Nelson \cite{nelson} diffusion coefficient.} 
\begin{eqnarray}
\label{ass24}
D=\frac{\hbar}{2m}.
\end{eqnarray}
Using the analogy with quantum mechanics and the fact that the Schr\"odinger-Kac operator in Eq. (\ref{ass20}) is Hermitian, we  can write the Green function of the Smoluchowski equation as
\begin{eqnarray}
\label{ass25}
P(x,t|x_0,t_0)&=&e^{-\beta(V(x)-V(x_0))/2} \nonumber\\
&\times& \sum_n \phi_n(x)
e^{-\lambda_n(t-t_0)}\phi^*_n(x_0),
\end{eqnarray}
where the eigenfunctions $\phi_n(x)$ and the eigenvalues $\lambda_n$ are determined by the differential equation
\begin{eqnarray}
\label{ass26}
D\frac{d^2\phi_n}{dx^2}-\frac{U(x)}{2Dm}\phi_n=-\lambda_n\phi_n.
\end{eqnarray}

The equilibrium
solution of the Smoluchowski equation (\ref{ass17}) is 
\begin{eqnarray}
\label{ass27}
P_{\rm eq}(x)=Ae^{-\beta V(x)},
\end{eqnarray}
where $A=1/\int e^{-\beta V(x)}\, dx$ is the normalisation factor. According
to Eqs. (\ref{ass19}) and (\ref{ass27}), the stationary solution of the
Schr\"odinger equation in imaginary time [Eq. (\ref{ass20})] is
\begin{eqnarray}
\label{ass28}
\psi_{\rm eq}(x)=A'e^{-\beta V(x)/2}.
\end{eqnarray}
We note that $P_{\rm eq}(x)\propto\psi_{\rm eq}^2(x)$. Since $\psi_{\rm
eq}(x)$ is the stationary solution of Eq. (\ref{ass20}), it satisfies
\begin{eqnarray}
\label{ass29}
D\frac{d^2\psi_{\rm eq}}{dx^2}-\frac{U(x)}{2Dm}\psi_{\rm eq}=0.
\end{eqnarray}
This is the eigenfunction of the Kac (or Schr\"odinger) equation (\ref{ass22}) or (\ref{ass23}) [see Eq. (\ref{ass26})] with eigenvalue $\lambda_0=E_0=0$ (ground state). 
We can check by direct substitution that  Eq. (\ref{ass28}) is indeed the
solution of 
Eq. (\ref{ass29}) provided that $V(x)$ and $U(x)$ are related to each other
by Eq. (\ref{ass21}). Actually, Eq.
(\ref{ass21}) can
be viewed as a Riccati equation for $V'(x)$. This is a nonlinear first order differential equation. It can be solved by making the
change of variables from Eq. (\ref{ass28}). Indeed, through this transformation, we find that $\psi_{\rm
eq}(x)$ is the solution of a second order linear differential equation given by 
Eq. (\ref{ass29}).

According to the previous discussion, if we know the solution of
the Smoluchowski equation (\ref{ass17}) with the potential $V(x)$, then
Eq. (\ref{ass19}) gives the solution of the Kac equation (\ref{ass23}) with the
potential $U(x)$ defined by Eq. (\ref{ass21}). Making the inverse Wick rotation $t\rightarrow i t$, we obtain the solution of the Schr\"odinger equation (\ref{ass22}) with the potential $U(x)$. Conversely,  if we know the  solution of the Schr\"odinger equation (\ref{ass22}) with the potential $U(x)$, or equivalently the solution of the  Kac equation (\ref{ass23}) with the
potential $U(x)$,  then Eq. (\ref{ass19}) gives the solution of
the Smoluchowski equation (\ref{ass17}) with the potential
$V(x)=-\frac{2}{\beta}\ln\psi_{\rm eq}(x)+C$ [see Eq. (\ref{ass28})] -- the
solution of the Riccati equation (\ref{ass21}) -- where $\psi_{\rm eq}(x)$ is 
the fundamental eigenfunction (ground state) of the Schr\"odinger equation 
(\ref{ass22}) with the potential $U(x)$ or the stationary solution of the Kac
equation  (\ref{ass23}) with the potential $U(x)$. It corresponds to a vanishing
energy (eigenvalue) $\lambda_0=E_0=0$. 
In conclusion, every soluble solution of the Smoluchowski (resp. Schr\"odinger)  equation provides a soluble solution of the Schr\"odinger (resp. Smoluchowski) equation for a certain potential. The potential $U(x)$ in the 
Schr\"odinger equation is determined by the potential $V(x)$ in the Smoluchowski equation through Eq. (\ref{ass21}).
Conversely, the potential $V(x)$ in the Smoluchowski equation is expressed by
the fundamental eigenfunction $\psi_{\rm eq}(x)$ of the Schr\"odinger equation
with the potential $U(x)$ as $V'(x)=-(2/\beta)\psi'_{\rm eq}(x)/\psi_{\rm
eq}(x)$. This supposes that the stationary solution of the Kac equation exists,
i.e., that the lowest eigenvalue $\lambda_0$ is zero.

In summary, the Smoluchowski equation  (\ref{ass17}) with a potential $V(x)$  is equivalent through Eq. (\ref{ass19}) to the Kac equation  (\ref{ass23})  with a potential $U(x)$, where the potentials $U$ and $V$  are related through the Riccati equation (\ref{ass21}). The stationary solution of the Kac equation with the potential $U(x)$ is given by the Boltzmann distribution (\ref{ass28}) with the potential $V(x)$ determined by the Riccati equation (\ref{ass21}).

{\it Remark:} Several authors like Ehrenfest \cite{ehrenfest},
M\'etadier
\cite{metadier},  Schr\"odinger
\cite{schrobrown,schrodingermarkoff}, F\"urth \cite{furth33}, F\'enyes
\cite{fenyes}, Weizel
\cite{weizel}, Saito and Namiki \cite{saito}, Kershaw \cite{kershaw} and Comisar \cite{comisar}  have
discussed the analogies and the differences between quantum mechanics
(Schr\"odinger's equation) and Brownian theory (Smoluchowski's
equation).\footnote{There is also a long list of works that
followed the seminal paper of Nelson \cite{nelson} on stochastic
quantization (see a detailed list of references in \cite{axioms}).} For
a free particle ($V=0$), the Schr\"odinger equation is
equivalent to the diffusion equation with an imaginary diffusion coefficient
${\cal D}=i\hbar/2m$ (this formula first appeared explicitly in the papers of
Ehrenfest \cite{ehrenfest} and F\"urth \cite{furth33}) or with a real diffusion coefficient $D=\hbar/2m$ and an imaginary time. On the other
hand, the formal transformation of the Smoluchowski equation into a
Schr\"odinger
equation in imaginary time seems to have been first introduced in statistical
mechanics by Risken and Vollmer \cite{rv1,rv2} and van Kampen
\cite{vk} (see also \cite{ch,mf}). Actually, it was already  
introduced by Uhlenbeck and Ornstein \cite{uo} in order to solve a kinetic equation introduced by Lord Rayleigh \cite{rayleigh} (now called the linear Fokker-Planck equation or the homogeneous Klein-Kramers equation) and by
Chandrasekhar \cite{evap}
in his
investigation of the escape of
stars from globular clusters.  This
transformation was also used by Chavanis \cite{planet} to evaluate the effect of
turbulent fluctuations on the  trapping of dust  by large-scale vortices in a
scenario of 
planet formation.

\subsection{From the Onsager-Machlup functional to the euclidean action}
\label{sec_assD}

According to Appendix \ref{sec_wom}, the transition probability from
$(x_0,t_0)$ to $(x,t)$ for an overdamped Brownian particle of mass $m$ in a
one-dimensional
potential $V(x)$  is given by the path integral
\begin{eqnarray}
\label{b14ap}
P(x,t|x_0,t_0)= \int {\cal D}x(t) J[x(t)] e^{-S[x(t)]/k_B T},
\end{eqnarray}
where 
\begin{eqnarray}
\label{omm}
S[x(t)]=\frac{\xi m}{4}\int_{t_0}^t \left\lbrack
{\dot
x}+\frac{V'(x(t))}{\xi m}\right\rbrack^2 \, {d} t
\end{eqnarray}
is the Onsager-Machlup functional and
\begin{eqnarray}
J[x(t)]=e^{\frac{1}{2\xi m}\int_{t_0}^t V''(x(t))\, dt}
\label{jac}
\end{eqnarray}
is the Jacobian \cite{graham} of the transformation from $\eta(t)$ to
$x(t)$.\footnote{The Jacobian was not explicitly written in Eq. (\ref{in3}) for
simplicity and because it is negligible in the limit $D\rightarrow 0$
considered in Appendix \ref{sec_wnl}.} The path integral is
explicitly given by
\begin{eqnarray}
\label{ass35}
P(x,t|x_0,t_0)&=& \int {\cal D}x(t)\nonumber\\
&\times&  e^{-\frac{1}{4D}\int_{t_0}^{t}\left\lbrace
\left\lbrack \dot
x+\frac{V'(x(t))}{\xi
m}\right\rbrack^2-\frac{2D}{\xi m}V''(x(t))\right\rbrace\, dt}.\nonumber\\
\end{eqnarray}
Developing the quadratic term in Eq. (\ref{ass35}) and using the identity
\begin{equation}
\label{ass36}
e^{-\frac{1}{2D\xi m}\int_{t_0}^{t} \dot x V'(x)\, dt}=e^{-\frac{1}{2D\xi
m}\int_{x_0}^{x} V'(x)\, dx}=e^{-\frac{V(x)-V(x_0)}{2k_B T}},
\end{equation}
we can  rewrite the  transition probability (\ref{ass35}) as
\begin{eqnarray}
\label{ass37}
P(x,t|x_0,t_0)&=&  e^{-\frac{V(x)-V(x_0)}{2k_B T}}\nonumber\\
&\times& \int {\cal D}x(t) \, 
e^{-\frac{1}{2Dm}\int_{t_0}^{t} \left\lbrack \frac{1}{2}m\dot
x^2+U(x(t))\right\rbrack\, dt},\nonumber\\
\end{eqnarray}
where $U(x)$ is the potential defined in terms of $V(x)$ by Eq.
(\ref{ass21}).
Provided
that we make the correspondance from Eq. (\ref{ass24}), the path integral in Eq.
(\ref{ass37}) corresponds  to the transition amplitude from
$(x_0,t_0)$ to $(x,t)$ for a quantum particle of mass $m$ in a
one-dimensional potential
$U(x)$. Indeed, according to Appendix
\ref{sec_ea}, this transition amplitude is given in imaginary time by the path
integral 
\begin{equation}
\label{ass38}
\psi(x,t|x_0,t_0)=\langle x|e^{-H(t-t_0)/\hbar}|x_0 \rangle=  \int {\cal D}x(t) \, 
e^{-S[x(t)]/\hbar},
\end{equation}
where 
\begin{eqnarray}
\label{ass39}
S[x(t)]=\int_{t_0}^{t} \left\lbrack \frac{1}{2}m\dot
x^2+U(x(t))\right\rbrack\, dt
\end{eqnarray}
is the Euclidean action associated with the potential $U(x)$.

Equation (\ref{ass37}) has been
obtained by starting from the path integral from Eq. (\ref{b14ap}) and
by expanding
the
quadratic term in the Onsager-Machlup functional from Eq. (\ref{omm}).
Alternatively, we can start from the Smoluchowski equation (\ref{ass17}),
use Eq. (\ref{ass19}) to transform it into a
Schr\"odinger equation with imaginary time (Kac equation) [Eq. (\ref{ass23})], use the
Feynman-Kac euclidean path integral
formulation of quantum mechanics to obtain Eqs. (\ref{ass38}) and
(\ref{ass39}), and finally use  Eq. (\ref{ass19}) to obtain Eq.
(\ref{ass37}). This equivalence provides a
direct way to prove that $P(x,t)$ given by Eq. (\ref{b14ap}) satisfies
the Smoluchowski equation (\ref{ass17}), knowing from the
studies of Feynman \cite{feynman} and Kac \cite{feynman,kac,kac49}
that $\psi(x,t)$ given by Eqs. (\ref{ass38}) and (\ref{ass39}) satisfies
the Schr\"odinger
equation in imaginary time [Eq. (\ref{ass23})]. It also provides the expression
of the Jacobian [Eq. (\ref{jac})] in an indirect manner.

{\it Remark:} If we account for the Jacobian
$J$ in the path integral from Eq. (\ref{b14ap}), we have to add a term
$-\frac{D}{2}V''(x)$ in the Lagrangian from Eq. (\ref{b16}) giving
\begin{eqnarray}
\label{b16n}
L(x,\dot x)=\frac{\xi m}{4}\left\lbrack
{\dot
x}+\frac{V'(x)}{\xi m}\right\rbrack^2-\frac{D}{2}V''(x).
\end{eqnarray}
In that case, repeating the calculations of Appendix \ref{sec_instwn}, we find
that
the equation of the thermal instanton associated with an overdamped Brownian
particle of
mass $m$ moving in a potential $V(x)$, which is obtained by minimizing the
action $S=\int
L\, dt$, is given by
\begin{eqnarray}
\label{b19n}
m\ddot x_c=U'(x_c)
\end{eqnarray}
with
\begin{eqnarray}
\label{b19nb}
U(x_c)=\frac{1}{2\xi^2m}\lbrack
V'(x_c)\rbrack^2-\frac{D}{\xi}V''(x_c).
\end{eqnarray}
This corresponds to the equation of the quantum
instanton associated with a particle of mass $m$ moving in the
potential $U(x)$
[see Eq. (\ref{q22})], which is obtained by minimizing the
euclidean action from Eq. (\ref{q17}). This correspondance enlightens the
analogy
between the Brownian problem and the quantum problem in imaginary
time discussed in Appendix
\ref{sec_assC} [see in particular Eq.
(\ref{ass21})].

\subsection{From the WKB formula to the Kramers formula}

We have shown in Appendix \ref{sec_assC} that the Smoluchowski equation with a
potential $V(x)$ could be transformed into a Schr\"odinger equation in imaginary
time (Kac equation) with a potential $U(x)$ related to $V(x)$ by Eq.
(\ref{ass21}).  On the other hand, we have shown
in Appendix \ref{sec_assD} that the Onsager-Machlup functional of a
Brownian particle moving in a potential $V(x)$ could be transformed
into the euclidean action of a quantum particle moving
in a potential $U(x)$. Finally, we have indicated at the end of
Appendix \ref{sec_assD} that the thermal instanton obtained by minimizing the
Onsager-Machlup functional with $V(x)$ is equivalent to the quantum instanton
obtained by minimizing the euclidean action with $U(x)$. From these analogies,
it should be
possible to derive the Kramers formula (\ref{gth}) for a Brownian particle in a potential
 $V(x)$ from the WKB formula (\ref{q54}) for a quantum particle in a potential 
$U(x)$.  Although a rigorous treatment of this correspondence requires a careful
consideration \cite{preparation}, we  give a brief sketch of its derivation
below.

We start from the WKB formula (\ref{q54}) that we rewrite in terms of
$U(x)$
and $D$ [using Eq. (\ref{ass24})] as 
\begin{eqnarray}
\label{kw1}
\Gamma\propto e^{-\frac{1}{mD}\int_{x'_M}^{x_M} \sqrt{2m\lbrack
U(x)-U(x_M)\rbrack}\,
{d}x}.
\end{eqnarray}
In the weak noise limit $D\rightarrow 0$, we can replace Eq. (\ref{ass21}) by
(see Appendix \ref{sec_wnl})
\begin{eqnarray}
\label{kw2}
U(x)=\frac{1}{2\xi^2m}V'(x)^2.
\end{eqnarray}
Substituting Eq. (\ref{kw2}) into Eq. (\ref{kw1}) and recalling that $V'(x_M)=0$
we get 
\begin{eqnarray}
\label{kw3}
\Gamma\propto e^{-\frac{1}{mD\xi}\int_{x_M}^{x_U} V'(x)\, dx}.
\end{eqnarray}
Recalling the Einstein relation from Eq. (\ref{b6b}), we finally obtain the
Kramers
formula 
\begin{eqnarray}
\label{kw4}
\Gamma\propto e^{-\frac{\Delta V}{k_B T}},
\end{eqnarray}
in agreement with Eq. (\ref{gth}). This provides a nice correspondence between two fundamental formulas in quantum mechanics and Brownian theory.

{\it Remark:} The exponential decay of a metastable equilibrium state due to quantum tunnel effect [Eq. (\ref{bib1})] or thermal activation [Eq. (\ref{bib})] is similar. The quantum rate is given by the WKB formula [Eq. (\ref{q54})] and the thermal rate is given by the Kramers formula [Eq. (\ref{gth})].  It is interesting to note that Kramers  was a
leading protagonist of the two theories (WKB formula in quantum mechanics \cite{kramers1926} and
Kramers formula \cite{kramers} in Brownian theory). However, he
apparently did not recognize the close formal connection between these two formulas. Indeed, we have seen that the Kramers formula can be written as a WKB formula for a different potential connected through the Riccati equation (\ref{ass21}).

\subsection{From the WKB formula to the Boltzmann formula}

In line with the previous section, we can make a link between the WKB formula for a quantum particle in a potential $U(x)$ and the Boltzmann formula for a Brownian particle in a potential $V(x)$.

The stationary solution of the Kac equation (\ref{ass23}) coincides with the stationary solution of the Schr\"odinger equation  (\ref{ass22})  with $E=0$. For $\hbar\rightarrow 0$, it is given by the WKB formula \cite{llquantique}
\begin{equation}
\label{bol1}
\psi_{\rm eq}\propto e^{\frac{i}{\hbar}S_{\rm class}}\propto e^{\frac{i}{\hbar}\int p_{\rm class}(x)\, dx}\propto e^{\pm\frac{i}{\hbar}\int \sqrt{2m(E-U(x))}\, dx}.
\end{equation}
For $E=0$ and $U(x)\ge 0$ we are always in the classically forbidden region and the wavefunction is 
\begin{eqnarray}
\label{bol2}
\psi_{\rm eq}(x) \propto e^{-\frac{1}{\hbar}\int \sqrt{2m U(x)}\, dx}.
\end{eqnarray}
It corresponds to an evanescent wave. On the other hand, for $\hbar\rightarrow 0$, the Riccati equation (\ref{ass21}) reduces to Eq. (\ref{kw2}). Substituting  Eq. (\ref{kw2}) into Eq. (\ref{bol2}), using Eq. (\ref{ass24}), and recalling the Einstein relation from Eq. (\ref{b6b}), we obtain for $T\sim \hbar\rightarrow 0$: 
\begin{eqnarray}
\label{bol3}
\psi_{\rm eq}(x) \propto e^{-\beta V(x)/2},
\end{eqnarray}
so that $P_{\rm eq}(x)\propto \psi_{\rm eq}^2(x)$ is the Boltzmann distribution from Eq. (\ref{ass27}) with the potential $V(x)$. This is the equilibrium solution of the Smoluchowski equation (\ref{ass17}).

More generally, the stationary solution of the Kac equation (\ref{ass23}) with the potential $U(x)$ is given by the Boltzmann distribution from Eq. (\ref{bol3}),  where $V(x)$ is related to $U(x)$ by the Riccati equation (\ref{ass21}). This solution is exact, i.e., it is valid for an arbitrary value of $\hbar$. If we consider the semiclassical limit $\hbar\rightarrow 0$, the Riccati equation (\ref{ass21}) reduces to Eq. (\ref{kw2}). After integration and substitution into Eq. (\ref{bol3}), we obtain the WKB solution (\ref{bol2}), corresponding to Eq. (\ref{bol1}) with $E=0$. Therefore, the semiclassical (WKB) solution of equilibrium Kac equation (\ref{ass23}) is given by Boltzmann distribution from Eq. (\ref{bol3}), where $V(x)$ is determined by the semiclassical Riccati equation (\ref{kw2}).\footnote{More generally, the stationary solution of the Schr\"odinger equation with any eigenenergy $E$ is equivalent to a Riccati equation. As a result, the semiclassical solution of the stationary Schr\"odinger equation is equivalent to the semiclassical solution of the corresponding Riccati equation. This is a manner to derive the WKB formula (see Sec. II.E of \cite{nottale2}). }

Conversely, the solution of the Riccati equation (\ref{ass21}) is given by $V(x)=-\frac{2}{\beta}\ln\psi(x)+C$, where $\psi(x)$ is the equilibrium solution of the Kac equation  (\ref{ass23}) with the potential $U(x)$. Therefore, the Riccati equation (\ref{ass21}) is equivalent to the equilibrium Kac equation  (\ref{ass23}) through the change of variables  $V(x)=-\frac{2}{\beta}\ln\psi(x)+C$ (i.e. $\psi(x)=A' e^{-\beta V(x)/2}$). In the limit $\hbar\rightarrow 0$, the reduced Riccati equation (\ref{kw2}) is equivalent to the WKB solution (\ref{bol2}).

\subsection{Early literature on the connection between the Onsager-Machlup and Feynman path integrals}

The connection between the probability distribution for fluctuation paths introduced by Onsager and Machlup \cite{onsagermachlup} in Brownian theory and the euclidean (Kac)  version \cite{kac49,kac} of the path integrals introduced by Feynman \cite{feynman} in quantum mechanics was first made by Onsager and Machlup \cite{onsagermachlup} (without knowning the works of Feynman and Kac), Hashitsume \cite{hashitsume}, Saito and Namiki \cite{saito}, Falkoff \cite{falkoff1956,falkoff1958}, Wiegel \cite{wiegel1,wiegel2}, Graham \cite{graham}, Moreau \cite{moreau}, Ito \cite{ito}, Caroli {\it et al.} \cite{ccr0,ccr}, and Weiss \cite{weiss}. They mentioned the similarity between the limits $T\rightarrow 0$ and $\hbar\rightarrow 0$, and the  similarity between the WKB formula in quantum mechanics  and the  Boltzmann or the  Kramers formula in Brownian theory.


\begin{thebibliography}{99}




\bibitem{pq}{\small R.D. Peccei, H.R. Quinn, Phys. Rev. Lett. {\bf 38}, 1440
(1977)}
\bibitem{marshrevue}{\small D. Marsh, Phys. Rep. {\bf 643}, 1 (2016)}
\bibitem{axiverse}{\small A. Arvanitaki, S. Dimopoulos, S.
Dubovsky, N. Kaloper,
J.  March-Russell,  Phys. Rev. D {\bf 81}, 123530 (2010)}
\bibitem{barkana}{\small W. Hu, R. Barkana, A. Gruzinov, Phys. Rev. Lett.
{\bf 85}, 1158 (2000)} 
\bibitem{sfdm}{\small M. Alcubierre, F.S. Guzm\'an, T. Matos, D. N\'u\~nez, L.A.
Ure\~na-L\'opez, P. Wiederhold, Class. Quantum Grav. {\bf 19}, 5017 (2002)} 
\bibitem{mocz}{\small P. Mocz {\it et al.}, Mon. Not. R. Astron. Soc.
{\bf 471}, 4559 (2017)}
\bibitem{moczmnras}{\small P. Mocz {\it
et al.}, Mon. Not. R. Astron. Soc. {\bf
494}, 2027 (2020)}
\bibitem{bohmer}{\small C.G. B\"ohmer, T. Harko, J. Cosmol. Astropart. Phys.
{\bf 06}, 025 (2007)} 
\bibitem{prd1}{\small P.H. Chavanis, Phys. Rev. D {\bf 84}, 043531 (2011)}
\bibitem{prd2}{\small P.H. Chavanis, L. Delfini, Phys. Rev. D {\bf 84}, 043532
(2011)}
\bibitem{tanja}{\small T. Rindler-Daller, P.R. Shapiro, Mon. Not. R. Astron.
Soc. {\bf 422}, 135 (2012)}
\bibitem{madelung}{\small E. Madelung, Z. Phys. {\bf 40}, 322 (1927)}
\bibitem{srm}{\small A. Su\'arez, V.H. Robles, T. Matos, Astrophys. Space
Sci. Proc. {\bf 38}, 107 (2014)}
\bibitem{rds}{\small T. Rindler-Daller, P.R. Shapiro, Astrophys. Space Sci.
Proc. {\bf 38}, 163 (2014)}
\bibitem{chavanisbook}{\small   P.H.  Chavanis, {\it Self-gravitating
Bose-Einstein condensates}, in Quantum Aspects of Black Holes, edited by X.
Calmet (Springer, 2015)}
\bibitem{leerevue}{\small J.W. Lee, EPJ Web of Conferences  {\bf 168}, 06005
(2018)}
\bibitem{braatenrevue}{\small E. Braaten, H. Zhang, Rev. Mod. Phys. {\bf 91},
041002 (2019)}
\bibitem{niemeyer}{\small J.C. Niemeyer, Prog. Part. Nucl. Phys. {\bf 113},
103787 (2020)}
\bibitem{mg16B}{\small P.H. Chavanis, {\it The maximum mass of dilute axion stars} in The sixteenth Marcel Grossmann meeting on recent developments in theoretical and experimental general relativity, astrophysics, and relativistic field theories. Editors R. Ruffini, and G. Vereshchagin, 2149–2173 (2023)}
\bibitem{ferreira}{\small E. Ferreira, Astron. Astrophys. Rev.
{\bf 29}, 7 (2021)}
\bibitem{huirevue}{\small L. Hui, Ann. Rev. Astron. Astrophys.  
{\bf 59}, 247 (2021)}
\bibitem{visinelliREVUE}  {\small L. Visinelli, Int. J. Mod.
Phys. D {\bf 30}, 2130006 (2021)}
\bibitem{khoury}  {\small J. Khoury, SciPost Phys. Lect. Notes  {\bf 42}, 1 (2022)}
\bibitem{mul}  {\small  T. Matos, L.A. Ure\~na-L\'opez, J.W.  Lee,  Frontiers in
Astronomy and Space Science {\bf 11}, 1347518 (2024)}
\bibitem{phgtmg}  {\small  L.E. Padilla, J.C. Hidalgo, T.D. Gomez-Aguilar, K.A.
Malik, G. German, Frontiers in Astronomy and Space Science {\bf 11}, 1361399
(2024)}
\bibitem{reviewBECDM}{\small P.H. Chavanis,  Front. Astron. Space Sci. {\bf 12}, 1538434 (2025)}
\bibitem{phi6}{\small P.H. Chavanis, Phys. Rev. D {\bf 98}, 023009
(2018)}
\bibitem{kaup}{\small D.J. Kaup, Phys. Rev. {\bf  172}, 1331 (1968)}
\bibitem{rb}{\small R. Ruffini, S. Bonazzola, Phys. Rev. {\bf  187}, 1767
(1969)}
\bibitem{colpi}{\small M. Colpi, S.L. Shapiro, I. Wasserman, Phys. Rev. Lett.
{\bf  57}, 2485 (1986)}
\bibitem{chavharko}{\small P.H. Chavanis, T. Harko, Phys. Rev. D {\bf 86},
064011 (2012)}
\bibitem{massmaxrel}{\small P.H. Chavanis, Phys. Rev. D {\bf 107}, 103503 
(2023)}
\bibitem{jeansapp}{\small P.H. Chavanis, Phys. Rev. D {\bf 103}, 123551 (2021)}
\bibitem{lb}{\small D. Lynden-Bell, Mon. Not. R. Astron. Soc. {\bf 136}, 101
(1967)}
\bibitem{seidel94}{\small E. Seidel, W.M. Suen, Phys. Rev. Lett.
{\bf  72}, 2516 (1994)}
\bibitem{gul0}{\small  F.S. Guzm\'an, L.A. Ure\~na-L\'opez, Phys. Rev. D {\bf
69}, 124033  (2004)}
\bibitem{gul}{\small  F.S. Guzm\'an, L.A. Ure\~na-L\'opez, Astrophys. J. {\bf
645}, 814  (2006)}
\bibitem{ch2}{\small H.Y. Schive, T. Chiueh, T. Broadhurst, Nature Physics {\bf
10}, 496 (2014)}
\bibitem{ch3}{\small H.Y. Schive {\it et al.}, Phys. Rev. Lett. {\bf 113},
261302 (2014)}
\bibitem{schwabe}{\small B. Schwabe, J. Niemeyer, J. Engels, Phys. Rev. D {\bf
94}, 043513 (2016)}
\bibitem{moczSV}{\small P. Mocz, L. Lancaster, A.Fialkov, F. Becerra, P.H.
Chavanis, Phys. Rev. D {\bf 97}, 083519 (2018)}
\bibitem{veltmaat}{\small J. Veltmaat, J.C. Niemeyer, B. Schwabe,
 Phys. Rev. D {\bf 98}, 043509 (2018)}
\bibitem{moczprl}{\small P. Mocz {\it et al.}, Phys. Rev. Lett.
{\bf 123}, 141301 (2019)}
\bibitem{veltmaat2}{\small J. Veltmaat, B. Schwabe, J.C.
Niemeyer, Phys. Rev. D {\bf 101}, 083518 (2020)}
\bibitem{moczatt}{\small P. Mocz {\it
et al.}, Mon. Not. R. Astron. Soc. {\bf 521}, 2608 (2023)}
\bibitem{modeldm}{\small P.H. Chavanis, Phys. Rev. D {\bf 100}, 083022  (2019)}
\bibitem{wignerPH}{\small P.H. Chavanis, Eur. Phys. J. B {\bf 95}, 48 (2022)}
\bibitem{nfw}{\small J.F. Navarro, C.S. Frenk, S.D.M. White, Mon. Not. R. astr.
Soc. {\bf 462}, 563 (1996)}
\bibitem{moore}{\small B. Moore, T. Quinn, F. Governato, J. Stadel, G.
Lake, Mon. Not. R. astr. Soc. {\bf  310}, 1147 (1999)}
\bibitem{mcmh}{\small P.H. Chavanis, Phys. Rev. D {\bf 100}, 123506 (2019)}
\bibitem{mcmhbh}{\small P.H. Chavanis, Phys. Rev. D {\bf 101}, 063532 (2020)}
\bibitem{levkov2}{\small D.G. Levkov, A.G.  Panin, I.I.
Tkachev, Phys. Rev. Lett. {\bf 121}, 051301 (2018)}
\bibitem{egg}{\small B. Eggemeier, J.C. Niemeyer, Phys. Rev. D {\bf 100},
063528 (2019)}
\bibitem{sgbne}{\small  B. Schwabe, M. Gosenca, C. Behrens, J.C.
Niemeyer, R. Easther, Phys. Rev. D {\bf 102},
083518 (2020)}
\bibitem{kmp}{\small K. Kirkpatrick, A.E. Mirasola, C. Prescod-Weinstein,
Phys. Rev. D {\bf 102}, 103012 (2020)}
\bibitem{cdlmn}{\small J. Chen, X. Du, E. Lentz, D. Marsh, J. Niemeyer, Phys.
Rev. D {\bf 104}, 083022 (2021)}
\bibitem{cdlm}{\small J. Chen, X. Du, E. Lentz, D. Marsh, Phys.
Rev. D {\bf 106}, 023009 (2022)}
\bibitem{kirk}{\small K. Kirkpatrick, A.E. Mirasola, C. Prescod-Weinstein, Phys.
Rev. D {\bf 106}, 043512 (2022)}
\bibitem{bectcoll}{\small P.H. Chavanis,  Phys. Rev. D {\bf 94},
083007 (2016)}
\bibitem{davidson}{\small S. Davidson, T. Schwetz, Phys. Rev. D {\bf 93},
123509 (2016)}
\bibitem{cotner}{\small E. Cotner, Phys. Rev. D {\bf 94}, 063503 (2016)}
\bibitem{tkachevprl}{\small D.G. Levkov, A.G.  Panin, I.I.
Tkachev, Phys. Rev. Lett. {\bf 118}, 011301 (2017)}
\bibitem{helfer}{\small T. Helfer {\it et al.}, JCAP {\bf 03}, 055 (2017)}
\bibitem{moss}{\small F. Michel, I.G. Moss, Phys. Lett.
B {\bf 785}, 9 (2018)}
\bibitem{braaten}{\small E. Braaten, A. Mohapatra, H. Zhang, Phy. Rev. Lett.
{\bf 117}, 121801 (2016)}
\bibitem{pt} {\small P.H. Chavanis, Phys. Rev. E {\bf 65}, 056123 (2002)}
\bibitem{ijmpb} {\small P.H. Chavanis, Int. J. Mod. Phys. B {\bf 20}, 3113
(2006)}
\bibitem{clm1}{\small P.H. Chavanis, M. Lemou, F. M\'ehats, Phys. Rev. D {\bf
91}, 063531 (2015)}
\bibitem{clm2}{\small  P.H. Chavanis, M. Lemou, F. M\'ehats,  Phys. Rev. D {\bf
92}, 123527 (2015)}
\bibitem{calettre}{\small P.H. Chavanis, G. Alberti, Phys. Lett. B {\bf
801}, 135155 (2020)}
\bibitem{acf}{\small G. Alberti, P.H. Chavanis, Eur. Phys. J. B  {\bf 93},
208 (2020)}
\bibitem{mg16F}{\small P.H. Chavanis, {\it The self-gravitating Fermi gas in Newtonian gravity and general relativity} in The sixteenth Marcel Grossmann meeting on recent developments in theoretical and experimental general relativity, astrophysics, and relativistic field theories. Editors R. Ruffini, and G. Vereshchagin, 2230–2251 (2023)}
\bibitem{htww}{\small B.K. Harrison, K.S. Thorne, M. Wakano, J.A. Wheeler {\it
Gravitation Theory and Gravitational Collapse} (University of Chicago
Press, 1965)}
\bibitem{shapiroteukolsky} {\small  S.L. Shapiro, S.A. Teukolsky
{\it Black Holes, White Dwarfs, and Neutron Stars} (Wiley Interscience, 1983)}
\bibitem{visinelli}{\small L. Visinelli, S. Baum, J. Redondo, K. Freese, F.
Wilczek, Phys. Lett. B {\bf 777}, 64 (2018)}
\bibitem{ebyexpansion}{\small J. Eby, P. Suranyi, L.C.R.
Wijewardhana, JCAP {\bf 04}, 038 (2018)}
\bibitem{ebyclass}{\small J. Eby, K. Mukaida, M. Takimoto, L.C.R.
Wijewardhana, M. Yamada, Phys. Rev. D {\bf 99}, 123503 (2019)}
\bibitem{elssw}{\small J. Eby, M. Leembruggen, L. Street, P. Suranyi,
L.C.R. Wijewardhana, Phys. Rev. D {\bf 100}, 063002 (2019)}
\bibitem{oscillon} {\small M. Gleiser, Phys.
Rev. D {\bf 49}, 2978 (1994)}
\bibitem{cgm} {\small E.J. Copeland, M. Gleiser, H.R. M\"uller, Phys.
Rev. D {\bf 52}, 1920 (1995)}
\bibitem{kt2}{\small E.W. Kolb, I.I. Tkachev, Phys. Rev. D {\bf 49} 5040 (1994)}
\bibitem{breather}{\small M.J. Ablowitz, D.J. Kaup, A.C. Newell, H. Segur, 
Phys. Rev. Lett. {\bf 30}, 1262 (1973)}
\bibitem{ebylifetime}{\small J. Eby, P. Suranyi, L.C.R.
Wijewardhana, Mod. Phys. Lett. {\bf 31}, 1650090 (2016)}
\bibitem{ebydecay}{\small J. Eby, M. Ma, P. Suranyi, L.C.R.
Wijewardhana, J. High Energ.
Phys. {\bf 01} 66 (2018)}
\bibitem{ebybh}{\small J. Eby, M. Leembruggen, P. Suranyi, L.C.R. Wijewardhana,
JCAP {\bf 10}, 058 (2018)}
\bibitem{braatenR}{\small E. Braaten, A. Mohapatra, H. Zhang, Phy. Rev. D
{\bf 96}, 031901(R) (2017)}
\bibitem{abrilph}{\small A. Su\'arez, P.H. Chavanis,  Phys. Rev. D {\bf 92},
023510 (2015)}
\bibitem{playa}{\small A. Su\'arez, P.H. Chavanis, J. Phys.: Conf. Series {\bf
654}, 012008 (2015)}
\bibitem{guerra}{\small D. Guerra, C.F.B. Macedo, P. Pani, JCAP
{\bf 09}, 061 (2019)}
\bibitem{tunnel}  {\small P.H. Chavanis, Phys. Rev. D {\bf 102}, 083531
(2020)}
\bibitem{wdsD}{\small  P.H. Chavanis, Phys. Rev. D {\bf 76}, 023004 (2007)}
\bibitem{llquantique}  {\small L.D. Landau, E.M. Lifshitz, {\it Quantum
mechanics} (Pergamon Press, 1958)}
\bibitem{kramers}  {\small H.A. Kramers,  Physica A {\bf 7}, 284 (1940)}
\bibitem{arrhenius}  {\small S. Arrhenius, Z. Phys. Chem. {\bf 4}, 226 (1889)}
\bibitem{preparation}{\small P.H. Chavanis, in preparation}
\bibitem{no}  {\small J.W. Negele, H. Orland, {\it Quantum
Many-Particle
Systems} (Addison-Wesley, Palo Alto, CA, 1988)}
\bibitem{stoof}{\small H.T.C. Stoof, J. Stat. Phys. {\bf 87}, 1353 (1997)}
\bibitem{leggett}{\small M. Ueda, A.J. Leggett,  Phys.
Rev. Lett. {\bf 80}, 1576 (1998)}
\bibitem{huepe}{\small C. Huepe, S. M\'etens, G. Dewel, P. Borckmans, M.E.
Brachet, Phys.
Rev. Lett. {\bf 82}, 1616 (1999)}
\bibitem{fa}{\small J.A. Freire, D.P. Arovas, Phys.
Rev. A {\bf 59}, 1461 (1999)}
\bibitem{skalski}{\small J. Skalski, Phys.
Rev. C {\bf 77}, 064610 (2008)}
\bibitem{lnp}{\small D. Levkov, E. Nugaev, A. Popescu, JHEP {\bf 12}, 131
(2017)}
\bibitem{stone1}{\small M. Stone, Phys. Rev. D {\bf 14}, 3568 (1976)}
\bibitem{stone2}{\small M. Stone, Phys. Lett. {\bf 67B}, 186 (1977)}
\bibitem{polyakov}{\small A.M. Polyakov, Nucl. Phys. B
{\bf 120}, 429 (1977)}
\bibitem{coleman}{\small S. Coleman, Phys. Rev. D {\bf 15},
2929 (1977)}
\bibitem{cc}{\small C.G. Callan, S. Coleman, Phys. Rev. D {\bf 16},
1762 (1977)}
\bibitem{abc}{A.I. Vainshtein, V.I. Zakharov, V.A. Novikov, M.A. Shifman, Sov.
Phys. Usp.  {\bf 25}, 195 (1982)}
\bibitem{brandenberger}{R.H. Brandenberger, Rev. Mod. Phys.  {\bf 57}, 1
(1985)}
\bibitem{gkw}{M. Gleiser, E.W. Kolb, R. Watkins, Nucl. Phys. B  {\bf 364}, 411
(1991)}
\bibitem{ggpp}{\small P.H. Chavanis, Eur. Phys. J. Plus {\bf
132}, 248 (2017)}
\bibitem{ll}  {\small  L. Landau, E. Lifshitz  {\it Fluid Mechanics} (Pergamon,
London 1959)}
\bibitem{brayprl}{\small A.J. Bray, A.J. McKane, Phys. Rev. Lett. {\bf 62}, 493
(1989)}
\bibitem{mk}{\small A.J. McKane, Phys. Rev. A {\bf 40}, 4050
(1989)}
\bibitem{bray1}{\small A.J. McKane, H.C. Luckock, A.J. Bray,  Phys. Rev. A {\bf
41}, 644 (1990)}
\bibitem{bray}{\small A.J. Bray, A.J. McKane, T.J. Newman, Phys. Rev. A {\bf
41}, 657 (1990)}
\bibitem{lmk}{\small H.C. Luckock, A.J. McKane, Phys. Rev. A {\bf 42}, 1982
(1990)}
\bibitem{nbk}{\small T.J. Newman, A.J. Bray, A.J. McKane, J. Stat. Phys. {\bf
59}, 357 (1990) }
\bibitem{katzokamoto}  {\small J. Katz, I. Okamoto, Mon. Not. R. astr.
Soc. {\bf 317}, 163
(2000)}
\bibitem{lifetime}  {\small P.H. Chavanis, Astron. Astrophys. {\bf 432}, 117
(2005)}
\bibitem{antonov}  {\small V.A. Antonov,  Vest. Leningr. Gos. Univ. {\bf 7}, 135
(1962)}
\bibitem{lbw}  {\small D. Lynden-Bell and R. Wood, Mon. Not. R. astr.
Soc. {\bf 138}, 495 (1968)}
\bibitem{langerspin}  {\small R.B. Griffiths, C.-Y. Weng, J.S. Langer, Phys.
Rev. {\bf 149}, 301 (1966)}
\bibitem{art}  {\small M. Antoni, S. Ruffo, A. Torcini, Europhys. Lett. {\bf
66}, 645 (2004)}
\bibitem{bmf}  {\small P.H. Chavanis, Eur. Phys. J. B {\bf 87}, 120 (2014)}
\bibitem{cdkramers}  {\small P.H. Chavanis, L. Delfini, Phys. Rev. E {\bf 89},
032139 (2014)}
\bibitem{crs}{\small  P. H. Chavanis, C. Rosier, C. Sire, Phys. Rev. E {\bf 66},
036105
(2002)}
\bibitem{ks}{\small E. Keller, L. A. Segel, J. Theor. Biol. {\bf 26}, 399
(1970)}
\bibitem{sks}{\small  P.H. Chavanis,  Commun. Nonlinear Sci. Numer. Simulat. 
{\bf 15}, 60 (2010)}
\bibitem{pre11}  {\small P.H. Chavanis, Phys. Rev. E {\bf 84}, 031101 (2011)}
\bibitem{back}{\small P.H. Chavanis, Symmetry {\bf 15}, 2195  (2023)}
\bibitem{nottale2}{\small P.H. Chavanis, in preparation}
\bibitem{gnz}{\small A. Griffin, T. Nikuni, E. Zaremba, {\it Bose-Condensed
Gases at Finite Temperatures} (Cambridge University Press, Cambridge, 2009)}
\bibitem{nottale}  {\small  L. Nottale,  {\it Scale Relativity and Fractal
Space-Time}  (Imperial College Press, 2011) }
\bibitem{epjpnottale}{\small P.H. Chavanis, Eur. Phys. J. Plus {\bf 132},
286 (2017)}
\bibitem{pdunottale}{\small P.H. Chavanis, Phys. Dark Univ. {\bf 22}, 80 (2018)}
\bibitem{wg}{\small P.H. Chavanis, in preparation}
\bibitem{holm}{\small D.D. Holm, J.E. Marsden, T. Ratiu, A. Weinstein, Phys.
Rep. {\bf 123}, 1 (1985)}
\bibitem{nelson} {\small E. Nelson, Phys. Rev.  {\bf 150}, 1079 (1966)}
\bibitem{axioms}  {\small P.H. Chavanis, Axioms  {\bf 13}, 606 (2024)}
\bibitem{madelungN} {\small E. Madelung, Naturwiss. {\bf 14}, 1004 (1926)}
\bibitem{weizsacker} {\small C.F. von Weizs\"acker, Z. f. Physik   {\bf 96}, 431
(1935)}   
\bibitem{fisher} {\small R.A. Fisher, Proc. Cambridge Philos. Soc. {\bf 22}, 700
(1925)}
\bibitem{vdwc}{\small J.D. van der Waals, Verhand. Kon. Akad. V Wetensch. Amst. sect. 1  (1893)}
\bibitem{korteweg}{\small D.J. Korteweg, Arch. Neerl. Sci Exactes Nat.  {\bf 6}, 1 (1901)}
\bibitem{poincare}  {\small H. Poincar\'e, Acta Math. {\bf 7}, 259 (1885)}
\bibitem{katzpoincare}  {\small J. Katz,  Mon. Not. R. Astron. Soc. {\bf 183},
765 (1978)}
\bibitem{whitney}  {\small  H. Whitney, Ann. Math. {\bf 62}, 374 (1955)}
\bibitem{post}  {\small  T. Poston, I. Stewart, {\it Catastrophe theory and its
applications} (Pitman, London, 1978)}
\bibitem{arnold}  {\small  V.I. Arnold, S.M. Gusein-Zade, A.N. Varchenko, {\it
Singularities of Differentiable Maps} (Birkh\"auser, Boston, 1985)}
\bibitem{derrick} {\small G.H. Derrick, J. Math. Phys. {\bf 5}, 1252 (1964)}
\bibitem{epjpbh}{\small P.H. Chavanis, Eur. Phys. J. Plus {\bf 134}, 352
(2019)}
\bibitem{glennon}{\small N. Glennon, C. Prescod-Weinstein, Phys. Rev. D {\bf
104}, 083532 (2021)}
\bibitem{opal1}{M. Smoluchowski,  Ann. Physik {\bf 330}, 205 (1908)}
\bibitem{opal2}{A. Einstein,  Ann. Physik {\bf 338}, 1275 (1910)}
\bibitem{paper5}  {\small P.H. Chavanis, Physica A {\bf 387},  5716 (2008)}
\bibitem{onsagermachlup}{\small L. Onsager, S. Machlup, Phys. Rev. {\bf
91}, 1505 (1953)}
\bibitem{maupertuis1}{\small P.L.M. de Maupertuis, ``Accord de diff\'erentes
loix de la nature qui avoient jusqu'ici paru incompatibles'', m\'emoire
pr\'esent\'e \`a l'Acad\'emie Royale des Sciences (1744)}
\bibitem{maupertuis2}{\small P.L.M. de Maupertuis, ``Les Loix du mouvement et du
repos d\'eduites d'un principe metaphysique'' (1746)}
\bibitem{hamilton}{\small W.R. Hamilton, Dublin University Review and Quarterly
Magazine {\bf 1}, 795 (1833)}
\bibitem{hoff}{\small  J.H. Van't Hoff, in Etudes de Dynamiques Chimiques (F. Muller and Co., Amsterdam), p. 114 (1884)}
\bibitem{hanggi}{\small  P. H\"anggi, P. Talkner, M. Borkovec, Rev. Mod. Phys. {\bf 62}, 251  (1990)}
\bibitem{dd}{\small T. Dreyfuss, H. Dym, Duke Math. J. {\bf 45}, 15
(1978)}
\bibitem{paddy} {\small T. Padmanabhan, Phys. Rep.  {\bf 188}, 285 (1990)}
\bibitem{found}  {\small J. Katz,  Found. Phys. {\bf 33}, 223 (2003)}
\bibitem{einsteinentropy}{\small A. Einstein, Ann. Phys. {\bf 33}, 1275 (1910)}
\bibitem{planck}{\small M. Planck,  Ann. Phys. {\bf 4}, 553 (1901)}
\bibitem{einsteinQT}{\small A. Einstein, Science {\bf 113}, 82 (1951)}
\bibitem{aaISO}{\small P.H. Chavanis,  Astron. Astrophys.  {\bf 381}, 340
(2002)}
\bibitem{n2d3}{\small P.H. Chavanis, Entropy {\bf 26}, 757 (2024)}
\bibitem{entropy1}{\small  P.H. Chavanis, Entropy {\bf 17}, 3205 (2015)}
\bibitem{entropy2}{\small  P.H. Chavanis, Entropy {\bf 21}, 1006 (2019)}
\bibitem{longshort}  {\small P.H. Chavanis, Physica A {\bf 390},  1546 (2011)}
\bibitem{pomeau0}{\small Y. Pomeau, M. Le Berre, P.H. Chavanis, B. Denet, Eur.
Phys. J. E {\bf 37}, 26 (2014)}
\bibitem{pomeau}{\small P.H. Chavanis, B. Denet, M. Le Berre, Y. Pomeau, Eur.
Phys. J. B {\bf 92}, 271 (2019)}
\bibitem{pomeauL}{\small P.H. Chavanis, B. Denet, M. Le Berre, Y. Pomeau,
Europhys. Lett. {\bf 129}, 30003 (2020)}
\bibitem{blowup}{\small P.H. Chavanis, C. Sire, Phys. Rev. E {\bf 70}, 026115
(2004)}
\bibitem{linde77}{\small A.D. Linde, Phys. Lett. B {\bf 70}, 306 (1977)}
\bibitem{linde80}{\small A.D. Linde, Phys. Lett. B {\bf 92}, 119 (1980)}
\bibitem{gw}{\small A.H. Guth, E.J. Weinberg, Phys. Rev. Lett.  {\bf 45}, 1131
(1980)}
\bibitem{linde81}{\small A.D. Linde, Phys. Lett. B {\bf 100}, 37 (1981)}
\bibitem{cook}{\small G.P. Cook, K.T. Mahanthappa, Phys. Rev. D  {\bf 23}, 1321
(1981)}
\bibitem{witten}{\small E. Witten, Nucl. Phys. B  {\bf 177}, 477 (1981)}
\bibitem{steinhardt81}{\small P.J. Steinhardt, Nucl. Phys. B  {\bf 179}, 492
(1981)}
\bibitem{billoire}{\small A. Billoire, K. Tamvakis, Nucl. Phys. B  {\bf 200},
329 (1982)}
\bibitem{hm}{\small S.W. Hawking, I.G. Moss, Phys. Lett. B  {\bf 110}, 35
(1982)}
\bibitem{linde83}{\small A.D. Linde, Nucl. Phys. B  {\bf 216}, 421 (1983)}
\bibitem{chh}{\small M. Claudson, L.J. Hall, I. Hinchliffe, Nucl. Phys. B  {\bf
228}, 501 (1983)}
\bibitem{gibbs}{\small  J.W. Gibbs, {\it Elementary Principles of Statistical
Mechanics} (Yale University, New Haven, CT, USA, 1902)}
\bibitem{einsteinfd}{\small A. Einstein,  Ann. Phys. {\bf 14}, 354 (1904)}
\bibitem{klein}{\small O. Klein, Ark. Mat. Och Fys. {\bf 16}, 1 (1921)} 
\bibitem{chandrarevue}{\small  S. Chandrasekhar, Rev. Mod. Phys. {\bf 15}, 1
(1943)}
\bibitem{uo}{\small G.E. Uhlenbeck, L.S. Ornstein, Phys. Rev.  {\bf
36}, 823 (1930)}
\bibitem{risken}{\small  H. Risken, {\it The Fokker-Planck Equation} (Springer:
Berlin/Heidelberg, Germany, 1989)}  
\bibitem{dean}  {\small D. Dean, J. Phys. A  {\bf 29},
L613 (1996)}
\bibitem{newdean}{\small P.H. Chavanis, in preparation}
\bibitem{fh}{\small R. Feynman, A.R. Hibbs, {\it Quantum
Mechanics and Path Integrals} (McGraw-Hill, New-York,
1965)}
\bibitem{dirac33}{\small P.A.M. Dirac, Physik. Zeits. Sowjetunion
{\bf 3}, 64 (1933)}
\bibitem{dirac35}{\small P.A.M. Dirac, {\it The Principles of Quantum
Mechanics} (The Clarendon Press, Oxford, 1935)}
\bibitem{dirac45}{\small P.A.M. Dirac, Rev. Mod. Phys.
{\bf 17}, 195 (1945)}
\bibitem{feynman}{\small R.P. Feynman, Rev. Mod. Phys.
{\bf 20}, 367 (1948)}
\bibitem{wiener}{\small N. Wiener, Acta Math. {\bf 55},
117 (1930)}
\bibitem{kac49}{M. Kac, Trans. Am. Math. Soc. {\bf 65}, 1 (1949)}
\bibitem{kac}{\small M. Kac, in Proceedings of the Second Berkeley
Symposium on Mathematical Statistics and Probability,  University of
California, 1951, p. 189}
\bibitem{montroll}{\small E. Montroll, Comm. Pure Appl. Math.  {\bf 5},
415 (1952)}
\bibitem{gy}{I.M. Gel'fand, A.M. Yaglom, Uspekhi Mat. Nauk {\bf 11}, 77
(1956); English translation: J. Math. Phys. {\bf
1}, 48 (1960)}
\bibitem{saito}{\small N. Saito, M. Namiki, Prog. Theor. Phys. Japan  {\bf
16}, 71 (1956)}
\bibitem{brush}{\small S.G. Brush, Rev. Mod. Phys.  {\bf
33}, 79 (1961)}
\bibitem{duru}{\small I.H. Duru, H. Kleinert, N. \"Unal, J. Low Temp. Phys. {\bf
42}, 137 (1981)}
\bibitem{merzbacher}{\small E. Merzbacher, Physics Today {\bf 55}, 44 (2002)}
\bibitem{gamow}{\small G. Gamow, Z. Phys.   {\bf 51}, 204 (1928)}
\bibitem{gurney1}{\small R.W. Gurney, E.U. Condon, Nature (London) {\bf 122},
439 (1928)}
\bibitem{gurney2}{\small R.W. Gurney, E.U. Condon, Phys. Rev.  {\bf 33}, 127
(1929)}
\bibitem{condon}{\small E.U. Condon, Am. J. Phys. {\bf 46}, 319
(1978)}
\bibitem{bornt}{\small M. Born, Z. Phys. {\bf 58}, 306 (1928)}
\bibitem{wentzel1926}{\small G. Wentzel, Z. Phys. {\bf 38}, 518 (1926)}
\bibitem{brillouin1926}{\small L. Brillouin, Compt. Rend. Acad.
Sci. Paris {\bf 183}, 24 (1926)}
\bibitem{kramers1926}{\small H.A. Kramers, Z. Phys. {\bf 39}, 828 (1926)}
\bibitem{hund1}{\small F. Hund, Z. Phys.   {\bf 40}, 742 (1927)}
\bibitem{hund2}{\small F. Hund, Z. Phys.   {\bf 43}, 805 (1927)}
\bibitem{opp1}{\small J.R. Oppenheimer, Phys. Rev.  {\bf 31}, 66 (1928)}
\bibitem{opp2}{\small J.R. Oppenheimer, Proc. Nat. Acad. Sci. USA {\bf 14}, 363
(1928)}
\bibitem{nordheim}{\small L. Nordheim, Z. Phys.   {\bf 46}, 833 (1927)}
\bibitem{fono}{\small R.H. Fowler, L. Nordheim, Proc. R. Soc. London, Ser. A
{\bf 119}, 173 (1928)}
\bibitem{mandelstam}{\small L. Mandelstam, M. Leontovich, Z. Phys.   {\bf 47}, 131 (1928)}
\bibitem{eckart}{\small C. Eckart, Phys. Rev.  {\bf 35}, 1303 (1930)}
\bibitem{rice}{\small O.K. Rice, Phys. Rev.  {\bf 35}, 1538 (1930)}
\bibitem{coll}{\small L.A. MacColl, Phys. Rev.  {\bf 40}, 621 (1932)}
\bibitem{bourgin}{\small D.G. Bourgin, Proc. Nat. Acad. Sci. USA {\bf 15}, 357
(1929)}
\bibitem{wignertunnel}{\small E.P. Wigner, Z. Phys. Chem. Abt. B {\bf 19}, 203
(1932)}
\bibitem{lw}{\small T.D. Lee, G.C. Wick, Phys. Rev. D {\bf 9}, 2291 (1974)}
\bibitem{voloshin}{\small M.B. Voloshin, I. Yu. Kobzarev, L.B. Okun', Sov.
J. Nucl. Phys. {\bf 20}, 644 (1975)}
\bibitem{cdl}{\small S. Coleman, F. De Luccia, Phys. Rev. D {\bf 21},
3305 (1980)}
\bibitem{banks}{\small T. Banks, C.M. Bender, T.T. Wu, Phys. Rev. D {\bf 8},
3346 (1973)}
\bibitem{benderwu}{\small C.M. Bender, T.T. Wu, Phys. Rev. D {\bf 7},
1620 (1973)}
\bibitem{hooft}{\small G. 't Hooft,  Phys. Rev. Lett. {\bf 37}, 8 (1976)}
\bibitem{polyakovPP}{A.M. Polyakov, Phys. Lett. B {\bf 59}, 82 (1975)}
\bibitem{langer}{\small J.S. Langer, Ann. Phys. {\bf 41},
108 (1967)}
\bibitem{langevin}  {\small P. Langevin, Comptes rendus {\bf 146}, 530 (1908)}
\bibitem{fokker}{\small A.D. Fokker, Ann. Physik  {\bf 43}, 810
(1914)}
\bibitem{planck1917}{\small M. Planck, Sitzber. Preuss. Akad. Wiss., p. 324
(1917)}
\bibitem{smoluchowski}  {\small M. von Smoluchowski, Ann. Physik  {\bf 48}, 1103
(1915)}
\bibitem{einstein}{\small A. Einstein, Ann. Physik  {\bf 17}, 549 (1905)}
\bibitem{instanton}  {\small R. Rajaraman, {\it Solitons and Instantons}
(North Holland, Amsterdam, 1982)}
\bibitem{farkas}{\small L. Farkas, Z. Phys. Chem. (Leipzig) {\bf 125}, 236
(1927)}
\bibitem{bedo}{\small  R. Becker, W. D\"oring, Ann. Phys. (Leipzig) {\bf 24},
719 (1935)}
\bibitem{powi}{\small M. Polanyi, E. Wigner, Z. Phys. Chem. Abt. A {\bf 139},
439 (1928)}
\bibitem{eyring}{\small  H. Eyring, J. Chem. Phys. {\bf 3}, 107 (1935)}
\bibitem{kais}{\small R. Kaischew, I. N. Stranski, Z. Phys. Chem. Abt. B {\bf
26}, 317 (1934)}
\bibitem{pont}{\small  L.A. Pontryagin, A. Andronov, A. Vitt, Zh. Eksp. Teor.
Fiz. {\bf 3}, 165 (1933)}
\bibitem{ccr0}{\small B. Caroli, C. Caroli, B. Roulet, J. Stat. Phys.
{\bf 21}, 415 (1979)}
\bibitem{ccr}{\small B. Caroli, C. Caroli, B. Roulet, J. Stat. Phys.
{\bf 26}, 83 (1981)}
\bibitem{weiss}{\small U. Weiss, Phys. Rev. A
{\bf 25}, 2444 (1982) }
\bibitem{fw}{\small  M. I. Freidlin, A. D. Wentzell, {\it Random Perturbations
of
Dynamical Systems} (Springer, New York, 1998)}
\bibitem{onsager1}{\small L. Onsager, Phys. Rev. {\bf 37}, 405 (1931)}
\bibitem{onsager2}{\small L. Onsager, Phys. Rev. {\bf 38}, 2265 (1931)}
\bibitem{rayleigh1}{\small Lord Rayleigh, Proc. Math. Soc. Lond. {\bf 4}, 357 (1873)}
\bibitem{rayleigh2}{\small Lord Rayleigh, Philos. Mag. {\bf 26}, 776 (1913)}
\bibitem{nfp}{\small P.H. Chavanis, Eur. Phys. J. B {\bf 62}, 179 (2008)}
\bibitem{gen}{\small P.H. Chavanis, Phys. Rev. E {\bf 68}, 036108 (2003)}
\bibitem{kingen}{\small P.H. Chavanis, Physica A {\bf 332}, 89 (2004)}
\bibitem{ehrenfest}{\small P. Ehrenfest, Zeits. Phys.
{\bf 45}, 455 (1927)}
\bibitem{metadier}{\small J. M\'etadier, Comptes Rendus
{\bf 193}, 1173 (1931)}
\bibitem{schrobrown}{\small E. Schr\"odinger, Sitzungsb. Preuss. Akad. Wiss.,
144 (1931)}
\bibitem{schrodingermarkoff}{\small E. Schr\"odinger, Ann. Inst. H. Poincar\'e
{\bf 4}, 269 (1932)}
\bibitem{furth33}{\small R. F\"urth, Zeits. Phys.
{\bf 81}, 143 (1933)}
\bibitem{fenyes} {\small I. F\'enyes, Z.  Phys.  {\bf 132}, 81 (1952)}
\bibitem{weizel} {\small W. Weizel, Z.  Phys.  {\bf 134}, 264 (1953)}
\bibitem{kershaw} {\small D. Kershaw, Phys. Rev.  {\bf 136}, 1850 (1964)}
\bibitem{comisar} {\small G.G. Comisar, Phys. Rev.  {\bf 138}, 1332 (1965)}
\bibitem{rv1}{\small H. Risken, H.D. Vollmer, Zeit. Phys.   {\bf
201}, 323 (1967)}
\bibitem{rv2}{\small H. Risken, H.D. Vollmer, Zeit. Phys.   {\bf
204}, 240 (1967)}
\bibitem{vk}{\small N.G. van Kampen, J. Stat. Phys.   {\bf
17}, 71 (1977)}
\bibitem{ch}{\small  R. Courant, D. Hilbert, {\it Methoden der Mathematischen
Physik},
Vol. I  (Springer, Berlin, 1931)}
\bibitem{mf}{\small  P.M. Morse, H. Feshbach, {\it Methods of Theoretical
Physics} (Mc
Graw-Hill, New-York, 1953)}
\bibitem{rayleigh}{\small Lord Rayleigh,  Phil. Mag. {\bf 32}, 424 (1891)}
\bibitem{evap}{\small S. Chandrasekhar, Astrophys. J.  {\bf 97}, 263 (1943)}
\bibitem{planet}{\small P.H. Chavanis, Astron. Astrophys.   {\bf
356}, 1089 (2000)}
\bibitem{graham}{\small R. Graham, {\it Statistical Theory of
Instabilities in Stationary Non-Equilibrium Systems}, Springer Tracts in Modern
Physics, Vol. 66, G. Hohler, ed. (Springer Verlag, Berlin, 1973)}
\bibitem{hashitsume}{\small N. Hashitsume, Prog. Theor. Phys. Japan  {\bf
15}, 369 (1956)}
\bibitem{falkoff1956}{\small D.L. Falkoff, Prog. Theor. Phys. Japan  {\bf
16}, 530 (1956)}
\bibitem{falkoff1958}{\small D.L. Falkoff, Ann. Phys.  {\bf
4}, 325 (1958)}
\bibitem{wiegel1}{\small F.W. Wiegel, Physica   {\bf
33}, 734 (1967)}
\bibitem{wiegel2}{\small F.W. Wiegel, Physica   {\bf
37}, 105 (1967)}
\bibitem{moreau}{\small M. Moreau, Physica A
{\bf 90}, 410 (1978)}
\bibitem{ito}{\small H. Ito, Prog. Theor. Phys.
{\bf 59}, 725 (1978)}


\end{thebibliography}
\end{document}